\documentclass[12pt]{iopart}
\usepackage{iopams,amsfonts,amssymb,amsthm} 

\pdfoutput=1
\usepackage{etex}
\usepackage{cite}
\usepackage{latexsym}
\usepackage{rawfonts}
\usepackage{graphicx}
\usepackage[usenames,dvipsnames,svgnames]{xcolor}
\usepackage{bbm}
\usepackage{latexsym}
\usepackage{multirow}
\usepackage{rotating}
\usepackage{lscape}
\usepackage{graphicx} %\graphicspath{{figs/}}
\usepackage[format=plain,labelfont=it,up,textfont=it,up]{caption}
\usepackage{subcaption}
\usepackage{xcolor}
\usepackage{fancybox}
\usepackage{ifmtarg}
\usepackage{fancyhdr}
\usepackage{wrapfig}
\usepackage{hyperref}

\usepackage{tikz}

\let\latexput\put

\input prepictex
\input pictex
\input postpictex

\def\2;{\;\;}

\def\RealN{{\mathbb R}}

\def\Ref#1{(\ref{#1})}

\def\C#1{{\mathcal #1}}

\def\U#1{\underline{#1}}

\def\Sfrac#1#2{\hbox{\large $\frac{#1}{#2}$}}
\def\sfrac#1#2{\hbox{\nor $\frac{#1}{#2}$}}

\def\LB{\left(}         \def\RB{\right)}
\def\LV{\left|}        \def\RV{\right|}
\def\LA{\left\langle}        \def\RA{\right\rangle}

\def\nor{\normalsize}

\def\svv{{\;\hbox{$|$}\;}}

\def\Var#1{\hbox{Var}(#1)}
\def\Cov#1#2{\hbox{Cov}(#1,#2)}

\def\tiny#1{\scalebox{0.75}{#1}}

\definecolor{blue}{rgb}{0,0.18,0.39}
\definecolor{RoyalBlue}{rgb}{0,0.2,0.7}

\def\axes#1#2#3#4#5#6#7{
\setplotarea x from #7 to #5, y from #2 to #6
\setplotarea x from #1 to #5, y from #2 to #6
\axis left shiftedto x=#3 
        ticks
        withvalues #6 #2 /
        at  #6 #2 /
 /
\axis bottom shiftedto y=#4
        ticks
        withvalues #1 #5 /
        at  #1 #5 /
/
\put {\footnotesize$\bullet$} at #3 #4
}

\def\axescenter#1#2#3#4#5#6#7{
\setplotarea x from #7 to #5, y from #2 to #6
\setplotarea x from #1 to #5, y from #2 to #6
\axis left shiftedto x=#3 
        ticks
        withvalues #6  /
        at  #6  /
 /
\axis bottom shiftedto y=#4
        ticks
        withvalues #1 #3 #5 /
        at  #1 #3 #5 /
/
\put {\footnotesize$\bullet$} at #3 #4
}

\def\axesnolabels#1#2#3#4#5#6#7{
\setplotarea x from #7 to #5, y from #2 to #6
\setplotarea x from #1 to #5, y from #2 to #6
\axis left shiftedto x=#3 
        ticks
        at  #6 #2 /
 /
\axis bottom shiftedto y=#4
        ticks
        at  #1 #5 /
/
\put {\footnotesize$\bullet$} at #3 #4
}

\definecolor{Maroon}{cmyk}{0,0.87,0.68,0.62}
\definecolor{Brown}{rgb}{0.7,0.3,0}
\definecolor{Navy}{rgb}{0.3,0.0,0.4}
\definecolor{Red}{cmyk}{0,1,1,0}
\definecolor{BrickRed}{cmyk}{0.16,0.89,0.61,0.02}
\definecolor{DarkRed}{cmyk}{0,1,1,0.5}
\definecolor{DarkBlue}{cmyk}{1,1,0,0.2}
\definecolor{DarkGreen}{cmyk}{1,0,1,0.4}
\definecolor{Green}{cmyk}{1,0,1,0}
\definecolor{DarkBrown}{cmyk}{0,0.81,1,0.6}
\definecolor{OrangeRed}{cmyk}{0,1,0.87,0}
\definecolor{RedOrange}{cmyk}{0,0.77,0.87,0}
\definecolor{Orange}{cmyk}{0,0.61,0.87,0}
\definecolor{Offwhite}{rgb}{.8,0.9,.8}
\definecolor{Offwhite2}{cmyk}{.04,.02,.01,0}
\definecolor{Tan}{rgb}{0.82,0.70,0.55}
\definecolor{Blue}{rgb}{0,0,1}
\definecolor{RoyalBlue}{rgb}{0.25,0.41,0.88}
\definecolor{Sepia}{rgb}{0.37,0.14,0.07}
\definecolor{myblue}{cmyk}{0.025,0.05,0,0}
\definecolor{Mahogany}{cmyk}{0.18,0.87,1,0.08}

\definecolor{green1}{cmyk}{0.25,0,0.76,0}
\definecolor{green2}{cmyk}{0.25,0,0.76,0.07}
\definecolor{green3}{cmyk}{0.25,0,0.76,0.20}
\definecolor{green4}{cmyk}{0.25,0,0.75,0.30}
\definecolor{green5}{cmyk}{0.25,0,0.75,0.40}
\definecolor{green6}{cmyk}{0.25,0,0.75,0.50}

\definecolor{B02}{cmyk}{0,0.14,0.22,0.12}
\definecolor{B03}{cmyk}{0,0.16,0.26,0.16}
\definecolor{B04}{cmyk}{0,0.19,0.28,0.19}
\definecolor{B05}{cmyk}{0,0.25,0.32,0.25}
\definecolor{B06}{cmyk}{0,0.31,0.36,0.31}
\definecolor{B07}{cmyk}{0,0.37,0.40,0.37}
\definecolor{B08}{cmyk}{0,0.46,0.46,0.46}
\definecolor{B09}{cmyk}{0,0.55,0.52,0.54}
\definecolor{B10}{cmyk}{0,0.69,0.61,0.62}
\definecolor{B11}{cmyk}{0,0.78,0.70,0.68}
\definecolor{B12}{cmyk}{0,0.93,0.85,0.60}
\definecolor{B13}{cmyk}{0.25,1,0.6,0.50}
\definecolor{B14}{cmyk}{0.5,1,0.30,0.40}
\definecolor{B15}{cmyk}{0.75,1,0,0.30}

\definecolor{C02}{cmyk}{0,0.22,0.14,0.12}
\definecolor{C03}{cmyk}{0,0.26,0.16,0.16}
\definecolor{C04}{cmyk}{0,0.28,0.19,0.19}
\definecolor{C05}{cmyk}{0,0.32,0.25,0.25}
\definecolor{C06}{cmyk}{0,0.36,0.31,0.31}
\definecolor{C07}{cmyk}{0,0.40,0.37,0.37}
\definecolor{C08}{cmyk}{0,0.46,0.46,0.46}
\definecolor{C09}{cmyk}{0,0.52,0.55,0.54}
\definecolor{C10}{cmyk}{0,0.61,0.69,0.62}
\definecolor{C11}{cmyk}{0,0.70,0.78,0.68}
\definecolor{C12}{cmyk}{0,0.85,0.93,0.60}
\definecolor{C13}{cmyk}{0.25,0.60,1,0.50}
\definecolor{C14}{cmyk}{0.5,0.30,1,0.40}
\definecolor{C15}{cmyk}{0.75,0,1,0.30}

\let\put\latexput

\begin{document}

\title[Phase diagrams of confined square lattice links]{Phase diagrams of confined square lattice links}

\author{EJ Janse van Rensburg$^1\dagger$ \& E Orlandini$^2\ddagger$}
\address{\sf$^1$Department of Mathematics and Statistics, 
York University, Toronto, Ontario M3J~1P3, Canada\\}
\address{\sf$^2$Dipartimento di Fisica e Astronomia ``Galileo Galilei'', 
Universit\'a degli studi di Padova, 8 - 35131 Padova, Italia\\}
\ead{$\dagger$\href{mailto:rensburg@yorku.ca}{rensburg@yorku.ca},
         $\ddagger$\href{mailto:orlandini@pd.infn.it}{orlandini@pd.infn.it}}
\vspace{10pt}
\begin{indented}
\item[]\today
\end{indented}

\begin{abstract}
We study by Monte Carlo simulations and scaling analysis two models of 
pairs of confined and dense ring polymers in two dimensions.  The pair of ring
polymers are modelled by squared lattice polygons confined within a square cavity 
and they are placed in relation to each other to be either unlinked or linked in the plane. 
The observed rich phase diagrams of the two models reveal several equilibrium 
phases separated by first order and continuous phase boundaries whose critical nature 
depend on this reciprocal placements.  We estimate numerically the critical exponents 
associated  with the phase boundaries and with the multicritical points where 
first order and continuous phase boundaries meet.
\end{abstract}

\vspace{2pc}

\submitto{J Phys A: Math Theor}

%\maketitle
% 
% For two-column output uncomment the next line and 
% choose [10pt] rather than [12pt] in the \documentclass declaration

\section{Introduction}

Entanglement is an important factor determining the physical properties
of polymers and biopolymers \cite{D62,deG84,KM91,BO12}.  Knotting and
linking of polymers aspects of their entanglement and have been 
studied extensively \cite{RCV93,SW93,GFR96} in lattice models
\cite{MW84,JvRW91,KM91,JvR02}, also in confining spaces \cite{MLY11}, 
or when adsorbed \cite{V95}, or when the polymer is beyond the
theta transition in the collapsed or dense phase \cite{TJvROSW94,OSV04}.
It is now known that knotting is surpressed in a lattice polygon stretched
in one direction, or stretched by a force \cite{JOTW07,JOTW07b}, 
but enhanced when the polygon is compressed in a confining space 
\cite{MLY11}.  It is also known that other thermodynamic quantities, such 
as pressure, is a function of knotting \cite{JvR07}, while recent studies
of compressed lattice knots show that osmotic pressure for finite length
polymers is a function of knotting \cite{JvR19} (see reference 
\cite{GJvR18} for results for linear compressed lattice polymers).   

\begin{figure}[h!]
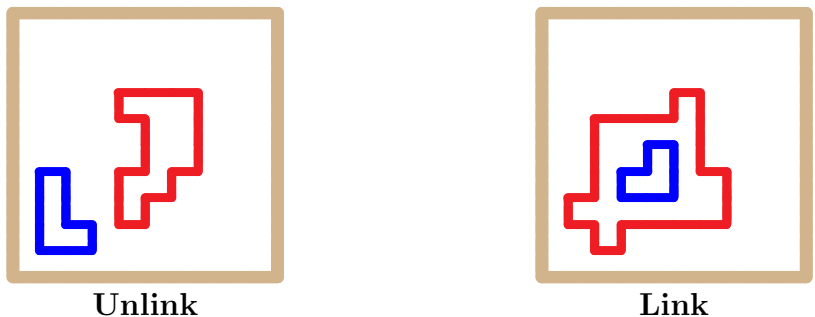

\beginpicture
\setcoordinatesystem units <1pt,1pt>
\setplotarea x from 0 to 250, y from 0 to 110
\setplotarea x from -50 to 250, y from 0 to 110

\color{Tan}
\setplotsymbol ({$\bullet$})
\plot 0 0 100 0 100 100 0 100 0 0  /
\plot 200 0 300 0 300 100 200 100 200 0  /

%Unlink
\setplotsymbol ({\scriptsize$\bullet$})
\color{Blue}
\plot 10 10 20 10 30 10 30 20 20 20 20 30 20 40 10 40 10 30 10 20 10 10 /
\color{Red}
\plot 50 20 50 30 60 30 60 40 70 40 70 50 70 60 70 70 60 70 50 70 40 70
40 60 50 60 50 50 50 40 40 40 40 30 40 20 50 20  /
\color{black}
\put {\textbf{Unlink}} at 50 -10 

%Link
\setplotsymbol ({\scriptsize$\bullet$})
\color{Blue}
\plot 230 30 240 30 250 30 250 40 250 50 240 50 240 40 230 40 230 30 /
\color{Red}
\plot 210 20 220 20 220 10 230 10 230 20 240 20 250 20 260 20 270 20
 270 30 270 40 260 40 260 50 260 60 260 70 250 70 250 60 240 60 
230 60 220 60 220 50 220 40 220 30 210 30 210 20 /
\color{black}
\put {\textbf{Link}} at 250 -10 

\normalcolor
\setlinear
\endpicture
\caption{Lattice models of the unlinked and linked pairs of
lattice polygons in a confining square of side-length $L$.}
\label{1}  %ZXZ[1]
\end{figure}

In this paper we determine the phase diagrams of a model of a pair of dense 
ring polymers in the plane and confined to a cavity (a convex region).  The model 
consists of two square lattice polygons placed inside a confining square in two 
topologically distinct ways. The splittable placement is called the 
\textit{unlinked} case, and the unsplittable placement the \textit{linked} 
case (see figure \ref{1}).  The lattice polygons are compressed by the 
confining square and so are models of dense ring polymers in two dimensions
\cite{deG79} (for example, ring polymers adsorbed on a surface and freely
fluctuating within a cavity).  

If the two polygons in figure \ref{1} are labelled by 1 and 2, such
that the outer polygon in the linked case carries label 1, then the partition functions
of the two models are given by
\begin{eqnarray}
\C{U}_L(\alpha,\beta) &=& \sum_{n1,n2} u_L(n_1,n_2)\,e^{\alpha n_1}\, e^{\beta n_2}, 
\label{1} \\
\C{L}_L(\alpha,\beta) &=& \sum_{n1,n2} \ell_L(n_1,n_2)\,e^{\alpha n_1}\, e^{\beta n_2} .
\label{2}
\end{eqnarray}
Here, $n_1$ and $n_2$ are the lengths of polygons 1 and 2 respectively while
the number of conformations (states) of two unlinked and linked polygons 
in a confining square of sidelength $L$ and area $L^2$ lattice sites are 
denoted $u_L(n_1,n_2)$ and $\ell_L(n_1,n_2)$, respectively.  The parameters
$(\alpha,\beta)$ are related to the chemical potentials of the monomers (vertices)
along the polygons.  If $\alpha\ll 0$ and $\beta\ll 0$ then short polygons dominate 
the partition functions.  This corresponds to a phase of short polygons exploring
the area of the confining square, and we call this the \textit{empty phase}.  
If either  $\alpha\gg 0$ or $\beta\gg 0$, or both, then the partition functions 
are dominated by one or two long polygons filling the square, and the model 
is in a \textit{dense phase}.  

If one considers our models as $L\times L$ systems with sites either occupied or
vacant subject to topological constraints (so that the occupied sites form
a pair of lattice polygons), then the (canonical) free energies of the two models 
are defined by
\begin{eqnarray}
f_L(\alpha,\beta) &=& \Sfrac{1}{L^2} \log \C{U}_L(\alpha,\beta),
\quad\hbox{and}\quad 
\label{3} \\
g_L(\alpha,\beta) &=& \Sfrac{1}{L^2} \log \C{L}_L(\alpha,\beta) .
\label{4}  
\end{eqnarray}
If the parameters $(\alpha,\beta)$ are considered to be chemical potentials of 
vertices along the polygons, then $f_L$ and $g_L$ are the corresponding grand
potentials of the models, and their first and second order derivatives 
are the concentrations and variances of the concentrations, respectively.
In this paper we shall refer to $f_L$ and $g_L$ as \textit{free energies},
but to their derivatives as \textit{mean concentrations} and
\textit{variances} (of the mean concentrations).
Notice that $f_L$ and $g_L$ are analytic functions of $(\alpha,\beta)$, 
and so phase boundaries will only exist in the $L\to\infty$ limit.  In small  
finite systems, finite size effects may be important, and phase boundaries 
are normally seen as peaks in the variances of concentrations, or as sharp 
changes in the concentrations.  In this paper we are able to choose $L$ large
enough that in some cases free energies can be fitted to (non-analytic) 
absolute value functions, concentrations are approximately step functions,
and the variances of concentrations have sharp spikes when crossing
phase boundaries.

\begin{figure}[h!]
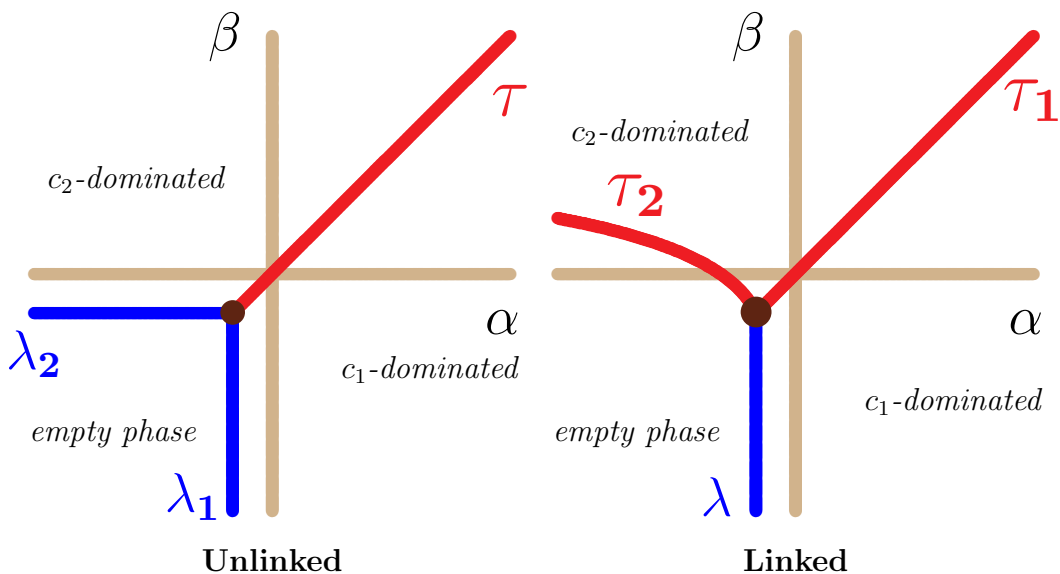

\beginpicture
%\setcoordinatesystem units <0.375pt,0.375pt>
\setcoordinatesystem units <0.6pt,0.6pt>
\setplotarea x from -150 to 150, y from -150 to 150

\color{Tan}
\setplotsymbol ({$\bullet$})
\plot -150 0 150 0 /
\plot  0 -150 0 150 /

\color{Red}
\plot -25 -25 150 150  /
\put {\scalebox{2.25}{$\mathbf{\tau}$}} at 150 110
\color{Blue}
\plot -150 -25 -25 -25 /
\plot -25 -150 -25 -25 /
\put {\scalebox{1.75}{$\mathbf{\lambda_1}$}} at -50 -140
\put {\scalebox{1.75}{$\mathbf{\lambda_2}$}} at -150 -50
\color{Sepia}
\put {\scalebox{2.0}{$\bullet$}} at -25 -25

\color{black}
\put {\LARGE$\alpha$} at 145 -30
\put {\LARGE$\beta$} at -30 150
\put {\textit{empty phase}} at -100 -100
\put {\textit{$c_1$-dominated}} at 100 -60
\put {\textit{$c_2$-dominated}} at -85 60

\setplotsymbol ({\scalebox{0.15}{$\bullet$}})
\setsolid

\normalcolor
\put {\textbf{Unlinked}} at 0 -180 

\setlinear

%\setcoordinatesystem units <0.375pt,0.375pt> point at -320 0 
\setcoordinatesystem units <0.6pt,0.6pt> point at -330 0 
\setplotarea x from -150 to 150, y from -150 to 150

\color{Tan}
\setplotsymbol ({$\bullet$})
\plot -150 0 150 0 /
\plot  0 -150 0 150 /

\color{Red}
\plot -25 -25 150 150  /
\setquadratic
\color{Red}
\plot -150 35 -65 10 -25 -25 /
\setlinear
\color{Blue}
\plot -25 -150 -25 -25 /
\color{Sepia}
\put {\scalebox{2.5}{$\bullet$}} at -25 -25

\color{black}
\put {\LARGE$\alpha$} at 145 -30
\put {\LARGE$\beta$} at -30 150
\put {\textit{empty phase}} at -100 -100
\put {\textit{$c_1$-dominated}} at 100 -80
\put {\textit{$c_2$-dominated}} at -85 90

\color{Red}
\put {\scalebox{2.25}{$\mathbf{\tau_1}$}} at 150 110
\put {\scalebox{2.25}{$\mathbf{\tau_2}$}} at -100 50
\color{Blue}
\put {\scalebox{1.75}{$\mathbf{\lambda}$}} at -50 -140

\color{black}
\setplotsymbol ({\scalebox{0.15}{$\bullet$}})
\setsolid

\normalcolor
\setlinear

\put {\textbf{Linked}} at 0 -180 
\endpicture
\caption{The phase diagrams of the unlinked (left) and linked (right) models
as determined in this paper.  In the unlinked model the multicritical point is 
located at  $(\alpha_c,\alpha_c)$, where $\alpha_c = -\log \mu_2$ in the $L\to\infty$ 
limit (and $\mu_2$ is the growth constant of square
lattice polygons \cite{H61A,HM54,HW62A}).  The location $(\alpha_c,\beta_c)$ of the 
multicritical point in the linked model is more difficult to determine, where
again $\alpha=-\log \mu_2$ in limit as $L\to\infty$.  The phase boundaries
$(\lambda_1,\lambda_2)$ in the unlinked model, and $\lambda$ in the
linked model are continuous transitions, while the $\tau$ phase boundaries
in both models are first order transitions.  These phase boundaries separate
empty phases from dense phase, namely a $c_1$-dominated phase when
the first polygon (or the outer polygon in the linked model) is dense in the
confining square, or a $c_2$-dominated phase when the second (or the inner
polygon in the linked model) is dense in the confining square.}
\label{2}  %ZXZ[2]
\end{figure}

Our finite size data show that the phase diagrams of these models each include
a multicritical point where three phase boundaries meet.  In the unlinked
model the multicritical point is the nexus of two continuous and one first
order phase boundary, while in the linked model, it is the meeting point
of one continuous and two first order phase boundaries, as shown in figure \ref{2}.
The locations of the multicritical points are obtained by examining the polygon
generating function $P(t)$ given by
\begin{equation}
P(t) = \sum_{n=0}^\infty p_n\, t^n,
\label{5}
\end{equation} 
where $p_n$ is the number of square lattice polygons of length $n$.
The radius of convergence of $P(t)$ is $t_c = 1/\mu_2$ \cite{H61A,HM54,HW62A}
(where $\mu_2$ is the growth constant of square lattice self-avoiding walks), 
so that if $L=\infty$, then the two polygons contributing to the partition
function in equation \Ref{1} become independent and the locations
of the critical lines $\lambda_1$ and $\lambda_2$ are along vertical
and horisontal lines with $\alpha_c=\beta_c=-\log \mu_2$.  Similarly,
the location of the critical line $\lambda$ in the linked model is
along the vertical line with $\alpha_c=-\log \mu_2$ since for $\beta<-\log \mu_2$
the inner polygon is short and the outer polygon can expand when 
$\alpha>-\log \mu_2$. 

\begin{figure}[h!]
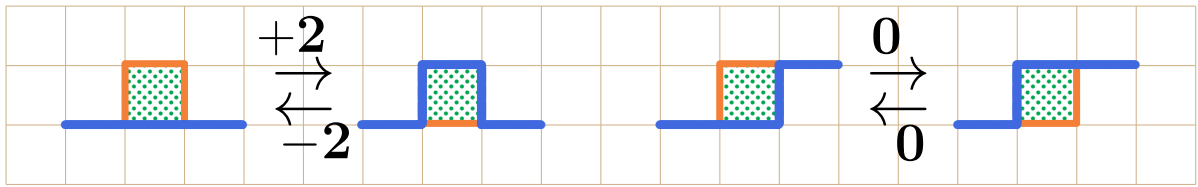

\beginpicture
\setcoordinatesystem units <2.25pt,2.25pt> point at 0 0 
\setplotarea x from -10 to 190, y from -10 to 20
\color{Tan}
\put {\beginpicture \grid 20 3 \endpicture} at 0 0 

\color{Green}
\setshadegrid span <2pt>
\setshadesymbol ({\scalebox{1.2}{$\cdot$}})
\vshade 10 0 10 <z,z,z,z> 20 0 10 /
\vshade 60 0 10 <z,z,z,z> 70 0 10 /
\vshade 110 0 10 <z,z,z,z> 120 0 10 /
\vshade 160 0 10 <z,z,z,z> 170 0 10 /

\setplotsymbol ({\scalebox{2.0}{$\cdot$}})
\color{Orange}
\plot 10 0 10 10 20 10 20 0 /
\plot 60 0 70 0 /
\plot 110 0 110 10 120 10 /
\plot 160 0 170 0 170 10 /

\setplotsymbol ({\scriptsize$\bullet$})

\color{RoyalBlue}
\plot 0 0 10 0 20 0 30 0 /
\plot  50 0 60 0 60 10 70 10 70 0 80 0 /
\plot 100 0 110 0 120 0 120 10 130 10 /
\plot 150 0 160 0 160 10 170 10 180 10 /
\color{blue}
%\multiput {$\bullet$} at 0 0 10 0 20 0 30 0 /
%\multiput {$\bullet$} at 50 0 60 0 60 10 70 10 70 0 80 0 /
%\multiput {$\bullet$} at 100 0 110 0 120 0 120 10 130 10 /
%\multiput {$\bullet$} at 150 0 160 0 160 10 170 10 180 10 /

\color{black}
\multiput {\scalebox{2.0}{$\mathbf{\rightarrow}$}} at 40 8 140 8 /
\multiput {\scalebox{2.0}{$\mathbf{\leftarrow}$}} at 40 2 140 2 /
\put {\LARGE$\mathbf{{+}2}$} at 38 15
\put {\LARGE$\mathbf{{-}2}$} at 42 -3
\multiput {\LARGE$\mathbf{0}$} at 138 15 142 -3 /

%\multiput {$\bullet$} at 50 0 60 0 60 10 70 10 70 0 80 0 /

\color{black}
\normalcolor
\endpicture
\caption{The BFACF elementary moves in the square lattice \cite{BF81,ACF83}.
On the left is the positive move (left to right), and the negative move
(right to left).  On the right is the neutral move.}
\label{3}  %ZXZ[3]
\end{figure}

In this paper our aim is to examine the phase diagrams in figure \ref{2} 
numerically by simulating a $20\times 20$ system using Monte Carlo methods.
Our data show that this finite size system is large enough to enable us to 
determine the phase diagram of the system with good accuracy, and to obtain 
reasonable estimates of the critical exponents associated with each phase boundary.

\begin{figure}[h!]
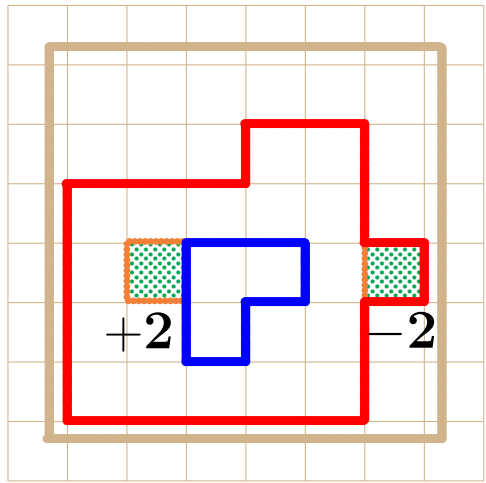

\beginpicture
\setcoordinatesystem units <2.25pt,2.25pt> point at 0 0 
\setplotarea x from -70 to 70, y from -10 to 70
\setplotarea x from -10 to 70, y from -10 to 70

\color{Tan}
\put {\beginpicture \grid 8 8 \endpicture} at 0 0 

\color{Green}
\setshadegrid span <1.75pt>
\setshadesymbol ({\scalebox{1.2}{$\cdot$}})
\vshade 10 20 30 <z,z,z,z> 20 20 30 /
\vshade 50 20 30 <z,z,z,z> 60 20 30 /

\color{Orange}
\setplotsymbol ({\scalebox{2.0}{$\cdot$}})
\setdots <2pt>
\plot 20 20 10 20 10 30 20 30 /
\plot 50 20 50 30 /

\setsolid
\setplotsymbol ({\scriptsize$\bullet$})

\color{Tan}
\plot -3 -3 -3 63 63 63 63 -3 -3 -3 /

\color{Blue}
\plot 20 10 20 20 20 30 30 30 40 30 40 20 30 20 30 10 20 10 / 
\color{red}
\plot 0 0 10 0 20 0 30 0 40 0 50 0 50 10 50 20 60 20 60 30 
 50 30 50 40 50 50 40 50 30 50 30 40 20 40 10 40 0 40 
 0 30 0 20 0 10 0 0  /

\color{black}
\put {\LARGE$\mathbf{{+}2}$} at 12 15 
\put {\LARGE$\mathbf{{-}2}$} at 56 15

\color{black}
\normalcolor
\endpicture
\caption{A neutral move on the pair of polygons in the linked model
consisting of a positive (${+}2$) move on the inner polygon, and 
a negative (${-}2$) move on the outer polygon.  The overall length 
of the two polygons remains unchanged, but they exchange edges
to increase and decrease their lengths respectively.}
\label{4}  %ZXZ[4]
\end{figure}

Sampling self-avoiding walks or polygons in a dense phase is a notoriously difficult
numerical problem \cite{TJvROW96} but for the unlinked model a parallel 
implementation \cite{CJvR20} of the GAS algorithm \cite{JvRR11A} proved effective.
In the linked model we instead used the GARM \cite{RJvR08} algorithm after the GAS
algorithm failed to converge.   These algorithms were implemented with 
BFACF elementary  moves \cite{BF81,ACF83}, (see figure \ref{3}) 
to approximately enumerate states in the models giving estimates of $u_L(n_1,n_2)$ 
and $\ell_L(n_1,n_2)$ in equations \Ref{1} and \Ref{2}. Convergence in
the linked model was improved dramatically by including 
the neutral (total length preserving) move shown in figure \ref{4}, 
in addition to the standard BFACF elementary moves.  Additionally, we used a multiple 
Markov Chain \cite{TJvROW96} implementation of the BFACF algorithm 
\cite{BF81,ACF83} to sample metric and other properties of the polygons 
across phase boundaries to further examine the nature of the transitions 
shown in figure \ref{2}.

In the next section we review the critical behaviour in the phase diagrams
in figure \ref{2}, before we discuss our numerical results, and examine the
behaviour of serveral observables in these models.

\section{Criticality in models of unlinked and linked dense polygons}
\label{section2}

It is not known that the thermodynamic limits (in the $L\to\infty$ limit) 
exist in our models (the methods of references \cite{M95} and \cite{JvR99} 
can be used to show that some models, similar to those in figure \ref{1}, 
have thermodynamic limits). Assuming existence of the thermodynamic limits,
define
\begin{eqnarray}
\upsilon(\alpha,\beta) &=& \lim_{L\to\infty} f_L(\alpha,\beta) , 
\quad\hbox{and}\quad 
\label{6} %ZXZ[6]
\\
\omega(\alpha,\beta) &=& \lim_{L\to\infty} g_L(\alpha,\beta) .
\label{7} %ZXZ[7]
\end{eqnarray}
These are the limiting free energies of the models, and we use the
theory in reference \cite{LS84} to determine the nature of the transitions along
the critical lines in figure \ref{2}.

\begin{figure}[h!]
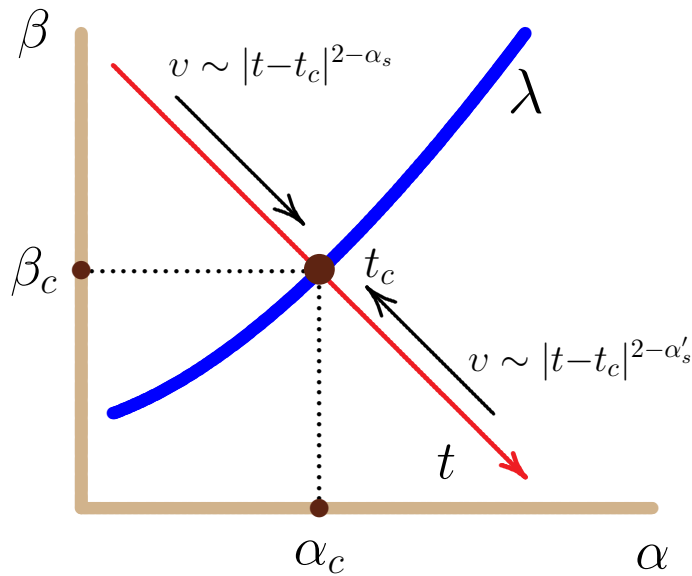

\beginpicture
\setcoordinatesystem units <1.2pt,1.2pt>
\setplotarea x from 0 to 250, y from 0 to 150

\color{Tan}
\setplotsymbol ({$\bullet$})
\plot 0 150 0 0 180 0 /

\setquadratic
\color{Blue}
\plot 10 30 70 70 140 150 /

\setlinear
\setplotsymbol ({\scalebox{0.35}{$\bullet$}})
\color{Red}
\arrow <10pt> [.2,.67] from 10 140 to 140 10 

\color{black}
\setplotsymbol ({\scalebox{0.35}{$\bullet$}})
\setdots <4pt>
\plot 0 75 75 75 75 0  /

\color{Sepia}
\put {\scalebox{2.5}{$\bullet$}} at 75 75
\put {\scalebox{1.5}{$\bullet$}} at 75 0
\put {\scalebox{1.5}{$\bullet$}} at 0 75

\color{black}
\put {\LARGE$\alpha$} at 180 -15
\put {\LARGE$\alpha_c$} at 75 -15
\put {\LARGE$\beta$} at -15 150
\put {\LARGE$\beta_c$} at -15 75
\put {\LARGE$t$} at 115 15
\put {\Large$t_c$} at 94 76.5
\put {\scalebox{2.0}{$\lambda$}} at 140 132

\setplotsymbol ({\scalebox{0.25}{$\bullet$}})
\setsolid
\put {\large$\upsilon \sim |t{-}t_c|^{2-\alpha_s}$} at 63 140
\arrow <12pt> [.2,.67] from 30 130 to 70 90 

\put {\large$\upsilon \sim |t{-}t_c|^{2-\alpha_s^\prime}$} at 157.5 47.5
\arrow <12pt> [.2,.67] from 130 30 to 90 70 

\normalcolor
\setlinear
\endpicture
\caption{A schematic drawing of the transverse scaling axis crossing 
a phase boundary when $t=t_c$ in the phase diagram of the unlinked 
or linked models.  The free energy $\upsilon(\alpha,\beta)$ is singular 
when $t=t_c$.  If the transition is continuous, then there are
critical exponents $(\alpha_s,\alpha_s^\prime)$ such that 
$\upsilon(\alpha,\beta) \sim |t{-}t_c|^{2-\alpha_s}$ as $t\to t_c^-$
and $\upsilon(\alpha,\beta) \sim |t{-}t_c|^{2-\alpha_s^\prime}$ as 
$t\to t_c^+$.}
\label{5}  %ZXZ[5]
\end{figure}

Along an axis in the phase diagram, parametrized by $t$ and crossing a critical
curve at $t=t_c$ (see figure \ref{5}), the singular part of $\upsilon(\alpha,\beta)$, 
denoted by $\upsilon_s(\alpha,\beta)$, is expected to exhibit powerlaw behaviour 
given by
\begin{equation}
\upsilon_s(\alpha,\beta) \sim
\cases{
 |t-t_c|^{2-\alpha_s}, & \hbox{if $t\to t_c^-$}; \cr
 |t-t_c|^{2-\alpha_s^\prime}, & \hbox{if $t\to t_c^+$}, 
}
\label{8}    %ZXZ[8]
\end{equation}
where $\alpha_s$ and $\alpha_s^\prime$ are critical exponents 
characteristic of the transition.  Similar expressions hold for the limiting 
free energy $\omega(\alpha,\beta)$ of the linked model.

The functions $\upsilon(\alpha,\beta)$ and $\omega(\alpha,\beta)$ are
approximated by $f_L(\alpha,\beta)$ and $g_L(\alpha,\beta)$ in equation \Ref{3} 
when $L$ is finite.  In this case there are finite size corrections to scaling 
which modifies equation \Ref{5} by introducing a \textit{finite size} 
crossover exponent $\phi_s$. For large $L$ the finite size corrections are small 
and are confined to a region close to the critical point.  In this case it may be
possible to extract critical exponents directly from numerical data.  In this
paper we are able to do this for $L=20$, where our data suggest that the 
free energies $f_L$ and $g_L$ or their derivatives are closely modelled by
absolute value or step functions, or closely follow singular behaviour given by
figure \Ref{8} or its derivatives outside small finite size regions.

\begin{figure}[h!]
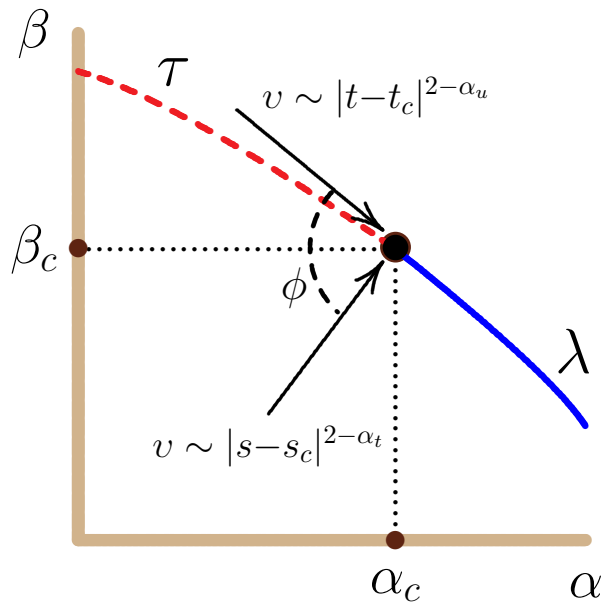

\beginpicture
\setcoordinatesystem units <2.4pt,2.4pt>
\setplotarea x from 0 to 80, y from 0 to 80

\color{Tan}
\setplotsymbol ({$\bullet$})
\plot 0 80 0 0 80 0 /

\color{black}
\put {\LARGE$\alpha$} at 80 -7
\put {\LARGE$\beta$} at -7 80
\put {\LARGE$\alpha_c$} at 50 -7
\put {\LARGE$\beta_c$} at -7 46

\color{black}
\setplotsymbol ({\scalebox{0.35}{$\bullet$}})
\setdots <4pt>
\plot 0 46 50 46 50 0  /

\color{Sepia}
\put {\scalebox{2.5}{$\bullet$}} at 50 46
\put {\scalebox{1.5}{$\bullet$}} at 50 0 
\put {\scalebox{1.5}{$\bullet$}} at 0 46

\setplotsymbol ({\scalebox{0.5}{$\bullet$}})
\setquadratic
\color{Red}
\setdashes <5pt>
\plot 0 74 20 65 50 46 /
\color{Blue}
\setsolid
\plot 50 46 70 29 80 18  /
\setlinear
\color{black}
\put {\huge $\bullet$} at 50 46 

\normalcolor
\put {\scalebox{2.0}{$\tau$}} at 15 74
\put {\scalebox{2.0}{$\lambda$}} at 78 30

\setplotsymbol ({\scalebox{0.25}{$\bullet$}})
\setsolid
\put {\large $\upsilon \sim |t{-}t_c|^{2-\alpha_u}$} at 47 70
\arrow <12pt> [.2,.67] from 25 68 to 48 49 

\put {\large $\upsilon \sim |s{-}s_c|^{2-\alpha_t}$} at 30 15
\arrow <12pt> [.2,.67] from 30 20 to 48 44 

\setplotsymbol ({\scalebox{0.35}{$\bullet$}})
\setdashes <5pt>
\circulararc 90 degrees from 40 55 center at 50 46 
\put {\Large $\phi$} at 34 40

\endpicture
\caption{A tricritical point where a critical curve of first order transitions 
($\tau$) meets a line of continuous transitions ($\lambda$).  Approaching 
the tricritical point along $\tau$ the free energy scales as $\upsilon(\alpha,\beta) 
\sim |t{-}t_c|^{2-\alpha_u}$, where $t$ is a coordinate along 
$\tau$.  Along a second axis through the tricritical point and 
transverse to $\lambda$, the free energy scales as 
$\upsilon(\alpha,\beta) \sim |s{-}s_c|^{2-\alpha_t}$, where $s$ 
is a coordinate transverse to $\lambda$.  The scaling of 
$\upsilon(\alpha,\beta)$ close to the tricritical point is consistent 
provided that their exists a crossover exponent $\phi$ relating to the 
exponents $\alpha_u$ and $\alpha_t$ by 
$\phi = (2-\alpha_t)/(2-\alpha_u)$.}
\label{6}  %ZXZ[6]
\end{figure}

The phase diagrams in figure \ref{2} include multicritical points where curves 
of first order and continuous transitions meet. This is shown schematically in 
figure \ref{6}, where a curve of first order transitions ($\tau$) meets a curve of 
continuous transitions ($\lambda$) at the \textit{tricritical point} $(\alpha_c,\beta_c)$.
Approaching the tricritical point along the $\tau$ phase boundary, the singular part 
of $\upsilon(\alpha,\beta)$ scales as 
\begin{equation}
\upsilon_s(\alpha,\beta) \sim |t-t_c|^{2-\alpha_u}, 
\label{9}   %ZXZ[9]
\end{equation}
where $t$ is the coordinate of a scaling axis along $\tau$. Transverse to the 
$\lambda$ phase boundary at the multicritical point,
\begin{equation}
\upsilon_s(\alpha,\beta) \sim |s-s_c|^{2-\alpha_t},
\label{10}   %ZXZ[10]
\end{equation}
where $s$ is the coordinate of the transverse scaling axis.  Since 
$\upsilon(\alpha,\beta)$ is a function of two variables, consistency of these
scaling laws requires that there exists a \textit{crossover exponent} $\phi$
and scaling function $F$ such that $F(x) \sim x^{2-\alpha_u}$ and
\begin{eqnarray}
&  &\upsilon_s (\alpha,\beta) 
\sim |s-s_c|^{2-\alpha_t} F(|s-s_c|^{-\phi}\, |t-t_c|) \nonumber \\
& & =  |t-t_c|^{2-\alpha_u} \LB |s-s_c|^{-(2-\alpha_t)/(2-\alpha_u)}\, 
 |t-t_c|\RB^{\alpha_u-2}  \nonumber \\
& & \hspace{2cm}\times F(|s-s_c|^{-\phi}\, |t-t_c|) \nonumber \\
& &= |t-t_c|^{2-\alpha_u}F_1 (|s-s_c|^{-\phi}\, |t-t_c|) , 
\end{eqnarray}
where $F_1(x) = x^{2-\alpha_u}\,F(x)$ and
provided that the crossover exponent is given by
\begin{equation}
\phi = \Sfrac{2-\alpha_t}{2-\alpha_u}\, .
\label{12}    %ZXZ[12]
\end{equation}
Similar expressions hold for $w(\alpha,\beta)$ of the linked model, with 
its exponents $\alpha_u$, $\alpha_t$ and crossover exponent $\phi$.

\begin{figure}[h!]
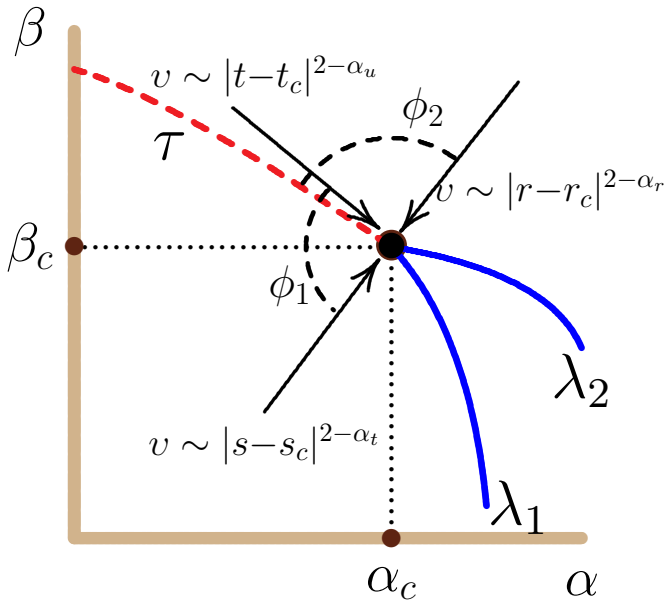

\beginpicture
\setcoordinatesystem units <2.4pt,2.4pt>
\setplotarea x from 0 to 80, y from 0 to 80

\color{Tan}
\setplotsymbol ({$\bullet$})
\plot 0 80 0 0 80 0 /

\color{black}
\put {\LARGE$\alpha$} at 80 -7
\put {\LARGE$\beta$} at -7 80
\put {\LARGE$\alpha_c$} at 50 -7
\put {\LARGE$\beta_c$} at -7 46

\color{black}
\setplotsymbol ({\scalebox{0.35}{$\bullet$}})
\setdots <4pt>
\plot 0 46 50 46 50 0  /

\color{Sepia}
\put {\scalebox{2.5}{$\bullet$}} at 50 46
\put {\scalebox{1.5}{$\bullet$}} at 50 0 
\put {\scalebox{1.5}{$\bullet$}} at 0 46

\setplotsymbol ({\scalebox{0.5}{$\bullet$}})
\setquadratic
\color{Red}
\setdashes <5pt>
\plot 0 74 20 65 50 46 /
\color{Blue}
\setsolid
\plot 50 46 60 29 65 5  /
\plot 50 46 70 40 80 30 /
\setlinear
\color{black}
\put {\huge $\bullet$} at 50 46 

\normalcolor
\put {\scalebox{2.0}{$\tau$}} at 15 62
\put {\scalebox{2.0}{$\lambda_1$}} at 70 5
\put {\scalebox{2.0}{$\lambda_2$}} at 80 25

\setplotsymbol ({\scalebox{0.25}{$\bullet$}})
\setsolid
\put {\large $\upsilon \sim |t{-}t_c|^{2-\alpha_u}$} at 30 73
\arrow <12pt> [.2,.67] from 25 68 to 48 49 

\put {\large $\upsilon \sim |s{-}s_c|^{2-\alpha_t}$} at 30 15
\arrow <12pt> [.2,.67] from 30 20 to 48 44 

\put {\large $\upsilon \sim |r{-}r_c|^{2-\alpha_r}$} at 75 55
\arrow <12pt> [.2,.67] from 70 72 to 51.5 48.5 

\setplotsymbol ({\scalebox{0.35}{$\bullet$}})
\setdashes <5pt>
\circulararc 90 degrees from 40 55 center at 50 46 
\put {\Large $\phi_1$} at 34 40
\circulararc 90 degrees from 60 60 center at 50 46 
\put {\Large $\phi_2$} at 55 68

\endpicture
\caption{A multicritical point where a critical curve of first order transitions 
($\tau$) meets two curves of continuous transitions ($\lambda_1$ and 
$\lambda_2$).  Approaching the multicritical point along $\tau$ the 
free energy scales as $\upsilon(\alpha,\beta) \sim |t{-}t_c|^{2-\alpha_u}$, 
where $t$ is a coordinate along $\tau$.  Along a second axis through the 
multicritical point and transverse to $\lambda_1$, the free energy scales as 
$\upsilon(\alpha,\beta) \sim |s{-}s_c|^{2-\alpha_t}$, where $s$ 
is a coordinate transverse to $\lambda_1$.  The scaling of 
$\upsilon(\alpha,\beta)$ between $\tau$ and $\lambda_1$ close to the 
multicritical point is consistent provided that their exists a crossover 
exponent $\phi_1$ relating the exponents $\alpha_u$ and $\alpha_t$ by 
$\phi_1 = (2-\alpha_t)/(2-\alpha_u)$.  Similarly, transverse to $\lambda_2$
the free energy scales as $\upsilon(\alpha,\beta) \sim |r{-}r_c|^{2-\alpha_t}$, 
where $r$ is a coordinate transverse to $\lambda_2$.  This defines a 
crossover exponent $\phi_2$ relating the exponents $\alpha_u$ and
$\alpha_r$ by $\phi_2 = (2-\alpha_r)/(2-\alpha_u)$.}
\label{7}  %ZXZ[7]
\end{figure}

In our models the phase diagrams of the both the unlinked and linked
models will be shown to include a multicritical point, generalising tricritical
scaling as shown in figure \ref{6} by the introduction of additional
crossover exponents relating scaling along first order and continuous 
phase boundaries.  In figure \ref{7} we show the case where two 
continuous phase boundaries $\lambda_1$ and $\lambda_2$ meets
a curve of first phase transitions $\tau$ in a multicritical point.  In this case 
there are crossover behaviour on either side of the $\tau$ phase boundary, 
and each of the $\lambda_1$ and $\lambda_2$ phase boundaries.  On each
side a crossover exponent $\phi_1$ and $\phi_2$ controls crossover
scaling.  We shall show that the phase boundary of the unlinked model has
the geometry shown in figure \ref{7}, albeit symmetric so that $\phi_1=\phi_2$
(as shown in figure \ref{2}). The phase diagram of the linked model, in contrast
has a multicritical point where two curves $\tau_1$ and $\tau_2$ of first order 
phase transitions meet a curve $\lambda$ of continuous transitions.  In this
model the situation is similar to that shown in figure \ref{7}, but now with
scaling along parallel axes on each of the $\tau_1$ and $\tau_2$ phase boundaries,
and transverse to the $\lambda$ phase boundary.  This gives crossover
scaling between $\tau_1$ and $\lambda$ with crossover exponent $\phi_1$,
and similarly between $\tau_2$ and $\lambda$ with crossover exponent $\phi_2$.

\section{Numerical results}

Simulations on the unlinked model with $L=20$ were performed using the
GAS algorithm \cite{JvRR11A} with BFACF elementary moves \cite{BF81,ACF83} 
(figure \ref{3}).  A parallel implementation (similar to the implementation 
of PERM in reference \cite{CJvR20}) using $8$ threads and shared data structures 
was run for $2,\!000$ started sequences (or \textit{tours}) each of length $10^7$
iterations. This amounts to  a total of $1.6\times 10^{11}$ sampled configurations 
(iterations). The simulation converged in reasonable time to very stable estimates
of $u_L(n_1,n_2)$ (see equation \Ref{1}).  The linked model proved far more
difficult and did not converge successfully, notable because the GAS algorithm
failed to effectively sample states with $n_1 < n_2$ (that is, a short outer polygon with
enclosed area filled with a long inner polygon as expected in the $c_2$-dominated
phase).  Using the GARM algorithm instead, implemented with BFACF 
elementary moves and an additional neutral move (see figure \ref{4}), proved
effective, and a parallel implementation with $12$ parallel threads, sampling
along GARM sequences for a total of $12\times 4.03\times 10^5$ tours, converged
in reasonable time (see reference \cite{CJvR20}).

The GAS and GARM algorithms are approximate enumeration algorithms producing
estimates of the microcanonical partitions $u_L(n_1,n_2)$ and $\ell_L(n_1,n_2)$
in equations \Ref{1} and \Ref{2}.  This gives estimates of the free energies
$f_L$ and $g_L$ (equations \Ref{3} and \Ref{4}) and in the event that $L$ is
large enough, we obtain accurate approximations to the limiting free energies
$\nu(\alpha,\beta)$ and $\omega(\alpha,\beta)$ in equations \Ref{6} and \Ref{7}.
Our simulations show that $L=20$ are sufficiently large, and that it will be 
challenging to perform simulations for $L>20$, in particular for the linked model.

\begin{figure}[h!]
\begin{subfigure}{0.33\textwidth}
	\caption{Free Energy $f_L(\alpha,\beta)$}
	\includegraphics[width=1.0\textwidth]{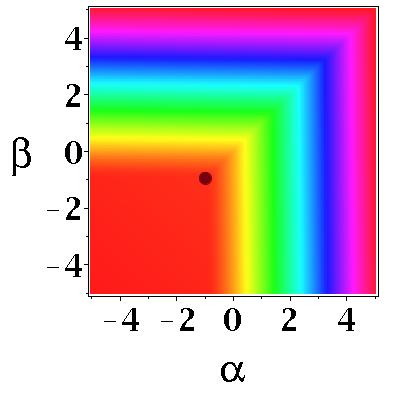}
\end{subfigure}
\begin{subfigure}{0.33\textwidth}
	\caption{$\LA c_1 \RA$}
	\includegraphics[width=1.0\textwidth]{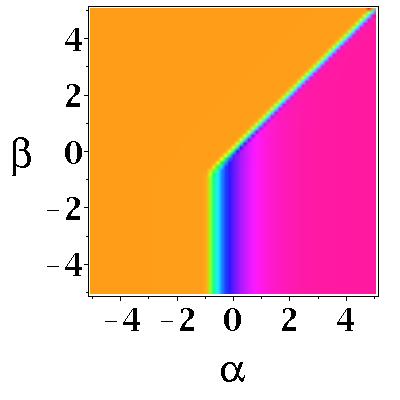}
\end{subfigure}
\begin{subfigure}{0.33\textwidth}
	\caption{$\LA c_2 \RA$}
	\includegraphics[width=1.0\textwidth]{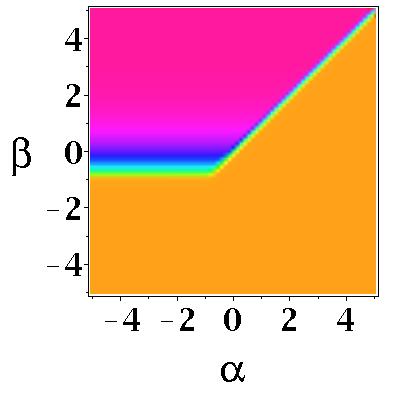}
\end{subfigure}
\begin{subfigure}{0.33\textwidth}
	\caption{$L^2\,\hbox{Cov}(c_1,c_2)$}
	\includegraphics[width=1.0\textwidth]{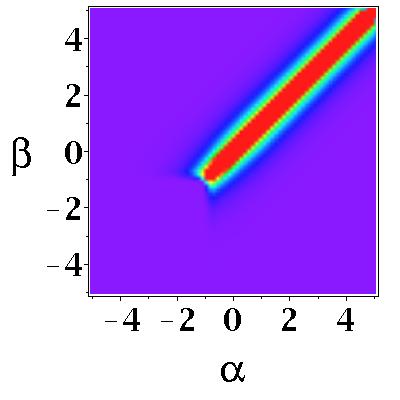}
\end{subfigure}
\begin{subfigure}{0.33\textwidth}
	\caption{$L^2\,\hbox{Var}(c_1)$}
	\includegraphics[width=1.0\textwidth]{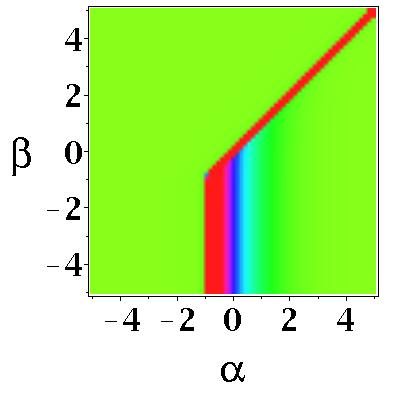}
\end{subfigure}
\begin{subfigure}{0.33\textwidth}
	\caption{$L^2\,\hbox{Var}(c_2)$}
	\includegraphics[width=1.0\textwidth]{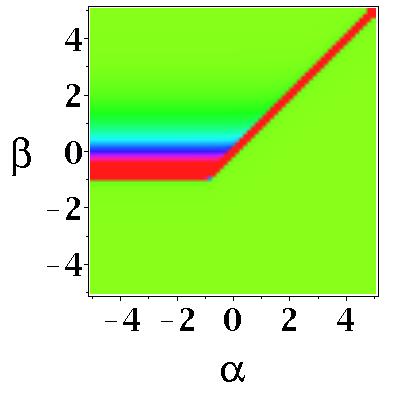}
\end{subfigure}
\caption{The free energy $f_L (\alpha,\beta)$ of the unlinked model
for $L=20$ (top left), and its first and second derivatives.  Proceeding 
clockwise from the top middle, the energies $\LA c_1 \RA$ 
and $\LA c_2 \RA$, the specific heats (variances) $\hbox{Var}(c_2)$ 
and $\hbox{Var}(c_1)$, and the covariance of $c_1$ and $c_2$ 
$\hbox{Cov}(c_1,c_2)$.  The concentrations $c_1$ and $c_2$ 
are negatively correlated, and shows a sharp transition 
along the main diagonal for $\alpha=\beta > \alpha_c$.  The location
of the multicritical point in the model is denoted by a bullet in (a).}
\label{8}   %ZXZ[8]
\end{figure}

\subsection{The phase diagram of two unlinked confined square lattice polygons}

In figure \ref{8}(a) a density plot of $f_L(\alpha,\beta)$ for $L=20$
is shown, with the left bottom colour low, and the right and top colours high.  
The free energy shows clear signs of transitions, and this is also confirmed by plotting 
other observables related to the first and second derivatives of $f_L(\alpha,\beta)$.
The mean concentrations of vertices in each polygon are related to the mean lengths
of the polygons by 
\begin{equation}
\LA c_1 \RA 
= \LA n_1 \RA / L^2 = \sfrac{d}{d\alpha} \,f_L(\alpha,\beta).
\label{13}   %ZXZ[13]
\end{equation}
The concentration of the second polygon is similarly given by
$\LA c_2 \RA = \LA n_2 \RA / L^2 = \sfrac{d}{d\beta} \,f_L(\alpha,\beta)$.
Density plots of $\LA c_1 \RA$ and $\LA c_2 \RA$ are shown in figures
\ref{8}(b) and (c).  These plots show rapid change in the mean
concentrations of one, or both, the polygons when phase boundaries are 
crossed.

The variance of the concentration $c_1$ is given by
\begin{equation}
\Var{c_1} = \Sfrac{1}{L^4}\Var{n_1}
= \Sfrac{1}{L^2}\Sfrac{\partial^2}{\partial \alpha^2}\,f_L(\alpha,\beta) 
\label{14}    %ZXZ[14]
\end{equation}
and in figure \ref{8}(e) a density plot of $L^2\,\Var{c_1}$ is shown, and this
shows a phase boundary (where $\Var{c_1}$ is large) separating
phases where $\LA c_1 \RA$ is small from a phase where $\LA c_1 \RA$ is
large.  A similar plot of $L^2\,\Var{c_2}$ is shown in figure \ref{8}(f) for
the concentration $c_2$ of the second polygon.  The covariance of
$c_1$ and $c_2$ is given by
\begin{eqnarray}
\Cov{c_1}{c_2} &=& \Sfrac{1}{L^4}\LB \LA n_1 n_2 \RA - \LA n_1 \RA \LA n_2 \RA \RB
\nonumber \\
&=& \Sfrac{1}{L^2}\Sfrac{\partial^2}{\partial \alpha\, \partial\beta}\,f_L(\alpha,\beta) ,
\label{15}   %ZXZ[15]
\end{eqnarray}
and $L^2\, \Cov{c_1}{c_2}$ is plotted in figure \ref{8}(d).  The covariance is 
small negative in most of the plot, but spikes to large negative along a
phase boundary which runs along the diagonal with points $(\alpha,\beta)$
where $\alpha=\beta > \alpha_c$ and starting in a multicritical point
$(\alpha_c,\alpha_c)$ since the phase diagram is symmetric under
exchange $\alpha \leftrightarrow \beta$.

These results numerically confirm the phase diagram of the unlinked model
shown in figures \ref{2} and figure \ref{9}. The multicritical point 
$(\alpha_c,\alpha_c)$ is situated at the meeting of three phase boundaries.  
The phase transition along the main diagonal, denoted by $\tau$, is a 
line of transitions starting in the multicritical point running into 
the first quadrant.  Up to numerical accuracy, above and to 
the left of $\tau$ the free energy is only a function of $\beta$, 
and to the right and below it, only a function of $\alpha$.  This shows 
that $\tau$ is a first order phase boundary separating a phase where 
$\LA c_2\RA$ is large from a phase where $\LA c_1\RA$ is large.  These 
are the \textit{$c_2$-dominated} and \textit{$c_1$-dominated} phases, respectively.

\begin{figure}[h!]
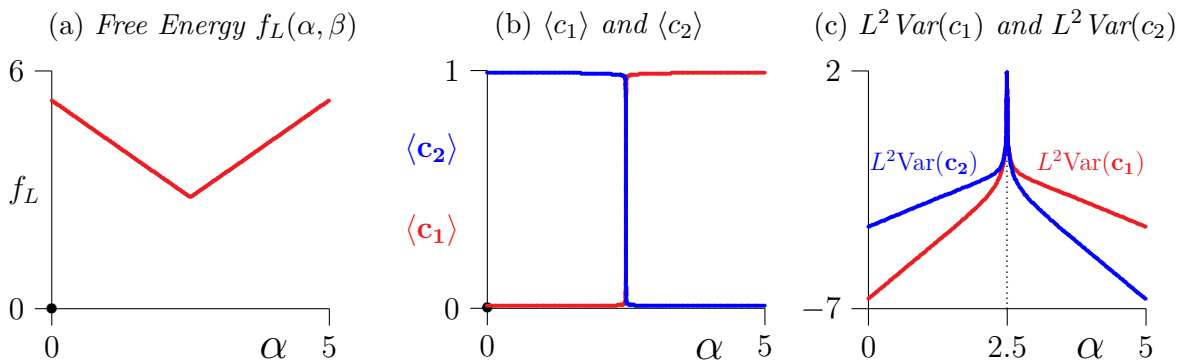

\beginpicture
\setcoordinatesystem units <0.75pt,0.75pt>
\setplotarea x from -150 to 150, y from -150 to 165

\color{Tan}
\setplotsymbol ({$\bullet$})
\plot -150 0 150 0 /
\plot  0 -150 0 150 /

\color{Black}
\setplotsymbol ({\scalebox{0.25}{$\bullet$}})
\circulararc 135 degrees from 18 18 center at -25 -25
\circulararc 135 degrees from -25 -80 center at -25 -25

\color{Red}
\setplotsymbol ({$\bullet$})
\plot -25 -25 150 150  /
\put {\scalebox{2.25}{$\mathbf{\tau}$}} at 150 130
\color{Blue}
\plot -150 -25 -25 -25 /
\plot -25 -150 -25 -25 /
\put {\scalebox{1.5}{$\mathbf{\lambda_1}$}} at -40 -140
\put {\scalebox{1.5}{$\mathbf{\lambda_2}$}} at -150 -40
\color{Sepia}
\put {\scalebox{2.0}{$\bullet$}} at -25 -25

\color{black}
\put {\LARGE$\alpha$} at 145 -20
\put {\LARGE$\beta$} at -20 150
\put {\textit{\Large empty phase}} at -100 -80
\put {\textit{\Large $c_1$-dominated}} at 90 -80
\put {\textit{\Large $c_2$-dominated}} at -85 90

\setsolid
\setplotsymbol (.)
\arrow <10pt> [.2,.67] from -125 -110 to -25 -110 
\put {\large$\alpha_s=2$} at -110 -120
\arrow <10pt> [.2,.67] from 90 -110 to -25 -110 
\put {\large$\alpha_s^\prime=0.26(3)$} at 80 -125

\normalcolor
\put {\textbf{Unlinked model phase diagram}} at 0 -180 

\setlinear

\put {\large $\phi=1.00(7)$} at 80 -33
\put {\large $\phi=1.00(7)$} at -50 47

\endpicture
\caption{Crossover exponents between $\tau$ and the $\lambda_1$ 
and $\lambda_2$ critical curves around the multicritical point in 
the phase diagrams of the unlinked models.  This phase diagram is symmetric 
over the main diagonal, so that the crossover exponents $\phi$ on either 
side have the same value.}
\label{9}  %ZXZ[9]
\end{figure}

Two additional phase boundaries are meeting at the 
multicritical point, namely a line of transitions $\lambda_1$ running 
vertically into the multicritical point and a line of transitions 
$\lambda_2$ running horisontally into the multicritical point.  The 
vertical phase boundary separates a phase where both 
$\LA c_1\RA$ and $\LA c_2 \RA$ are small (this is the \textit{empty phase}) 
from a $c_1$-dominated phase.  Similarly, the horisontal phase boundary
separates the empty phase from a $c_2$-dominated phase.  

Note that, since $f_L(\alpha,\beta)$ is an analytic function of $(\alpha,\beta)$, 
these phases and phase boundaries exist only in the $L\to\infty$ limit.
In our data the phase boundaries manifest themselves as peaks in the variances
or sharp changes in the mean concentrations.  
The location of the multicritical point is obtained by observing  that 
the radius of convergence of the polygon partition function $P(t)$ 
in equation \Ref{2} is $t_c = 1/\mu_2$, so that if $L=\infty$, then the 
two polygons contributing to equation \Ref{3} become independent and 
the critical points are determined by the polygon generating function.
This gives $\alpha_c=\beta_c=-\log \mu_2 \approx -0.97$. 

\begin{figure}[h!]
\begin{subfigure}{0.33\textwidth}
\normalcolor
\color{black}
	\caption{Free Energy $f_L(\alpha,\beta)$}
\beginpicture
\setcoordinatesystem units <21pt,15pt>
\axes{0}{0}{0}{0}{5}{6}{0}
\put {\Large$\alpha$} at 4.0 -1
\setplotsymbol ({\scalebox{0.3}{$\bullet$}})
\color{Red}
\plot 0.00 5.253  0.05 5.204  0.10 5.155  0.15 5.106  0.20 5.057  0.25 5.008  0.30 4.959  0.35 4.910  0.40 4.861  0.45 4.812  0.50 4.763  0.55 4.714  0.60 4.665  0.65 4.616  0.70 4.567  0.75 4.518  0.80 4.469  0.85 4.420  0.90 4.371  0.95 4.322  1.00 4.273  1.05 4.224  1.10 4.175  1.15 4.126  1.20 4.077  1.25 4.028  1.30 3.979  1.35 3.930  1.40 3.881  1.45 3.832  1.50 3.783  1.55 3.734  1.60 3.685  1.65 3.636  1.70 3.588  1.75 3.539  1.80 3.490  1.85 3.441  1.90 3.392  1.95 3.343  2.00 3.294  2.05 3.246  2.10 3.197  2.15 3.148  2.20 3.099  2.25 3.051  2.30 3.002  2.35 2.954  2.40 2.905  2.45 2.857  2.50 2.811  2.55 2.857  2.60 2.905  2.65 2.954  2.70 3.002  2.75 3.051  2.80 3.099  2.85 3.148  2.90 3.197  2.95 3.246  3.00 3.294  3.05 3.343  3.10 3.392  3.15 3.441  3.20 3.490  3.25 3.539  3.30 3.588  3.35 3.636  3.40 3.685  3.45 3.734  3.50 3.783  3.55 3.832  3.60 3.881  3.65 3.930  3.70 3.979  3.75 4.028  3.80 4.077  3.85 4.126  3.90 4.175  3.95 4.224  4.00 4.273  4.05 4.322  4.10 4.371  4.15 4.420  4.20 4.469  4.25 4.518  4.30 4.567  4.35 4.616  4.40 4.665  4.45 4.714  4.50 4.763  4.55 4.812  4.60 4.861  4.65 4.910  4.70 4.959  4.75 5.008  4.80 5.057  4.85 5.106  4.90 5.155  4.95 5.204  5.00 5.253  /
\color{black}
\put {$f_L$} at -0.5 3
\endpicture
\end{subfigure}
\begin{subfigure}{0.33\textwidth}
\normalcolor
\color{black}
	\caption{$\LA c_1 \RA$ and $\LA c_2 \RA$}
\beginpicture
\setcoordinatesystem units <21pt,90pt>
\axes{0}{0}{0}{0}{5}{1}{0}
\put {\Large$\alpha$} at 4.0 -0.17
\setplotsymbol ({\scalebox{0.3}{$\bullet$}})
\color{Red}
\plot 0.00 0.0100  0.15 0.0100  0.29 0.0100  0.42 0.0100  0.55 0.0100  0.67 0.0100  0.79 0.0100  0.90 0.0100  1.01 0.0100  1.11 0.0100  1.21 0.0100  1.30 0.0100  1.39 0.0100  1.48 0.0100  1.56 0.0101  1.63 0.0101  1.70 0.0101  1.77 0.0101  1.83 0.0102  1.89 0.0102  1.95 0.0103  2.00 0.0104  2.05 0.0105  2.10 0.0106  2.14 0.0107  2.18 0.0109  2.21 0.0111  2.25 0.0113  2.28 0.0115  2.31 0.0118  2.33 0.0121  2.36 0.0124  2.38 0.0128  2.40 0.0133  2.41 0.0139  2.43 0.0146  2.44 0.0153  2.45 0.0163  2.46 0.0174  2.47 0.0188  2.48 0.0206  2.48 0.0229  2.49 0.0261  2.49 0.0326  2.49 0.0546  2.50 0.1212  2.50 0.2408  2.50 0.3636  2.50 0.4452  2.50 0.4828  2.50 0.4923  2.50 0.4923  2.50 0.5018  2.50 0.5395  2.50 0.6217  2.50 0.7464  2.50 0.8691  2.51 0.9378  2.51 0.9603  2.51 0.9669  2.52 0.9701  2.52 0.9724  2.53 0.9742  2.54 0.9757  2.55 0.9770  2.56 0.9780  2.57 0.9790  2.59 0.9798  2.60 0.9805  2.62 0.9812  2.64 0.9818  2.67 0.9824  2.69 0.9829  2.72 0.9835  2.75 0.9840  2.79 0.9844  2.82 0.9849  2.86 0.9853  2.90 0.9858  2.95 0.9862  3.00 0.9865  3.05 0.9869  3.11 0.9872  3.17 0.9876  3.23 0.9878  3.30 0.9881  3.37 0.9884  3.44 0.9886  3.52 0.9888  3.61 0.9890  3.70 0.9892  3.79 0.9893  3.89 0.9894  3.99 0.9895  4.10 0.9896  4.21 0.9897  4.33 0.9898  4.45 0.9898  4.58 0.9899  4.71 0.9899  4.85 0.9899  5.00 0.9899   /
\put {$\mathbf{\LA c_1 \RA}$} at -1 0.33
\color{Blue}
\plot 0.00 0.9899  0.15 0.9899  0.29 0.9899  0.42 0.9899  0.55 0.9898  0.67 0.9898  0.79 0.9897  0.90 0.9896  1.01 0.9895  1.11 0.9894  1.21 0.9893  1.30 0.9892  1.39 0.9890  1.48 0.9888  1.56 0.9886  1.63 0.9884  1.70 0.9881  1.77 0.9878  1.83 0.9875  1.89 0.9872  1.95 0.9869  2.00 0.9865  2.05 0.9861  2.10 0.9857  2.14 0.9853  2.18 0.9849  2.21 0.9844  2.25 0.9839  2.28 0.9834  2.31 0.9829  2.33 0.9823  2.36 0.9818  2.38 0.9811  2.40 0.9805  2.41 0.9797  2.43 0.9789  2.44 0.9780  2.45 0.9769  2.46 0.9756  2.47 0.9741  2.48 0.9723  2.48 0.9699  2.49 0.9667  2.49 0.9601  2.49 0.9381  2.50 0.8714  2.50 0.7518  2.50 0.6290  2.50 0.5474  2.50 0.5097  2.50 0.5003  2.50 0.5003  2.50 0.4908  2.50 0.4530  2.50 0.3709  2.50 0.2462  2.50 0.1235  2.51 0.0549  2.51 0.0324  2.51 0.0258  2.52 0.0227  2.52 0.0204  2.53 0.0187  2.54 0.0173  2.55 0.0162  2.56 0.0153  2.57 0.0145  2.59 0.0138  2.60 0.0133  2.62 0.0128  2.64 0.0124  2.67 0.0120  2.69 0.0117  2.72 0.0115  2.75 0.0112  2.79 0.0110  2.82 0.0109  2.86 0.0107  2.90 0.0106  2.95 0.0105  3.00 0.0104  3.05 0.0103  3.11 0.0102  3.17 0.0102  3.23 0.0101  3.30 0.0101  3.37 0.0101  3.44 0.0101  3.52 0.0100  3.61 0.0100  3.70 0.0100  3.79 0.0100  3.89 0.0100  3.99 0.0100  4.10 0.0100  4.21 0.0100  4.33 0.0100  4.45 0.0100  4.58 0.0100  4.71 0.0100  4.85 0.0100  5.00 0.0100   /
\put {$\mathbf{\LA c_2 \RA}$} at -1 0.67
\endpicture
\end{subfigure}
\begin{subfigure}{0.33\textwidth}
	\caption{$L^2\Var{c_1}$ and $L^2\Var{c_2}$}
\beginpicture
\normalcolor
\color{black}
\setcoordinatesystem units <21pt,10pt>
\setplotarea x from 0 to 5, y from -7 to 2
\axis left shiftedto x=0
        ticks withvalues $-7$ $2$ /
        at -7 2 /
/
\axis bottom 
        ticks withvalues $0$ $2.5$ $5$ /
        at 0 2.5 5 /
/
\setdots <2pt>
\plot 2.5 -7 2.5 2 /
\setsolid
\put {\Large$\alpha$} at 4 -8.5
\setplotsymbol ({\scalebox{0.3}{$\bullet$}})
\color{Red}
\plot 0.00 -6.628  0.15 -6.375  0.29 -6.132  0.42 -5.898  0.55 -5.675  0.67 -5.460  0.79 -5.255  0.90 -5.059  1.01 -4.871  1.11 -4.692  1.21 -4.520  1.30 -4.357  1.39 -4.200  1.48 -4.051  1.56 -3.909  1.63 -3.773  1.70 -3.644  1.77 -3.519  1.83 -3.400  1.89 -3.286  1.95 -3.176  2.00 -3.069  2.05 -2.965  2.10 -2.864  2.14 -2.765  2.18 -2.667  2.21 -2.570  2.25 -2.472  2.28 -2.373  2.31 -2.272  2.33 -2.169  2.36 -2.063  2.38 -1.953  2.40 -1.838  2.41 -1.718  2.43 -1.591  2.44 -1.457  2.45 -1.314  2.46 -1.160  2.47 -0.994  2.48 -0.811  2.48 -0.602  2.49 -0.313  2.49 0.215  2.49 0.897  2.50 1.442  2.50 1.750  2.50 1.872  2.50 1.905  2.50 1.910  2.50 1.910  2.50 1.910  2.50 1.910  2.50 1.906  2.50 1.877  2.50 1.759  2.50 1.454  2.51 0.910  2.51 0.223  2.51 -0.309  2.52 -0.590  2.52 -0.784  2.53 -0.949  2.54 -1.095  2.55 -1.223  2.56 -1.337  2.57 -1.437  2.59 -1.525  2.60 -1.602  2.62 -1.671  2.64 -1.732  2.67 -1.787  2.69 -1.838  2.72 -1.885  2.75 -1.929  2.79 -1.973  2.82 -2.015  2.86 -2.057  2.90 -2.100  2.95 -2.144  3.00 -2.189  3.05 -2.236  3.11 -2.285  3.17 -2.336  3.23 -2.390  3.30 -2.447  3.37 -2.507  3.44 -2.571  3.52 -2.638  3.61 -2.709  3.70 -2.784  3.79 -2.864  3.89 -2.947  3.99 -3.035  4.10 -3.127  4.21 -3.224  4.33 -3.325  4.45 -3.432  4.58 -3.543  4.71 -3.659  4.85 -3.780  5.00 -3.906  /
\put {\scalebox{0.8}{$L^2\mathbf{\Var{c_1}}$}} at 4 -1.5
\color{Blue}
\plot 0.00 -3.904  0.15 -3.778  0.29 -3.657  0.42 -3.541  0.55 -3.430  0.67 -3.323  0.79 -3.222  0.90 -3.125  1.01 -3.033  1.11 -2.945  1.21 -2.862  1.30 -2.782  1.39 -2.707  1.48 -2.636  1.56 -2.569  1.63 -2.505  1.70 -2.445  1.77 -2.388  1.83 -2.334  1.89 -2.283  1.95 -2.234  2.00 -2.187  2.05 -2.142  2.10 -2.098  2.14 -2.056  2.18 -2.014  2.21 -1.971  2.25 -1.928  2.28 -1.884  2.31 -1.836  2.33 -1.786  2.36 -1.731  2.38 -1.669  2.40 -1.600  2.41 -1.523  2.43 -1.435  2.44 -1.335  2.45 -1.221  2.46 -1.092  2.47 -0.946  2.48 -0.779  2.48 -0.581  2.49 -0.302  2.49 0.218  2.49 0.898  2.50 1.442  2.50 1.750  2.50 1.872  2.50 1.905  2.50 1.910  2.50 1.910  2.50 1.910  2.50 1.910  2.50 1.906  2.50 1.877  2.50 1.759  2.50 1.454  2.51 0.909  2.51 0.220  2.51 -0.321  2.52 -0.612  2.52 -0.819  2.53 -0.999  2.54 -1.165  2.55 -1.318  2.56 -1.462  2.57 -1.596  2.59 -1.724  2.60 -1.844  2.62 -1.959  2.64 -2.070  2.67 -2.176  2.69 -2.279  2.72 -2.380  2.75 -2.479  2.79 -2.578  2.82 -2.676  2.86 -2.775  2.90 -2.875  2.95 -2.977  3.00 -3.081  3.05 -3.189  3.11 -3.299  3.17 -3.414  3.23 -3.534  3.30 -3.659  3.37 -3.789  3.44 -3.925  3.52 -4.067  3.61 -4.216  3.70 -4.373  3.79 -4.536  3.89 -4.708  3.99 -4.887  4.10 -5.075  4.21 -5.271  4.33 -5.477  4.45 -5.691  4.58 -5.915  4.71 -6.148  4.85 -6.391  5.00 -6.644   /
\put {\scalebox{0.8}{$L^2\mathbf{\Var{c_2}}$}} at 1 -1.5
\normalcolor
\color{black}
\endpicture
\end{subfigure}
\normalcolor
\color{black}
\caption{The free energy, mean concentrations and variances of the unlinked
model as a function of $\alpha$ along the line crossing the $\tau$ phase boundary
between the points $(0,5)$ to $(5,0)$ in the phase diagram in figure \ref{9}.  
These data were calculated for $L=20$, but shows that the free energy is
to numerical accuracy an absolute function, non-analytic on the $\tau$
phase boundary.  The mean concentrations $\LA c_1\RA$ and $\LA c_2\RA$ are
approximately step functions at the critical line, while the variances,
plotted on a logarithm vertical scale, develope sharp spikes at the critical
point, while having very small values (less than $10^{-3}$) away from 
the critical point.}
\label{10}   %ZXZ[10]
\end{figure}

\subsubsection{Critical scaling in the unlinked model}

{\bf The $\tau$ phase boundary:}
In figure \ref{10} the free energy, mean concentrations and variances 
for $L=20$ are plotted as a function of $\alpha$ from the point $(0,5)$ 
along the line segment to $(0,5)$.  The critical point (where the $\tau$
phase boundary is crossed) corresponds to $\alpha = 2.5$, and this
is seen in the plot of the free energy $f_L$ in figure \ref{10}(a).  The
phase transition along the $\tau$ phase boundary is a first order
transition to numerical accuracy, as the limiting free energy in equation
\Ref{6} is accurately approximated by $f_L$ for $L=20$, and $f_L$
can be directly modelled by an absolutely value function.  Indeed,
using the $5$-parameter model
\begin{equation}
\upsilon (\alpha,5{-}\alpha) 
= a_0 + a_1\, |\alpha-a_c|^{2-\alpha_s} + a_2\, (\alpha-a_c)^2
\label{16}   %ZXZ[16]
\end{equation}
a nonlinear fit (done using the \textit{NonlinearFit} function in Maple
\cite{Maple17}) can be performed to obtain estimates of the exponent
$2{-}\alpha_s$.  If we first ignore the quadratic term (by putting $a_2=0$)
to obtain
\begin{equation}
%\hspace{-2.5cm}
\upsilon(\alpha,5{-}\alpha) \approx % f_L ( \alpha,5{-}\alpha) = 
2.810 + 0.975\,|\alpha-2.498|^{1.003} .
\label{17}   %ZXZ[17]
\end{equation} 
Repeating the fit with $a_2$ as an additional  parameter gives instead
\begin{eqnarray}
\hspace{-1.5cm}
\upsilon(\alpha,5{-}\alpha) 
&\approx& %f_L ( \alpha,5{-}\alpha) = 
2.810 + 0.976\,|\alpha-2.499|^{1.006} \nonumber \\
& & \hspace{1cm} +\, 0.00153\, (\alpha-2.499)^2 .
\label{18}  %ZXZ[18]
\end{eqnarray} 
Notice the very small coefficient of the quadratic term, showing that 
the free energy is very closely approximated by the absolute value term.
These fits give the critical point at $a_c=2.498$ (compared to its exact value
$2.5$ by symmetry) and the estimate of  the exponent $2{-}\alpha_s 
= 2{-}\alpha_s^\prime \approx 1.003$ (see equation \Ref{5}).
Notice that the coefficient of the quadratic term is very small, showing
that even in this finite size model, with $L=20$, the free energy is
well approximated by a non-analytic function, showing a sharp transition
at the critical point. 

As in figure \ref{5} one may now estimate the scaling exponent associated
with the $\tau$ phase boundary.  Taking as error bar the difference in
the estimates in equations \Ref{17} and \Ref{18},
\begin{equation}
2-\alpha_s = 2-\alpha_s^\prime = 1.003 \pm 0.003  . 
\label{19} %ZXZ[19]
\end{equation} 
This is consistent with the transition across the $\tau$ phase boundary 
being first order.

\begin{figure}[h!]
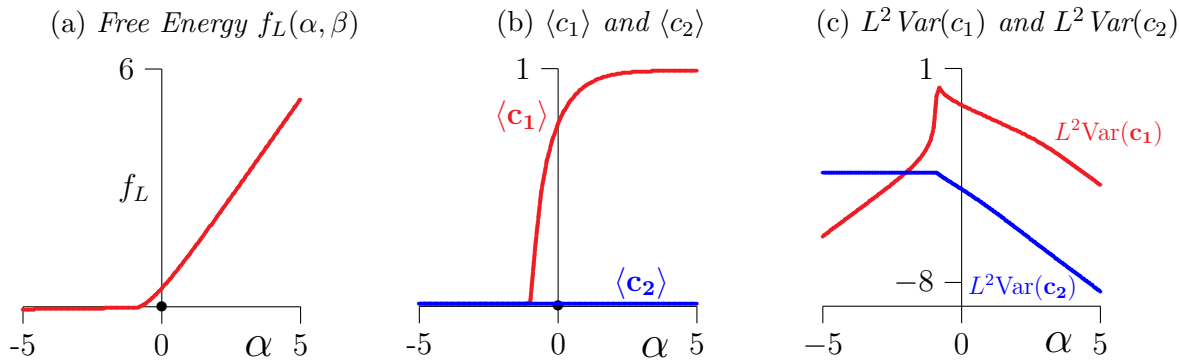

\begin{subfigure}{0.33\textwidth}
	\caption{Free Energy $f_L(\alpha,\beta)$}
\beginpicture
\normalcolor
\color{black}
\setcoordinatesystem units <10.5pt,15pt>
\axescenter{-5}{0}{0}{0}{5}{6}{0}
\put {\Large$\alpha$} at 3.5 -1
\setplotsymbol ({\scalebox{0.3}{$\bullet$}})
\color{Red}
\plot -5.00000 -0.0699098275  -4.80000 -0.0679097226  -4.60000 -0.0659095662  -4.40000 -0.0639093328  -4.20000 -0.0619089845  -4.00000 -0.0599084647  -3.80000 -0.0579076887  -3.60000 -0.0559065299  -3.40000 -0.0539047987  -3.20000 -0.0519022106  -3.00000 -0.0498983370  -2.80000 -0.0478925302  -2.60000 -0.0458838046  -2.40000 -0.0438706444  -2.20000 -0.0418506845  -2.00000 -0.0398201479  -1.80000 -0.0377727853  -1.60000 -0.0356976631  -1.40000 -0.0335737936  -1.20000 -0.0313534133  -1.00000 -0.0288709893  -0.80000 -0.0128754725  -0.60000 0.0647447300  -0.40000 0.1749706407  -0.20000 0.3065481510  0.00000 0.4535126438  0.20000 0.6120206201  0.40000 0.7793672984  0.60000 0.9535627964  0.80000 1.1330990236  1.00000 1.3168115087  1.20000 1.5037928560  1.40000 1.6933309573  1.60000 1.8848654890  1.80000 2.0779563637  2.00000 2.2722588347  2.20000 2.4675025442  2.40000 2.6634740611  2.60000 2.8600030007  2.80000 3.0569523789  3.00000 3.2542127775  3.20000 3.4516985838  3.40000 3.6493446868  3.60000 3.8471029875  3.80000 4.0449388499  4.00000 4.2428278359  4.20000 4.4407529650  4.40000 4.6387025691  4.60000 4.8366686930  4.80000 5.0346459422  5.00000 5.2326306726  /
\color{black}
\put {$f_L$} at -1 3
\endpicture
\end{subfigure}
\begin{subfigure}{0.33\textwidth}
	\caption{$\LA c_1 \RA$ and $\LA c_2 \RA$}
\beginpicture
\normalcolor
\color{black}
\setcoordinatesystem units <10.5pt,90pt>
\axescenter{-5}{0}{0}{0}{5}{1}{0}
\put {\Large$\alpha$} at 3.5 -0.17
\setplotsymbol ({\scalebox{0.3}{$\bullet$}})
\color{Red}
\plot -5.00000 0.0100004263  -4.80000 0.0100006361  -4.60000 0.0100009490  -4.40000 0.0100014161  -4.20000 0.0100021133  -4.00000 0.0100031544  -3.80000 0.0100047094  -3.60000 0.0100070337  -3.40000 0.0100105112  -3.20000 0.0100157215  -3.00000 0.0100235446  -2.80000 0.0100353287  -2.60000 0.0100531655  -2.40000 0.0100803629  -2.20000 0.0101223015  -2.00000 0.0101881068  -1.80000 0.0102942474  -1.60000 0.0104733911  -1.40000 0.0108009165  -1.20000 0.0115055904  -1.00000 0.0140199550  -0.80000 0.2717653427  -0.60000 0.4830037150  -0.40000 0.6108038726  -0.20000 0.7001957494  0.00000 0.7663149811  0.20000 0.8165431061  0.40000 0.8552839211  0.60000 0.8854256956  0.80000 0.9089729435  1.00000 0.9274008705  1.20000 0.9418227337  1.40000 0.9530938941  1.60000 0.9618867325  1.80000 0.9687364698  2.00000 0.9740643953  2.20000 0.9781960113  2.40000 0.9813779693  2.60000 0.9837977347  2.80000 0.9856051579  3.00000 0.9869277522  3.20000 0.9878763453  3.40000 0.9885449025  3.60000 0.9890095165  3.80000 0.9893289718  4.00000 0.9895469156  4.20000 0.9896947841  4.40000 0.9897947224  4.60000 0.9898620875  4.80000 0.9899074139  5.00000 0.9899378743  /
\put {$\mathbf{\LA c_1 \RA}$} at -1.3 0.8
\color{Blue}
\plot -5.00000 0.0101882450  -4.80000 0.0101882449  -4.60000 0.0101882449  -4.40000 0.0101882449  -4.20000 0.0101882448  -4.00000 0.0101882448  -3.80000 0.0101882447  -3.60000 0.0101882445  -3.40000 0.0101882442  -3.20000 0.0101882439  -3.00000 0.0101882433  -2.80000 0.0101882425  -2.60000 0.0101882412  -2.40000 0.0101882392  -2.20000 0.0101882362  -2.00000 0.0101882315  -1.80000 0.0101882238  -1.60000 0.0101882108  -1.40000 0.0101881868  -1.20000 0.0101881341  -1.00000 0.0101879376  -0.80000 0.0101533501  -0.60000 0.0101100450  -0.40000 0.0100792493  -0.20000 0.0100571526  0.00000 0.0100409418  0.20000 0.0100290918  0.40000 0.0100204781  0.60000 0.0100142534  0.80000 0.0100098194  1.00000 0.0100066985  1.20000 0.0100045294  1.40000 0.0100030425  1.60000 0.0100020343  1.80000 0.0100013554  2.00000 0.0100009006  2.20000 0.0100005973  2.40000 0.0100003958  2.60000 0.0100002625  2.80000 0.0100001742  3.00000 0.0100001159  3.20000 0.0100000772  3.40000 0.0100000515  3.60000 0.0100000344  3.80000 0.0100000230  4.00000 0.0100000154  4.20000 0.0100000103  4.40000 0.0100000069  4.60000 0.0100000046  4.80000 0.0100000031  5.00000 0.0100000021   /
\put {$\mathbf{\LA c_2 \RA}$} at 3 0.1
\endpicture
\end{subfigure}
\begin{subfigure}{0.33\textwidth}
	\caption{$L^2\Var{c_1}$ and $L^2\Var{c_2}$}
\beginpicture
\normalcolor
\color{black}
\setcoordinatesystem units <10.5pt,9pt>
\setplotarea x from -5 to 5, y from -9 to 1
\axis left shiftedto x=0
        ticks withvalues $-8$ $1$ /
        at -8 1 /
/
\axis bottom 
        ticks withvalues $-5$ $0$ $5$ /
        at -5 0 5 /
/
%\setdots <2pt>
%\plot 3.536 -7 3.536 2 /
\setsolid
\put {\Large$\alpha$} at 3.5 -10.5
\setplotsymbol ({\scalebox{0.3}{$\bullet$}})
\color{Red}
\plot -5.00000 -6.06915  -4.90000 -5.98226  -4.80000 -5.89534  -4.70000 -5.80841  -4.60000 -5.72148  -4.50000 -5.63453  -4.40000 -5.54757  -4.30000 -5.46057  -4.20000 -5.37355  -4.10000 -5.28648  -4.00000 -5.19937  -3.90000 -5.11221  -3.80000 -5.02499  -3.70000 -4.93767  -3.60000 -4.85026  -3.50000 -4.76273  -3.40000 -4.67504  -3.30000 -4.58717  -3.20000 -4.49908  -3.10000 -4.41071  -3.00000 -4.32200  -2.90000 -4.23288  -2.80000 -4.14325  -2.70000 -4.05299  -2.60000 -3.96196  -2.50000 -3.86998  -2.40000 -3.77683  -2.30000 -3.68221  -2.20000 -3.58578  -2.10000 -3.48709  -2.00000 -3.38553  -1.90000 -3.28035  -1.80000 -3.17052  -1.70000 -3.05466  -1.60000 -2.93080  -1.50000 -2.79608  -1.40000 -2.64603  -1.30000 -2.47314  -1.20000 -2.26302  -1.10000 -1.98207  -1.00000 -1.51425  -0.90000 -0.21086  -0.80000 0.18952  -0.70000 0.01489  -0.60000 -0.10634  -0.50000 -0.19929  -0.40000 -0.27990  -0.30000 -0.35254  -0.20000 -0.41945  -0.10000 -0.48260  0.00000 -0.54314  0.10000 -0.60162  0.20000 -0.65852  0.30000 -0.71418  0.40000 -0.76894  0.50000 -0.82304  0.60000 -0.87674  0.70000 -0.93022  0.80000 -0.98348  0.90000 -1.03663  1.00000 -1.08980  1.10000 -1.14307  1.20000 -1.19650  1.30000 -1.25013  1.40000 -1.30398  1.50000 -1.35800  1.60000 -1.41217  1.70000 -1.46646  1.80000 -1.52092  1.90000 -1.57557  2.00000 -1.63054  2.10000 -1.68598  2.20000 -1.74216  2.30000 -1.79939  2.40000 -1.85801  2.50000 -1.91837  2.60000 -1.98071  2.70000 -2.04524  2.80000 -2.11205  2.90000 -2.18112  3.00000 -2.25240  3.10000 -2.32575  3.20000 -2.40100  3.30000 -2.47795  3.40000 -2.55643  3.50000 -2.63622  3.60000 -2.71716  3.70000 -2.79908  3.80000 -2.88183  3.90000 -2.96528  4.00000 -3.04931  4.10000 -3.13384  4.20000 -3.21877  4.30000 -3.30405  4.40000 -3.38960  4.50000 -3.47539  4.60000 -3.56137  4.70000 -3.64750  4.80000 -3.73377  4.90000 -3.82014  5.00000 -3.90660  /
\put {\scalebox{0.8}{$L^2\mathbf{\Var{c_1}}$}} at 5.3 -1.7
\color{Blue}
\plot -5.00000 -3.38521  -4.90000 -3.38521  -4.80000 -3.38522  -4.70000 -3.38522  -4.60000 -3.38522  -4.50000 -3.38522  -4.40000 -3.38522  -4.30000 -3.38522  -4.20000 -3.38522  -4.10000 -3.38522  -4.00000 -3.38522  -3.90000 -3.38522  -3.80000 -3.38522  -3.70000 -3.38522  -3.60000 -3.38522  -3.50000 -3.38522  -3.40000 -3.38522  -3.30000 -3.38522  -3.20000 -3.38522  -3.10000 -3.38522  -3.00000 -3.38522  -2.90000 -3.38522  -2.80000 -3.38522  -2.70000 -3.38522  -2.60000 -3.38522  -2.50000 -3.38523  -2.40000 -3.38523  -2.30000 -3.38523  -2.20000 -3.38524  -2.10000 -3.38524  -2.00000 -3.38525  -1.90000 -3.38526  -1.80000 -3.38527  -1.70000 -3.38528  -1.60000 -3.38530  -1.50000 -3.38532  -1.40000 -3.38536  -1.30000 -3.38541  -1.20000 -3.38549  -1.10000 -3.38563  -1.00000 -3.38598  -0.90000 -3.38969  -0.80000 -3.48142  -0.70000 -3.55807  -0.60000 -3.63510  -0.50000 -3.71024  -0.40000 -3.78466  -0.30000 -3.85823  -0.20000 -3.93159  -0.10000 -4.00549  0.00000 -4.07993  0.10000 -4.15496  0.20000 -4.23071  0.30000 -4.30726  0.40000 -4.38480  0.50000 -4.46347  0.60000 -4.54322  0.70000 -4.62397  0.80000 -4.70572  0.90000 -4.78849  1.00000 -4.87225  1.10000 -4.95694  1.20000 -5.04245  1.30000 -5.12865  1.40000 -5.21544  1.50000 -5.30270  1.60000 -5.39036  1.70000 -5.47839  1.80000 -5.56676  1.90000 -5.65541  2.00000 -5.74434  2.10000 -5.83345  2.20000 -5.92270  2.30000 -6.01207  2.40000 -6.10138  2.50000 -6.19071  2.60000 -6.27992  2.70000 -6.36896  2.80000 -6.45780  2.90000 -6.54653  3.00000 -6.63507  3.10000 -6.72331  3.20000 -6.81135  3.30000 -6.89928  3.40000 -6.98716  3.50000 -7.07469  3.60000 -7.16241  3.70000 -7.24949  3.80000 -7.33724  3.90000 -7.42481  4.00000 -7.51145  4.10000 -7.59860  4.20000 -7.68613  4.30000 -7.77211  4.40000 -7.86012  4.50000 -7.94692  4.60000 -8.03152  4.70000 -8.11919  4.80000 -8.20761  4.90000 -8.29243  5.00000 -8.37675  /
\put {\scalebox{0.8}{$L^2\mathbf{\Var{c_2}}$}} at 2.2 -8.2
\normalcolor
\color{black}
\endpicture
\end{subfigure}
\caption{The free energy, mean concentrations and variances of the 
unlinked model as a function of $\alpha$ along the line from the point 
$(-5,-2)$ to $(5,-2)$ crossing $\lambda_1$ in the phase diagram 
in figure \ref{9}.  In this figure $L=20$.  The free energy
shows a transition at approximately $\alpha_c\approx -\log \mu_2 \approx -0.97$,
and this seen in the behaviour of the mean concentration $\LA c_1 \RA$.
Our analysis strongly suggests that this transition between the empty
and the $c_1$-dominated phase is continuous.
Notice that $\LA c_2 \RA$ appears to be unchanged in the middle graph,
but if the logarithm of variances are plotted, then $\Var{c_2}$ is singular 
at the critical point, going from constant into a steady decline.  
Since our model is symmetric in $\{c_1,c_2\}$,  these graphs 
will be unchanged (except with $c_1$ and $c_2$ interchanged)
if instead these quantities are calculated along the line from $(-2,-5)$ to $(-2,5)$ 
which crosses over the phase boundary $\lambda_2$
separating the empty and the $c_2$-dominated phases.}
\label{11}   %ZXZ[11]
\end{figure}

{\bf The $\lambda_1$ and $\lambda_2$ phase boundaries:}
Next, consider the transition across the $\lambda_1$ phase boundary 
by calculating the free energy along the line segment from $(-5,-2)$
to $(5,-2)$.  The results are plotted in figure \ref{11} as a function of
$\alpha$.  The graphs in figure \ref{11}(b) of the concentrations
$\LA c_1\RA$ and $\LA c_2 \RA$ show that the transition across 
the $\lambda_1$ phase boundary does not have a step-like increase,
unlike the case for the $\tau$ phase boundary.  Since the $\lambda_1$ 
phase boundary in figure \ref{9} is a vertical line and $\alpha_c \approx -0.97$, 
the transition across $\lambda_1$ is expected to occur at $\alpha_c$, 
and our data are consistent with this.  The critical exponent associated 
with the transition is defined by equation \Ref{8}, and since the 
free energy $f_L $ in figure \ref{11} is (up to numerical accuracy) 
a constant for $\alpha < \alpha_c$ we conclude that $2{-}\alpha_s = 0$ 
along the $\lambda_1$ phase boundary as $\alpha \to \alpha_c^-$.

If $\alpha > \alpha_c$ the concentration $\LA c_1 \RA$ increases sharply 
at the critical point (as shown in figure \ref{11}(b)).  As $\alpha$ increases
in size, one expects that the first polygon becomes dense and then fills 
the confining square, and $\LA c_1 \RA \to 1$.  In view of equation \Ref{8},
\begin{equation}
\LA c_1 \RA \sim |\alpha-\alpha_c|^{1-\alpha_s^\prime},\; 
\hbox{for $\alpha\to \alpha_c^+$ and $\beta=-2$}.
\label{20}    %ZXZ[20]
\end{equation}
The exponent can be estimated by plotting
$\log \LA c_1 \RA / \log |\alpha{-}\alpha_c|$ as a function of 
$1/\log  |\alpha{-}\alpha_c|$.  This is improved if we subtract from
$\LA c_1 \RA$ its (small but non-zero) value $\LA c_1 \RA\!\vert_{crit}$ at 
$\alpha = \alpha_c$, to zero it at the critical point.  This removes a
background term due to the finite size of our model since $L=20$. 
Defining
\begin{equation}
\log \LA c_1 \RA_s = \log (\LA c_1 \RA - \LA c_1 \RA\!\vert_{crit})
\label{21}  %ZXZ[21]
\end{equation}
we get the model
\begin{eqnarray}
\gamma &=& \log \LA c_1 \RA_s / \log |\alpha{-}\alpha_c| \nonumber \\
&=& (1{-}\alpha_s^\prime) + C/\log  |\alpha{-}\alpha_c| + \ldots
\label{22}  %ZXZ[22]
\end{eqnarray}
Choosing $\alpha_c=-0.97$ and then plotting 
$\gamma$ as a function of $1/\log  |\alpha{-}\alpha_c|$ for 
$-0.95 \leq \alpha \leq 0$ gives the graph in figure \ref{12}.  This graph is
almost a straight line, except at points where $\alpha$ approaches $\alpha_c$.
Fitting a linear function in $1/\log  |\alpha{-}\alpha_c|$ to the graph gives 
the estimate $1{-}\alpha_s^\prime \approx 0.545$, and fitting a quadratic 
instead gives $1{-}\alpha_s^\prime \approx 0.516$.  The difference in 
these two estimates is taken as a confidence interval in the estimate.

\begin{figure}[h!]
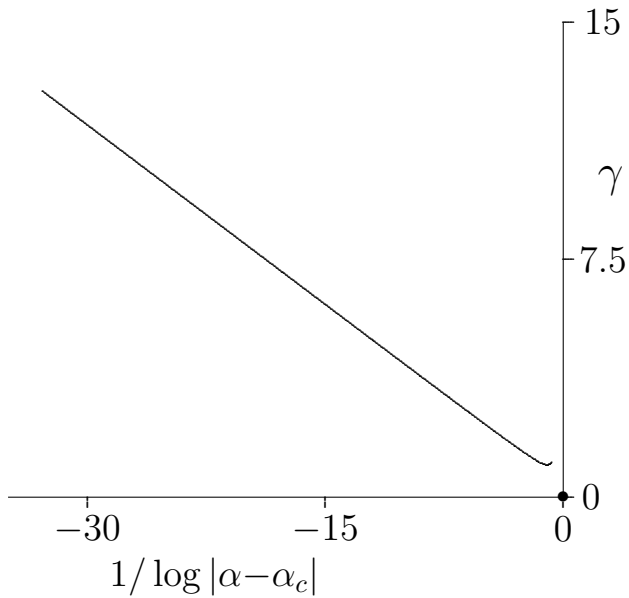

\color{black}
\normalcolor
\beginpicture
\setcoordinatesystem units <6pt,12pt>
\setplotarea x from -35 to 0, y from 0 to 15

\axis right shiftedto x=0 
        %ticks
        %withvalues 10  /
        %at  10  /
 /
\axis bottom shiftedto y=0
        %ticks
        %withvalues -30 -15 0 /
        %at  -30 -15 0 /
/
\put {\footnotesize$\bullet$} at 0 0 

\plot -30 -0.33 -30 0 / \plot -15 -0.33 -15 0 / \plot 0 -0.33 0 0 /
\put {\large${-}30$} at -30 -1
\put {\large${-}15$} at -15 -1
\put {\large$0$} at 0 -1
\plot 0 0 0.65 0 / \plot 0 7.5 0.65 7.5 / \plot 0 15 0.65 15 / 
\put {\large$0$} at 1.75 0  
\put {\large$7.5$} at 2.5 7.5  
\put {\large$15$} at 2.5 15 

\plot -0.7110 1.0888  -0.7637 1.0668  -0.8191 1.0476  -0.8776 1.0329  -0.9397 1.0235  -1.0058 1.0197  -1.0766 1.0214  -1.1527 1.0289  -1.2351 1.0420  -1.3245 1.0605  -1.4221 1.0843  -1.5292 1.1131  -1.6475 1.1471  -1.7790 1.1868  -1.9261 1.2329  -2.0919 1.2866  -2.2805 1.3494  -2.4970 1.4232  -2.7484 1.5107  -3.0441 1.6153  -3.3971 1.7420  -3.8261 1.8978  -4.3589 2.0931  -5.0390 2.3443  -5.9376 2.6783  -7.1807 3.1426  -9.0146 3.8303  -11.9931 4.9505  -17.6771 7.0931  -32.8308 12.8134 /

\put {\Large$\gamma$} at 3 10
\put {\large$1/\log|\alpha{-}\alpha_c|$} at -22 -2.5

\color{black}
\normalcolor
\endpicture
\caption{Plotting $\gamma 
= \log(\LA c_1 \RA- \LA c_1 \RA\!\vert_{crit})/ \log |\alpha{-}\alpha_c|$ as
a function of $1/\log|\alpha{-}\alpha_c|$ for the unlinked model along a 
line segment in the $c_1$-dominated phase starting in the $\lambda_1$ 
phase boundary with $\beta=-2$.   In this graph $-0.9 \leq \alpha < 0$.}
\label{12}  %ZXZ[12]
\end{figure}

On the empty phase side of $\lambda_1$ the free energy is constant to numerical
accuracy, and so one expects $\alpha_s=2$. Taken together,
\begin{equation}
\alpha_s = 2
\qquad\hbox{and}\qquad
\alpha_s^\prime \approx 0.45 \pm 0.03
\label{23}    %ZXZ[23]
\end{equation}
along the $\lambda_1$ phase boundary.  The stated error bar on 
$\alpha_s^\prime$ is the standard deviation of the estimates in the table
above.  

Thus, the transitions along the $\lambda_1$ phase boundary has exponents
$2{-}\alpha_s=0$ and $2{-}\alpha_s^\prime=1.55(3)$ in equation \Ref{8}.
Since the model is symmetric in $\{c_1,c_2\}$, these results 
will be the same if instead the $\lambda_2$ phase boundary between the 
empty and $c_2$-dominated phases is crossed.

\begin{figure}[h!]
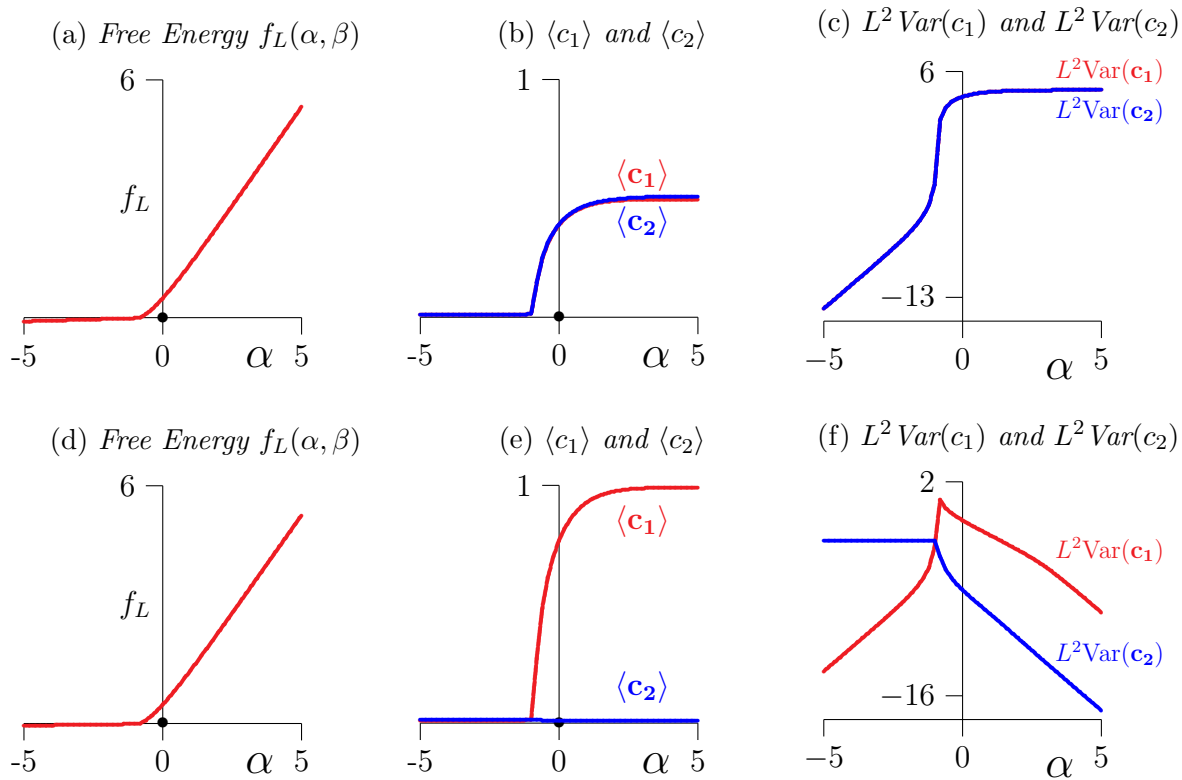

\begin{subfigure}{0.33\textwidth}
	\caption{Free Energy $f_L(\alpha,\beta)$}
\beginpicture
\normalcolor
\color{black}
\setcoordinatesystem units <10.5pt,15pt>
\axescenter{-5}{0}{0}{0}{5}{6}{0}
\put {\Large$\alpha$} at 3.5 -1
\setplotsymbol ({\scalebox{0.3}{$\bullet$}})
\color{Red}
\plot -5.00000 -0.0999995736  -4.80000 -0.0959993638  -4.60000 -0.0919990508  -4.40000 -0.0879985838  -4.20000 -0.0839978869  -4.00000 -0.0799968468  -3.80000 -0.0759952942  -3.60000 -0.0719929757  -3.40000 -0.0679895120  -3.20000 -0.0639843335  -3.00000 -0.0599765833  -2.80000 -0.0559649652  -2.60000 -0.0519475071  -2.40000 -0.0479211767  -2.20000 -0.0438812423  -2.00000 -0.0398201479  -1.80000 -0.0357253935  -1.60000 -0.0315751129  -1.40000 -0.0273273480  -1.20000 -0.0228866987  -1.00000 -0.0179240360  -0.80000 0.0025018404  -0.60000 0.0826667394  -0.40000 0.1952301401  -0.20000 0.3290391657  0.00000 0.4781629940  0.20000 0.6387746287  0.40000 0.8081831791  0.60000 0.9844079968  0.80000 1.1659505500  1.00000 1.3516539986  1.20000 1.5406160944  1.40000 1.7321292527  1.60000 1.9256364572  1.80000 2.1206995659  2.00000 2.3169748616  2.20000 2.5141927210  2.40000 2.7121404236  2.60000 2.9106482757  2.80000 3.1095798246  3.00000 3.3088258556  3.20000 3.5083005871  3.40000 3.7079384663  3.60000 3.9076908371  3.80000 4.1075225161  4.00000 4.3074085972  4.20000 4.5073317316  4.40000 4.7072799765  4.60000 4.9072451791  4.80000 5.1072218062  5.00000 5.3072061173 /
\color{black}
\put {$f_L$} at -1 3
\endpicture
\end{subfigure}
\begin{subfigure}{0.33\textwidth}
	\caption{$\LA c_1 \RA$ and $\LA c_2 \RA$}
\beginpicture
\normalcolor
\color{black}
\setcoordinatesystem units <10.5pt,90pt>
\axescenter{-5}{0}{0}{0}{5}{1}{0}
\put {\Large$\alpha$} at 3.5 -0.17
\setplotsymbol ({\scalebox{0.3}{$\bullet$}})
\color{Red}
\plot -5.00 0.0100004263  -4.80 0.0100006361  -4.60 0.0100009491  -4.40 0.0100014162  -4.20 0.0100021135  -4.00 0.0100031546  -3.80 0.0100047097  -3.60 0.0100070342  -3.40 0.0100105119  -3.20 0.0100157225  -3.00 0.0100235460  -2.80 0.0100353307  -2.60 0.0100531681  -2.40 0.0100803660  -2.20 0.0101223045  -2.00 0.0101881068  -1.80 0.0102942348  -1.60 0.0104733324  -1.40 0.0108006715  -1.20 0.0115042914  -1.00 0.0139969335  -0.80 0.1425861700  -0.60 0.2469430732  -0.40 0.3102317440  -0.20 0.3545739277  0.00 0.3871850043  0.20 0.4120348032  0.40 0.4312505332  0.60 0.4461918363  0.80 0.4578533431  1.00 0.4668994475  1.20 0.4739102626  1.40 0.4793881329  1.60 0.4836505053  1.80 0.4868957833  2.00 0.4892664799  2.20 0.4909009180  2.40 0.4919524687  2.60 0.4925805892  2.80 0.4929294069  3.00 0.4931104854  3.20 0.4931988562  3.40 0.4932394681  3.60 0.4932569088  3.80 0.4932637103  4.00 0.4932659077  4.20 0.4932662661  4.40 0.4932659888  4.60 0.4932655652  4.80 0.4932651727  5.00 0.4932648604  /
\put {$\mathbf{\LA c_1 \RA}$} at 3 0.6
\color{Blue}
\plot -5.00 0.0100004266  -4.80 0.0100006365  -4.60 0.0100009497  -4.40 0.0100014172  -4.20 0.0100021149  -4.00 0.0100031567  -3.80 0.0100047129  -3.60 0.0100070390  -3.40 0.0100105191  -3.20 0.0100157332  -3.00 0.0100235620  -2.80 0.0100353546  -2.60 0.0100532040  -2.40 0.0100804201  -2.20 0.0101223863  -2.00 0.0101882315  -1.80 0.0102944271  -1.60 0.0104736348  -1.40 0.0108011657  -1.20 0.0115051698  -1.00 0.0139991099  -0.80 0.1428156313  -0.60 0.2481073669  -0.40 0.3119406970  -0.20 0.3565871869  0.00 0.3897735307  0.20 0.4149137537  0.40 0.4342534300  0.60 0.4493151091  0.80 0.4611081047  1.00 0.4704266079  1.20 0.4777986437  1.40 0.4835729220  1.60 0.4880973734  1.80 0.4917026772  2.00 0.4946647656  2.20 0.4971703346  2.40 0.4993127050  2.60 0.5011195255  2.80 0.5025952513  3.00 0.5037539406  3.20 0.5046296845  3.40 0.5052705267  3.60 0.5057277539  3.80 0.5060478900  4.00 0.5062690245  4.20 0.5064203267  4.40 0.5065231696  4.60 0.5065927592  4.80 0.5066397035  5.00 0.5066713059   /
\put {$\mathbf{\LA c_2 \RA}$} at 3 0.4
\endpicture
\end{subfigure}
\begin{subfigure}{0.33\textwidth}
	\caption{$L^2\Var{c_1}$ and $L^2\Var{c_2}$}
\beginpicture
\normalcolor
\color{black}
\setcoordinatesystem units <10.5pt,4.5pt>
\setplotarea x from -5 to 5, y from -15 to 6
\axis left shiftedto x=0
        ticks withvalues $-13$ $6$ /
        at -13 6 /
/
\axis bottom 
        ticks withvalues $-5$ $0$ $5$ /
        at -5 0 5 /
/
%\setdots <2pt>
%\plot 3.536 -7 3.536 2 /
\setsolid
\put {\Large$\alpha$} at 3.5 -19
\setplotsymbol ({\scalebox{0.3}{$\bullet$}})
\color{Red}
\plot -5.00 -13.9746602676  -4.80 -13.5744509396  -4.60 -13.1741386186  -4.40 -12.7736726008  -4.20 -12.3729771841  -4.00 -11.9719392995  -3.80 -11.5703899657  -3.60 -11.1680764185  -3.40 -10.7646200691  -3.20 -10.3594527423  -3.00 -9.9517191831  -2.80 -9.5401261713  -2.60 -9.1227047459  -2.40 -8.6964252790  -2.20 -8.2565488805  -2.00 -7.7954676999  -1.80 -7.3004411976  -1.60 -6.7485634065  -1.40 -6.0931242652  -1.20 -5.2121468730  -1.00 -3.4991640001  -0.80 1.9022124719  -0.60 2.9491006110  -0.40 3.4040139270  -0.20 3.6766157323  0.00 3.8604415839  0.20 3.9921056698  0.40 4.0898205892  0.60 4.1639267807  0.80 4.2208494572  1.00 4.2649555347  1.20 4.2993558469  1.40 4.3262184507  1.60 4.3472195421  1.80 4.3636839073  2.00 4.3766413653  2.20 4.3868737205  2.40 4.3949509410  2.60 4.4012778156  2.80 4.4061539796  3.00 4.4098290502  3.20 4.4125323501  3.40 4.4144762954  3.60 4.4158478850  3.80 4.4168014007  4.00 4.4174570184  4.20 4.4179042566  4.40 4.4182076561  4.60 4.4184126885  4.80 4.4185508820  5.00 4.4186438592  /
\put {\scalebox{0.8}{$L^2\mathbf{\Var{c_1}}$}} at 5.3 6
\color{Blue}
\plot -5.00 -13.9739794543  -4.80 -13.5737701668  -4.60 -13.1734579062  -4.40 -12.7729919786  -4.20 -12.3722966966  -4.00 -11.9712590132  -3.80 -11.5697099800  -3.60 -11.1673968824  -3.40 -10.7639412065  -3.20 -10.3587748898  -3.00 -9.9510428492  -2.80 -9.5394521271  -2.60 -9.1220341658  -2.40 -8.6957599541  -2.20 -8.2558915296  -2.00 -7.7948223617  -1.80 -7.2998135406  -1.60 -6.7479604843  -1.40 -6.0925537439  -1.20 -5.2116244408  -1.00 -3.4985628705  -0.80 1.9026540136  -0.60 2.9493090685  -0.40 3.4040260933  -0.20 3.6766902057  0.00 3.8604845743  0.20 3.9921269221  0.40 4.0898261008  0.60 4.1639377102  0.80 4.2208620369  1.00 4.2649804576  1.20 4.2993787855  1.40 4.3262357874  1.60 4.3472380116  1.80 4.3637130963  2.00 4.3766874016  2.20 4.3869358497  2.40 4.3950224803  2.60 4.4013497688  2.80 4.4062188732  3.00 4.4098828429  3.20 4.4125742184  3.40 4.4145074262  3.60 4.4158702926  3.80 4.4168171681  4.00 4.4174679414  4.20 4.4179117432  4.40 4.4182127503  4.60 4.4184161379  4.80 4.4185532099  5.00 4.4186454268  /
\put {\scalebox{0.8}{$L^2\mathbf{\Var{c_2}}$}} at 5.3 2.7
\normalcolor
\color{black}
\endpicture
\end{subfigure}

\vspace{5mm}
\begin{subfigure}{0.33\textwidth}
	\caption{Free Energy $f_L(\alpha,\beta)$}
\beginpicture
\normalcolor
\color{black}
\setcoordinatesystem units <10.5pt,15pt>
\axescenter{-5}{0}{0}{0}{5}{6}{0}
\put {\Large$\alpha$} at 3.5 -1
\setplotsymbol ({\scalebox{0.3}{$\bullet$}})
\color{Red}
\plot -5.00 -0.0585237746  -4.80 -0.0565236699  -4.60 -0.0545235138  -4.40 -0.0525232808  -4.20 -0.0505229331  -4.00 -0.0485224143  -3.80 -0.0465216397  -3.60 -0.0445204830  -3.40 -0.0425187550  -3.20 -0.0405161716  -3.00 -0.0385123051  -2.80 -0.0365065092  -2.60 -0.0344977998  -2.40 -0.0324846645  -2.20 -0.0304647429  -2.00 -0.0284342665  -1.80 -0.0263870012  -1.60 -0.0243120437  -1.40 -0.0221884770  -1.20 -0.0199687557  -1.00 -0.0174887627  -0.80 -0.0018256415  -0.60 0.0755116536  -0.40 0.1855810757  -0.20 0.3170611611  0.00 0.4639603567  0.20 0.6224232779  0.40 0.7897384232  0.60 0.9639116971  0.80 1.1434323423  1.00 1.3271339689  1.20 1.5141078161  1.40 1.7036407954  1.60 1.8951718622  1.80 2.0882604076  2.00 2.2825613192  2.20 2.4778039894  2.40 2.6737748163  2.60 2.8703032991  2.80 3.0672523753  3.00 3.2645125740  3.20 3.4619982480  3.40 3.6596442630  3.60 3.8574025053  3.80 4.0552383287  4.00 4.2531272886  4.20 4.4510524003  4.40 4.6490019928  4.60 4.8469681088  4.80 5.0449453529  5.00 5.2429300797 /
\color{black}
\put {$f_L$} at -1 3
\endpicture
\end{subfigure}
\begin{subfigure}{0.33\textwidth}
	\caption{$\LA c_1 \RA$ and $\LA c_2 \RA$}
\beginpicture
\normalcolor
\color{black}
\setcoordinatesystem units <10.5pt,90pt>
\axescenter{-5}{0}{0}{0}{5}{1}{0}
\put {\Large$\alpha$} at 3.5 -0.17
\setplotsymbol ({\scalebox{0.3}{$\bullet$}})
\color{Red}
\plot -5.00 0.0100004255  -4.80 0.0100006349  -4.60 0.0100009473  -4.40 0.0100014135  -4.20 0.0100021095  -4.00 0.0100031486  -3.80 0.0100047008  -3.60 0.0100070208  -3.40 0.0100104920  -3.20 0.0100156926  -3.00 0.0100235012  -2.80 0.0100352631  -2.60 0.0100530658  -2.40 0.0100802101  -2.20 0.0101220642  -2.00 0.0101877299  -1.80 0.0102936278  -1.60 0.0104723109  -1.40 0.0107988200  -1.20 0.0115004490  -1.00 0.0139895285  -0.80 0.2696643746  -0.60 0.4819910727  -0.40 0.6102001807  -0.20 0.6998028771  0.00 0.7660462006  0.20 0.8163559561  0.40 0.8551522221  0.60 0.8853330295  0.80 0.9089081892  1.00 0.9273559218  1.20 0.9417918742  1.40 0.9530729370  1.60 0.9618726096  1.80 0.9687269978  2.00 0.9740580691  2.20 0.9781918040  2.40 0.9813751811  2.60 0.9837958906  2.80 0.9856039385  3.00 0.9869269450  3.20 0.9878758098  3.40 0.9885445465  3.60 0.9890092793  3.80 0.9893288136  4.00 0.9895468098  4.20 0.9896947133  4.40 0.9897946750  4.60 0.9898620558  4.80 0.9899073927  5.00 0.9899378601  /
\put {$\mathbf{\LA c_1 \RA}$} at 3 0.85
\color{Blue}
\plot -5.00 0.0152278118  -4.80 0.0152278095  -4.60 0.0152278062  -4.40 0.0152278012  -4.20 0.0152277938  -4.00 0.0152277827  -3.80 0.0152277661  -3.60 0.0152277413  -3.40 0.0152277043  -3.20 0.0152276488  -3.00 0.0152275654  -2.80 0.0152274398  -2.60 0.0152272497  -2.40 0.0152269598  -2.20 0.0152265129  -2.00 0.0152258116  -1.80 0.0152246803  -1.60 0.0152227704  -1.40 0.0152192757  -1.20 0.0152117416  -1.00 0.0151847363  -0.80 0.0126441849  -0.60 0.0113566305  -0.40 0.0108227469  -0.20 0.0105358339  0.00 0.0103600838  0.20 0.0102455185  0.40 0.0101682641  0.60 0.0101151390  0.80 0.0100784696  1.00 0.0100531687  1.20 0.0100358005  1.40 0.0100239839  1.60 0.0100160091  1.80 0.0100106555  2.00 0.0100070752  2.20 0.0100046901  2.40 0.0100031073  2.60 0.0100020598  2.80 0.0100013672  3.00 0.0100009091  3.20 0.0100006056  3.40 0.0100004041  3.60 0.0100002700  3.80 0.0100001805  4.00 0.0100001208  4.20 0.0100000809  4.40 0.0100000542  4.60 0.0100000363  4.80 0.0100000243  5.00 0.0100000163   /
\put {$\mathbf{\LA c_2 \RA}$} at 3 0.15
\endpicture
\end{subfigure}
\begin{subfigure}{0.33\textwidth}
	\caption{$L^2\Var{c_1}$ and $L^2\Var{c_2}$}
\beginpicture
\normalcolor
\color{black}
\setcoordinatesystem units <10.5pt,4.5pt>
\setplotarea x from -5 to 5, y from -18 to 2
\axis left shiftedto x=0
        ticks withvalues $-16$ $2$ /
        at -16 2 /
/
\axis bottom 
        ticks withvalues $-5$ $0$ $5$ /
        at -5 0 5 /
/
%\setdots <2pt>
%\plot 3.536 -7 3.536 2 /
\setsolid
\put {\Large$\alpha$} at 3.5 -22
\setplotsymbol ({\scalebox{0.3}{$\bullet$}})
\color{Red}
\plot -5.00 -13.9765534319  -4.80 -13.5763444398  -4.60 -13.1760326200  -4.40 -12.7755673501  -4.20 -12.3748730497  -4.00 -11.9738368312  -3.80 -11.5722899852  -3.60 -11.1699801544  -3.40 -10.7665293603  -3.20 -10.3613703457  -3.00 -9.9536492424  -2.80 -9.5420749383  -2.60 -9.1246817075  -2.40 -8.6984449589  -2.20 -8.2586338220  -2.00 -7.7976536557  -1.80 -7.3027868338  -1.60 -6.7511703282  -1.40 -6.0961840649  -1.20 -5.2160664776  -1.00 -3.5030365659  -0.80 0.4542508290  -0.60 -0.2409377988  -0.40 -0.6417735586  -0.20 -0.9637999296  0.00 -1.2488810623  0.20 -1.5147856997  0.40 -1.7692128043  0.60 -2.0175319788  0.80 -2.2634297233  1.00 -2.5083426840  1.20 -2.7541210155  1.40 -3.0016916933  1.60 -3.2509077666  1.80 -3.5014075424  2.00 -3.7539027745  2.20 -4.0109943646  2.40 -4.2778201470  2.60 -4.5603961435  2.80 -4.8628400137  3.00 -5.1860503046  3.20 -5.5282223604  3.40 -5.8861293582  3.60 -6.2562471179  3.80 -6.6354096540  4.00 -7.0210603345  4.20 -7.4112641779  4.40 -7.8046159472  4.60 -8.2001222181  4.80 -8.5970930465  5.00 -8.9950548676 /
\put {\scalebox{0.8}{$L^2\mathbf{\Var{c_1}}$}} at 5.3 -3.8
\color{Blue}
\plot -5.00 -2.9343961031  -4.80 -2.9343971241  -4.60 -2.9343986477  -4.40 -2.9344009215  -4.20 -2.9344043156  -4.00 -2.9344093832  -3.80 -2.9344169530  -3.60 -2.9344282674  -3.40 -2.9344451947  -3.20 -2.9344705553  -3.00 -2.9345086313  -2.80 -2.9345659801  -2.60 -2.9346527723  -2.40 -2.9347850835  -2.20 -2.9349890438  -2.00 -2.9353089250  -1.80 -2.9358245237  -1.60 -2.9366938680  -1.40 -2.9382807622  -1.20 -2.9416841532  -1.00 -2.9536034770  -0.80 -4.3014758087  -0.60 -5.3948791267  -0.40 -6.0942913564  -0.20 -6.6381365882  0.00 -7.1071832045  0.20 -7.5363514480  0.40 -7.9442242665  0.60 -8.3428674622  0.80 -8.7385520027  1.00 -9.1354978700  1.20 -9.5358259650  1.40 -9.9394425360  1.60 -10.3456069317  1.80 -10.7539316563  2.00 -11.1642129420  2.20 -11.5758626557  2.40 -11.9879028685  2.60 -12.3992985677  2.80 -12.8092656875  3.00 -13.2174170258  3.20 -13.6237314804  3.40 -14.0284228878  3.60 -14.4318021873  3.80 -14.8341827262  4.00 -15.2358335850  4.20 -15.6369660349  4.40 -16.0377370910  4.60 -16.4382594213  4.80 -16.8386120428  5.00 -17.2388495421  /
\put {\scalebox{0.8}{$L^2\mathbf{\Var{c_2}}$}} at 5.3 -12.5
\normalcolor
\color{black}
\endpicture
\end{subfigure}
\caption{
\underbar{Top panels}: 
The free energy, mean concentrations and variances of the unlinked model
as a function of $\alpha=\beta$ along the line from the point $(-5,-5)$ to 
$(5,5)$ through the multicritical point in the phase diagram in figure \ref{f9}. 
The free energy (top left) shows a transition at approximately 
$\alpha_c = \beta_c \approx -\log \mu_2 \approx -0.97$ (as also seen in the 
graphs of the mean concentrations $\LA c_i \RA$).  These critical curves
show a transition as the multicritical point is crossed into the dense phase
where the polygons fill the confining square.  The mean concentrations 
show a singular point in the free energy at the critical point, and the 
variances (plotted on a logarithmic scale) show a sharp increase at 
this point.\newline
\underbar{Bottom panels}:
The free energy, mean concentrations and variances as a function of
$\alpha$ on the line thougth the multicritical point and with
endpoints $(-5,\beta_c)$ and $(5,\beta_c)$ (where 
$\beta_c=-\log \mu_2 \approx -0.97$).  These
data shown a continuous transition, similar to the transition seen
in figure \ref{f8} between the empty phase, and the $c_1$-dominated
phase.}
\label{13}   %ZXZ[13]
\end{figure}

{\bf Scaling around the multicritical point:}
Scaling axes throught the multicritical point are set up consistent with
figure \ref{7}.  The parallel scaling axis (parallel to the $\tau$-phase boundary) 
runs along the $\tau$ phase boundary on the main diagonal from $(-5,-5)$ 
to $(5,5)$.  The transverse axis (transverse to the $\tau$-line) runs along 
the $\lambda_2$ phase boundary and crosses the multicritical point into 
the $c_1$-dominated phase.  Similarly, a second transverse axis runs along 
the $\lambda_1$ phase boundary into the $c_2$-dominated phase.

Estimates of the free energy $f_L$ and its derivatives are plotted for
$L=20$ on the axes throught the multicritical point in figure \ref{13}.
The three top panels show results along the parallel scaling axis,
while the bottom panels display results along the transverse scaling
axis running along $\lambda_2$ and crossing into the $c_1$-dominated
phase.  

Since the phase diagram is symmetric about the main diagonal in the
$\alpha\beta$-plane, and since the free energy is not a function of
$\beta$ in the $c_1$-dominated phase, and of $\alpha$ in the
$c_2$-dominated phase, it follows that the scaling of $\LA c_1 \RA$
and $\LA c_2 \RA$ along these axes is the same as given in equation
\Ref{20} with $\alpha^\prime_s$ and $\alpha_s$ having the values
in equation \Ref{23}.  This follows by symmetry, since 
$f_L(\alpha,\alpha_c) = f_L(\alpha,\beta)$ for any $\beta<\alpha_c$, 
and $f_L(\alpha,\alpha) = f_L(\alpha,\alpha_c)$ in the dense phases
(that is, for $\alpha>\alpha_c$).  Thus, it follows that the exponents 
$\alpha_u$ and $\alpha_t$ in equations \Ref{9} and \Ref{10} have 
the same values as in equation \Ref{23}, namely
\begin{equation}
\alpha_u = \alpha_t \approx 0.45\pm 0.03.
\label{24}  %ZXZ[24]
\end{equation}
By equation \Ref{12}, the crossover exponent in this model is $\phi=1$.  
Taking into account the uncertainty of the estimate in equation \Ref{24}, 
the estimate of the crossover exponent of the unlinked model is given by 
\begin{equation}
\phi = 1.00 \pm 0.04.
\label{25}  %ZXZ[25]
\end{equation}

\begin{figure}[h!]
\begin{subfigure}{0.33\textwidth}
	\caption{Free Energy $g_L(\alpha,\beta)$}
	\includegraphics[width=1.0\textwidth]{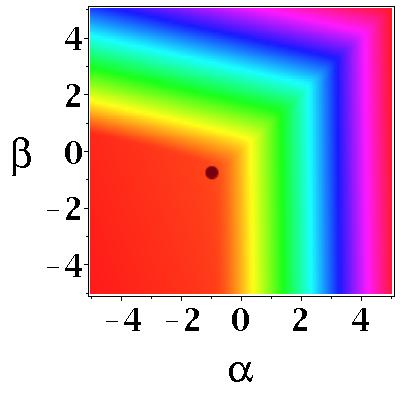}
\end{subfigure}
\begin{subfigure}{0.33\textwidth}
	\caption{$\LA c_1 \RA$}
	\includegraphics[width=1.0\textwidth]{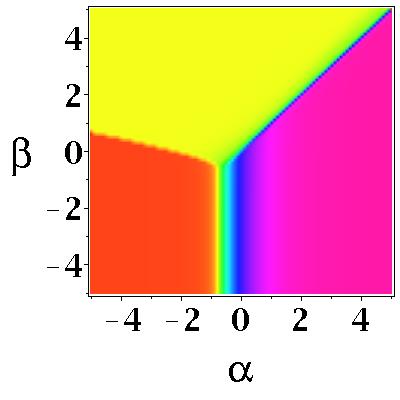}
\end{subfigure}
\begin{subfigure}{0.33\textwidth}
	\caption{$\LA c_2 \RA$}
	\includegraphics[width=1.0\textwidth]{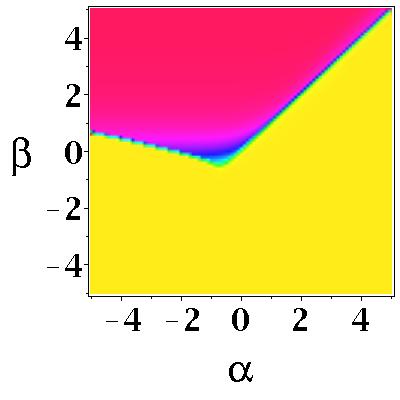}
\end{subfigure}
\begin{subfigure}{0.33\textwidth}
	\caption{$L^2\,\hbox{Cov}(c_1,c_2)$}
	\includegraphics[width=1.0\textwidth]{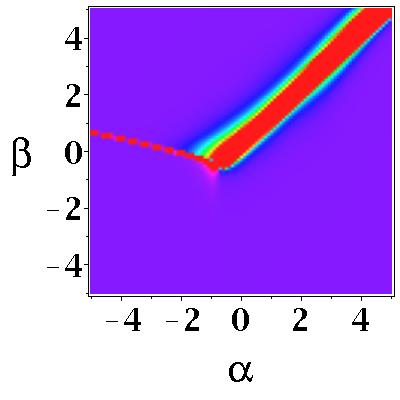}
\end{subfigure}
\begin{subfigure}{0.33\textwidth}
	\caption{$L^2\,\hbox{Var}(c_1)$}
	\includegraphics[width=1.0\textwidth]{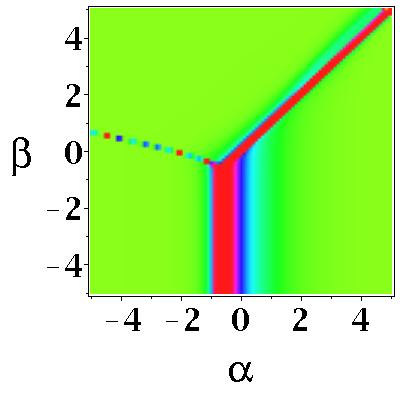}
\end{subfigure}
\begin{subfigure}{0.33\textwidth}
	\caption{$L^2\,\hbox{Var}(c_2)$}
	\includegraphics[width=1.0\textwidth]{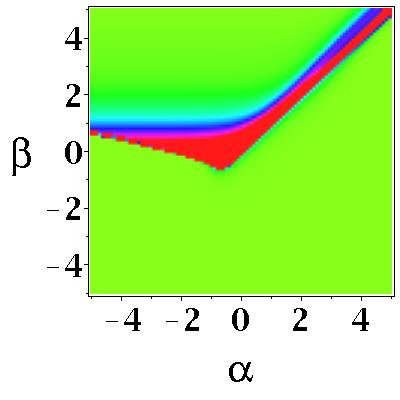}
\end{subfigure}
\caption{The free energy $g_L (\alpha,\beta)$ of the linked model
for $L=20$ (top left), and its first and second derivatives.  Proceeding 
clockwise from the top middle,  the mean concentrations of the outer 
polygon $\LA c_1 \RA$ and the inner polygon $\LA c_2 \RA$, their variances
$\hbox{Var}(c_2)$ and $\hbox{Var}(c_1)$, and the covariance
of $c_1$ and $c_2$, $\hbox{Cov}(c_1,c_2)$.  The concentrations $c_1$ 
and $c_2$ are negatively correlated, and the covariance shows a sharp 
transition close to the main diagonal for $\alpha=\beta > \alpha_c$. In 
addition, a second, apparently weaker, phase change appears
along a curve starting in the  multicritical point and runs along along 
a curve into the second quadrant, and a third transition runs vertically
into the multicritical point from below.  The approximate location of the 
multicritical point in the model is denoted by a bullet in (a).}
\label{14}   %ZXZ[14]
\end{figure}

\subsection{The phase diagram of two linked confined square lattice polygons}

In figure \ref{14}(a) is a density plot of $g_L(\alpha,\beta)$ 
(for $L=20$) with the left bottom colour low, and the right and top colours high.  
The free energy shows clear signs of phase transitions, and this is also seen 
in plots of the first and second derivaties of $g_L(\alpha,\beta)$ shown 
in figures \ref{14}(b)-(f).

The concentrations $\LA c_1\RA$ and $\LA c_2\RA$ (see equation \Ref{13})
are plotted in figures \ref{14}(b) and (c).  These plots show a rapid change in 
the mean concentration of one, or both, $\LA c_1\RA$ and $\LA c_2\RA$ when 
phase boundaries are crossed.  Note that these plots are different from those in 
figure \ref{8},  suggesting that the different topology of this model (namely that 
the two polygons are linked in $\RealN^2$) has a relevant impact on its 
thermodynamic properties.

The variances (equation \Ref{14}) and covariance (equation \Ref{15})
are plotted in figure \ref{14}(d), (e) and (f).    The covariance shows 
(as expected) strong negative correlations between $c_1$ and $c_2$ induced by
the confining square, except along two phase boundaries, where the correlation
is positive.  $\Var{c_1}$ in figure \ref{14}(e), considered with figure 
\ref{14}(b), shows a phase boundary $\tau_1$ close to or on the main 
diagonal for $\alpha>\alpha_c$ and $\beta>\beta_c$ 
where $(\alpha_c,\beta_c)$ is the location of a multicritical point.  
 
\begin{figure}[h!]
\centering
\begin{subfigure}{0.67\textwidth}
	\includegraphics[width=1.0\textwidth]{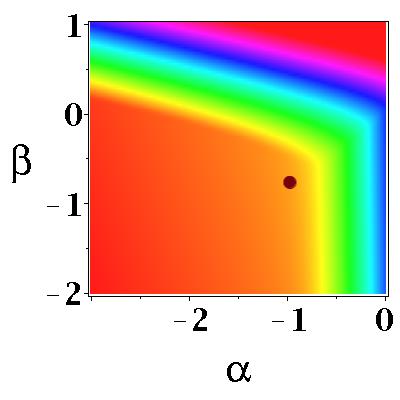}
\end{subfigure}
\caption{A density plot the free energy $g_L(\alpha,\beta)$ of 
the linked model in the vicinity of the multicritical point.  The 
location of the multicritical point is estimated to be
$(\alpha,\beta)=(-0.97,-0.77)$ and is denoted by the bullet.}
\label{15}   %ZXZ[15]
\end{figure}

Magnifying figure \ref{14}(a) in the vicinity of the multicritical point
gives the density plot in figure \ref{15}.  This plot suggest that the
a \textit{curved} phase boundary $\tau_2$ separates the empty phase 
from a $c_2$-dominated phase for large positive $\beta$ (in this phase
the inner polygon is dense, and the outer polygon is expanded to near
the perimeter of the confining square).

The rest of the phase diagram, as suggested by figures \ref{14} and \ref{15}, is
shown in figure \ref{16}.  A vertical phase boundary $\alpha=\alpha_c$ runs  
from the multicritical point for $\beta<\beta_c$.  This phase boundary, denoted
by $\lambda$, separates an empty phase, for large negative $\alpha$
and $\beta$, from a $c_1$-dominated phase for large positive $\alpha$.
This transition is similar to the transition across the $\lambda_1$ phase 
boundary in the unlinked model, and the critical value of $\alpha$ is given 
by $\alpha_c=-\log \mu_2$ (using an argument similar to the one used 
for $\lambda_1$ in the unlinked model).  Crossing $\lambda$ takes the model from 
its empty phase into a phase where the outer polygon is dense, while 
the inner polygon remains small.

\begin{figure}[h!]
\beginpicture
\setcoordinatesystem units <0.75pt,0.75pt> point at -320 0 
\setplotarea x from -150 to 150, y from -150 to 165

\color{Tan}
\setplotsymbol ({$\bullet$})
\plot -150 0 150 0 /
\plot  0 -150 0 150 /

\setplotsymbol (.)
\color{black}
\arrow <10pt> [.2,.67] from -100 -130 to -25 -130
\arrow <10pt> [.2,.67] from 75 -130 to -25 -130

\put {\large$\alpha_t^\prime = 2$} at -100 -145
\put {\large$\alpha_t = -0.53(4)$} at 75 -145

\color{Black}
\setplotsymbol ({\scalebox{0.25}{$\bullet$}})
\circulararc 135 degrees from -60 10 center at -25 -25
\circulararc 135 degrees from -25 -80 center at -25 -25

\color{Red}
\setplotsymbol ({$\bullet$})
\plot -25 -25 150 150  /
\setquadratic
\color{Red}
\plot -150 35 -65 10 -25 -25 /
\setlinear
\color{Blue}
\plot -25 -150 -25 -25 /
\color{Sepia}
\put {\scalebox{2.5}{$\bullet$}} at -25 -25

\color{black}
\put {\LARGE$\alpha$} at 145 -15
\put {\LARGE$\beta$} at -20 150
\put {\textit{\Large empty phase}} at -100 -90
\put {\textit{\Large $c_1$-dominated}} at 100 -80
\put {\textit{\Large $c_2$-dominated}} at -85 90

\color{Red}
\put {\hbox{\LARGE$\tau_1$}} at 150 120
\put {\hbox{\LARGE$\tau_2$}} at -100 40
\color{Blue}
\put {\hbox{\LARGE$\lambda$}} at -40 -150

\color{black}
\put {\large $\phi_1=1.3(2)$} at 75 -30
\put {\large $\phi_2=1$} at -100 -50

\normalcolor
\setlinear

\put {\textbf{Linked model phase diagram}} at 0 -180 
\endpicture
\caption{Crossover exponents between the $\tau_1$ and $\lambda$,
and $\tau_2$ and $\lambda$ critical curves around the multicritical point 
in the phase diagram of the linked model.}
\label{16}  %ZXZ[16]
\end{figure}

Since the multicritical point $(\alpha_c,\beta_c)$ is the intersection of 
the three phase boundaries $\lambda$, $\tau_1$ and $\tau_2$, with
the $\lambda$ phase boundary running vertically and separating the
empty phase from the $c_1$-dominated phase, we conclude that
$\alpha_c = -\log \mu_2 \approx -0.97$, but, as will be seen below,
it appears from our numerical data that $\beta_c > -0.97$, although
the resolution of our data was not good enough to confirm this.

To determine the trajectories of $\tau_1$ and $\tau_2$, we estimated
from our data sets of points along them.  In the case of $\tau_1$ a
linear least squares fit using a quadratic model shows that
\begin{eqnarray}
\tau_1(\alpha) &\approx& 0.0824 + 0.9824\,\alpha + 0.00268\,\alpha^2 ,
\nonumber \\
& & \hspace{3cm}\hbox{for $\alpha>-0.97$}.
\label{26} %ZXZ[26]
\end{eqnarray}
Since along the critical curve $\lambda$ the value of $\alpha=\alpha_c 
\approx - 0.97$, it follows that this predicts that the multicritical point appears 
to be close to $(-0.97,-0.868)$.

In the case of $\tau_2$ a $3$-parameter non-linear fit using 
the \textit{NonlinearFit} function in the Statistics package of 
Maple17 \cite{Maple17} was used along points on the phase boundary.
Assuming a powerlaw shape gives the trajectory
\begin{eqnarray}
\tau_2(\alpha) &\approx& -0.6680 + 0.5184 \LV -0.97 - \alpha \RV^{0.64},
\nonumber \\ 
& & \hspace{3cm}\hbox{for $\alpha<-0.97$}.
 \label{27}   %ZXZ[27]
\end{eqnarray}
This predicts that the multicritical point is close to $(-0.97,-0.668)$.  Taking
the average of this estimate, and the estimate obtained from the trajectory
of the $\tau_1$ phase boundary, gives our best estimate of the location of
the multicritical point for the linked case: 
\begin{equation}
(\alpha_c,\beta_c) \approx (-0.97,-0.77) ,
\label{28}   %ZXZ[28]
\end{equation}
assuming, of course, that finite size corrections at $L=20$ is negligible.
Note that the estimate of $\beta_c$ is larger than the one reported for the 
unlinked model (namely, $\beta_c \approx -0.97$).  This estimate may have 
a large confidence interval,  since numerical estimates in the vicinity of the 
multicritical point are strongly affected by the environment (namely, it is the 
intersection of three phase boundaries) and the resolution of our simulations 
is not good enough to determine a more accurate location.

\begin{figure}[h!]
\begin{subfigure}{0.33\textwidth}
\normalcolor
\color{black}
	\caption{Free Energy $f_L(\alpha,\beta)$}
\beginpicture
\setcoordinatesystem units <21pt,15pt>
\axes{0}{0}{0}{0}{5}{6}{0}
\put {\Large$\alpha$} at 4.0 -1
\setplotsymbol ({\scalebox{0.3}{$\bullet$}})
\color{Red}
\plot 0.000 4.2651  0.098 4.2044  0.196 4.1436  0.294 4.0828  0.392 4.0221  0.490 3.9613  0.588 3.9006  0.686 3.8399  0.784 3.7791  0.882 3.7184  0.980 3.6577  1.078 3.5971  1.176 3.5364  1.274 3.4758  1.373 3.4153  1.471 3.3548  1.569 3.2944  1.667 3.2342  1.765 3.1741  1.863 3.1143  1.961 3.0549  2.059 2.9960  2.157 2.9378  2.255 2.8809  2.353 2.8269  2.451 2.7845  2.549 2.8438  2.647 2.9382  2.745 3.0332  2.843 3.1285  2.941 3.2239  3.039 3.3194  3.137 3.4151  3.235 3.5108  3.333 3.6066  3.431 3.7024  3.529 3.7982  3.627 3.8942  3.725 3.9901  3.823 4.0861  3.921 4.1820  4.020 4.2780  4.118 4.3741  4.216 4.4701  4.314 4.5661  4.412 4.6622  4.510 4.7582  4.608 4.8543  4.706 4.9503  4.804 5.0464  4.902 5.1424  5.000 5.2385 /
\color{black}
\put {$g_L$} at -0.5 3
\endpicture
\end{subfigure}
\begin{subfigure}{0.33\textwidth}
\normalcolor
\color{black}
	\caption{$\LA c_1 \RA$ and $\LA c_2 \RA$}
\beginpicture
\setcoordinatesystem units <21pt,90pt>
\axes{0}{0}{0}{0}{5}{1}{0}
\put {\Large$\alpha$} at 4.0 -0.17
\setplotsymbol ({\scalebox{0.3}{$\bullet$}})
\color{Red}
\plot 0.000 0.1900  0.278 0.1900  0.534 0.1900  0.770 0.1900  0.986 0.1900  1.183 0.1900  1.362 0.1900  1.525 0.1900  1.671 0.1900  1.801 0.1901  1.918 0.1901  2.020 0.1902  2.110 0.1902  2.188 0.1904  2.254 0.1905  2.311 0.1908  2.358 0.1912  2.397 0.1919  2.427 0.1928  2.451 0.1942  2.469 0.1962  2.482 0.1991  2.491 0.2035  2.496 0.2115  2.499 0.2371  2.500 0.3613  2.500 0.9744  2.500 0.9803  2.501 0.9824  2.504 0.9837  2.509 0.9848  2.518 0.9856  2.530 0.9863  2.548 0.9869  2.573 0.9874  2.603 0.9878  2.642 0.9882  2.689 0.9885  2.745 0.9888  2.812 0.9890  2.890 0.9891  2.980 0.9893  3.082 0.9894  3.198 0.9895  3.329 0.9896  3.475 0.9897  3.638 0.9897  3.817 0.9898  4.014 0.9898  4.230 0.9898  4.466 0.9899  4.722 0.9899  5.000 0.9899
   /
\put {$\mathbf{\LA c_1 \RA}$} at -0.7 0.22
\color{Blue}
\plot 0.000 0.8099  0.278 0.8098  0.534 0.8096  0.770 0.8094  0.986 0.8090  1.183 0.8086  1.362 0.8079  1.525 0.8069  1.671 0.8055  1.801 0.8036  1.918 0.8011  2.020 0.7980  2.110 0.7942  2.188 0.7895  2.254 0.7830  2.311 0.7721  2.358 0.7537  2.397 0.7258  2.427 0.6867  2.451 0.6295  2.469 0.5232  2.482 0.2365  2.491 0.0787  2.496 0.0500  2.499 0.0426  2.500 0.0406  2.500 0.0404  2.500 0.0402  2.501 0.0384  2.504 0.0346  2.509 0.0296  2.518 0.0248  2.530 0.0208  2.548 0.0178  2.573 0.0156  2.603 0.0141  2.642 0.0131  2.689 0.0124  2.745 0.0118  2.812 0.0114  2.890 0.0111  2.980 0.0108  3.082 0.0105  3.198 0.0104  3.329 0.0102  3.475 0.0101  3.638 0.0101  3.817 0.0100  4.014 0.0100  4.230 0.0100  4.466 0.0100  4.722 0.0100  5.000 0.0100 
   /
\put {$\mathbf{\LA c_2 \RA}$} at -0.7 0.76
\endpicture
\end{subfigure}
\begin{subfigure}{0.33\textwidth}
	\caption{$L^2\Var{c_1}$ and $L^2\Var{c_2}$}
\beginpicture
\normalcolor
\color{black}
\setcoordinatesystem units <21pt,4.506pt>
\setplotarea x from 0 to 5, y from -15 to 5
\axis left shiftedto x=0
        ticks withvalues $-10$  $5$ /
        at -10  5 /
/
\axis bottom 
        ticks withvalues $0$ $2.5$ $5$ /
        at 0 2.5 5 /
/
\put {\Large$\alpha$} at 4 -19.0
\setplotsymbol ({\scalebox{0.3}{$\bullet$}})
\color{Red}
\plot 0.000 -12.3870  0.278 -11.2717  0.534 -10.2389  0.770 -9.2845  0.986 -8.4040  1.183 -7.5939  1.362 -6.8511  1.525 -6.1746  1.671 -5.5663  1.801 -5.0301  1.918 -4.5694  2.020 -4.1767  2.110 -3.8158  2.188 -3.3911  2.254 -2.7781  2.311 -2.0295  2.358 -1.3381  2.397 -0.7407  2.427 -0.1796  2.451 0.5463  2.469 1.6866  2.482 2.6701  2.491 1.4392  2.496 0.5295  2.499 0.1275  2.500 -0.0046  2.500 -0.0189  2.500 -0.0331  2.501 -0.1571  2.504 -0.4685  2.509 -0.9572  2.518 -1.5236  2.530 -2.0828  2.548 -2.6141  2.573 -3.1071  2.603 -3.5374  2.642 -3.8873  2.689 -4.1625  2.745 -4.3821  2.812 -4.5654  2.890 -4.7291  2.980 -4.8873  3.082 -5.0512  3.198 -5.2303  3.329 -5.4326  3.475 -5.6653  3.638 -5.9346  3.817 -6.2453  4.014 -6.6009  4.230 -7.0032  4.466 -7.4531  4.722 -7.9508  5.000 -8.4964 
  /
\put {\scalebox{0.8}{$L^2\mathbf{\Var{c_1}}$}} at 4.2 -3
\color{Blue}
\plot 0.000 -8.2578  0.278 -7.7167  0.534 -7.2233  0.770 -6.7740  0.986 -6.3619  1.183 -5.9757  1.362 -5.6026  1.525 -5.2336  1.671 -4.8683  1.801 -4.5164  1.918 -4.1901  2.020 -3.8919  2.110 -3.5944  2.188 -3.2158  2.254 -2.6523  2.311 -1.9560  2.358 -1.3006  2.397 -0.7203  2.427 -0.1663  2.451 0.5547  2.469 1.6888  2.482 2.6728  2.491 1.4406  2.496 0.5273  2.499 0.1223  2.500 -0.0111  2.500 -0.0256  2.500 -0.0399  2.501 -0.1653  2.504 -0.4812  2.509 -0.9807  2.518 -1.5705  2.530 -2.1760  2.548 -2.7907  2.573 -3.4180  2.603 -4.0352  2.642 -4.6050  2.689 -5.1033  2.745 -5.5277  2.812 -5.8915  2.890 -6.2199  2.980 -6.5438  3.082 -6.8915  3.198 -7.2847  3.329 -7.7381  3.475 -8.2612  3.638 -8.8601  3.817 -9.5383  4.014 -10.2982  4.230 -11.1416  4.466 -12.0701  4.722 -13.0855  5.000 -14.1898
   /
\put {\scalebox{0.8}{$L^2\mathbf{\Var{c_2}}$}} at 3.0 -12
\normalcolor
\color{black}
\endpicture
\end{subfigure}

\vspace{5mm}
\begin{subfigure}{0.33\textwidth}
	\caption{Free Energy $f_L(\alpha,\beta)$}
\beginpicture
\normalcolor
\color{black}
\setcoordinatesystem units <10.5pt,22.5pt>
\axescenter{-5}{0}{0}{0}{5}{4}{0}
\put {\Large$\beta$} at 3.5 -0.667
\setplotsymbol ({\scalebox{0.3}{$\bullet$}})
\color{Red}
\plot -5.000 -0.1089  -4.804 -0.1069  -4.608 -0.1050  -4.412 -0.1030  -4.216 -0.1010  -4.020 -0.0991  -3.824 -0.0971  -3.627 -0.0952  -3.431 -0.0932  -3.235 -0.0912  -3.039 -0.0893  -2.843 -0.0873  -2.647 -0.0854  -2.451 -0.0834  -2.255 -0.0814  -2.059 -0.0795  -1.863 -0.0775  -1.667 -0.0756  -1.471 -0.0736  -1.275 -0.0716  -1.078 -0.0697  -0.882 -0.0677  -0.686 -0.0657  -0.490 -0.0637  -0.294 -0.0618  -0.098 -0.0586  0.098 0.0540  0.294 0.1797  0.490 0.3135  0.686 0.4531  0.882 0.5971  1.078 0.7445  1.275 0.8946  1.471 1.0468  1.667 1.2006  1.863 1.3558  2.059 1.5119  2.255 1.6687  2.451 1.8259  2.647 1.9834  2.843 2.1412  3.039 2.2992  3.235 2.4574  3.431 2.6158  3.627 2.7742  3.824 2.9328  4.020 3.0914  4.216 3.2501  4.412 3.4088  4.608 3.5676  4.804 3.7264  5.000 3.8851
  /
\color{black}
\put {$g_L$} at -1 2
\endpicture
\end{subfigure}
\begin{subfigure}{0.33\textwidth}
	\caption{$\LA c_1 \RA$ and $\LA c_2 \RA$}
\beginpicture
\normalcolor
\color{black}
\setcoordinatesystem units <10.5pt,90pt>
\axescenter{-5}{0}{0}{0}{5}{1}{0}
\put {\Large$\beta$} at 3.5 -0.17
\setplotsymbol ({\scalebox{0.3}{$\bullet$}})
\color{Red}
\plot -5.000 0.0323  -4.804 0.0323  -4.608 0.0323  -4.412 0.0323  -4.216 0.0323  -4.020 0.0323  -3.824 0.0323  -3.627 0.0323  -3.431 0.0323  -3.235 0.0323  -3.039 0.0323  -2.843 0.0323  -2.647 0.0323  -2.451 0.0323  -2.255 0.0323  -2.059 0.0323  -1.863 0.0323  -1.667 0.0323  -1.471 0.0323  -1.275 0.0323  -1.078 0.0323  -0.882 0.0324  -0.686 0.0324  -0.490 0.0324  -0.294 0.0325  -0.098 0.0911  0.098 0.1915  0.294 0.1909  0.490 0.1906  0.686 0.1904  0.882 0.1902  1.078 0.1901  1.275 0.1901  1.471 0.1901  1.667 0.1900  1.863 0.1900  2.059 0.1900  2.255 0.1900  2.451 0.1900  2.647 0.1900  2.843 0.1900  3.039 0.1900  3.235 0.1900  3.431 0.1900  3.627 0.1900  3.824 0.1900  4.020 0.1900  4.216 0.1900  4.412 0.1900  4.608 0.1900  4.804 0.1900  5.000 0.1900
  /
\put {$\mathbf{\LA c_1 \RA}$} at 3 0.1
\color{Blue}
\plot -5.000 0.0100  -4.804 0.0100  -4.608 0.0100  -4.412 0.0100  -4.216 0.0100  -4.020 0.0100  -3.824 0.0100  -3.627 0.0100  -3.431 0.0100  -3.235 0.0100  -3.039 0.0100  -2.843 0.0100  -2.647 0.0100  -2.451 0.0100  -2.255 0.0100  -2.059 0.0100  -1.863 0.0100  -1.667 0.0100  -1.471 0.0100  -1.275 0.0100  -1.078 0.0100  -0.882 0.0100  -0.686 0.0101  -0.490 0.0101  -0.294 0.0101  -0.098 0.2086  0.098 0.6168  0.294 0.6638  0.490 0.6986  0.686 0.7242  0.882 0.7438  1.078 0.7590  1.275 0.7712  1.471 0.7808  1.667 0.7884  1.863 0.7940  2.059 0.7980  2.255 0.8008  2.451 0.8027  2.647 0.8042  2.843 0.8054  3.039 0.8063  3.235 0.8072  3.431 0.8078  3.627 0.8084  3.824 0.8088  4.020 0.8092  4.216 0.8094  4.412 0.8096  4.608 0.8097  4.804 0.8098  5.000 0.8099
 /
\put {$\mathbf{\LA c_2 \RA}$} at 3 0.9
\endpicture
\end{subfigure}
\begin{subfigure}{0.33\textwidth}
	\caption{$L^2\Var{c_1}$ and $L^2\Var{c_2}$}
\beginpicture
\normalcolor
\color{black}
\setcoordinatesystem units <10.5pt,3.6pt>
\setplotarea x from -5 to 5, y from -20 to 5
\axis left shiftedto x=0
        ticks withvalues $-15$ $5$ /
        at -15 5 /
/
\axis bottom 
        ticks withvalues $-5$ $0$ $5$ /
        at -5 0 5 /
/
\setsolid
\put {\Large$\beta$} at 3.5 -24.0
\setplotsymbol ({\scalebox{0.3}{$\bullet$}})
\color{Red}
\plot -5.000 -5.2963  -4.804 -5.2963  -4.608 -5.2963  -4.412 -5.2963  -4.216 -5.2963  -4.020 -5.2963  -3.824 -5.2962  -3.627 -5.2962  -3.431 -5.2962  -3.235 -5.2961  -3.039 -5.2960  -2.843 -5.2959  -2.647 -5.2957  -2.451 -5.2954  -2.255 -5.2950  -2.059 -5.2944  -1.863 -5.2935  -1.667 -5.2921  -1.471 -5.2900  -1.275 -5.2870  -1.078 -5.2823  -0.882 -5.2753  -0.686 -5.2646  -0.490 -5.2478  -0.294 -5.2207  -0.098 0.8644  0.098 -5.7927  0.294 -6.2653  0.490 -6.7455  0.686 -7.2230  0.882 -7.6937  1.078 -8.1590  1.275 -8.6184  1.471 -9.0668  1.667 -9.4971  1.863 -9.9086  2.059 -10.3082  2.255 -10.7043  2.451 -11.1025  2.647 -11.5055  2.843 -11.9140  3.039 -12.3270  3.235 -12.7426  3.431 -13.1586  3.627 -13.5730  3.824 -13.9842  4.020 -14.3918  4.216 -14.7958  4.412 -15.1968  4.608 -15.5953  4.804 -15.9919  5.000 -16.3872
  /
\put {\scalebox{0.8}{$L^2\mathbf{\Var{c_1}}$}} at -3.5 -3
\color{Blue}
\plot -5.000 -17.8011  -4.804 -17.4089  -4.608 -17.0167  -4.412 -16.6246  -4.216 -16.2324  -4.020 -15.8402  -3.824 -15.4480  -3.627 -15.0558  -3.431 -14.6635  -3.235 -14.2711  -3.039 -13.8787  -2.843 -13.4862  -2.647 -13.0935  -2.451 -12.7005  -2.255 -12.3070  -2.059 -11.9130  -1.863 -11.5181  -1.667 -11.1219  -1.471 -10.7237  -1.275 -10.3225  -1.078 -9.9169  -0.882 -9.5047  -0.686 -9.0823  -0.490 -8.6441  -0.294 -8.1799  -0.098 3.3095  0.098 -1.3135  0.294 -1.5728  0.490 -1.8922  0.686 -2.1758  0.882 -2.4367  1.078 -2.6738  1.275 -2.8965  1.471 -3.1279  1.667 -3.3998  1.863 -3.7258  2.059 -4.0845  2.255 -4.4389  2.451 -4.7563  2.647 -5.0184  2.843 -5.2263  3.039 -5.4002  3.235 -5.5689  3.431 -5.7565  3.627 -5.9766  3.824 -6.2328  4.020 -6.5225  4.216 -6.8398  4.412 -7.1787  4.608 -7.5334  4.804 -7.8997  5.000 -8.2741
  /
\put {\scalebox{0.8}{$L^2\mathbf{\Var{c_2}}$}} at -4 -10
\normalcolor
\color{black}
\endpicture
\end{subfigure}

\vspace{5mm}
\begin{subfigure}{0.33\textwidth}
	\caption{Free Energy $f_L(\alpha,\beta)$}
\beginpicture
\normalcolor
\color{black}
\setcoordinatesystem units <10.5pt,15pt>
\axescenter{-5}{0}{0}{0}{5}{6}{0}
\put {\Large$\alpha$} at 3.5 -1
\setplotsymbol ({\scalebox{0.3}{$\bullet$}})
\color{Red}
\plot -5.000 -0.1700  -4.804 -0.1641  -4.608 -0.1582  -4.412 -0.1523  -4.216 -0.1465  -4.020 -0.1406  -3.824 -0.1347  -3.627 -0.1288  -3.431 -0.1229  -3.235 -0.1170  -3.039 -0.1110  -2.843 -0.1051  -2.647 -0.0991  -2.451 -0.0931  -2.255 -0.0870  -2.059 -0.0808  -1.863 -0.0744  -1.667 -0.0677  -1.471 -0.0606  -1.275 -0.0526  -1.078 -0.0426  -0.882 -0.0247  -0.686 0.0226  -0.490 0.1128  -0.294 0.2316  -0.098 0.3688  0.098 0.5191  0.294 0.6791  0.490 0.8464  0.686 1.0195  0.882 1.1972  1.078 1.3784  1.275 1.5625  1.471 1.7489  1.667 1.9371  1.863 2.1267  2.059 2.3173  2.255 2.5087  2.451 2.7008  2.647 2.8933  2.843 3.0863  3.039 3.2795  3.235 3.4730  3.431 3.6666  3.627 3.8604  3.824 4.0543  4.020 4.2482  4.216 4.4422  4.412 4.6363  4.608 4.8303  4.804 5.0244  5.000 5.2185
  /
\color{black}
\put {$g_L$} at -1 3
\endpicture
\end{subfigure}
\begin{subfigure}{0.33\textwidth}
	\caption{$\LA c_1 \RA$ and $\LA c_2 \RA$}
\beginpicture
\normalcolor
\color{black}
\setcoordinatesystem units <10.5pt,90pt>
\axescenter{-5}{0}{0}{0}{5}{1}{0}
\put {\Large$\alpha$} at 3.5 -0.17
\setplotsymbol ({\scalebox{0.3}{$\bullet$}})
\color{Red}
\plot -5.000 0.0300  -4.804 0.0300  -4.608 0.0300  -4.412 0.0300  -4.216 0.0300  -4.020 0.0300  -3.824 0.0301  -3.627 0.0301  -3.431 0.0301  -3.235 0.0302  -3.039 0.0303  -2.843 0.0304  -2.647 0.0306  -2.451 0.0309  -2.255 0.0313  -2.059 0.0320  -1.863 0.0331  -1.667 0.0349  -1.471 0.0380  -1.275 0.0442  -1.078 0.0612  -0.882 0.1437  -0.686 0.3547  -0.490 0.5462  -0.294 0.6586  -0.098 0.7364  0.098 0.7935  0.294 0.8363  0.490 0.8693  0.686 0.8953  0.882 0.9159  1.078 0.9322  1.275 0.9452  1.471 0.9556  1.667 0.9636  1.863 0.9697  2.059 0.9744  2.255 0.9780  2.451 0.9808  2.647 0.9830  2.843 0.9848  3.039 0.9861  3.235 0.9872  3.431 0.9880  3.627 0.9886  3.824 0.9890  4.020 0.9893  4.216 0.9895  4.412 0.9897  4.608 0.9898  4.804 0.9898  5.000 0.9899
  /
\put {$\mathbf{\LA c_1 \RA}$} at -1.3 0.8
\color{Blue}
\plot -5.000 0.0100  -4.804 0.0100  -4.608 0.0100  -4.412 0.0100  -4.216 0.0100  -4.020 0.0100  -3.824 0.0100  -3.627 0.0100  -3.431 0.0100  -3.235 0.0100  -3.039 0.0100  -2.843 0.0100  -2.647 0.0100  -2.451 0.0100  -2.255 0.0100  -2.059 0.0100  -1.863 0.0100  -1.667 0.0100  -1.471 0.0100  -1.275 0.0100  -1.078 0.0101  -0.882 0.0101  -0.686 0.0101  -0.490 0.0101  -0.294 0.0101  -0.098 0.0100  0.098 0.0100  0.294 0.0100  0.490 0.0100  0.686 0.0100  0.882 0.0100  1.078 0.0100  1.275 0.0100  1.471 0.0100  1.667 0.0100  1.863 0.0100  2.059 0.0100  2.255 0.0100  2.451 0.0100  2.647 0.0100  2.843 0.0100  3.039 0.0100  3.235 0.0100  3.431 0.0100  3.627 0.0100  3.824 0.0100  4.020 0.0100  4.216 0.0100  4.412 0.0100  4.608 0.0100  4.804 0.0100  5.000 0.0100  
   /
\put {$\mathbf{\LA c_2 \RA}$} at 3 0.1
\endpicture
\end{subfigure}
\begin{subfigure}{0.33\textwidth}
	\caption{$L^2\Var{c_1}$ and $L^2\Var{c_2}$}
\beginpicture
\normalcolor
\color{black}
\setcoordinatesystem units <10.5pt,3.444pt>
\setplotarea x from -5 to 5, y from -25 to 1
\axis left shiftedto x=0
        ticks withvalues $-20$ $0$ /
        at -20 0 /
/
\axis bottom 
        ticks withvalues $-5$ $0$ $5$ /
        at -5 0 5 /
/
\setsolid
\put {\Large$\alpha$} at 3.5 -30.0
\setplotsymbol ({\scalebox{0.3}{$\bullet$}})
\color{Red}
\plot -5.000 -11.4504  -4.804 -11.0580  -4.608 -10.6656  -4.412 -10.2731  -4.216 -9.8804  -4.020 -9.4874  -3.824 -9.0940  -3.627 -8.7000  -3.431 -8.3052  -3.235 -7.9090  -3.039 -7.5109  -2.843 -7.1098  -2.647 -6.7044  -2.451 -6.2922  -2.255 -5.8696  -2.059 -5.4305  -1.863 -4.9644  -1.667 -4.4515  -1.471 -3.8507  -1.275 -3.0646  -1.078 -1.8386  -0.882 -0.2293  -0.686 0.2138  -0.490 -0.3488  -0.294 -0.7553  -0.098 -1.0885  0.098 -1.3850  0.294 -1.6590  0.490 -1.9060  0.686 -2.1416  0.882 -2.3732  1.078 -2.5996  1.275 -2.8278  1.471 -3.0705  1.667 -3.3312  1.863 -3.6002  2.059 -3.8661  2.255 -4.1235  2.451 -4.3722  2.647 -4.6147  2.843 -4.8560  3.039 -5.1042  3.235 -5.3678  3.431 -5.6523  3.627 -5.9593  3.824 -6.2869  4.020 -6.6318  4.216 -6.9902  4.412 -7.3587  4.608 -7.7344  4.804 -8.1152  5.000 -8.4996
  /
\put {\scalebox{0.8}{$L^2\mathbf{\Var{c_1}}$}} at 3.0 0.25
\color{Blue}
\plot -5.000 -17.9376  -4.804 -17.5453  -4.608 -17.1529  -4.412 -16.7604  -4.216 -16.3677  -4.020 -15.9747  -3.824 -15.5813  -3.627 -15.1874  -3.431 -14.7926  -3.235 -14.3966  -3.039 -13.9987  -2.843 -13.5980  -2.647 -13.1932  -2.451 -12.7822  -2.255 -12.3621  -2.059 -11.9284  -1.863 -11.4743  -1.667 -10.9896  -1.471 -10.4580  -1.275 -9.8514  -1.078 -9.1199  -0.882 -8.3835  -0.686 -8.3977  -0.490 -8.6160  -0.294 -8.9224  -0.098 -9.2636  0.098 -9.6224  0.294 -9.9883  0.490 -10.3316  0.686 -10.6543  0.882 -10.9687  1.078 -11.2821  1.275 -11.5984  1.471 -11.9209  1.667 -12.2512  1.863 -12.5880  2.059 -12.9265  2.255 -13.2605  2.451 -13.5858  2.647 -13.9027  2.843 -14.2164  3.039 -14.5334  3.235 -14.8591  3.431 -15.1964  3.627 -15.5457  3.824 -15.9057  4.020 -16.2745  4.216 -16.6500  4.412 -17.0306  4.608 -17.4147  4.804 -17.8013  5.000 -18.1897
  /
\put {\scalebox{0.8}{$L^2\mathbf{\Var{c_2}}$}} at 3 -20
\normalcolor
\color{black}
\endpicture
\end{subfigure}
\normalcolor
\color{black}
\caption{
\underbar{Top row}:
The free energy, mean concentrations and variances of the linked model
as a function $\alpha$ along the line segment from $(0,5)$ to $(5,0)$ across 
the $\tau_1$ phase boundary.   
The variances are plotted on a logarithmic vertical scale. \newline
\underbar{Middle row}:
The free energy, mean concentrations and variances as a function 
of $\beta$ on the line segment from the point $(-2,-5)$
to the point $(-2,5)$ across the $\tau_2$ phase boundary. \newline
\underbar{Bottom row}:
The free energy, mean concentrations and variances as a function 
of $\alpha$ on the line segment across the $\lambda$
phase boundary from $(-5,-2)$ to $(5,-2)$.}
\label{17}    %ZXZ[17]
\color{black}
\normalcolor
\end{figure}

\subsubsection{Critical scaling in the linked model}

{\bf The $\tau_1$ phase boundary:}
In the top row of figure \ref{17} the free energy and its derivatives are plotted 
as a function of $\alpha$ along the line segment from $(0,5)$ to $(5,0)$ in the 
phase diagram. As in the unlinked model, the free energy can be fitted to an 
absolute value function using the model 
\begin{eqnarray}
w(\alpha,5{-}\alpha) 
&=& a_0 + a_1\, \alpha + a_2\,|\alpha-\alpha_c|^{2-\alpha_s} \nonumber \\
& & \hspace{2cm} +\, a_3\,(\alpha-\alpha_c)^2 .
\label{29}   %ZXZ[29]
\end{eqnarray}
The curve in figure \ref{17}(a) appears to be rotated about its minimum, and to 
account for this, a linear term in $\alpha$ was included in equation \Ref{29}.  
A fit with $a_3=0$ (so that there is no quadratic correction) gives
\begin{eqnarray}
%\hspace{-2.5cm}
w(\alpha,5{-}\alpha) 
&\approx& 2.3172+ 0.1823\,\alpha \nonumber \\
& & \hspace{0cm} +\, 0.7747\,|\alpha-2.4592|^{1.018}  ,
\label{30}    %ZXZ[30]
\end{eqnarray}
while including the quadratic term in the model gives
\begin{eqnarray}
\hspace{-2.5cm}
& & \hspace{-1cm}w(\alpha,5{-}\alpha) 
\approx 2.3259 + 0.1819\,\alpha \nonumber \\
& & \hspace{-1cm} +\, 0.7737\,|\alpha-2.4583|^{1.057} 
 - 0.0124\, (\alpha-2.4583)^2 .
\label{31}    %ZXZ[31]
\end{eqnarray}
We note again the relative small coefficient of the quadratic term,
suggesting minor curvature of the $\tau_1$ phase boundary for
$-0.97 \leq \alpha \leq 5$.  Equation \Ref{27} shows that the $\tau_1$ 
phase boundary is crossed approximately at the point $(2.4592,2.5408)$
by the line segment from $(0,5)$ to $(5,0)$ in the phase diagram, consistent with 
the observation that the $\tau_1$ phase boundary is slightly off-set from 
the main diagonal in the $\alpha\beta$-plane.

The critical exponents $2{-}\alpha_s$ and $2{-}\alpha_s^\prime$ 
(see equation \Ref{8}) can also be read from the fits in equations
\Ref{30} and \Ref{31}:
\begin{equation}
2-\alpha_s = 2-\alpha_s^\prime = 1.02 \pm 0.04. 
\label{32}  %ZXZ[32]
\end{equation} 
The error bar is the absolute difference between the estimates
in equations \Ref{30} and \Ref{31}.  These results are consistent with 
a first order transition along $\tau_1$ (which is seen in the jump 
discontinuities in $\LA c_1\RA$ and $\LA c_2 \RA$ shown in the 
top middle graph in figure \ref{17}).

{\bf The $\tau_2$ phase boundary:}
The free energy, concentrations and variances across this phase boundary 
are plotted in figure \ref{17} along a line segment from $(-2,-5)$ to $(-2,5)$.  
The jump discontinuities in $\LA c_1\RA$ and $\LA c_2 \RA$ are consistent 
with this also being a first order transition. Fits similar to equations 
\Ref{30} and \Ref{31} give the approximations
\begin{eqnarray}
w(-2,\beta) 
&\approx& -0.0392 +  0.4013\,\beta \nonumber \\
& & +\, 0.3517\,|\beta-0.0517|^{1.059}  .
\label{33}    %ZXZ[33]
\end{eqnarray}
and
\begin{eqnarray}
& & \hspace{-1cm} w(-2,\beta) 
\approx -0.0188 + 0.4017\,\beta \nonumber \\
& & \hspace{-1cm} +\, 0.3400\,|\beta-0.0567|^{1.182} 
 - 0.0151\, (\beta-0.0567)^2 .
\label{34}    %ZXZ[34]
\end{eqnarray}
Comparison of these fits with equation \Ref{8} give the estimated 
critical exponents
\begin{equation}
2-\alpha_s = 2-\alpha_s^\prime = 1.06 \pm 0.12 
\label{35}  %ZXZ[35]
\end{equation}
and critical point $(-2,\tau_2(-2)) \approx (-2,-0.52(5))$ on the
$\tau_2$ phase boundary.  These values are again consistent with 
$\tau_2$ being a curve of first order transitions.

{\bf The $\lambda$ phase boundary:}
The data along the $\lambda$ phase boundary are plotted in the bottom panels 
of figure \ref{17}.  The concentration $\LA c_1 \RA$ plotted in figure \ref{17}(h) 
is obtained from estimates along the line segment between the points $(-5,-2)$ 
and $(5,-2)$ in the phase diagram. In this graph $\LA c_1 \RA$ has a profile 
consistent with a continuous transition.

As in the unlinked model, the scaling exponent $\alpha_s^\prime$  in 
equation \Ref{8} can be estimated by plotting 
$\log \LA c_1 \RA_s / \log |\alpha{-}\alpha_c|$ (equation \Ref{22}) as a function 
of $1/\log  |\alpha{-}\alpha_c|$.  Plotting $\gamma$ in equation \Ref{22}
as a function of $1/\log  |\alpha{-}\alpha_c|$ for $-0.95 \leq \alpha \leq 0$ 
gives the graph in figure \ref{18} .  Fitting a linear function in 
$1/\log  |\alpha{-}\alpha_c|$ to the graph gives the estimate
$1{-}\alpha_s^\prime \approx 0.57$, and fitting a quadratic instead gives
$1{-}\alpha_s^\prime \approx 0.63$.  The difference in these two estimates
is taken as a confidence interval in the estimate.

\begin{figure}[h!]
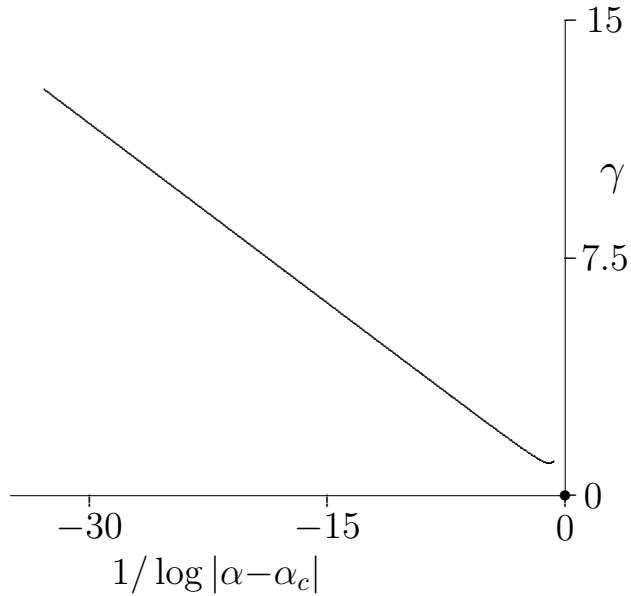

\beginpicture
\color{black}
\normalcolor
\setcoordinatesystem units <6pt,12pt>
\setplotarea x from -35 to 0, y from 0 to 15

\axis right shiftedto x=0 
        %ticks
        %withvalues 10  /
        %at  10  /
 /
\axis bottom shiftedto y=0
        %ticks
        %withvalues -30 -15 0 /
        %at  -30 -15 0 /
/
\put {\footnotesize$\bullet$} at 0 0 

\plot -30 -0.33 -30 0 / \plot -15 -0.33 -15 0 / \plot 0 -0.33 0 0 /
\put {\large${-}30$} at -30 -1
\put {\large${-}15$} at -15 -1
\put {\large$0$} at 0 -1
\plot 0 0 0.65 0 / \plot 0 7.5 0.65 7.5 / \plot 0 15 0.65 15 / 
\put {\large$0$} at 1.75 0  
\put {\large$7.5$} at 2.5 7.5  
\put {\large$15$} at 2.5 15 

\plot -0.7110 1.0888  -0.7637 1.0668  -0.8191 1.0476  -0.8776 1.0329  -0.9397 1.0235  -1.0058 1.0197  -1.0766 1.0214  -1.1527 1.0289  -1.2351 1.0420  -1.3245 1.0605  -1.4221 1.0843  -1.5292 1.1131  -1.6475 1.1471  -1.7790 1.1868  -1.9261 1.2329  -2.0919 1.2866  -2.2805 1.3494  -2.4970 1.4232  -2.7484 1.5107  -3.0441 1.6153  -3.3971 1.7420  -3.8261 1.8978  -4.3589 2.0931  -5.0390 2.3443  -5.9376 2.6783  -7.1807 3.1426  -9.0146 3.8303  -11.9931 4.9505  -17.6771 7.0931  -32.8308 12.8134 /

\put {\Large$\gamma$} at 3 10
\put {\large$1/\log|\alpha{-}\alpha_c|$} at -22 -2.5

\color{black}
\normalcolor
\endpicture
\caption{Plotting $\gamma 
= \log(\LA c_1 \RA- \LA c_1 \RA\!\vert_{crit})/ \log |\alpha{-}\alpha_c|$ as
a function of $1/\log|\alpha{-}\alpha_c|$ for the linked model in the 
$c_1$-dominated phase along a line segment starting in the $\lambda$ 
phase boundary, with $\beta=-2$.   In this graph $-0.9 \leq \alpha \leq 0$.}
\label{18}  %%ZXZ[18]
\end{figure}

On the empty phase side of $\lambda$ the free energy is constant to numerical
accuracy, and so one expects $\alpha_s=2$. Taken together,
\begin{equation}
\alpha_s = 2
\qquad\hbox{and}\qquad
\alpha_s^\prime \approx 0.43 \pm 0.06
\label{36}    %ZXZ[36]
\end{equation}
This shows that $2{-}\alpha_s=0$ on the empty phase side of $\lambda$, and
$2{-}\alpha_s^\prime=1.57(6)$ on the $c_1$-dominated phase side.

\begin{figure}[h!]
\begin{subfigure}{0.33\textwidth}
\normalcolor
\color{black}
	\caption{Free Energy $f_L(\alpha,\beta)$}
\beginpicture
\setcoordinatesystem units <10.5pt,15pt>
\axescenter{-5}{0}{0}{0}{5}{6}{0}
\put {\Large$\alpha$} at 3.0 -1
\setplotsymbol ({\scalebox{0.3}{$\bullet$}})
\color{Red}
\plot -5.000 -0.2000  -4.804 -0.1922  -4.608 -0.1843  -4.412 -0.1765  -4.216 -0.1686  -4.020 -0.1608  -3.824 -0.1529  -3.627 -0.1451  -3.431 -0.1372  -3.235 -0.1293  -3.039 -0.1214  -2.843 -0.1135  -2.647 -0.1056  -2.451 -0.0976  -2.255 -0.0895  -2.059 -0.0814  -1.863 -0.0730  -1.667 -0.0644  -1.471 -0.0553  -1.275 -0.0453  -1.078 -0.0333  -0.882 -0.0129  -0.686 0.0369  -0.490 0.1296  -0.294 0.2508  -0.098 0.3902  0.098 0.5426  0.294 0.7046  0.490 0.8740  0.686 1.0492  0.882 1.2289  1.078 1.4123  1.275 1.5985  1.471 1.7870  1.667 1.9772  1.863 2.1688  2.059 2.3614  2.255 2.5549  2.451 2.7489  2.647 2.9435  2.843 3.1384  3.039 3.3336  3.235 3.5291  3.431 3.7248  3.627 3.9205  3.824 4.1164  4.020 4.3123  4.216 4.5083  4.412 4.7043  4.608 4.9004  4.804 5.0964  5.000 5.2925 
  /
\color{black}
\put {$g_L$} at -1 3
\endpicture
\end{subfigure}
\begin{subfigure}{0.33\textwidth}
\normalcolor
\color{black}
	\caption{$\LA c_1 \RA$ and $\LA c_2 \RA$}
\beginpicture
\setcoordinatesystem units <10.5pt,150pt>
\axescenter{-5}{0}{0}{0}{5}{0.6}{0}
\put {\Large$\alpha$} at 3.0 -0.1
\setplotsymbol ({\scalebox{0.3}{$\bullet$}})
\color{Red}
\plot -5.000 0.0300  -4.804 0.0300  -4.608 0.0300  -4.412 0.0300  -4.216 0.0300  -4.020 0.0300  -3.824 0.0301  -3.627 0.0301  -3.431 0.0301  -3.235 0.0302  -3.039 0.0303  -2.843 0.0304  -2.647 0.0306  -2.451 0.0309  -2.255 0.0314  -2.059 0.0320  -1.863 0.0331  -1.667 0.0349  -1.471 0.0381  -1.275 0.0445  -1.078 0.0622  -0.882 0.1463  -0.686 0.2858  -0.490 0.3149  -0.294 0.3494  -0.098 0.3824  0.098 0.3786  0.294 0.4259  0.490 0.4427  0.686 0.4547  0.882 0.4636  1.078 0.4732  1.275 0.4779  1.471 0.4836  1.667 0.4872  1.863 0.4909  2.059 0.4930  2.255 0.4949  2.451 0.4961  2.647 0.4986  2.843 0.4983  3.039 0.4821  3.235 0.4987  3.431 0.5012  3.627 0.5001  3.824 0.4994  4.020 0.4997  4.216 0.4989  4.412 0.5004  4.608 0.4970  4.804 0.5027  5.000 0.5032
   /
\put {$\mathbf{\LA c_1 \RA}$} at -2 0.23
\color{Blue}
\plot -5.000 0.0100  -4.804 0.0100  -4.608 0.0100  -4.412 0.0100  -4.216 0.0100  -4.020 0.0100  -3.824 0.0100  -3.627 0.0100  -3.431 0.0100  -3.235 0.0100  -3.039 0.0100  -2.843 0.0100  -2.647 0.0100  -2.451 0.0100  -2.255 0.0100  -2.059 0.0100  -1.863 0.0100  -1.667 0.0100  -1.471 0.0101  -1.275 0.0102  -1.078 0.0106  -0.882 0.0135  -0.686 0.2270  -0.490 0.3152  -0.294 0.3555  -0.098 0.3834  0.098 0.4400  0.294 0.4290  0.490 0.4434  0.686 0.4561  0.882 0.4667  1.078 0.4726  1.275 0.4801  1.471 0.4840  1.667 0.4879  1.863 0.4902  2.059 0.4926  2.255 0.4942  2.451 0.4958  2.647 0.4954  2.843 0.4973  3.039 0.5147  3.235 0.4991  3.431 0.4973  3.627 0.4988  3.824 0.4999  4.020 0.4998  4.216 0.5007  4.412 0.4993  4.608 0.5028  4.804 0.4972  5.000 0.4967
   /
\put {$\mathbf{\LA c_2 \RA}$} at 3 0.4
\endpicture
\end{subfigure}
\begin{subfigure}{0.33\textwidth}
	\caption{$L^2\Var{c_1}$ and $L^2\Var{c_2}$}
\beginpicture
\normalcolor
\color{black}
\setcoordinatesystem units <10.5pt,4.0912pt>
\setplotarea x from -5 to 5, y from -20 to 2
\axis left shiftedto x=0
        ticks withvalues $-15$ $2$ /
        at -15 2 /
/
\axis bottom 
        ticks withvalues $-5$ $0$ $5$ /
        at -5 0 5 /
/
\put {\Large$\alpha$} at 3 -22
\setplotsymbol ({\scalebox{0.3}{$\bullet$}})
\color{Red}
\plot -5.000 -11.4519  -4.804 -11.0595  -4.608 -10.6671  -4.412 -10.2746  -4.216 -9.8819  -4.020 -9.4889  -3.824 -9.0955  -3.627 -8.7015  -3.431 -8.3066  -3.235 -7.9104  -3.039 -7.5122  -2.843 -7.1111  -2.647 -6.7054  -2.451 -6.2930  -2.255 -5.8700  -2.059 -5.4302  -1.863 -4.9627  -1.667 -4.4473  -1.471 -3.8417  -1.275 -3.0459  -1.078 -1.8067  -0.882 -0.2655  -0.686 0.2159  -0.490 1.0421  -0.294 0.9224  -0.098 0.9235  0.098 0.9886  0.294 1.0740  0.490 1.1659  0.686 1.2604  0.882 1.3560  1.078 1.4525  1.275 1.5495  1.471 1.6461  1.667 1.7396  1.863 1.8262  2.059 1.9024  2.255 1.9666  2.451 2.0178  2.647 2.0561  2.843 2.0819  3.039 2.0959  3.235 2.0992  3.431 2.0924  3.627 2.0760  3.824 2.0504  4.020 2.0155  4.216 1.9716  4.412 1.9189  4.608 1.8578  4.804 1.7889  5.000 1.7130 
  /
\put {\scalebox{0.8}{$L^2\mathbf{\Var{c_1}}$}} at -3.75 -2
\color{Blue}
\plot -5.000 -23.4623  -4.804 -22.6950  -4.608 -21.9273  -4.412 -21.1590  -4.216 -20.3902  -4.020 -19.6207  -3.824 -18.8504  -3.627 -18.0792  -3.431 -17.3067  -3.235 -16.5325  -3.039 -15.7561  -2.843 -14.9766  -2.647 -14.1926  -2.451 -13.4018  -2.255 -12.6011  -2.059 -11.7850  -1.863 -10.9447  -1.667 -10.0643  -1.471 -9.1128  -1.275 -8.0160  -1.078 -6.5489  -0.882 -4.2348  -0.686 -0.6377  -0.490 1.3540  -0.294 1.0968  -0.098 1.0469  0.098 1.0790  0.294 1.1408  0.490 1.2149  0.686 1.2962  0.882 1.3819  1.078 1.4712  1.275 1.5631  1.471 1.6560  1.667 1.7470  1.863 1.8317  2.059 1.9066  2.255 1.9695  2.451 2.0197  2.647 2.0572  2.843 2.0823  3.039 2.0960  3.235 2.0990  3.431 2.0921  3.627 2.0758  3.824 2.0502  4.020 2.0154  4.216 1.9715  4.412 1.9188  4.608 1.8577  4.804 1.7888  5.000 1.7129
   /
\put {\scalebox{0.8}{$L^2\mathbf{\Var{c_2}}$}} at 3.75 -2
\normalcolor
\color{black}
\endpicture
\end{subfigure}

\vspace{5mm}
\begin{subfigure}{0.33\textwidth}
	\caption{Free Energy $f_L(\alpha,\beta)$}
\beginpicture
\normalcolor
\color{black}
\setcoordinatesystem units <10.5pt,22.5pt>
\axescenter{-5}{0}{0}{0}{5}{4}{0}
\put {\Large$\beta$} at 3.5 -0.667
\setplotsymbol ({\scalebox{0.3}{$\bullet$}})
\color{Red}
\plot -5.000 -0.0647  -4.804 -0.0628  -4.608 -0.0608  -4.412 -0.0588  -4.216 -0.0569  -4.020 -0.0549  -3.824 -0.0530  -3.627 -0.0510  -3.431 -0.0490  -3.235 -0.0471  -3.039 -0.0451  -2.843 -0.0431  -2.647 -0.0412  -2.451 -0.0392  -2.255 -0.0372  -2.059 -0.0353  -1.863 -0.0333  -1.667 -0.0313  -1.471 -0.0293  -1.275 -0.0273  -1.078 -0.0252  -0.882 -0.0231  -0.686 -0.0207  -0.490 -0.0163  -0.294 0.0488  -0.098 0.1431  0.098 0.2549  0.294 0.3787  0.490 0.5112  0.686 0.6500  0.882 0.7936  1.078 0.9407  1.275 1.0906  1.471 1.2427  1.667 1.3965  1.863 1.5516  2.059 1.7077  2.255 1.8644  2.451 2.0216  2.647 2.1791  2.843 2.3369  3.039 2.4950  3.235 2.6531  3.431 2.8115  3.627 2.9699  3.824 3.1285  4.020 3.2871  4.216 3.4458  4.412 3.6045  4.608 3.7633  4.804 3.9221  5.000 4.0808
  /
\color{black}
\put {$g_L$} at -1 2
\endpicture
\end{subfigure}
\begin{subfigure}{0.33\textwidth}
	\caption{$\LA c_1 \RA$ and $\LA c_2 \RA$}
\beginpicture
\normalcolor
\color{black}
\setcoordinatesystem units <10.5pt,90pt>
\axescenter{-5}{0}{0}{0}{5}{1}{0}
\put {\Large$\beta$} at 3.5 -0.17
\setplotsymbol ({\scalebox{0.3}{$\bullet$}})
\color{Red}
\plot -5.000 0.0901  -4.804 0.0901  -4.608 0.0901  -4.412 0.0901  -4.216 0.0901  -4.020 0.0901  -3.824 0.0901  -3.627 0.0901  -3.431 0.0901  -3.235 0.0901  -3.039 0.0901  -2.843 0.0901  -2.647 0.0901  -2.451 0.0901  -2.255 0.0902  -2.059 0.0902  -1.863 0.0903  -1.667 0.0904  -1.471 0.0906  -1.275 0.0908  -1.078 0.0913  -0.882 0.0923  -0.686 0.0947  -0.490 0.1453  -0.294 0.2201  -0.098 0.2100  0.098 0.2021  0.294 0.1975  0.490 0.1946  0.686 0.1929  0.882 0.1918  1.078 0.1911  1.275 0.1907  1.471 0.1905  1.667 0.1903  1.863 0.1902  2.059 0.1901  2.255 0.1901  2.451 0.1901  2.647 0.1900  2.843 0.1900  3.039 0.1900  3.235 0.1900  3.431 0.1900  3.627 0.1900  3.824 0.1900  4.020 0.1900  4.216 0.1900  4.412 0.1900  4.608 0.1900  4.804 0.1900  5.000 0.1900
 /
\put {$\mathbf{\LA c_1 \RA}$} at 3 0.12
\color{Blue}
\plot -5.000 0.0100  -4.804 0.0100  -4.608 0.0100  -4.412 0.0100  -4.216 0.0100  -4.020 0.0100  -3.824 0.0100  -3.627 0.0100  -3.431 0.0100  -3.235 0.0100  -3.039 0.0100  -2.843 0.0100  -2.647 0.0100  -2.451 0.0100  -2.255 0.0100  -2.059 0.0101  -1.863 0.0101  -1.667 0.0102  -1.471 0.0103  -1.275 0.0104  -1.078 0.0107  -0.882 0.0113  -0.686 0.0129  -0.490 0.0962  -0.294 0.4246  -0.098 0.5309  0.098 0.6044  0.294 0.6559  0.490 0.6936  0.686 0.7212  0.882 0.7419  1.078 0.7579  1.275 0.7704  1.471 0.7804  1.667 0.7881  1.863 0.7938  2.059 0.7979  2.255 0.8007  2.451 0.8027  2.647 0.8042  2.843 0.8053  3.039 0.8063  3.235 0.8071  3.431 0.8078  3.627 0.8084  3.824 0.8088  4.020 0.8092  4.216 0.8094  4.412 0.8096  4.608 0.8097  4.804 0.8098  5.000 0.8099
 /
\put {$\mathbf{\LA c_2 \RA}$} at 3 0.90
\endpicture
\end{subfigure}
\begin{subfigure}{0.33\textwidth}
	\caption{$L^2\Var{c_1}$ and $L^2\Var{c_2}$}
\beginpicture
\normalcolor
\color{black}
\setcoordinatesystem units <10.5pt,4.864pt>
\setplotarea x from -5 to 5, y from -17 to 1
\axis left shiftedto x=0
        ticks withvalues $-15$ $1$ /
        at -15 1 /
/
\axis bottom 
        ticks withvalues $-5$ $0$ $5$ /
        at -5 0 5 /
/
\setsolid
\put {\Large$\beta$} at 3.5 -20
\setplotsymbol ({\scalebox{0.3}{$\bullet$}})
\color{Red}
\plot -5.000 -0.8605  -4.804 -0.8605  -4.608 -0.8605  -4.412 -0.8605  -4.216 -0.8605  -4.020 -0.8605  -3.824 -0.8605  -3.627 -0.8605  -3.431 -0.8605  -3.235 -0.8604  -3.039 -0.8604  -2.843 -0.8604  -2.647 -0.8604  -2.451 -0.8603  -2.255 -0.8602  -2.059 -0.8601  -1.863 -0.8599  -1.667 -0.8595  -1.471 -0.8589  -1.275 -0.8576  -1.078 -0.8551  -0.882 -0.8483  -0.686 -0.8203  -0.490 0.3380  -0.294 -2.7330  -0.098 -3.1787  0.098 -3.6850  0.294 -4.1893  0.490 -4.6821  0.686 -5.1632  0.882 -5.6335  1.078 -6.0977  1.275 -6.5568  1.471 -7.0054  1.667 -7.4361  1.863 -7.8481  2.059 -8.2481  2.255 -8.6443  2.451 -9.0424  2.647 -9.4454  2.843 -9.8538  3.039 -10.2667  3.235 -10.6824  3.431 -11.0984  3.627 -11.5128  3.824 -11.9241  4.020 -12.3317  4.216 -12.7357  4.412 -13.1367  4.608 -13.5352  4.804 -13.9319  5.000 -14.3272
  /
\put {\scalebox{0.8}{$L^2\mathbf{\Var{c_1}}$}} at -4 -3
\color{Blue}
\plot -5.000 -14.7514  -4.804 -14.3592  -4.608 -13.9669  -4.412 -13.5745  -4.216 -13.1820  -4.020 -12.7893  -3.824 -12.3964  -3.627 -12.0032  -3.431 -11.6094  -3.235 -11.2148  -3.039 -10.8191  -2.843 -10.4217  -2.647 -10.0216  -2.451 -9.6178  -2.255 -9.2082  -2.059 -8.7901  -1.863 -8.3588  -1.667 -7.9072  -1.471 -7.4237  -1.275 -6.8875  -1.078 -6.2580  -0.882 -5.4382  -0.686 -4.1001  -0.490 1.4739  -0.294 -0.4441  -0.098 -0.7776  0.098 -1.1919  0.294 -1.4922  0.490 -1.8128  0.686 -2.1117  0.882 -2.3861  1.078 -2.6341  1.275 -2.8652  1.471 -3.1033  1.667 -3.3804  1.863 -3.7098  2.059 -4.0705  2.255 -4.4259  2.451 -4.7443  2.647 -5.0076  2.843 -5.2172  3.039 -5.3929  3.235 -5.5630  3.431 -5.7517  3.627 -5.9727  3.824 -6.2296  4.020 -6.5196  4.216 -6.8373  4.412 -7.1763  4.608 -7.5312  4.804 -7.8976  5.000 -8.2720
  /
\put {\scalebox{0.8}{$L^2\mathbf{\Var{c_2}}$}} at 4 -3
\normalcolor
\color{black}
\endpicture
\end{subfigure}

\vspace{5mm}
\begin{subfigure}{0.33\textwidth}
	\caption{Free Energy $f_L(\alpha,\beta)$}
\beginpicture
\normalcolor
\color{black}
\setcoordinatesystem units <10.5pt,15pt>
\axescenter{-5}{0}{0}{0}{5}{6}{0}
\put {\Large$\alpha$} at 3.5 -1
\setplotsymbol ({\scalebox{0.3}{$\bullet$}})
\color{Red}
\plot -5.000 -0.1597  -4.804 -0.1538  -4.608 -0.1479  -4.412 -0.1420  -4.216 -0.1362  -4.020 -0.1303  -3.824 -0.1244  -3.627 -0.1185  -3.431 -0.1126  -3.235 -0.1067  -3.039 -0.1007  -2.843 -0.0948  -2.647 -0.0888  -2.451 -0.0828  -2.255 -0.0767  -2.059 -0.0705  -1.863 -0.0641  -1.667 -0.0574  -1.471 -0.0503  -1.275 -0.0422  -1.078 -0.0321  -0.882 -0.0139  -0.686 0.0334  -0.490 0.1234  -0.294 0.2422  -0.098 0.3793  0.098 0.5295  0.294 0.6895  0.490 0.8568  0.686 1.0299  0.882 1.2075  1.078 1.3887  1.275 1.5729  1.471 1.7592  1.667 1.9474  1.863 2.1370  2.059 2.3276  2.255 2.5190  2.451 2.7111  2.647 2.9036  2.843 3.0966  3.039 3.2898  3.235 3.4833  3.431 3.6769  3.627 3.8707  3.824 4.0646  4.020 4.2585  4.216 4.4525  4.412 4.6466  4.608 4.8406  4.804 5.0347  5.000 5.2288
  /
\color{black}
\put {$g_L$} at -1 3
\endpicture
\end{subfigure}
\begin{subfigure}{0.33\textwidth}
	\caption{$\LA c_1 \RA$ and $\LA c_2 \RA$}
\beginpicture
\normalcolor
\color{black}
\setcoordinatesystem units <10.5pt,90pt>
\axescenter{-5}{0}{0}{0}{5}{1}{0}
\put {\Large$\alpha$} at 3.5 -0.17
\setplotsymbol ({\scalebox{0.3}{$\bullet$}})
\color{Red}
\plot -5.000 0.0300  -4.804 0.0300  -4.608 0.0300  -4.412 0.0300  -4.216 0.0300  -4.020 0.0300  -3.824 0.0301  -3.627 0.0301  -3.431 0.0301  -3.235 0.0302  -3.039 0.0303  -2.843 0.0304  -2.647 0.0306  -2.451 0.0309  -2.255 0.0314  -2.059 0.0321  -1.863 0.0332  -1.667 0.0350  -1.471 0.0382  -1.275 0.0445  -1.078 0.0622  -0.882 0.1447  -0.686 0.3542  -0.490 0.5457  -0.294 0.6581  -0.098 0.7361  0.098 0.7933  0.294 0.8362  0.490 0.8693  0.686 0.8952  0.882 0.9158  1.078 0.9322  1.275 0.9452  1.471 0.9555  1.667 0.9636  1.863 0.9697  2.059 0.9744  2.255 0.9780  2.451 0.9808  2.647 0.9830  2.843 0.9848  3.039 0.9861  3.235 0.9872  3.431 0.9880  3.627 0.9886  3.824 0.9890  4.020 0.9893  4.216 0.9895  4.412 0.9897  4.608 0.9898  4.804 0.9898  5.000 0.9899 
  /
\put {$\mathbf{\LA c_1 \RA}$} at 3 0.875
\color{Blue}
\plot -5.000 0.0100  -4.804 0.0100  -4.608 0.0100  -4.412 0.0100  -4.216 0.0100  -4.020 0.0100  -3.824 0.0100  -3.627 0.0100  -3.431 0.0100  -3.235 0.0100  -3.039 0.0100  -2.843 0.0100  -2.647 0.0100  -2.451 0.0100  -2.255 0.0100  -2.059 0.0100  -1.863 0.0100  -1.667 0.0101  -1.471 0.0101  -1.275 0.0102  -1.078 0.0105  -0.882 0.0114  -0.686 0.0113  -0.490 0.0109  -0.294 0.0106  -0.098 0.0104  0.098 0.0103  0.294 0.0102  0.490 0.0101  0.686 0.0101  0.882 0.0101  1.078 0.0100  1.275 0.0100  1.471 0.0100  1.667 0.0100  1.863 0.0100  2.059 0.0100  2.255 0.0100  2.451 0.0100  2.647 0.0100  2.843 0.0100  3.039 0.0100  3.235 0.0100  3.431 0.0100  3.627 0.0100  3.824 0.0100  4.020 0.0100  4.216 0.0100  4.412 0.0100  4.608 0.0100  4.804 0.0100  5.000 0.0100
   /
\put {$\mathbf{\LA c_2 \RA}$} at 3 0.1
\endpicture
\end{subfigure}
\begin{subfigure}{0.33\textwidth}
	\caption{$L^2\Var{c_1}$ and $L^2\Var{c_2}$}
\beginpicture
\normalcolor
\color{black}
\setcoordinatesystem units <10.5pt,4.524pt>
\setplotarea x from -5 to 5, y from -20 to 1
\axis left shiftedto x=0
        ticks withvalues $-15$ $1$ /
        at -15 1 /
/
\axis bottom 
        ticks withvalues $-5$ $0$ $5$ /
        at -5 0 5 /
/
\setsolid
\put {\Large$\alpha$} at 3.5 -24
\setplotsymbol ({\scalebox{0.3}{$\bullet$}})
\color{Red}
\plot -5.000 -11.4400  -4.804 -11.0476  -4.608 -10.6552  -4.412 -10.2627  -4.216 -9.8699  -4.020 -9.4769  -3.824 -9.0835  -3.627 -8.6895  -3.431 -8.2945  -3.235 -7.8982  -3.039 -7.4999  -2.843 -7.0986  -2.647 -6.6927  -2.451 -6.2798  -2.255 -5.8563  -2.059 -5.4157  -1.863 -4.9474  -1.667 -4.4312  -1.471 -3.8254  -1.275 -3.0325  -1.078 -1.8063  -0.882 -0.2484  -0.686 0.2123  -0.490 -0.3484  -0.294 -0.7537  -0.098 -1.0869  0.098 -1.3832  0.294 -1.6572  0.490 -1.9046  0.686 -2.1406  0.882 -2.3724  1.078 -2.5988  1.275 -2.8271  1.471 -3.0699  1.667 -3.3305  1.863 -3.5996  2.059 -3.8655  2.255 -4.1229  2.451 -4.3717  2.647 -4.6142  2.843 -4.8556  3.039 -5.1039  3.235 -5.3675  3.431 -5.6520  3.627 -5.9589  3.824 -6.2865  4.020 -6.6314  4.216 -6.9898  4.412 -7.3583  4.608 -7.7340  4.804 -8.1148  5.000 -8.4992
  /
\put {\scalebox{0.8}{$L^2\mathbf{\Var{c_1}}$}} at -4 -3.5
\color{Blue}
\plot -5.000 -15.8775  -4.804 -15.4852  -4.608 -15.0927  -4.412 -14.7000  -4.216 -14.3072  -4.020 -13.9140  -3.824 -13.5203  -3.627 -13.1258  -3.431 -12.7303  -3.235 -12.3331  -3.039 -11.9335  -2.843 -11.5303  -2.647 -11.1217  -2.451 -10.7051  -2.255 -10.2763  -2.059 -9.8291  -1.863 -9.3536  -1.667 -8.8339  -1.471 -8.2412  -1.275 -7.5181  -1.078 -6.5355  -0.882 -5.3839  -0.686 -5.4291  -0.490 -5.9292  -0.294 -6.4334  -0.098 -6.9246  0.098 -7.3837  0.294 -7.8128  0.490 -8.1973  0.686 -8.5467  0.882 -8.8782  1.078 -9.2026  1.275 -9.5262  1.471 -9.8537  1.667 -10.1874  1.863 -10.5262  2.059 -10.8660  2.255 -11.2007  2.451 -11.5262  2.647 -11.8433  2.843 -12.1569  3.039 -12.4738  3.235 -12.7994  3.431 -13.1366  3.627 -13.4859  3.824 -13.8458  4.020 -14.2146  4.216 -14.5901  4.412 -14.9706  4.608 -15.3547  4.804 -15.7413  5.000 -16.1297
  /
\put {\scalebox{0.8}{$L^2\mathbf{\Var{c_2}}$}} at 4 -18
\normalcolor
\color{black}
\endpicture
\end{subfigure}
\normalcolor
\color{black}
\caption{
\underbar{Top row}:
The free energy, mean concentrations and variances of the linked model
as a function of $\alpha$ from the empty phase and then along the 
$\tau_1$ phase boundary.  The variances are plotted on a logarithmic 
vertical scale.\newline
\underbar{Middle row}:
The free energy, mean concentrations and variances as a function
of $\beta$ along the $\lambda$ phase boundary and tangent to the
$\tau_2$ phase boundary at the multicritical point.  The data is
plotted against $\beta$ along the line segment from $(\alpha_c,-5)$ 
to the point $(\alpha_c,5)$.\newline
\underbar{Bottom row}:
The free energy, mean concentrations and variances as a function
of $\alpha$ transverse to the $\lambda$ phase boundary
thought the multicritical point along the line segment from
$(-5,\beta_c)$ to $(5,\beta_c)$ into the $c_1$-dominated phase.}
\label{19}    %ZXZ[19]
\normalcolor
\color{black}
\end{figure}

{\bf Scaling around the multicritical point:}
In the phase diagram in figure \ref{16} a transverse scaling axis
is set up to run horisontally through the multicritical point, while two
parallel scaling axes are set up to run along the first phase order
phase boundaries.

The free energies, mean concentrations and variances of the concentrations 
are plotted along the parallel and transverse scaling axes in figure \ref{19}.  
Along the transverse axis (shown in figure \ref{19}(h)) our data show that
the concentrations remain unchanged when compared to figure \ref{17}(h).
Thus, the transverse scaling exponents are the same as those found along
the $\lambda$ phase boundary given in equation \Ref{35}, namely
$\alpha_t=2$ from the $\tau_2$ side of the phase boundary, and
$\alpha_t=0.43(6)$ on the $\tau_1$ side.

Since the trajectory of the $\tau_1$ phase boundary and location of the
multicritical point are not known exactly, it is more challenging to estimate 
the scaling exponent $\alpha_u$.  Putting $\ell(\alpha)=(\alpha,\tau_1(\alpha))$ 
and denoting the distance along $\tau_1$ between $\ell(\alpha_c)$ and $\ell(\alpha)$ in 
the $\alpha\beta$-plane by $d(\alpha_c,\alpha)$, the exponent $\alpha_u$ can 
be estimated by calculating the ratio $\log \LA c_1 \RA_s / \log d(\alpha_c,\alpha)$ 
with $\alpha_c = -0.970$ along $\tau_1$ (see equations \Ref{21} and \Ref{22}).  
By the same arguments leading to equation \Ref{22}, we plot
\begin{equation}
\gamma = \log\LA c_1 \RA_s / \log d(\alpha_c,\alpha)
\label{37}  %ZXZ[37]
\end{equation}
as a function of $1/ \log d(\alpha_c,\alpha)$ and estimate the exponent
$1-\alpha_u$ by a linear fit. This gives the graph in figure \ref{20}.

\begin{figure}[h!]
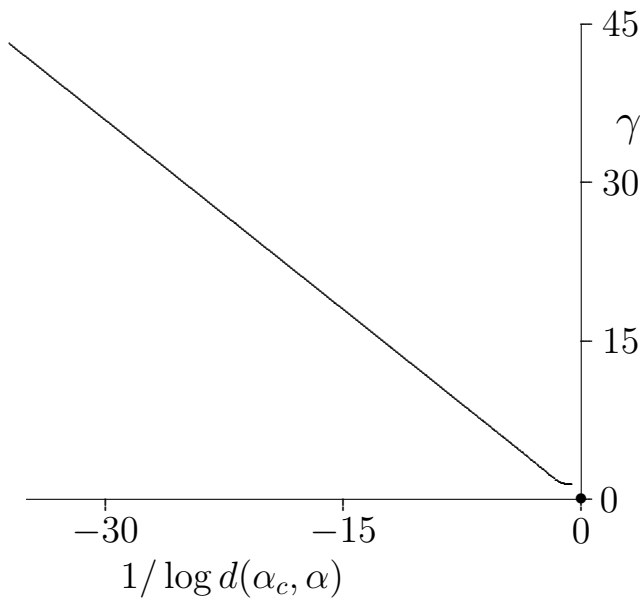

\beginpicture
\color{black}
\normalcolor
\setcoordinatesystem units <6pt,4pt>
\setplotarea x from -35 to 0, y from 0 to 45

\axis right shiftedto x=0 
        %ticks
        %withvalues 10  /
        %at  10  /
 /
\axis bottom shiftedto y=0
        %ticks
        %withvalues -30 -15 0 /
        %at  -30 -15 0 /
/
\put {\footnotesize$\bullet$} at 0 0 

\plot -30 -1 -30 0 / \plot -15 -1 -15 0 / \plot 0 -1 0 0 /
\put {\large${-}30$} at -30 -3
\put {\large${-}15$} at -15 -3
\put {\large$0$} at 0 -3
\plot 0 0 0.65 0 / \plot 0 15 0.65 15 / \plot 0 30 0.65 30 /  \plot 0 45 0.65 45 / 
\put {\large$0$} at 1.75 0  
\put {\large$15$} at 2.5 15  
\put {\large$30$} at 2.5 30 
\put {\large$45$} at 2.5 45 

\plot -0.5932 1.4192  -0.6267 1.4246  -0.6610 1.4295  -0.6961 1.4338  -0.7323 1.4378  -0.7696 1.4412  -0.8083 1.4444  -0.8484 1.4474  -0.8902 1.4508  -0.9337 1.4551  -0.9792 1.4611  -1.0269 1.4697  -1.0770 1.4820  -1.1297 1.4992  -1.1853 1.5225  -1.2440 1.5533  -1.3063 1.5927  -1.3725 1.6421  -1.4431 1.7023  -1.5184 1.7740  -1.5991 1.8577  -1.6859 1.9537  -1.7794 2.0621  -1.8805 2.1833  -1.9904 2.3176  -2.1101 2.4658  -2.2412 2.6290  -2.3854 2.8088  -2.5448 3.0073  -2.7221 3.2273  -2.9205 3.4725  -3.1441 3.7476  -3.3981 4.0586  -3.6892 4.4137  -4.0262 4.8232  -4.4213 5.3015  -4.8907 5.8683  -5.4579 6.5514  -6.1571 7.3919  -7.0408 8.4524  -8.1932 9.8336  -9.7594 11.7089  -12.0114 14.4033  -15.5268 18.6071  -21.7849 26.0878  -36.0526 43.1387
 /

\put {\Large$\gamma$} at 3 35
\put {\large$1/\log d(\alpha_c,\alpha)$} at -22 -7.5

\color{black}
\normalcolor
\endpicture
\caption{Plotting $\gamma 
= \log(\LA c_1 \RA- \LA c_1 \RA\!\vert_{crit})/ \log d(\alpha_c,\alpha)$ as
a function of $1/\log d(\alpha_c,\alpha)$ for the linked model along the trajectory 
of the $\tau_1$ phase boundary, starting in the multicritical point.  In this
graph $-0.9 \leq \alpha \leq -0.2$.}
\label{20}  %%ZXZ[20]
\end{figure}

A linear fit to the data in figure \ref{20} gives $1{-}\alpha_u \approx 0.172$ and
a quadratic fit gives $1{-}\alpha_u \approx 0.242$.  Taking the difference as an 
error bar gives the estimate $1{-}\alpha_u = 0.17 \pm 0.07$. Thus, our estimate
is
\begin{equation}
\alpha_u = 0.83 \pm 0.07,  \quad\hbox{along $\tau_1$}.
\label{38}    %ZXZ[38]
\end{equation} 
Comparing this with the estimate of $\alpha_t$ on the $c_1$-dominated
phase side of the $\lambda$ gives the estimated crossover exponent 
$\phi_1 = 1.3 \pm 0.2$, by equation \Ref{12}. 

Along the $\tau_2$ phase boundary the situation is simpler.  This phase
boundary is between the empty phase and the $c_2$-dominated phase,
and the free energy is continuous along $\tau_2$.  Since the free energy 
in the scaling limit a constant in the empty phase, it is also, by continuity,
a constant along the $\tau_2$ phase boundary in the scaling limit.  Thus,
one concludes that $\alpha_u=2$ along $\tau_2$, whereas the transverse 
exponent on the empty phase side of $\lambda$ is $\alpha_t=2$ as well.  
This gives the crossover exponent $\phi=1$ between the $\lambda$ 
and $\tau_2$ phase boundaries. 

\subsection{Finite size effects}

In the previous sections we have considered the phase diagrams for $L=20$.
In this section we briefly consider finite size effects in our data.  We do this
only for the linked model; finite size effects in the unlinked model are
smaller and phase boundaries do not move with increasing $L$.

\begin{figure}[h!]
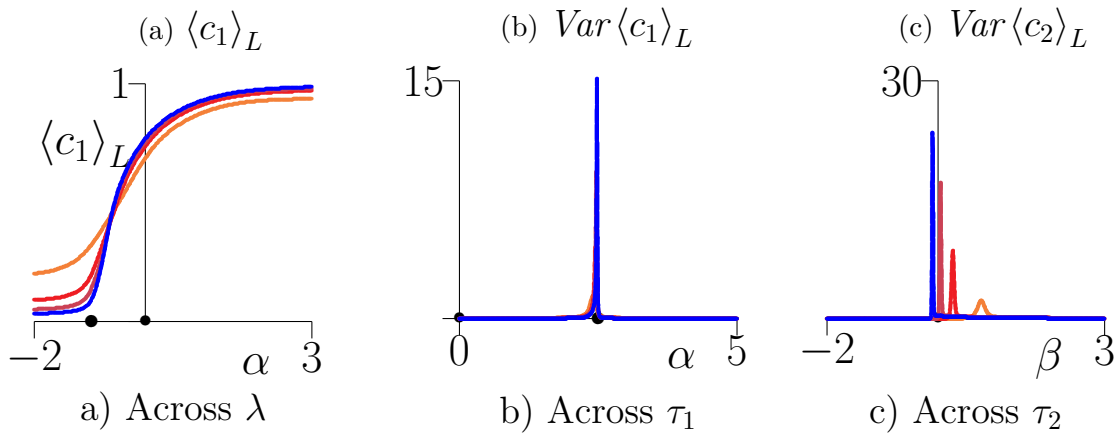

\begin{subfigure}{0.33\textwidth}
\normalcolor
\color{black}
	\caption{\large$\LA c_1 \RA_L$}
\beginpicture
\setcoordinatesystem units <21pt,90pt>
\axesnolabels{-2}{0}{0}{0}{3}{1}{0}
\put {\Large$-2$} at -2 -0.15
\put {\Large$3$} at 3 -0.15
\put {\Large$1$} at -0.45 1
\put {\Large$\alpha$} at 2.0 -0.18
\setplotsymbol ({\scalebox{0.3}{$\bullet$}})
\put {$\bullet$} at -0.97 0 
\color{Orange}
\plot -2.0000 0.2003  -1.9750 0.2010  -1.9500 0.2017  -1.9250 0.2025  -1.9000 0.2033  -1.8750 0.2042  -1.8500 0.2051  -1.8250 0.2060  -1.8000 0.2071  -1.7750 0.2082  -1.7500 0.2093  -1.7250 0.2105  -1.7000 0.2118  -1.6750 0.2132  -1.6500 0.2147  -1.6250 0.2162  -1.6000 0.2179  -1.5750 0.2196  -1.5500 0.2215  -1.5250 0.2235  -1.5000 0.2256  -1.4750 0.2279  -1.4500 0.2303  -1.4250 0.2328  -1.4000 0.2356  -1.3750 0.2385  -1.3500 0.2416  -1.3250 0.2449  -1.3000 0.2485  -1.2750 0.2522  -1.2500 0.2562  -1.2250 0.2605  -1.2000 0.2651  -1.1750 0.2699  -1.1500 0.2751  -1.1250 0.2806  -1.1000 0.2863  -1.0750 0.2924  -1.0500 0.2989  -1.0250 0.3056  -1.0000 0.3127  -0.9750 0.3200  -0.9500 0.3277  -0.9250 0.3356  -0.9000 0.3437  -0.8750 0.3521  -0.8500 0.3606  -0.8250 0.3693  -0.8000 0.3782  -0.7750 0.3872  -0.7500 0.3962  -0.7250 0.4053  -0.7000 0.4145  -0.6750 0.4237  -0.6500 0.4330  -0.6250 0.4424  -0.6000 0.4517  -0.5750 0.4612  -0.5500 0.4707  -0.5250 0.4802  -0.5000 0.4899  -0.4750 0.4996  -0.4500 0.5094  -0.4250 0.5193  -0.4000 0.5293  -0.3750 0.5394  -0.3500 0.5495  -0.3250 0.5597  -0.3000 0.5699  -0.2750 0.5802  -0.2500 0.5904  -0.2250 0.6006  -0.2000 0.6107  -0.1750 0.6207  -0.1500 0.6306  -0.1250 0.6403  -0.1000 0.6499  -0.0750 0.6592  -0.0500 0.6683  -0.0250 0.6772  0.0000 0.6858  0.0250 0.6942  0.0500 0.7023  0.0750 0.7102  0.1000 0.7178  0.1250 0.7251  0.1500 0.7322  0.1750 0.7391  0.2000 0.7457  0.2250 0.7520  0.2500 0.7582  0.2750 0.7641  0.3000 0.7698  0.3250 0.7754  0.3500 0.7807  0.3750 0.7859  0.4000 0.7909  0.4250 0.7957  0.4500 0.8004  0.4750 0.8049  0.5000 0.8093  0.5250 0.8135  0.5500 0.8176  0.5750 0.8216  0.6000 0.8255  0.6250 0.8292  0.6500 0.8328  0.6750 0.8364  0.7000 0.8398  0.7250 0.8431  0.7500 0.8463  0.7750 0.8495  0.8000 0.8525  0.8250 0.8555  0.8500 0.8583  0.8750 0.8611  0.9000 0.8638  0.9250 0.8665  0.9500 0.8690  0.9750 0.8715  1.0000 0.8739  1.0250 0.8763  1.0500 0.8786  1.0750 0.8808  1.1000 0.8829  1.1250 0.8850  1.1500 0.8870  1.1750 0.8890  1.2000 0.8909  1.2250 0.8927  1.2500 0.8945  1.2750 0.8963  1.3000 0.8979  1.3250 0.8995  1.3500 0.9011  1.3750 0.9026  1.4000 0.9040  1.4250 0.9055  1.4500 0.9068  1.4750 0.9081  1.5000 0.9094  1.5250 0.9106  1.5500 0.9117  1.5750 0.9129  1.6000 0.9139  1.6250 0.9150  1.6500 0.9160  1.6750 0.9169  1.7000 0.9178  1.7250 0.9187  1.7500 0.9196  1.7750 0.9204  1.8000 0.9211  1.8250 0.9219  1.8500 0.9226  1.8750 0.9233  1.9000 0.9239  1.9250 0.9246  1.9500 0.9252  1.9750 0.9257  2.0000 0.9263  2.0250 0.9268  2.0500 0.9273  2.0750 0.9278  2.1000 0.9282  2.1250 0.9287  2.1500 0.9291  2.1750 0.9295  2.2000 0.9299  2.2250 0.9302  2.2500 0.9306  2.2750 0.9309  2.3000 0.9312  2.3250 0.9315  2.3500 0.9318  2.3750 0.9321  2.4000 0.9323  2.4250 0.9326  2.4500 0.9328  2.4750 0.9330  2.5000 0.9332  2.5250 0.9335  2.5500 0.9336  2.5750 0.9338  2.6000 0.9340  2.6250 0.9342  2.6500 0.9343  2.6750 0.9345  2.7000 0.9346  2.7250 0.9348  2.7500 0.9349  2.7750 0.9350  2.8000 0.9351  2.8250 0.9353  2.8500 0.9354  2.8750 0.9355  2.9000 0.9356  2.9250 0.9357  2.9500 0.9358  2.9750 0.9358  3.0000 0.9359  /
\color{Red}
\plot -2.0000 0.0895  -1.9750 0.0898  -1.9500 0.0902  -1.9250 0.0906  -1.9000 0.0910  -1.8750 0.0914  -1.8500 0.0919  -1.8250 0.0924  -1.8000 0.0929  -1.7750 0.0934  -1.7500 0.0940  -1.7250 0.0947  -1.7000 0.0953  -1.6750 0.0960  -1.6500 0.0968  -1.6250 0.0976  -1.6000 0.0985  -1.5750 0.0995  -1.5500 0.1005  -1.5250 0.1015  -1.5000 0.1027  -1.4750 0.1040  -1.4500 0.1054  -1.4250 0.1069  -1.4000 0.1085  -1.3750 0.1102  -1.3500 0.1122  -1.3250 0.1143  -1.3000 0.1166  -1.2750 0.1192  -1.2500 0.1221  -1.2250 0.1253  -1.2000 0.1289  -1.1750 0.1329  -1.1500 0.1375  -1.1250 0.1427  -1.1000 0.1487  -1.0750 0.1555  -1.0500 0.1633  -1.0250 0.1722  -1.0000 0.1823  -0.9750 0.1937  -0.9500 0.2064  -0.9250 0.2203  -0.9000 0.2352  -0.8750 0.2508  -0.8500 0.2670  -0.8250 0.2835  -0.8000 0.3001  -0.7750 0.3167  -0.7500 0.3334  -0.7250 0.3501  -0.7000 0.3670  -0.6750 0.3840  -0.6500 0.4012  -0.6250 0.4186  -0.6000 0.4361  -0.5750 0.4536  -0.5500 0.4710  -0.5250 0.4881  -0.5000 0.5049  -0.4750 0.5211  -0.4500 0.5368  -0.4250 0.5520  -0.4000 0.5665  -0.3750 0.5805  -0.3500 0.5939  -0.3250 0.6068  -0.3000 0.6191  -0.2750 0.6310  -0.2500 0.6424  -0.2250 0.6534  -0.2000 0.6639  -0.1750 0.6741  -0.1500 0.6839  -0.1250 0.6933  -0.1000 0.7025  -0.0750 0.7112  -0.0500 0.7197  -0.0250 0.7279  0.0000 0.7358  0.0250 0.7435  0.0500 0.7509  0.0750 0.7580  0.1000 0.7649  0.1250 0.7715  0.1500 0.7780  0.1750 0.7842  0.2000 0.7902  0.2250 0.7961  0.2500 0.8017  0.2750 0.8072  0.3000 0.8125  0.3250 0.8176  0.3500 0.8225  0.3750 0.8273  0.4000 0.8320  0.4250 0.8365  0.4500 0.8408  0.4750 0.8450  0.5000 0.8491  0.5250 0.8531  0.5500 0.8569  0.5750 0.8606  0.6000 0.8642  0.6250 0.8677  0.6500 0.8711  0.6750 0.8744  0.7000 0.8775  0.7250 0.8806  0.7500 0.8836  0.7750 0.8865  0.8000 0.8893  0.8250 0.8920  0.8500 0.8946  0.8750 0.8972  0.9000 0.8997  0.9250 0.9021  0.9500 0.9044  0.9750 0.9066  1.0000 0.9088  1.0250 0.9109  1.0500 0.9130  1.0750 0.9150  1.1000 0.9169  1.1250 0.9188  1.1500 0.9206  1.1750 0.9224  1.2000 0.9241  1.2250 0.9258  1.2500 0.9274  1.2750 0.9289  1.3000 0.9304  1.3250 0.9319  1.3500 0.9333  1.3750 0.9347  1.4000 0.9361  1.4250 0.9374  1.4500 0.9386  1.4750 0.9399  1.5000 0.9410  1.5250 0.9422  1.5500 0.9433  1.5750 0.9444  1.6000 0.9454  1.6250 0.9465  1.6500 0.9474  1.6750 0.9484  1.7000 0.9493  1.7250 0.9502  1.7500 0.9511  1.7750 0.9519  1.8000 0.9527  1.8250 0.9535  1.8500 0.9542  1.8750 0.9550  1.9000 0.9557  1.9250 0.9564  1.9500 0.9570  1.9750 0.9576  2.0000 0.9583  2.0250 0.9588  2.0500 0.9594  2.0750 0.9600  2.1000 0.9605  2.1250 0.9610  2.1500 0.9615  2.1750 0.9619  2.2000 0.9624  2.2250 0.9628  2.2500 0.9632  2.2750 0.9636  2.3000 0.9640  2.3250 0.9644  2.3500 0.9647  2.3750 0.9651  2.4000 0.9654  2.4250 0.9657  2.4500 0.9660  2.4750 0.9663  2.5000 0.9665  2.5250 0.9668  2.5500 0.9670  2.5750 0.9673  2.6000 0.9675  2.6250 0.9677  2.6500 0.9679  2.6750 0.9681  2.7000 0.9683  2.7250 0.9685  2.7500 0.9687  2.7750 0.9689  2.8000 0.9690  2.8250 0.9692  2.8500 0.9693  2.8750 0.9694  2.9000 0.9696  2.9250 0.9697  2.9500 0.9698  2.9750 0.9699  3.0000 0.9700 /
\color{BrickRed}
\plot -2.0000 0.0504  -1.9750 0.0506  -1.9500 0.0508  -1.9250 0.0511  -1.9000 0.0513  -1.8750 0.0516  -1.8500 0.0518  -1.8250 0.0521  -1.8000 0.0524  -1.7750 0.0527  -1.7500 0.0531  -1.7250 0.0535  -1.7000 0.0539  -1.6750 0.0543  -1.6500 0.0547  -1.6250 0.0552  -1.6000 0.0557  -1.5750 0.0563  -1.5500 0.0569  -1.5250 0.0576  -1.5000 0.0583  -1.4750 0.0590  -1.4500 0.0598  -1.4250 0.0607  -1.4000 0.0617  -1.3750 0.0628  -1.3500 0.0640  -1.3250 0.0653  -1.3000 0.0668  -1.2750 0.0684  -1.2500 0.0703  -1.2250 0.0724  -1.2000 0.0749  -1.1750 0.0777  -1.1500 0.0810  -1.1250 0.0850  -1.1000 0.0898  -1.0750 0.0957  -1.0500 0.1029  -1.0250 0.1117  -1.0000 0.1223  -0.9750 0.1348  -0.9500 0.1489  -0.9250 0.1643  -0.9000 0.1806  -0.8750 0.1974  -0.8500 0.2147  -0.8250 0.2328  -0.8000 0.2521  -0.7750 0.2731  -0.7500 0.2960  -0.7250 0.3208  -0.7000 0.3470  -0.6750 0.3737  -0.6500 0.4000  -0.6250 0.4250  -0.6000 0.4485  -0.5750 0.4703  -0.5500 0.4906  -0.5250 0.5095  -0.5000 0.5274  -0.4750 0.5443  -0.4500 0.5603  -0.4250 0.5756  -0.4000 0.5900  -0.3750 0.6038  -0.3500 0.6169  -0.3250 0.6294  -0.3000 0.6413  -0.2750 0.6527  -0.2500 0.6636  -0.2250 0.6741  -0.2000 0.6841  -0.1750 0.6938  -0.1500 0.7032  -0.1250 0.7122  -0.1000 0.7209  -0.0750 0.7294  -0.0500 0.7375  -0.0250 0.7454  0.0000 0.7530  0.0250 0.7603  0.0500 0.7674  0.0750 0.7743  0.1000 0.7810  0.1250 0.7874  0.1500 0.7937  0.1750 0.7997  0.2000 0.8056  0.2250 0.8112  0.2500 0.8167  0.2750 0.8220  0.3000 0.8272  0.3250 0.8322  0.3500 0.8370  0.3750 0.8417  0.4000 0.8462  0.4250 0.8506  0.4500 0.8549  0.4750 0.8590  0.5000 0.8630  0.5250 0.8669  0.5500 0.8706  0.5750 0.8742  0.6000 0.8778  0.6250 0.8812  0.6500 0.8845  0.6750 0.8877  0.7000 0.8908  0.7250 0.8938  0.7500 0.8968  0.7750 0.8996  0.8000 0.9024  0.8250 0.9051  0.8500 0.9077  0.8750 0.9102  0.9000 0.9127  0.9250 0.9151  0.9500 0.9174  0.9750 0.9197  1.0000 0.9219  1.0250 0.9240  1.0500 0.9261  1.0750 0.9281  1.1000 0.9300  1.1250 0.9319  1.1500 0.9338  1.1750 0.9356  1.2000 0.9373  1.2250 0.9390  1.2500 0.9406  1.2750 0.9422  1.3000 0.9437  1.3250 0.9452  1.3500 0.9466  1.3750 0.9479  1.4000 0.9493  1.4250 0.9505  1.4500 0.9517  1.4750 0.9529  1.5000 0.9541  1.5250 0.9551  1.5500 0.9562  1.5750 0.9572  1.6000 0.9581  1.6250 0.9591  1.6500 0.9599  1.6750 0.9608  1.7000 0.9616  1.7250 0.9624  1.7500 0.9631  1.7750 0.9638  1.8000 0.9645  1.8250 0.9652  1.8500 0.9658  1.8750 0.9664  1.9000 0.9670  1.9250 0.9675  1.9500 0.9681  1.9750 0.9686  2.0000 0.9691  2.0250 0.9695  2.0500 0.9700  2.0750 0.9704  2.1000 0.9708  2.1250 0.9712  2.1500 0.9716  2.1750 0.9720  2.2000 0.9724  2.2250 0.9727  2.2500 0.9731  2.2750 0.9734  2.3000 0.9737  2.3250 0.9740  2.3500 0.9743  2.3750 0.9746  2.4000 0.9749  2.4250 0.9751  2.4500 0.9754  2.4750 0.9756  2.5000 0.9759  2.5250 0.9761  2.5500 0.9764  2.5750 0.9766  2.6000 0.9768  2.6250 0.9770  2.6500 0.9773  2.6750 0.9775  2.7000 0.9777  2.7250 0.9779  2.7500 0.9781  2.7750 0.9783  2.8000 0.9784  2.8250 0.9786  2.8500 0.9788  2.8750 0.9790  2.9000 0.9792  2.9250 0.9793  2.9500 0.9795  2.9750 0.9796  3.0000 0.9798 /
\color{Blue}
\plot -2.0000 0.0323  -1.9750 0.0324  -1.9500 0.0326  -1.9250 0.0327  -1.9000 0.0329  -1.8750 0.0330  -1.8500 0.0332  -1.8250 0.0334  -1.8000 0.0336  -1.7750 0.0338  -1.7500 0.0340  -1.7250 0.0343  -1.7000 0.0345  -1.6750 0.0348  -1.6500 0.0351  -1.6250 0.0354  -1.6000 0.0358  -1.5750 0.0361  -1.5500 0.0365  -1.5250 0.0370  -1.5000 0.0374  -1.4750 0.0379  -1.4500 0.0385  -1.4250 0.0391  -1.4000 0.0397  -1.3750 0.0404  -1.3500 0.0412  -1.3250 0.0421  -1.3000 0.0431  -1.2750 0.0442  -1.2500 0.0454  -1.2250 0.0468  -1.2000 0.0484  -1.1750 0.0503  -1.1500 0.0525  -1.1250 0.0550  -1.1000 0.0581  -1.0750 0.0618  -1.0500 0.0663  -1.0250 0.0720  -1.0000 0.0791  -0.9750 0.0882  -0.9500 0.0995  -0.9250 0.1135  -0.9000 0.1303  -0.8750 0.1496  -0.8500 0.1712  -0.8250 0.1948  -0.8000 0.2201  -0.7750 0.2472  -0.7500 0.2761  -0.7250 0.3065  -0.7000 0.3376  -0.6750 0.3686  -0.6500 0.3984  -0.6250 0.4267  -0.6000 0.4531  -0.5750 0.4774  -0.5500 0.4998  -0.5250 0.5204  -0.5000 0.5392  -0.4750 0.5567  -0.4500 0.5732  -0.4250 0.5887  -0.4000 0.6034  -0.3750 0.6175  -0.3500 0.6309  -0.3250 0.6436  -0.3000 0.6558  -0.2750 0.6674  -0.2500 0.6785  -0.2250 0.6890  -0.2000 0.6992  -0.1750 0.7089  -0.1500 0.7182  -0.1250 0.7272  -0.1000 0.7358  -0.0750 0.7441  -0.0500 0.7520  -0.0250 0.7597  0.0000 0.7671  0.0250 0.7742  0.0500 0.7810  0.0750 0.7876  0.1000 0.7940  0.1250 0.8001  0.1500 0.8060  0.1750 0.8117  0.2000 0.8172  0.2250 0.8226  0.2500 0.8277  0.2750 0.8327  0.3000 0.8375  0.3250 0.8421  0.3500 0.8466  0.3750 0.8510  0.4000 0.8552  0.4250 0.8593  0.4500 0.8632  0.4750 0.8671  0.5000 0.8708  0.5250 0.8744  0.5500 0.8779  0.5750 0.8813  0.6000 0.8846  0.6250 0.8878  0.6500 0.8910  0.6750 0.8940  0.7000 0.8969  0.7250 0.8998  0.7500 0.9025  0.7750 0.9052  0.8000 0.9078  0.8250 0.9103  0.8500 0.9128  0.8750 0.9152  0.9000 0.9175  0.9250 0.9197  0.9500 0.9219  0.9750 0.9240  1.0000 0.9261  1.0250 0.9281  1.0500 0.9301  1.0750 0.9320  1.1000 0.9338  1.1250 0.9356  1.1500 0.9373  1.1750 0.9390  1.2000 0.9406  1.2250 0.9422  1.2500 0.9438  1.2750 0.9453  1.3000 0.9467  1.3250 0.9481  1.3500 0.9495  1.3750 0.9508  1.4000 0.9521  1.4250 0.9534  1.4500 0.9546  1.4750 0.9558  1.5000 0.9569  1.5250 0.9580  1.5500 0.9591  1.5750 0.9601  1.6000 0.9611  1.6250 0.9620  1.6500 0.9630  1.6750 0.9639  1.7000 0.9647  1.7250 0.9656  1.7500 0.9664  1.7750 0.9672  1.8000 0.9679  1.8250 0.9687  1.8500 0.9694  1.8750 0.9701  1.9000 0.9707  1.9250 0.9714  1.9500 0.9720  1.9750 0.9726  2.0000 0.9731  2.0250 0.9737  2.0500 0.9742  2.0750 0.9748  2.1000 0.9753  2.1250 0.9757  2.1500 0.9762  2.1750 0.9767  2.2000 0.9771  2.2250 0.9775  2.2500 0.9780  2.2750 0.9784  2.3000 0.9787  2.3250 0.9791  2.3500 0.9795  2.3750 0.9798  2.4000 0.9802  2.4250 0.9805  2.4500 0.9808  2.4750 0.9811  2.5000 0.9814  2.5250 0.9817  2.5500 0.9820  2.5750 0.9823  2.6000 0.9826  2.6250 0.9828  2.6500 0.9831  2.6750 0.9833  2.7000 0.9836  2.7250 0.9838  2.7500 0.9840  2.7750 0.9842  2.8000 0.9844  2.8250 0.9846  2.8500 0.9848  2.8750 0.9850  2.9000 0.9852  2.9250 0.9854  2.9500 0.9856  2.9750 0.9857  3.0000 0.9859 /
\color{Black}
\put {\Large$\LA c_1 \RA_L$} at -1.1 0.75
\put {\large a) Across $\lambda$} at 0.5 -0.375
\endpicture
\end{subfigure}
\normalcolor
\color{Black}
\begin{subfigure}{0.33\textwidth}
\normalcolor
\color{black}
	\caption{\large$\hbox{Var}\LA c_1 \RA_L$}
\beginpicture
\setcoordinatesystem units <21pt,6pt>
\normalcolor
\color{Black}
\axesnolabels{0}{0}{0}{0}{5}{15}{0}
\put {\Large$0$} at 0 -2
\put {\Large$5$} at 5 -2
\put {\Large$15$} at -0.65 15
\put {\Large$\alpha$} at 4.0 -2.5
\setplotsymbol ({\scalebox{0.3}{$\bullet$}})
\put {$\bullet$} at 2.5 0 
\color{black}
\normalcolor
\color{Orange}
\plot 0.0000 0.0000  0.0100 0.0000  0.0200 0.0000  0.0300 0.0000  0.0400 0.0000  0.0500 0.0000  0.0600 0.0000  0.0700 0.0000  0.0800 0.0000  0.0900 0.0000  0.1000 0.0000  0.1100 0.0000  0.1200 0.0000  0.1300 0.0000  0.1400 0.0000  0.1500 0.0000  0.1600 0.0000  0.1700 0.0000  0.1800 0.0000  0.1900 0.0000  0.2000 0.0000  0.2100 0.0000  0.2200 0.0000  0.2300 0.0000  0.2400 0.0000  0.2500 0.0000  0.2600 0.0000  0.2700 0.0000  0.2800 0.0000  0.2900 0.0000  0.3000 0.0000  0.3100 0.0000  0.3200 0.0000  0.3300 0.0000  0.3400 0.0000  0.3500 0.0000  0.3600 0.0000  0.3700 0.0000  0.3800 0.0000  0.3900 0.0000  0.4000 0.0000  0.4100 0.0000  0.4200 0.0000  0.4300 0.0001  0.4400 0.0001  0.4500 0.0001  0.4600 0.0001  0.4700 0.0001  0.4800 0.0001  0.4900 0.0001  0.5000 0.0001  0.5100 0.0001  0.5200 0.0001  0.5300 0.0001  0.5400 0.0001  0.5500 0.0001  0.5600 0.0001  0.5700 0.0001  0.5800 0.0001  0.5900 0.0001  0.6000 0.0001  0.6100 0.0001  0.6200 0.0001  0.6300 0.0001  0.6400 0.0001  0.6500 0.0001  0.6600 0.0001  0.6700 0.0001  0.6800 0.0001  0.6900 0.0001  0.7000 0.0002  0.7100 0.0002  0.7200 0.0002  0.7300 0.0002  0.7400 0.0002  0.7500 0.0002  0.7600 0.0002  0.7700 0.0002  0.7800 0.0002  0.7900 0.0002  0.8000 0.0002  0.8100 0.0002  0.8200 0.0002  0.8300 0.0003  0.8400 0.0003  0.8500 0.0003  0.8600 0.0003  0.8700 0.0003  0.8800 0.0003  0.8900 0.0003  0.9000 0.0003  0.9100 0.0004  0.9200 0.0004  0.9300 0.0004  0.9400 0.0004  0.9500 0.0004  0.9600 0.0004  0.9700 0.0004  0.9800 0.0005  0.9900 0.0005  1.0000 0.0005  1.0100 0.0005  1.0200 0.0005  1.0300 0.0006  1.0400 0.0006  1.0500 0.0006  1.0600 0.0006  1.0700 0.0007  1.0800 0.0007  1.0900 0.0007  1.1000 0.0008  1.1100 0.0008  1.1200 0.0008  1.1300 0.0008  1.1400 0.0009  1.1500 0.0009  1.1600 0.0010  1.1700 0.0010  1.1800 0.0010  1.1900 0.0011  1.2000 0.0011  1.2100 0.0012  1.2200 0.0012  1.2300 0.0013  1.2400 0.0013  1.2500 0.0014  1.2600 0.0014  1.2700 0.0015  1.2800 0.0016  1.2900 0.0016  1.3000 0.0017  1.3100 0.0018  1.3200 0.0018  1.3300 0.0019  1.3400 0.0020  1.3500 0.0021  1.3600 0.0022  1.3700 0.0022  1.3800 0.0023  1.3900 0.0024  1.4000 0.0025  1.4100 0.0026  1.4200 0.0028  1.4300 0.0029  1.4400 0.0030  1.4500 0.0031  1.4600 0.0033  1.4700 0.0034  1.4800 0.0035  1.4900 0.0037  1.5000 0.0038  1.5100 0.0040  1.5200 0.0042  1.5300 0.0044  1.5400 0.0045  1.5500 0.0047  1.5600 0.0049  1.5700 0.0051  1.5800 0.0054  1.5900 0.0056  1.6000 0.0058  1.6100 0.0061  1.6200 0.0063  1.6300 0.0066  1.6400 0.0069  1.6500 0.0072  1.6600 0.0075  1.6700 0.0079  1.6800 0.0082  1.6900 0.0086  1.7000 0.0089  1.7100 0.0093  1.7200 0.0098  1.7300 0.0102  1.7400 0.0107  1.7500 0.0111  1.7600 0.0116  1.7700 0.0122  1.7800 0.0127  1.7900 0.0133  1.8000 0.0139  1.8100 0.0146  1.8200 0.0153  1.8300 0.0160  1.8400 0.0167  1.8500 0.0175  1.8600 0.0184  1.8700 0.0192  1.8800 0.0202  1.8900 0.0212  1.9000 0.0222  1.9100 0.0233  1.9200 0.0245  1.9300 0.0257  1.9400 0.0270  1.9500 0.0284  1.9600 0.0299  1.9700 0.0315  1.9800 0.0331  1.9900 0.0349  2.0000 0.0368  2.0100 0.0388  2.0200 0.0410  2.0300 0.0433  2.0400 0.0458  2.0500 0.0485  2.0600 0.0513  2.0700 0.0544  2.0800 0.0577  2.0900 0.0613  2.1000 0.0652  2.1100 0.0694  2.1200 0.0740  2.1300 0.0790  2.1400 0.0844  2.1500 0.0904  2.1600 0.0970  2.1700 0.1043  2.1800 0.1123  2.1900 0.1212  2.2000 0.1312  2.2100 0.1424  2.2200 0.1551  2.2300 0.1695  2.2400 0.1860  2.2500 0.2051  2.2600 0.2274  2.2700 0.2537  2.2800 0.2849  2.2900 0.3225  2.3000 0.3682  2.3100 0.4240  2.3200 0.4924  2.3300 0.5759  2.3400 0.6765  2.3500 0.7948  2.3600 0.9284  2.3700 1.0704  2.3800 1.2077  2.3900 1.3224  2.4000 1.3950  2.4100 1.4105  2.4200 1.3642  2.4300 1.2638  2.4400 1.1265  2.4500 0.9723  2.4600 0.8188  2.4700 0.6779  2.4800 0.5556  2.4900 0.4535  2.5000 0.3703  2.5100 0.3036  2.5200 0.2506  2.5300 0.2086  2.5400 0.1753  2.5500 0.1487  2.5600 0.1275  2.5700 0.1103  2.5800 0.0963  2.5900 0.0848  2.6000 0.0752  2.6100 0.0673  2.6200 0.0605  2.6300 0.0548  2.6400 0.0499  2.6500 0.0457  2.6600 0.0420  2.6700 0.0387  2.6800 0.0359  2.6900 0.0334  2.7000 0.0311  2.7100 0.0291  2.7200 0.0273  2.7300 0.0256  2.7400 0.0241  2.7500 0.0227  2.7600 0.0215  2.7700 0.0203  2.7800 0.0193  2.7900 0.0183  2.8000 0.0174  2.8100 0.0166  2.8200 0.0158  2.8300 0.0151  2.8400 0.0144  2.8500 0.0137  2.8600 0.0131  2.8700 0.0126  2.8800 0.0121  2.8900 0.0116  2.9000 0.0111  2.9100 0.0106  2.9200 0.0102  2.9300 0.0098  2.9400 0.0095  2.9500 0.0091  2.9600 0.0088  2.9700 0.0084  2.9800 0.0081  2.9900 0.0078  3.0000 0.0075  3.0100 0.0073  3.0200 0.0070  3.0300 0.0068  3.0400 0.0065  3.0500 0.0063  3.0600 0.0061  3.0700 0.0059  3.0800 0.0057  3.0900 0.0055  3.1000 0.0053  3.1100 0.0052  3.1200 0.0050  3.1300 0.0048  3.1400 0.0047  3.1500 0.0045  3.1600 0.0044  3.1700 0.0043  3.1800 0.0041  3.1900 0.0040  3.2000 0.0039  3.2100 0.0038  3.2200 0.0037  3.2300 0.0036  3.2400 0.0034  3.2500 0.0033  3.2600 0.0032  3.2700 0.0032  3.2800 0.0031  3.2900 0.0030  3.3000 0.0029  3.3100 0.0028  3.3200 0.0027  3.3300 0.0027  3.3400 0.0026  3.3500 0.0025  3.3600 0.0024  3.3700 0.0024  3.3800 0.0023  3.3900 0.0022  3.4000 0.0022  3.4100 0.0021  3.4200 0.0021  3.4300 0.0020  3.4400 0.0020  3.4500 0.0019  3.4600 0.0019  3.4700 0.0018  3.4800 0.0018  3.4900 0.0017  3.5000 0.0017  3.5100 0.0016  3.5200 0.0016  3.5300 0.0015  3.5400 0.0015  3.5500 0.0015  3.5600 0.0014  3.5700 0.0014  3.5800 0.0014  3.5900 0.0013  3.6000 0.0013  3.6100 0.0013  3.6200 0.0012  3.6300 0.0012  3.6400 0.0012  3.6500 0.0011  3.6600 0.0011  3.6700 0.0011  3.6800 0.0011  3.6900 0.0010  3.7000 0.0010  3.7100 0.0010  3.7200 0.0010  3.7300 0.0009  3.7400 0.0009  3.7500 0.0009  3.7600 0.0009  3.7700 0.0008  3.7800 0.0008  3.7900 0.0008  3.8000 0.0008  3.8100 0.0008  3.8200 0.0008  3.8300 0.0007  3.8400 0.0007  3.8500 0.0007  3.8600 0.0007  3.8700 0.0007  3.8800 0.0007  3.8900 0.0006  3.9000 0.0006  3.9100 0.0006  3.9200 0.0006  3.9300 0.0006  3.9400 0.0006  3.9500 0.0006  3.9600 0.0005  3.9700 0.0005  3.9800 0.0005  3.9900 0.0005  4.0000 0.0005  4.0100 0.0005  4.0200 0.0005  4.0300 0.0005  4.0400 0.0005  4.0500 0.0004  4.0600 0.0004  4.0700 0.0004  4.0800 0.0004  4.0900 0.0004  4.1000 0.0004  4.1100 0.0004  4.1200 0.0004  4.1300 0.0004  4.1400 0.0004  4.1500 0.0004  4.1600 0.0003  4.1700 0.0003  4.1800 0.0003  4.1900 0.0003  4.2000 0.0003  4.2100 0.0003  4.2200 0.0003  4.2300 0.0003  4.2400 0.0003  4.2500 0.0003  4.2600 0.0003  4.2700 0.0003  4.2800 0.0003  4.2900 0.0003  4.3000 0.0003  4.3100 0.0003  4.3200 0.0002  4.3300 0.0002  4.3400 0.0002  4.3500 0.0002  4.3600 0.0002  4.3700 0.0002  4.3800 0.0002  4.3900 0.0002  4.4000 0.0002  4.4100 0.0002  4.4200 0.0002  4.4300 0.0002  4.4400 0.0002  4.4500 0.0002  4.4600 0.0002  4.4700 0.0002  4.4800 0.0002  4.4900 0.0002  4.5000 0.0002  4.5100 0.0002  4.5200 0.0002  4.5300 0.0002  4.5400 0.0002  4.5500 0.0002  4.5600 0.0001  4.5700 0.0001  4.5800 0.0001  4.5900 0.0001  4.6000 0.0001  4.6100 0.0001  4.6200 0.0001  4.6300 0.0001  4.6400 0.0001  4.6500 0.0001  4.6600 0.0001  4.6700 0.0001  4.6800 0.0001  4.6900 0.0001  4.7000 0.0001  4.7100 0.0001  4.7200 0.0001  4.7300 0.0001  4.7400 0.0001  4.7500 0.0001  4.7600 0.0001  4.7700 0.0001  4.7800 0.0001  4.7900 0.0001  4.8000 0.0001  4.8100 0.0001  4.8200 0.0001  4.8300 0.0001  4.8400 0.0001  4.8500 0.0001  4.8600 0.0001  4.8700 0.0001  4.8800 0.0001  4.8900 0.0001  4.9000 0.0001  4.9100 0.0001  4.9200 0.0001  4.9300 0.0001  4.9400 0.0001  4.9500 0.0001  4.9600 0.0001  4.9700 0.0001  4.9800 0.0001  4.9900 0.0001  5.0000 0.0001     /
\color{black}
\normalcolor
\color{Red}
\plot 0.0000 0.0000  0.0100 0.0000  0.0200 0.0000  0.0300 0.0000  0.0400 0.0000  0.0500 0.0000  0.0600 0.0000  0.0700 0.0000  0.0800 0.0000  0.0900 0.0000  0.1000 0.0000  0.1100 0.0000  0.1200 0.0000  0.1300 0.0000  0.1400 0.0000  0.1500 0.0000  0.1600 0.0000  0.1700 0.0000  0.1800 0.0000  0.1900 0.0000  0.2000 0.0000  0.2100 0.0000  0.2200 0.0000  0.2300 0.0000  0.2400 0.0000  0.2500 0.0000  0.2600 0.0000  0.2700 0.0000  0.2800 0.0000  0.2900 0.0000  0.3000 0.0000  0.3100 0.0000  0.3200 0.0000  0.3300 0.0000  0.3400 0.0000  0.3500 0.0000  0.3600 0.0000  0.3700 0.0000  0.3800 0.0000  0.3900 0.0000  0.4000 0.0000  0.4100 0.0000  0.4200 0.0000  0.4300 0.0000  0.4400 0.0000  0.4500 0.0000  0.4600 0.0000  0.4700 0.0000  0.4800 0.0000  0.4900 0.0000  0.5000 0.0000  0.5100 0.0000  0.5200 0.0001  0.5300 0.0001  0.5400 0.0001  0.5500 0.0001  0.5600 0.0001  0.5700 0.0001  0.5800 0.0001  0.5900 0.0001  0.6000 0.0001  0.6100 0.0001  0.6200 0.0001  0.6300 0.0001  0.6400 0.0001  0.6500 0.0001  0.6600 0.0001  0.6700 0.0001  0.6800 0.0001  0.6900 0.0001  0.7000 0.0001  0.7100 0.0001  0.7200 0.0001  0.7300 0.0001  0.7400 0.0001  0.7500 0.0001  0.7600 0.0001  0.7700 0.0001  0.7800 0.0001  0.7900 0.0001  0.8000 0.0002  0.8100 0.0002  0.8200 0.0002  0.8300 0.0002  0.8400 0.0002  0.8500 0.0002  0.8600 0.0002  0.8700 0.0002  0.8800 0.0002  0.8900 0.0002  0.9000 0.0002  0.9100 0.0002  0.9200 0.0003  0.9300 0.0003  0.9400 0.0003  0.9500 0.0003  0.9600 0.0003  0.9700 0.0003  0.9800 0.0003  0.9900 0.0003  1.0000 0.0003  1.0100 0.0004  1.0200 0.0004  1.0300 0.0004  1.0400 0.0004  1.0500 0.0004  1.0600 0.0004  1.0700 0.0005  1.0800 0.0005  1.0900 0.0005  1.1000 0.0005  1.1100 0.0005  1.1200 0.0006  1.1300 0.0006  1.1400 0.0006  1.1500 0.0006  1.1600 0.0007  1.1700 0.0007  1.1800 0.0007  1.1900 0.0007  1.2000 0.0008  1.2100 0.0008  1.2200 0.0008  1.2300 0.0009  1.2400 0.0009  1.2500 0.0010  1.2600 0.0010  1.2700 0.0010  1.2800 0.0011  1.2900 0.0011  1.3000 0.0012  1.3100 0.0012  1.3200 0.0013  1.3300 0.0013  1.3400 0.0014  1.3500 0.0014  1.3600 0.0015  1.3700 0.0015  1.3800 0.0016  1.3900 0.0017  1.4000 0.0017  1.4100 0.0018  1.4200 0.0019  1.4300 0.0020  1.4400 0.0021  1.4500 0.0021  1.4600 0.0022  1.4700 0.0023  1.4800 0.0024  1.4900 0.0025  1.5000 0.0026  1.5100 0.0027  1.5200 0.0029  1.5300 0.0030  1.5400 0.0031  1.5500 0.0032  1.5600 0.0034  1.5700 0.0035  1.5800 0.0037  1.5900 0.0038  1.6000 0.0040  1.6100 0.0041  1.6200 0.0043  1.6300 0.0045  1.6400 0.0047  1.6500 0.0049  1.6600 0.0051  1.6700 0.0053  1.6800 0.0055  1.6900 0.0058  1.7000 0.0060  1.7100 0.0063  1.7200 0.0066  1.7300 0.0069  1.7400 0.0072  1.7500 0.0075  1.7600 0.0078  1.7700 0.0081  1.7800 0.0085  1.7900 0.0089  1.8000 0.0093  1.8100 0.0097  1.8200 0.0101  1.8300 0.0106  1.8400 0.0111  1.8500 0.0116  1.8600 0.0121  1.8700 0.0127  1.8800 0.0133  1.8900 0.0139  1.9000 0.0146  1.9100 0.0153  1.9200 0.0160  1.9300 0.0168  1.9400 0.0176  1.9500 0.0185  1.9600 0.0195  1.9700 0.0204  1.9800 0.0215  1.9900 0.0226  2.0000 0.0238  2.0100 0.0251  2.0200 0.0264  2.0300 0.0279  2.0400 0.0294  2.0500 0.0311  2.0600 0.0329  2.0700 0.0348  2.0800 0.0369  2.0900 0.0391  2.1000 0.0415  2.1100 0.0441  2.1200 0.0470  2.1300 0.0501  2.1400 0.0535  2.1500 0.0571  2.1600 0.0612  2.1700 0.0656  2.1800 0.0705  2.1900 0.0759  2.2000 0.0819  2.2100 0.0885  2.2200 0.0959  2.2300 0.1041  2.2400 0.1134  2.2500 0.1238  2.2600 0.1356  2.2700 0.1490  2.2800 0.1643  2.2900 0.1819  2.3000 0.2022  2.3100 0.2258  2.3200 0.2535  2.3300 0.2861  2.3400 0.3249  2.3500 0.3714  2.3600 0.4280  2.3700 0.4978  2.3800 0.5857  2.3900 0.7003  2.4000 0.8583  2.4100 1.0960  2.4200 1.4941  2.4300 2.2015  2.4400 3.3332  2.4500 4.4806  2.4600 4.5263  2.4700 3.3035  2.4800 1.9491  2.4900 1.0749  2.5000 0.6099  2.5100 0.3715  2.5200 0.2448  2.5300 0.1732  2.5400 0.1299  2.5500 0.1019  2.5600 0.0829  2.5700 0.0693  2.5800 0.0592  2.5900 0.0514  2.6000 0.0454  2.6100 0.0405  2.6200 0.0365  2.6300 0.0332  2.6400 0.0304  2.6500 0.0280  2.6600 0.0259  2.6700 0.0241  2.6800 0.0226  2.6900 0.0212  2.7000 0.0199  2.7100 0.0188  2.7200 0.0178  2.7300 0.0169  2.7400 0.0161  2.7500 0.0153  2.7600 0.0146  2.7700 0.0140  2.7800 0.0134  2.7900 0.0128  2.8000 0.0123  2.8100 0.0118  2.8200 0.0114  2.8300 0.0109  2.8400 0.0105  2.8500 0.0102  2.8600 0.0098  2.8700 0.0095  2.8800 0.0091  2.8900 0.0088  2.9000 0.0085  2.9100 0.0083  2.9200 0.0080  2.9300 0.0078  2.9400 0.0075  2.9500 0.0073  2.9600 0.0071  2.9700 0.0069  2.9800 0.0067  2.9900 0.0065  3.0000 0.0063  3.0100 0.0061  3.0200 0.0059  3.0300 0.0058  3.0400 0.0056  3.0500 0.0055  3.0600 0.0053  3.0700 0.0052  3.0800 0.0050  3.0900 0.0049  3.1000 0.0048  3.1100 0.0047  3.1200 0.0045  3.1300 0.0044  3.1400 0.0043  3.1500 0.0042  3.1600 0.0041  3.1700 0.0040  3.1800 0.0039  3.1900 0.0038  3.2000 0.0037  3.2100 0.0036  3.2200 0.0035  3.2300 0.0034  3.2400 0.0034  3.2500 0.0033  3.2600 0.0032  3.2700 0.0031  3.2800 0.0031  3.2900 0.0030  3.3000 0.0029  3.3100 0.0028  3.3200 0.0028  3.3300 0.0027  3.3400 0.0026  3.3500 0.0026  3.3600 0.0025  3.3700 0.0025  3.3800 0.0024  3.3900 0.0024  3.4000 0.0023  3.4100 0.0023  3.4200 0.0022  3.4300 0.0022  3.4400 0.0021  3.4500 0.0021  3.4600 0.0020  3.4700 0.0020  3.4800 0.0019  3.4900 0.0019  3.5000 0.0018  3.5100 0.0018  3.5200 0.0018  3.5300 0.0017  3.5400 0.0017  3.5500 0.0016  3.5600 0.0016  3.5700 0.0016  3.5800 0.0015  3.5900 0.0015  3.6000 0.0015  3.6100 0.0014  3.6200 0.0014  3.6300 0.0014  3.6400 0.0014  3.6500 0.0013  3.6600 0.0013  3.6700 0.0013  3.6800 0.0012  3.6900 0.0012  3.7000 0.0012  3.7100 0.0012  3.7200 0.0011  3.7300 0.0011  3.7400 0.0011  3.7500 0.0011  3.7600 0.0010  3.7700 0.0010  3.7800 0.0010  3.7900 0.0010  3.8000 0.0010  3.8100 0.0009  3.8200 0.0009  3.8300 0.0009  3.8400 0.0009  3.8500 0.0009  3.8600 0.0008  3.8700 0.0008  3.8800 0.0008  3.8900 0.0008  3.9000 0.0008  3.9100 0.0008  3.9200 0.0007  3.9300 0.0007  3.9400 0.0007  3.9500 0.0007  3.9600 0.0007  3.9700 0.0007  3.9800 0.0007  3.9900 0.0006  4.0000 0.0006  4.0100 0.0006  4.0200 0.0006  4.0300 0.0006  4.0400 0.0006  4.0500 0.0006  4.0600 0.0006  4.0700 0.0005  4.0800 0.0005  4.0900 0.0005  4.1000 0.0005  4.1100 0.0005  4.1200 0.0005  4.1300 0.0005  4.1400 0.0005  4.1500 0.0005  4.1600 0.0005  4.1700 0.0004  4.1800 0.0004  4.1900 0.0004  4.2000 0.0004  4.2100 0.0004  4.2200 0.0004  4.2300 0.0004  4.2400 0.0004  4.2500 0.0004  4.2600 0.0004  4.2700 0.0004  4.2800 0.0004  4.2900 0.0003  4.3000 0.0003  4.3100 0.0003  4.3200 0.0003  4.3300 0.0003  4.3400 0.0003  4.3500 0.0003  4.3600 0.0003  4.3700 0.0003  4.3800 0.0003  4.3900 0.0003  4.4000 0.0003  4.4100 0.0003  4.4200 0.0003  4.4300 0.0003  4.4400 0.0003  4.4500 0.0003  4.4600 0.0002  4.4700 0.0002  4.4800 0.0002  4.4900 0.0002  4.5000 0.0002  4.5100 0.0002  4.5200 0.0002  4.5300 0.0002  4.5400 0.0002  4.5500 0.0002  4.5600 0.0002  4.5700 0.0002  4.5800 0.0002  4.5900 0.0002  4.6000 0.0002  4.6100 0.0002  4.6200 0.0002  4.6300 0.0002  4.6400 0.0002  4.6500 0.0002  4.6600 0.0002  4.6700 0.0002  4.6800 0.0002  4.6900 0.0002  4.7000 0.0002  4.7100 0.0001  4.7200 0.0001  4.7300 0.0001  4.7400 0.0001  4.7500 0.0001  4.7600 0.0001  4.7700 0.0001  4.7800 0.0001  4.7900 0.0001  4.8000 0.0001  4.8100 0.0001  4.8200 0.0001  4.8300 0.0001  4.8400 0.0001  4.8500 0.0001  4.8600 0.0001  4.8700 0.0001  4.8800 0.0001  4.8900 0.0001  4.9000 0.0001  4.9100 0.0001  4.9200 0.0001  4.9300 0.0001  4.9400 0.0001  4.9500 0.0001  4.9600 0.0001  4.9700 0.0001  4.9800 0.0001  4.9900 0.0001  5.0000 0.0001 /
\color{BrickRed}
\plot 0.0000 0.0000  0.0100 0.0000  0.0200 0.0000  0.0300 0.0000  0.0400 0.0000  0.0500 0.0000  0.0600 0.0000  0.0700 0.0000  0.0800 0.0000  0.0900 0.0000  0.1000 0.0000  0.1100 0.0000  0.1200 0.0000  0.1300 0.0000  0.1400 0.0000  0.1500 0.0000  0.1600 0.0000  0.1700 0.0000  0.1800 0.0000  0.1900 0.0000  0.2000 0.0000  0.2100 0.0000  0.2200 0.0000  0.2300 0.0000  0.2400 0.0000  0.2500 0.0000  0.2600 0.0000  0.2700 0.0000  0.2800 0.0000  0.2900 0.0000  0.3000 0.0000  0.3100 0.0000  0.3200 0.0000  0.3300 0.0000  0.3400 0.0000  0.3500 0.0000  0.3600 0.0000  0.3700 0.0000  0.3800 0.0000  0.3900 0.0000  0.4000 0.0000  0.4100 0.0000  0.4200 0.0000  0.4300 0.0000  0.4400 0.0000  0.4500 0.0000  0.4600 0.0000  0.4700 0.0000  0.4800 0.0000  0.4900 0.0000  0.5000 0.0000  0.5100 0.0000  0.5200 0.0000  0.5300 0.0000  0.5400 0.0000  0.5500 0.0000  0.5600 0.0000  0.5700 0.0000  0.5800 0.0000  0.5900 0.0000  0.6000 0.0000  0.6100 0.0000  0.6200 0.0001  0.6300 0.0001  0.6400 0.0001  0.6500 0.0001  0.6600 0.0001  0.6700 0.0001  0.6800 0.0001  0.6900 0.0001  0.7000 0.0001  0.7100 0.0001  0.7200 0.0001  0.7300 0.0001  0.7400 0.0001  0.7500 0.0001  0.7600 0.0001  0.7700 0.0001  0.7800 0.0001  0.7900 0.0001  0.8000 0.0001  0.8100 0.0001  0.8200 0.0001  0.8300 0.0001  0.8400 0.0001  0.8500 0.0001  0.8600 0.0001  0.8700 0.0001  0.8800 0.0001  0.8900 0.0002  0.9000 0.0002  0.9100 0.0002  0.9200 0.0002  0.9300 0.0002  0.9400 0.0002  0.9500 0.0002  0.9600 0.0002  0.9700 0.0002  0.9800 0.0002  0.9900 0.0002  1.0000 0.0002  1.0100 0.0002  1.0200 0.0003  1.0300 0.0003  1.0400 0.0003  1.0500 0.0003  1.0600 0.0003  1.0700 0.0003  1.0800 0.0003  1.0900 0.0003  1.1000 0.0004  1.1100 0.0004  1.1200 0.0004  1.1300 0.0004  1.1400 0.0004  1.1500 0.0004  1.1600 0.0004  1.1700 0.0005  1.1800 0.0005  1.1900 0.0005  1.2000 0.0005  1.2100 0.0006  1.2200 0.0006  1.2300 0.0006  1.2400 0.0006  1.2500 0.0007  1.2600 0.0007  1.2700 0.0007  1.2800 0.0007  1.2900 0.0008  1.3000 0.0008  1.3100 0.0008  1.3200 0.0009  1.3300 0.0009  1.3400 0.0009  1.3500 0.0010  1.3600 0.0010  1.3700 0.0011  1.3800 0.0011  1.3900 0.0012  1.4000 0.0012  1.4100 0.0013  1.4200 0.0013  1.4300 0.0014  1.4400 0.0014  1.4500 0.0015  1.4600 0.0016  1.4700 0.0016  1.4800 0.0017  1.4900 0.0018  1.5000 0.0019  1.5100 0.0019  1.5200 0.0020  1.5300 0.0021  1.5400 0.0022  1.5500 0.0023  1.5600 0.0024  1.5700 0.0025  1.5800 0.0026  1.5900 0.0027  1.6000 0.0029  1.6100 0.0030  1.6200 0.0031  1.6300 0.0032  1.6400 0.0034  1.6500 0.0035  1.6600 0.0037  1.6700 0.0039  1.6800 0.0040  1.6900 0.0042  1.7000 0.0044  1.7100 0.0046  1.7200 0.0048  1.7300 0.0051  1.7400 0.0053  1.7500 0.0055  1.7600 0.0058  1.7700 0.0060  1.7800 0.0063  1.7900 0.0066  1.8000 0.0069  1.8100 0.0072  1.8200 0.0076  1.8300 0.0079  1.8400 0.0083  1.8500 0.0087  1.8600 0.0091  1.8700 0.0095  1.8800 0.0100  1.8900 0.0105  1.9000 0.0109  1.9100 0.0115  1.9200 0.0120  1.9300 0.0126  1.9400 0.0132  1.9500 0.0138  1.9600 0.0145  1.9700 0.0152  1.9800 0.0159  1.9900 0.0167  2.0000 0.0176  2.0100 0.0184  2.0200 0.0194  2.0300 0.0203  2.0400 0.0214  2.0500 0.0225  2.0600 0.0236  2.0700 0.0249  2.0800 0.0262  2.0900 0.0276  2.1000 0.0291  2.1100 0.0308  2.1200 0.0326  2.1300 0.0345  2.1400 0.0366  2.1500 0.0390  2.1600 0.0415  2.1700 0.0444  2.1800 0.0476  2.1900 0.0512  2.2000 0.0554  2.2100 0.0601  2.2200 0.0656  2.2300 0.0721  2.2400 0.0796  2.2500 0.0886  2.2600 0.0993  2.2700 0.1121  2.2800 0.1274  2.2900 0.1457  2.3000 0.1673  2.3100 0.1927  2.3200 0.2220  2.3300 0.2550  2.3400 0.2915  2.3500 0.3306  2.3600 0.3718  2.3700 0.4152  2.3800 0.4627  2.3900 0.5198  2.4000 0.5978  2.4100 0.7181  2.4200 0.9192  2.4300 1.2655  2.4400 1.8597  2.4500 2.9732  2.4600 5.7422  2.4700 9.2834  2.4800 5.4292  2.4900 1.7034  2.5000 0.5709  2.5100 0.2509  2.5200 0.1401  2.5300 0.0920  2.5400 0.0671  2.5500 0.0524  2.5600 0.0429  2.5700 0.0363  2.5800 0.0316  2.5900 0.0280  2.6000 0.0252  2.6100 0.0230  2.6200 0.0211  2.6300 0.0196  2.6400 0.0184  2.6500 0.0173  2.6600 0.0163  2.6700 0.0155  2.6800 0.0148  2.6900 0.0142  2.7000 0.0136  2.7100 0.0131  2.7200 0.0127  2.7300 0.0122  2.7400 0.0119  2.7500 0.0115  2.7600 0.0112  2.7700 0.0109  2.7800 0.0106  2.7900 0.0103  2.8000 0.0101  2.8100 0.0099  2.8200 0.0097  2.8300 0.0095  2.8400 0.0093  2.8500 0.0091  2.8600 0.0089  2.8700 0.0088  2.8800 0.0086  2.8900 0.0084  2.9000 0.0083  2.9100 0.0082  2.9200 0.0080  2.9300 0.0079  2.9400 0.0078  2.9500 0.0077  2.9600 0.0075  2.9700 0.0074  2.9800 0.0073  2.9900 0.0072  3.0000 0.0071  3.0100 0.0070  3.0200 0.0069  3.0300 0.0068  3.0400 0.0067  3.0500 0.0066  3.0600 0.0066  3.0700 0.0065  3.0800 0.0064  3.0900 0.0063  3.1000 0.0062  3.1100 0.0061  3.1200 0.0061  3.1300 0.0060  3.1400 0.0059  3.1500 0.0058  3.1600 0.0058  3.1700 0.0057  3.1800 0.0056  3.1900 0.0055  3.2000 0.0055  3.2100 0.0054  3.2200 0.0053  3.2300 0.0053  3.2400 0.0052  3.2500 0.0051  3.2600 0.0051  3.2700 0.0050  3.2800 0.0049  3.2900 0.0049  3.3000 0.0048  3.3100 0.0048  3.3200 0.0047  3.3300 0.0046  3.3400 0.0046  3.3500 0.0045  3.3600 0.0045  3.3700 0.0044  3.3800 0.0043  3.3900 0.0043  3.4000 0.0042  3.4100 0.0042  3.4200 0.0041  3.4300 0.0041  3.4400 0.0040  3.4500 0.0039  3.4600 0.0039  3.4700 0.0038  3.4800 0.0038  3.4900 0.0037  3.5000 0.0037  3.5100 0.0036  3.5200 0.0036  3.5300 0.0035  3.5400 0.0035  3.5500 0.0034  3.5600 0.0034  3.5700 0.0033  3.5800 0.0033  3.5900 0.0032  3.6000 0.0032  3.6100 0.0031  3.6200 0.0031  3.6300 0.0030  3.6400 0.0030  3.6500 0.0030  3.6600 0.0029  3.6700 0.0029  3.6800 0.0028  3.6900 0.0028  3.7000 0.0027  3.7100 0.0027  3.7200 0.0027  3.7300 0.0026  3.7400 0.0026  3.7500 0.0025  3.7600 0.0025  3.7700 0.0025  3.7800 0.0024  3.7900 0.0024  3.8000 0.0023  3.8100 0.0023  3.8200 0.0023  3.8300 0.0022  3.8400 0.0022  3.8500 0.0022  3.8600 0.0021  3.8700 0.0021  3.8800 0.0020  3.8900 0.0020  3.9000 0.0020  3.9100 0.0019  3.9200 0.0019  3.9300 0.0019  3.9400 0.0019  3.9500 0.0018  3.9600 0.0018  3.9700 0.0018  3.9800 0.0017  3.9900 0.0017  4.0000 0.0017  4.0100 0.0016  4.0200 0.0016  4.0300 0.0016  4.0400 0.0016  4.0500 0.0015  4.0600 0.0015  4.0700 0.0015  4.0800 0.0015  4.0900 0.0014  4.1000 0.0014  4.1100 0.0014  4.1200 0.0014  4.1300 0.0013  4.1400 0.0013  4.1500 0.0013  4.1600 0.0013  4.1700 0.0012  4.1800 0.0012  4.1900 0.0012  4.2000 0.0012  4.2100 0.0012  4.2200 0.0011  4.2300 0.0011  4.2400 0.0011  4.2500 0.0011  4.2600 0.0011  4.2700 0.0010  4.2800 0.0010  4.2900 0.0010  4.3000 0.0010  4.3100 0.0010  4.3200 0.0009  4.3300 0.0009  4.3400 0.0009  4.3500 0.0009  4.3600 0.0009  4.3700 0.0009  4.3800 0.0008  4.3900 0.0008  4.4000 0.0008  4.4100 0.0008  4.4200 0.0008  4.4300 0.0008  4.4400 0.0008  4.4500 0.0007  4.4600 0.0007  4.4700 0.0007  4.4800 0.0007  4.4900 0.0007  4.5000 0.0007  4.5100 0.0007  4.5200 0.0007  4.5300 0.0006  4.5400 0.0006  4.5500 0.0006  4.5600 0.0006  4.5700 0.0006  4.5800 0.0006  4.5900 0.0006  4.6000 0.0006  4.6100 0.0005  4.6200 0.0005  4.6300 0.0005  4.6400 0.0005  4.6500 0.0005  4.6600 0.0005  4.6700 0.0005  4.6800 0.0005  4.6900 0.0005  4.7000 0.0005  4.7100 0.0005  4.7200 0.0004  4.7300 0.0004  4.7400 0.0004  4.7500 0.0004  4.7600 0.0004  4.7700 0.0004  4.7800 0.0004  4.7900 0.0004  4.8000 0.0004  4.8100 0.0004  4.8200 0.0004  4.8300 0.0004  4.8400 0.0004  4.8500 0.0003  4.8600 0.0003  4.8700 0.0003  4.8800 0.0003  4.8900 0.0003  4.9000 0.0003  4.9100 0.0003  4.9200 0.0003  4.9300 0.0003  4.9400 0.0003  4.9500 0.0003  4.9600 0.0003  4.9700 0.0003  4.9800 0.0003  4.9900 0.0003  5.0000 0.0003 /
\color{Blue}
\plot 0.0000 0.0000  0.0100 0.0000  0.0200 0.0000  0.0300 0.0000  0.0400 0.0000  0.0500 0.0000  0.0600 0.0000  0.0700 0.0000  0.0800 0.0000  0.0900 0.0000  0.1000 0.0000  0.1100 0.0000  0.1200 0.0000  0.1300 0.0000  0.1400 0.0000  0.1500 0.0000  0.1600 0.0000  0.1700 0.0000  0.1800 0.0000  0.1900 0.0000  0.2000 0.0000  0.2100 0.0000  0.2200 0.0000  0.2300 0.0000  0.2400 0.0000  0.2500 0.0000  0.2600 0.0000  0.2700 0.0000  0.2800 0.0000  0.2900 0.0000  0.3000 0.0000  0.3100 0.0000  0.3200 0.0000  0.3300 0.0000  0.3400 0.0000  0.3500 0.0000  0.3600 0.0000  0.3700 0.0000  0.3800 0.0000  0.3900 0.0000  0.4000 0.0000  0.4100 0.0000  0.4200 0.0000  0.4300 0.0000  0.4400 0.0000  0.4500 0.0000  0.4600 0.0000  0.4700 0.0000  0.4800 0.0000  0.4900 0.0000  0.5000 0.0000  0.5100 0.0000  0.5200 0.0000  0.5300 0.0000  0.5400 0.0000  0.5500 0.0000  0.5600 0.0000  0.5700 0.0000  0.5800 0.0000  0.5900 0.0000  0.6000 0.0000  0.6100 0.0000  0.6200 0.0001  0.6300 0.0001  0.6400 0.0001  0.6500 0.0001  0.6600 0.0001  0.6700 0.0001  0.6800 0.0001  0.6900 0.0001  0.7000 0.0001  0.7100 0.0001  0.7200 0.0001  0.7300 0.0001  0.7400 0.0001  0.7500 0.0001  0.7600 0.0001  0.7700 0.0001  0.7800 0.0001  0.7900 0.0001  0.8000 0.0001  0.8100 0.0001  0.8200 0.0001  0.8300 0.0001  0.8400 0.0001  0.8500 0.0001  0.8600 0.0001  0.8700 0.0001  0.8800 0.0001  0.8900 0.0002  0.9000 0.0002  0.9100 0.0002  0.9200 0.0002  0.9300 0.0002  0.9400 0.0002  0.9500 0.0002  0.9600 0.0002  0.9700 0.0002  0.9800 0.0002  0.9900 0.0002  1.0000 0.0002  1.0100 0.0002  1.0200 0.0003  1.0300 0.0003  1.0400 0.0003  1.0500 0.0003  1.0600 0.0003  1.0700 0.0003  1.0800 0.0003  1.0900 0.0003  1.1000 0.0004  1.1100 0.0004  1.1200 0.0004  1.1300 0.0004  1.1400 0.0004  1.1500 0.0004  1.1600 0.0005  1.1700 0.0005  1.1800 0.0005  1.1900 0.0005  1.2000 0.0005  1.2100 0.0006  1.2200 0.0006  1.2300 0.0006  1.2400 0.0006  1.2500 0.0007  1.2600 0.0007  1.2700 0.0007  1.2800 0.0008  1.2900 0.0008  1.3000 0.0008  1.3100 0.0009  1.3200 0.0009  1.3300 0.0009  1.3400 0.0010  1.3500 0.0010  1.3600 0.0010  1.3700 0.0011  1.3800 0.0011  1.3900 0.0012  1.4000 0.0012  1.4100 0.0013  1.4200 0.0013  1.4300 0.0014  1.4400 0.0015  1.4500 0.0015  1.4600 0.0016  1.4700 0.0017  1.4800 0.0017  1.4900 0.0018  1.5000 0.0019  1.5100 0.0020  1.5200 0.0020  1.5300 0.0021  1.5400 0.0022  1.5500 0.0023  1.5600 0.0024  1.5700 0.0025  1.5800 0.0026  1.5900 0.0027  1.6000 0.0028  1.6100 0.0030  1.6200 0.0031  1.6300 0.0032  1.6400 0.0034  1.6500 0.0035  1.6600 0.0037  1.6700 0.0038  1.6800 0.0040  1.6900 0.0041  1.7000 0.0043  1.7100 0.0045  1.7200 0.0047  1.7300 0.0049  1.7400 0.0051  1.7500 0.0053  1.7600 0.0055  1.7700 0.0058  1.7800 0.0060  1.7900 0.0062  1.8000 0.0065  1.8100 0.0068  1.8200 0.0070  1.8300 0.0073  1.8400 0.0076  1.8500 0.0079  1.8600 0.0083  1.8700 0.0086  1.8800 0.0089  1.8900 0.0093  1.9000 0.0097  1.9100 0.0101  1.9200 0.0105  1.9300 0.0109  1.9400 0.0113  1.9500 0.0117  1.9600 0.0122  1.9700 0.0127  1.9800 0.0132  1.9900 0.0137  2.0000 0.0142  2.0100 0.0148  2.0200 0.0153  2.0300 0.0159  2.0400 0.0166  2.0500 0.0172  2.0600 0.0179  2.0700 0.0186  2.0800 0.0194  2.0900 0.0202  2.1000 0.0211  2.1100 0.0220  2.1200 0.0230  2.1300 0.0241  2.1400 0.0254  2.1500 0.0267  2.1600 0.0282  2.1700 0.0300  2.1800 0.0319  2.1900 0.0342  2.2000 0.0368  2.2100 0.0399  2.2200 0.0435  2.2300 0.0478  2.2400 0.0529  2.2500 0.0590  2.2600 0.0663  2.2700 0.0750  2.2800 0.0854  2.2900 0.0977  2.3000 0.1123  2.3100 0.1295  2.3200 0.1497  2.3300 0.1732  2.3400 0.2006  2.3500 0.2326  2.3600 0.2701  2.3700 0.3144  2.3800 0.3668  2.3900 0.4292  2.4000 0.5039  2.4100 0.5958  2.4200 0.7155  2.4300 0.8865  2.4400 1.1567  2.4500 1.6297  2.4600 2.6019  2.4700 5.7117  2.4800 15.1179  2.4900 5.0913  2.5000 0.9813  2.5100 0.3540  2.5200 0.1928  2.5300 0.1269  2.5400 0.0920  2.5500 0.0707  2.5600 0.0566  2.5700 0.0468  2.5800 0.0396  2.5900 0.0343  2.6000 0.0303  2.6100 0.0271  2.6200 0.0245  2.6300 0.0225  2.6400 0.0208  2.6500 0.0194  2.6600 0.0182  2.6700 0.0172  2.6800 0.0163  2.6900 0.0155  2.7000 0.0148  2.7100 0.0142  2.7200 0.0137  2.7300 0.0132  2.7400 0.0127  2.7500 0.0123  2.7600 0.0119  2.7700 0.0116  2.7800 0.0113  2.7900 0.0110  2.8000 0.0107  2.8100 0.0105  2.8200 0.0102  2.8300 0.0100  2.8400 0.0098  2.8500 0.0096  2.8600 0.0094  2.8700 0.0092  2.8800 0.0090  2.8900 0.0088  2.9000 0.0087  2.9100 0.0085  2.9200 0.0084  2.9300 0.0082  2.9400 0.0081  2.9500 0.0079  2.9600 0.0078  2.9700 0.0077  2.9800 0.0075  2.9900 0.0074  3.0000 0.0073  3.0100 0.0072  3.0200 0.0071  3.0300 0.0070  3.0400 0.0068  3.0500 0.0067  3.0600 0.0066  3.0700 0.0065  3.0800 0.0064  3.0900 0.0063  3.1000 0.0062  3.1100 0.0061  3.1200 0.0060  3.1300 0.0059  3.1400 0.0059  3.1500 0.0058  3.1600 0.0057  3.1700 0.0056  3.1800 0.0055  3.1900 0.0054  3.2000 0.0053  3.2100 0.0053  3.2200 0.0052  3.2300 0.0051  3.2400 0.0050  3.2500 0.0049  3.2600 0.0049  3.2700 0.0048  3.2800 0.0047  3.2900 0.0046  3.3000 0.0046  3.3100 0.0045  3.3200 0.0044  3.3300 0.0044  3.3400 0.0043  3.3500 0.0042  3.3600 0.0042  3.3700 0.0041  3.3800 0.0040  3.3900 0.0040  3.4000 0.0039  3.4100 0.0038  3.4200 0.0038  3.4300 0.0037  3.4400 0.0037  3.4500 0.0036  3.4600 0.0036  3.4700 0.0035  3.4800 0.0034  3.4900 0.0034  3.5000 0.0033  3.5100 0.0033  3.5200 0.0032  3.5300 0.0032  3.5400 0.0031  3.5500 0.0031  3.5600 0.0030  3.5700 0.0030  3.5800 0.0029  3.5900 0.0029  3.6000 0.0028  3.6100 0.0028  3.6200 0.0027  3.6300 0.0027  3.6400 0.0026  3.6500 0.0026  3.6600 0.0025  3.6700 0.0025  3.6800 0.0025  3.6900 0.0024  3.7000 0.0024  3.7100 0.0023  3.7200 0.0023  3.7300 0.0023  3.7400 0.0022  3.7500 0.0022  3.7600 0.0021  3.7700 0.0021  3.7800 0.0021  3.7900 0.0020  3.8000 0.0020  3.8100 0.0020  3.8200 0.0019  3.8300 0.0019  3.8400 0.0019  3.8500 0.0018  3.8600 0.0018  3.8700 0.0018  3.8800 0.0017  3.8900 0.0017  3.9000 0.0017  3.9100 0.0016  3.9200 0.0016  3.9300 0.0016  3.9400 0.0016  3.9500 0.0015  3.9600 0.0015  3.9700 0.0015  3.9800 0.0014  3.9900 0.0014  4.0000 0.0014  4.0100 0.0014  4.0200 0.0013  4.0300 0.0013  4.0400 0.0013  4.0500 0.0013  4.0600 0.0012  4.0700 0.0012  4.0800 0.0012  4.0900 0.0012  4.1000 0.0012  4.1100 0.0011  4.1200 0.0011  4.1300 0.0011  4.1400 0.0011  4.1500 0.0011  4.1600 0.0010  4.1700 0.0010  4.1800 0.0010  4.1900 0.0010  4.2000 0.0010  4.2100 0.0009  4.2200 0.0009  4.2300 0.0009  4.2400 0.0009  4.2500 0.0009  4.2600 0.0009  4.2700 0.0008  4.2800 0.0008  4.2900 0.0008  4.3000 0.0008  4.3100 0.0008  4.3200 0.0008  4.3300 0.0008  4.3400 0.0007  4.3500 0.0007  4.3600 0.0007  4.3700 0.0007  4.3800 0.0007  4.3900 0.0007  4.4000 0.0007  4.4100 0.0006  4.4200 0.0006  4.4300 0.0006  4.4400 0.0006  4.4500 0.0006  4.4600 0.0006  4.4700 0.0006  4.4800 0.0006  4.4900 0.0006  4.5000 0.0005  4.5100 0.0005  4.5200 0.0005  4.5300 0.0005  4.5400 0.0005  4.5500 0.0005  4.5600 0.0005  4.5700 0.0005  4.5800 0.0005  4.5900 0.0005  4.6000 0.0004  4.6100 0.0004  4.6200 0.0004  4.6300 0.0004  4.6400 0.0004  4.6500 0.0004  4.6600 0.0004  4.6700 0.0004  4.6800 0.0004  4.6900 0.0004  4.7000 0.0004  4.7100 0.0004  4.7200 0.0004  4.7300 0.0003  4.7400 0.0003  4.7500 0.0003  4.7600 0.0003  4.7700 0.0003  4.7800 0.0003  4.7900 0.0003  4.8000 0.0003  4.8100 0.0003  4.8200 0.0003  4.8300 0.0003  4.8400 0.0003  4.8500 0.0003  4.8600 0.0003  4.8700 0.0003  4.8800 0.0003  4.8900 0.0003  4.9000 0.0002  4.9100 0.0002  4.9200 0.0002  4.9300 0.0002  4.9400 0.0002  4.9500 0.0002  4.9600 0.0002  4.9700 0.0002  4.9800 0.0002  4.9900 0.0002  5.0000 0.0002 /
\normalcolor
\color{black}
%\put {\Large$\hbox{Var}\LA c_1 \RA_L$} at -0.95 0.75
\put {\large b) Across $\tau_1$} at 2.5 -6
\endpicture
\end{subfigure}
\normalcolor
\color{Black}
\begin{subfigure}{0.33\textwidth}
\normalcolor
\color{black}
	\caption{\large$\hbox{Var}\LA c_2 \RA_L$}
\beginpicture
\setcoordinatesystem units <21pt,3pt>
\normalcolor
\color{Black}
\axesnolabels{-2}{0}{0}{0}{3}{30}{0}
\put {\Large$-2$} at -2 -4
\put {\Large$3$} at 3 -4
\put {\Large$30$} at -0.65 30
\put {\Large$\beta$} at 2.0 -5
\setplotsymbol ({\scalebox{0.3}{$\bullet$}})
\color{Orange}
\plot -2.0000 0.0000  -1.9900 0.0000  -1.9800 0.0000  -1.9700 0.0000  -1.9600 0.0000  -1.9500 0.0000  -1.9400 0.0000  -1.9300 0.0000  -1.9200 0.0000  -1.9100 0.0000  -1.9000 0.0001  -1.8900 0.0001  -1.8800 0.0001  -1.8700 0.0001  -1.8600 0.0001  -1.8500 0.0001  -1.8400 0.0001  -1.8300 0.0001  -1.8200 0.0001  -1.8100 0.0001  -1.8000 0.0001  -1.7900 0.0001  -1.7800 0.0001  -1.7700 0.0001  -1.7600 0.0001  -1.7500 0.0001  -1.7400 0.0001  -1.7300 0.0001  -1.7200 0.0001  -1.7100 0.0001  -1.7000 0.0001  -1.6900 0.0001  -1.6800 0.0001  -1.6700 0.0001  -1.6600 0.0001  -1.6500 0.0001  -1.6400 0.0001  -1.6300 0.0001  -1.6200 0.0001  -1.6100 0.0001  -1.6000 0.0001  -1.5900 0.0001  -1.5800 0.0001  -1.5700 0.0001  -1.5600 0.0001  -1.5500 0.0001  -1.5400 0.0001  -1.5300 0.0001  -1.5200 0.0001  -1.5100 0.0001  -1.5000 0.0001  -1.4900 0.0001  -1.4800 0.0001  -1.4700 0.0001  -1.4600 0.0001  -1.4500 0.0001  -1.4400 0.0001  -1.4300 0.0001  -1.4200 0.0001  -1.4100 0.0001  -1.4000 0.0001  -1.3900 0.0001  -1.3800 0.0001  -1.3700 0.0001  -1.3600 0.0001  -1.3500 0.0002  -1.3400 0.0002  -1.3300 0.0002  -1.3200 0.0002  -1.3100 0.0002  -1.3000 0.0002  -1.2900 0.0002  -1.2800 0.0002  -1.2700 0.0002  -1.2600 0.0002  -1.2500 0.0002  -1.2400 0.0002  -1.2300 0.0002  -1.2200 0.0002  -1.2100 0.0002  -1.2000 0.0002  -1.1900 0.0002  -1.1800 0.0002  -1.1700 0.0002  -1.1600 0.0002  -1.1500 0.0002  -1.1400 0.0002  -1.1300 0.0002  -1.1200 0.0002  -1.1100 0.0002  -1.1000 0.0003  -1.0900 0.0003  -1.0800 0.0003  -1.0700 0.0003  -1.0600 0.0003  -1.0500 0.0003  -1.0400 0.0003  -1.0300 0.0003  -1.0200 0.0003  -1.0100 0.0003  -1.0000 0.0003  -0.9900 0.0003  -0.9800 0.0003  -0.9700 0.0003  -0.9600 0.0003  -0.9500 0.0003  -0.9400 0.0004  -0.9300 0.0004  -0.9200 0.0004  -0.9100 0.0004  -0.9000 0.0004  -0.8900 0.0004  -0.8800 0.0004  -0.8700 0.0004  -0.8600 0.0004  -0.8500 0.0004  -0.8400 0.0004  -0.8300 0.0004  -0.8200 0.0005  -0.8100 0.0005  -0.8000 0.0005  -0.7900 0.0005  -0.7800 0.0005  -0.7700 0.0005  -0.7600 0.0005  -0.7500 0.0005  -0.7400 0.0005  -0.7300 0.0006  -0.7200 0.0006  -0.7100 0.0006  -0.7000 0.0006  -0.6900 0.0006  -0.6800 0.0006  -0.6700 0.0006  -0.6600 0.0006  -0.6500 0.0007  -0.6400 0.0007  -0.6300 0.0007  -0.6200 0.0007  -0.6100 0.0007  -0.6000 0.0007  -0.5900 0.0008  -0.5800 0.0008  -0.5700 0.0008  -0.5600 0.0008  -0.5500 0.0008  -0.5400 0.0008  -0.5300 0.0009  -0.5200 0.0009  -0.5100 0.0009  -0.5000 0.0009  -0.4900 0.0009  -0.4800 0.0010  -0.4700 0.0010  -0.4600 0.0010  -0.4500 0.0010  -0.4400 0.0010  -0.4300 0.0011  -0.4200 0.0011  -0.4100 0.0011  -0.4000 0.0011  -0.3900 0.0012  -0.3800 0.0012  -0.3700 0.0012  -0.3600 0.0013  -0.3500 0.0013  -0.3400 0.0013  -0.3300 0.0013  -0.3200 0.0014  -0.3100 0.0014  -0.3000 0.0014  -0.2900 0.0015  -0.2800 0.0015  -0.2700 0.0016  -0.2600 0.0016  -0.2500 0.0016  -0.2400 0.0017  -0.2300 0.0017  -0.2200 0.0018  -0.2100 0.0018  -0.2000 0.0018  -0.1900 0.0019  -0.1800 0.0019  -0.1700 0.0020  -0.1600 0.0020  -0.1500 0.0021  -0.1400 0.0021  -0.1300 0.0022  -0.1200 0.0022  -0.1100 0.0023  -0.1000 0.0024  -0.0900 0.0024  -0.0800 0.0025  -0.0700 0.0026  -0.0600 0.0026  -0.0500 0.0027  -0.0400 0.0028  -0.0300 0.0028  -0.0200 0.0029  -0.0100 0.0030  0.0000 0.0031  0.0100 0.0032  0.0200 0.0033  0.0300 0.0034  0.0400 0.0035  0.0500 0.0036  0.0600 0.0037  0.0700 0.0038  0.0800 0.0039  0.0900 0.0040  0.1000 0.0041  0.1100 0.0043  0.1200 0.0044  0.1300 0.0045  0.1400 0.0047  0.1500 0.0048  0.1600 0.0050  0.1700 0.0052  0.1800 0.0053  0.1900 0.0055  0.2000 0.0057  0.2100 0.0059  0.2200 0.0061  0.2300 0.0064  0.2400 0.0066  0.2500 0.0069  0.2600 0.0072  0.2700 0.0075  0.2800 0.0078  0.2900 0.0082  0.3000 0.0085  0.3100 0.0089  0.3200 0.0094  0.3300 0.0099  0.3400 0.0104  0.3500 0.0110  0.3600 0.0116  0.3700 0.0123  0.3800 0.0131  0.3900 0.0140  0.4000 0.0150  0.4100 0.0162  0.4200 0.0175  0.4300 0.0190  0.4400 0.0207  0.4500 0.0227  0.4600 0.0250  0.4700 0.0277  0.4800 0.0309  0.4900 0.0347  0.5000 0.0391  0.5100 0.0445  0.5200 0.0509  0.5300 0.0586  0.5400 0.0678  0.5500 0.0790  0.5600 0.0926  0.5700 0.1092  0.5800 0.1293  0.5900 0.1539  0.6000 0.1839  0.6100 0.2204  0.6200 0.2648  0.6300 0.3187  0.6400 0.3838  0.6500 0.4621  0.6600 0.5553  0.6700 0.6653  0.6800 0.7933  0.6900 0.9397  0.7000 1.1037  0.7100 1.2823  0.7200 1.4704  0.7300 1.6597  0.7400 1.8399  0.7500 1.9984  0.7600 2.1226  0.7700 2.2012  0.7800 2.2270  0.7900 2.1976  0.8000 2.1166  0.8100 1.9923  0.8200 1.8363  0.8300 1.6614  0.8400 1.4795  0.8500 1.3007  0.8600 1.1322  0.8700 0.9787  0.8800 0.8426  0.8900 0.7242  0.9000 0.6231  0.9100 0.5376  0.9200 0.4662  0.9300 0.4069  0.9400 0.3578  0.9500 0.3173  0.9600 0.2839  0.9700 0.2563  0.9800 0.2335  0.9900 0.2146  1.0000 0.1988  1.0100 0.1855  1.0200 0.1742  1.0300 0.1646  1.0400 0.1564  1.0500 0.1492  1.0600 0.1429  1.0700 0.1373  1.0800 0.1323  1.0900 0.1278  1.1000 0.1237  1.1100 0.1200  1.1200 0.1165  1.1300 0.1133  1.1400 0.1103  1.1500 0.1075  1.1600 0.1048  1.1700 0.1023  1.1800 0.0999  1.1900 0.0976  1.2000 0.0954  1.2100 0.0933  1.2200 0.0913  1.2300 0.0894  1.2400 0.0876  1.2500 0.0858  1.2600 0.0840  1.2700 0.0824  1.2800 0.0807  1.2900 0.0792  1.3000 0.0777  1.3100 0.0762  1.3200 0.0747  1.3300 0.0733  1.3400 0.0720  1.3500 0.0707  1.3600 0.0694  1.3700 0.0681  1.3800 0.0669  1.3900 0.0657  1.4000 0.0645  1.4100 0.0634  1.4200 0.0623  1.4300 0.0612  1.4400 0.0601  1.4500 0.0591  1.4600 0.0581  1.4700 0.0571  1.4800 0.0561  1.4900 0.0551  1.5000 0.0542  1.5100 0.0533  1.5200 0.0524  1.5300 0.0515  1.5400 0.0506  1.5500 0.0498  1.5600 0.0489  1.5700 0.0481  1.5800 0.0473  1.5900 0.0465  1.6000 0.0457  1.6100 0.0449  1.6200 0.0442  1.6300 0.0434  1.6400 0.0427  1.6500 0.0420  1.6600 0.0413  1.6700 0.0406  1.6800 0.0399  1.6900 0.0392  1.7000 0.0386  1.7100 0.0379  1.7200 0.0373  1.7300 0.0366  1.7400 0.0360  1.7500 0.0354  1.7600 0.0348  1.7700 0.0342  1.7800 0.0336  1.7900 0.0330  1.8000 0.0325  1.8100 0.0319  1.8200 0.0314  1.8300 0.0308  1.8400 0.0303  1.8500 0.0298  1.8600 0.0293  1.8700 0.0288  1.8800 0.0283  1.8900 0.0278  1.9000 0.0273  1.9100 0.0268  1.9200 0.0263  1.9300 0.0259  1.9400 0.0254  1.9500 0.0250  1.9600 0.0245  1.9700 0.0241  1.9800 0.0237  1.9900 0.0233  2.0000 0.0229  2.0100 0.0224  2.0200 0.0220  2.0300 0.0217  2.0400 0.0213  2.0500 0.0209  2.0600 0.0205  2.0700 0.0201  2.0800 0.0198  2.0900 0.0194  2.1000 0.0191  2.1100 0.0187  2.1200 0.0184  2.1300 0.0181  2.1400 0.0177  2.1500 0.0174  2.1600 0.0171  2.1700 0.0168  2.1800 0.0165  2.1900 0.0162  2.2000 0.0159  2.2100 0.0156  2.2200 0.0153  2.2300 0.0150  2.2400 0.0147  2.2500 0.0145  2.2600 0.0142  2.2700 0.0139  2.2800 0.0137  2.2900 0.0134  2.3000 0.0132  2.3100 0.0129  2.3200 0.0127  2.3300 0.0125  2.3400 0.0122  2.3500 0.0120  2.3600 0.0118  2.3700 0.0116  2.3800 0.0113  2.3900 0.0111  2.4000 0.0109  2.4100 0.0107  2.4200 0.0105  2.4300 0.0103  2.4400 0.0101  2.4500 0.0099  2.4600 0.0097  2.4700 0.0096  2.4800 0.0094  2.4900 0.0092  2.5000 0.0090  2.5100 0.0089  2.5200 0.0087  2.5300 0.0085  2.5400 0.0084  2.5500 0.0082  2.5600 0.0081  2.5700 0.0079  2.5800 0.0077  2.5900 0.0076  2.6000 0.0075  2.6100 0.0073  2.6200 0.0072  2.6300 0.0070  2.6400 0.0069  2.6500 0.0068  2.6600 0.0066  2.6700 0.0065  2.6800 0.0064  2.6900 0.0063  2.7000 0.0061  2.7100 0.0060  2.7200 0.0059  2.7300 0.0058  2.7400 0.0057  2.7500 0.0056  2.7600 0.0055  2.7700 0.0054  2.7800 0.0053  2.7900 0.0052  2.8000 0.0051  2.8100 0.0050  2.8200 0.0049  2.8300 0.0048  2.8400 0.0047  2.8500 0.0046  2.8600 0.0045  2.8700 0.0044  2.8800 0.0043  2.8900 0.0042  2.9000 0.0042  2.9100 0.0041  2.9200 0.0040  2.9300 0.0039  2.9400 0.0038  2.9500 0.0038  2.9600 0.0037  2.9700 0.0036  2.9800 0.0036  2.9900 0.0035  3.0000 0.0034 /
\color{Red}
\plot -2.0000 0.0000  -1.9900 0.0000  -1.9800 0.0000  -1.9700 0.0000  -1.9600 0.0000  -1.9500 0.0000  -1.9400 0.0000  -1.9300 0.0000  -1.9200 0.0000  -1.9100 0.0000  -1.9000 0.0000  -1.8900 0.0000  -1.8800 0.0000  -1.8700 0.0000  -1.8600 0.0000  -1.8500 0.0000  -1.8400 0.0000  -1.8300 0.0000  -1.8200 0.0000  -1.8100 0.0000  -1.8000 0.0000  -1.7900 0.0000  -1.7800 0.0000  -1.7700 0.0000  -1.7600 0.0000  -1.7500 0.0000  -1.7400 0.0000  -1.7300 0.0000  -1.7200 0.0000  -1.7100 0.0000  -1.7000 0.0000  -1.6900 0.0000  -1.6800 0.0000  -1.6700 0.0000  -1.6600 0.0000  -1.6500 0.0000  -1.6400 0.0000  -1.6300 0.0000  -1.6200 0.0000  -1.6100 0.0000  -1.6000 0.0000  -1.5900 0.0000  -1.5800 0.0000  -1.5700 0.0000  -1.5600 0.0000  -1.5500 0.0000  -1.5400 0.0001  -1.5300 0.0001  -1.5200 0.0001  -1.5100 0.0001  -1.5000 0.0001  -1.4900 0.0001  -1.4800 0.0001  -1.4700 0.0001  -1.4600 0.0001  -1.4500 0.0001  -1.4400 0.0001  -1.4300 0.0001  -1.4200 0.0001  -1.4100 0.0001  -1.4000 0.0001  -1.3900 0.0001  -1.3800 0.0001  -1.3700 0.0001  -1.3600 0.0001  -1.3500 0.0001  -1.3400 0.0001  -1.3300 0.0001  -1.3200 0.0001  -1.3100 0.0001  -1.3000 0.0001  -1.2900 0.0001  -1.2800 0.0001  -1.2700 0.0001  -1.2600 0.0001  -1.2500 0.0001  -1.2400 0.0001  -1.2300 0.0001  -1.2200 0.0001  -1.2100 0.0001  -1.2000 0.0001  -1.1900 0.0001  -1.1800 0.0001  -1.1700 0.0001  -1.1600 0.0001  -1.1500 0.0001  -1.1400 0.0001  -1.1300 0.0001  -1.1200 0.0001  -1.1100 0.0001  -1.1000 0.0001  -1.0900 0.0001  -1.0800 0.0001  -1.0700 0.0001  -1.0600 0.0001  -1.0500 0.0001  -1.0400 0.0001  -1.0300 0.0001  -1.0200 0.0001  -1.0100 0.0002  -1.0000 0.0002  -0.9900 0.0002  -0.9800 0.0002  -0.9700 0.0002  -0.9600 0.0002  -0.9500 0.0002  -0.9400 0.0002  -0.9300 0.0002  -0.9200 0.0002  -0.9100 0.0002  -0.9000 0.0002  -0.8900 0.0002  -0.8800 0.0002  -0.8700 0.0002  -0.8600 0.0002  -0.8500 0.0002  -0.8400 0.0002  -0.8300 0.0002  -0.8200 0.0002  -0.8100 0.0002  -0.8000 0.0002  -0.7900 0.0002  -0.7800 0.0002  -0.7700 0.0003  -0.7600 0.0003  -0.7500 0.0003  -0.7400 0.0003  -0.7300 0.0003  -0.7200 0.0003  -0.7100 0.0003  -0.7000 0.0003  -0.6900 0.0003  -0.6800 0.0003  -0.6700 0.0003  -0.6600 0.0003  -0.6500 0.0003  -0.6400 0.0003  -0.6300 0.0003  -0.6200 0.0003  -0.6100 0.0004  -0.6000 0.0004  -0.5900 0.0004  -0.5800 0.0004  -0.5700 0.0004  -0.5600 0.0004  -0.5500 0.0004  -0.5400 0.0004  -0.5300 0.0004  -0.5200 0.0004  -0.5100 0.0004  -0.5000 0.0005  -0.4900 0.0005  -0.4800 0.0005  -0.4700 0.0005  -0.4600 0.0005  -0.4500 0.0005  -0.4400 0.0005  -0.4300 0.0005  -0.4200 0.0005  -0.4100 0.0006  -0.4000 0.0006  -0.3900 0.0006  -0.3800 0.0006  -0.3700 0.0006  -0.3600 0.0006  -0.3500 0.0006  -0.3400 0.0007  -0.3300 0.0007  -0.3200 0.0007  -0.3100 0.0007  -0.3000 0.0007  -0.2900 0.0007  -0.2800 0.0008  -0.2700 0.0008  -0.2600 0.0008  -0.2500 0.0008  -0.2400 0.0008  -0.2300 0.0009  -0.2200 0.0009  -0.2100 0.0009  -0.2000 0.0009  -0.1900 0.0010  -0.1800 0.0010  -0.1700 0.0010  -0.1600 0.0010  -0.1500 0.0011  -0.1400 0.0011  -0.1300 0.0011  -0.1200 0.0011  -0.1100 0.0012  -0.1000 0.0012  -0.0900 0.0012  -0.0800 0.0013  -0.0700 0.0013  -0.0600 0.0014  -0.0500 0.0014  -0.0400 0.0014  -0.0300 0.0015  -0.0200 0.0015  -0.0100 0.0016  0.0000 0.0016  0.0100 0.0017  0.0200 0.0017  0.0300 0.0018  0.0400 0.0018  0.0500 0.0019  0.0600 0.0020  0.0700 0.0021  0.0800 0.0023  0.0900 0.0025  0.1000 0.0028  0.1100 0.0033  0.1200 0.0042  0.1300 0.0059  0.1400 0.0088  0.1500 0.0144  0.1600 0.0249  0.1700 0.0448  0.1800 0.0833  0.1900 0.1575  0.2000 0.3010  0.2100 0.5763  0.2200 1.0946  0.2300 2.0277  0.2400 3.5627  0.2500 5.6791  0.2600 7.7415  0.2700 8.5391  0.2800 7.4776  0.2900 5.3732  0.3000 3.3857  0.3100 2.0087  0.3200 1.1952  0.3300 0.7519  0.3400 0.5191  0.3500 0.3978  0.3600 0.3335  0.3700 0.2979  0.3800 0.2767  0.3900 0.2627  0.4000 0.2525  0.4100 0.2442  0.4200 0.2371  0.4300 0.2306  0.4400 0.2246  0.4500 0.2189  0.4600 0.2135  0.4700 0.2084  0.4800 0.2035  0.4900 0.1988  0.5000 0.1943  0.5100 0.1899  0.5200 0.1858  0.5300 0.1817  0.5400 0.1779  0.5500 0.1741  0.5600 0.1705  0.5700 0.1670  0.5800 0.1637  0.5900 0.1604  0.6000 0.1573  0.6100 0.1542  0.6200 0.1513  0.6300 0.1484  0.6400 0.1456  0.6500 0.1429  0.6600 0.1403  0.6700 0.1378  0.6800 0.1353  0.6900 0.1329  0.7000 0.1305  0.7100 0.1283  0.7200 0.1261  0.7300 0.1239  0.7400 0.1218  0.7500 0.1197  0.7600 0.1178  0.7700 0.1158  0.7800 0.1139  0.7900 0.1121  0.8000 0.1102  0.8100 0.1085  0.8200 0.1068  0.8300 0.1051  0.8400 0.1034  0.8500 0.1018  0.8600 0.1002  0.8700 0.0987  0.8800 0.0972  0.8900 0.0957  0.9000 0.0943  0.9100 0.0928  0.9200 0.0915  0.9300 0.0901  0.9400 0.0888  0.9500 0.0875  0.9600 0.0862  0.9700 0.0850  0.9800 0.0837  0.9900 0.0825  1.0000 0.0814  1.0100 0.0802  1.0200 0.0791  1.0300 0.0780  1.0400 0.0769  1.0500 0.0758  1.0600 0.0747  1.0700 0.0737  1.0800 0.0727  1.0900 0.0717  1.1000 0.0707  1.1100 0.0698  1.1200 0.0688  1.1300 0.0679  1.1400 0.0670  1.1500 0.0661  1.1600 0.0652  1.1700 0.0644  1.1800 0.0635  1.1900 0.0627  1.2000 0.0619  1.2100 0.0611  1.2200 0.0603  1.2300 0.0595  1.2400 0.0587  1.2500 0.0580  1.2600 0.0572  1.2700 0.0565  1.2800 0.0558  1.2900 0.0551  1.3000 0.0544  1.3100 0.0537  1.3200 0.0530  1.3300 0.0524  1.3400 0.0517  1.3500 0.0511  1.3600 0.0504  1.3700 0.0498  1.3800 0.0492  1.3900 0.0486  1.4000 0.0480  1.4100 0.0474  1.4200 0.0468  1.4300 0.0462  1.4400 0.0456  1.4500 0.0451  1.4600 0.0445  1.4700 0.0440  1.4800 0.0434  1.4900 0.0429  1.5000 0.0424  1.5100 0.0418  1.5200 0.0413  1.5300 0.0408  1.5400 0.0403  1.5500 0.0398  1.5600 0.0393  1.5700 0.0388  1.5800 0.0384  1.5900 0.0379  1.6000 0.0374  1.6100 0.0369  1.6200 0.0365  1.6300 0.0360  1.6400 0.0356  1.6500 0.0351  1.6600 0.0347  1.6700 0.0342  1.6800 0.0338  1.6900 0.0334  1.7000 0.0330  1.7100 0.0325  1.7200 0.0321  1.7300 0.0317  1.7400 0.0313  1.7500 0.0309  1.7600 0.0305  1.7700 0.0301  1.7800 0.0297  1.7900 0.0293  1.8000 0.0289  1.8100 0.0286  1.8200 0.0282  1.8300 0.0278  1.8400 0.0274  1.8500 0.0271  1.8600 0.0267  1.8700 0.0263  1.8800 0.0260  1.8900 0.0256  1.9000 0.0253  1.9100 0.0249  1.9200 0.0246  1.9300 0.0242  1.9400 0.0239  1.9500 0.0236  1.9600 0.0232  1.9700 0.0229  1.9800 0.0226  1.9900 0.0223  2.0000 0.0219  2.0100 0.0216  2.0200 0.0213  2.0300 0.0210  2.0400 0.0207  2.0500 0.0204  2.0600 0.0201  2.0700 0.0198  2.0800 0.0195  2.0900 0.0192  2.1000 0.0189  2.1100 0.0187  2.1200 0.0184  2.1300 0.0181  2.1400 0.0178  2.1500 0.0176  2.1600 0.0173  2.1700 0.0170  2.1800 0.0168  2.1900 0.0165  2.2000 0.0162  2.2100 0.0160  2.2200 0.0157  2.2300 0.0155  2.2400 0.0153  2.2500 0.0150  2.2600 0.0148  2.2700 0.0145  2.2800 0.0143  2.2900 0.0141  2.3000 0.0138  2.3100 0.0136  2.3200 0.0134  2.3300 0.0132  2.3400 0.0130  2.3500 0.0128  2.3600 0.0125  2.3700 0.0123  2.3800 0.0121  2.3900 0.0119  2.4000 0.0117  2.4100 0.0115  2.4200 0.0113  2.4300 0.0112  2.4400 0.0110  2.4500 0.0108  2.4600 0.0106  2.4700 0.0104  2.4800 0.0102  2.4900 0.0101  2.5000 0.0099  2.5100 0.0097  2.5200 0.0096  2.5300 0.0094  2.5400 0.0092  2.5500 0.0091  2.5600 0.0089  2.5700 0.0088  2.5800 0.0086  2.5900 0.0085  2.6000 0.0083  2.6100 0.0082  2.6200 0.0080  2.6300 0.0079  2.6400 0.0077  2.6500 0.0076  2.6600 0.0075  2.6700 0.0073  2.6800 0.0072  2.6900 0.0071  2.7000 0.0069  2.7100 0.0068  2.7200 0.0067  2.7300 0.0066  2.7400 0.0065  2.7500 0.0063  2.7600 0.0062  2.7700 0.0061  2.7800 0.0060  2.7900 0.0059  2.8000 0.0058  2.8100 0.0057  2.8200 0.0056  2.8300 0.0055  2.8400 0.0054  2.8500 0.0053  2.8600 0.0052  2.8700 0.0051  2.8800 0.0050  2.8900 0.0049  2.9000 0.0048  2.9100 0.0047  2.9200 0.0046  2.9300 0.0045  2.9400 0.0045  2.9500 0.0044  2.9600 0.0043  2.9700 0.0042  2.9800 0.0041  2.9900 0.0041  3.0000 0.0040 /
\color{BrickRed}
\plot -2.0000 0.0000  -1.9900 0.0000  -1.9800 0.0000  -1.9700 0.0000  -1.9600 0.0000  -1.9500 0.0000  -1.9400 0.0000  -1.9300 0.0000  -1.9200 0.0000  -1.9100 0.0000  -1.9000 0.0000  -1.8900 0.0000  -1.8800 0.0000  -1.8700 0.0000  -1.8600 0.0000  -1.8500 0.0000  -1.8400 0.0000  -1.8300 0.0000  -1.8200 0.0000  -1.8100 0.0000  -1.8000 0.0000  -1.7900 0.0000  -1.7800 0.0000  -1.7700 0.0000  -1.7600 0.0000  -1.7500 0.0000  -1.7400 0.0000  -1.7300 0.0000  -1.7200 0.0000  -1.7100 0.0000  -1.7000 0.0000  -1.6900 0.0000  -1.6800 0.0000  -1.6700 0.0000  -1.6600 0.0000  -1.6500 0.0000  -1.6400 0.0000  -1.6300 0.0000  -1.6200 0.0000  -1.6100 0.0000  -1.6000 0.0000  -1.5900 0.0000  -1.5800 0.0000  -1.5700 0.0000  -1.5600 0.0000  -1.5500 0.0000  -1.5400 0.0000  -1.5300 0.0000  -1.5200 0.0000  -1.5100 0.0000  -1.5000 0.0000  -1.4900 0.0000  -1.4800 0.0000  -1.4700 0.0000  -1.4600 0.0000  -1.4500 0.0000  -1.4400 0.0000  -1.4300 0.0000  -1.4200 0.0000  -1.4100 0.0000  -1.4000 0.0000  -1.3900 0.0000  -1.3800 0.0000  -1.3700 0.0000  -1.3600 0.0000  -1.3500 0.0000  -1.3400 0.0000  -1.3300 0.0000  -1.3200 0.0000  -1.3100 0.0000  -1.3000 0.0000  -1.2900 0.0000  -1.2800 0.0001  -1.2700 0.0001  -1.2600 0.0001  -1.2500 0.0001  -1.2400 0.0001  -1.2300 0.0001  -1.2200 0.0001  -1.2100 0.0001  -1.2000 0.0001  -1.1900 0.0001  -1.1800 0.0001  -1.1700 0.0001  -1.1600 0.0001  -1.1500 0.0001  -1.1400 0.0001  -1.1300 0.0001  -1.1200 0.0001  -1.1100 0.0001  -1.1000 0.0001  -1.0900 0.0001  -1.0800 0.0001  -1.0700 0.0001  -1.0600 0.0001  -1.0500 0.0001  -1.0400 0.0001  -1.0300 0.0001  -1.0200 0.0001  -1.0100 0.0001  -1.0000 0.0001  -0.9900 0.0001  -0.9800 0.0001  -0.9700 0.0001  -0.9600 0.0001  -0.9500 0.0001  -0.9400 0.0001  -0.9300 0.0001  -0.9200 0.0001  -0.9100 0.0001  -0.9000 0.0001  -0.8900 0.0001  -0.8800 0.0001  -0.8700 0.0001  -0.8600 0.0001  -0.8500 0.0001  -0.8400 0.0001  -0.8300 0.0001  -0.8200 0.0001  -0.8100 0.0001  -0.8000 0.0001  -0.7900 0.0001  -0.7800 0.0001  -0.7700 0.0001  -0.7600 0.0001  -0.7500 0.0002  -0.7400 0.0002  -0.7300 0.0002  -0.7200 0.0002  -0.7100 0.0002  -0.7000 0.0002  -0.6900 0.0002  -0.6800 0.0002  -0.6700 0.0002  -0.6600 0.0002  -0.6500 0.0002  -0.6400 0.0002  -0.6300 0.0002  -0.6200 0.0002  -0.6100 0.0002  -0.6000 0.0002  -0.5900 0.0002  -0.5800 0.0002  -0.5700 0.0002  -0.5600 0.0002  -0.5500 0.0002  -0.5400 0.0002  -0.5300 0.0002  -0.5200 0.0003  -0.5100 0.0003  -0.5000 0.0003  -0.4900 0.0003  -0.4800 0.0003  -0.4700 0.0003  -0.4600 0.0003  -0.4500 0.0003  -0.4400 0.0003  -0.4300 0.0003  -0.4200 0.0003  -0.4100 0.0003  -0.4000 0.0003  -0.3900 0.0003  -0.3800 0.0004  -0.3700 0.0004  -0.3600 0.0004  -0.3500 0.0004  -0.3400 0.0004  -0.3300 0.0004  -0.3200 0.0004  -0.3100 0.0004  -0.3000 0.0004  -0.2900 0.0004  -0.2800 0.0004  -0.2700 0.0005  -0.2600 0.0005  -0.2500 0.0005  -0.2400 0.0005  -0.2300 0.0005  -0.2200 0.0005  -0.2100 0.0005  -0.2000 0.0005  -0.1900 0.0006  -0.1800 0.0006  -0.1700 0.0006  -0.1600 0.0006  -0.1500 0.0006  -0.1400 0.0006  -0.1300 0.0007  -0.1200 0.0007  -0.1100 0.0007  -0.1000 0.0007  -0.0900 0.0007  -0.0800 0.0008  -0.0700 0.0008  -0.0600 0.0009  -0.0500 0.0011  -0.0400 0.0017  -0.0300 0.0038  -0.0200 0.0113  -0.0100 0.0385  0.0000 0.1380  0.0100 0.5036  0.0200 1.8119  0.0300 5.9467  0.0400 14.2771  0.0500 17.1490  0.0600 9.2766  0.0700 3.3077  0.0800 1.1870  0.0900 0.5836  0.1000 0.4176  0.1100 0.3673  0.1200 0.3469  0.1300 0.3343  0.1400 0.3239  0.1500 0.3144  0.1600 0.3053  0.1700 0.2967  0.1800 0.2885  0.1900 0.2807  0.2000 0.2733  0.2100 0.2662  0.2200 0.2596  0.2300 0.2532  0.2400 0.2473  0.2500 0.2416  0.2600 0.2362  0.2700 0.2311  0.2800 0.2263  0.2900 0.2217  0.3000 0.2173  0.3100 0.2131  0.3200 0.2091  0.3300 0.2052  0.3400 0.2015  0.3500 0.1979  0.3600 0.1945  0.3700 0.1911  0.3800 0.1879  0.3900 0.1847  0.4000 0.1817  0.4100 0.1787  0.4200 0.1758  0.4300 0.1730  0.4400 0.1702  0.4500 0.1675  0.4600 0.1649  0.4700 0.1623  0.4800 0.1598  0.4900 0.1573  0.5000 0.1549  0.5100 0.1525  0.5200 0.1502  0.5300 0.1479  0.5400 0.1457  0.5500 0.1435  0.5600 0.1413  0.5700 0.1392  0.5800 0.1371  0.5900 0.1351  0.6000 0.1331  0.6100 0.1311  0.6200 0.1292  0.6300 0.1273  0.6400 0.1254  0.6500 0.1236  0.6600 0.1218  0.6700 0.1201  0.6800 0.1184  0.6900 0.1167  0.7000 0.1150  0.7100 0.1134  0.7200 0.1118  0.7300 0.1103  0.7400 0.1087  0.7500 0.1072  0.7600 0.1057  0.7700 0.1043  0.7800 0.1029  0.7900 0.1014  0.8000 0.1001  0.8100 0.0987  0.8200 0.0974  0.8300 0.0960  0.8400 0.0947  0.8500 0.0935  0.8600 0.0922  0.8700 0.0910  0.8800 0.0898  0.8900 0.0886  0.9000 0.0874  0.9100 0.0862  0.9200 0.0851  0.9300 0.0839  0.9400 0.0828  0.9500 0.0817  0.9600 0.0806  0.9700 0.0796  0.9800 0.0785  0.9900 0.0775  1.0000 0.0765  1.0100 0.0755  1.0200 0.0745  1.0300 0.0735  1.0400 0.0725  1.0500 0.0716  1.0600 0.0707  1.0700 0.0697  1.0800 0.0688  1.0900 0.0679  1.1000 0.0671  1.1100 0.0662  1.1200 0.0654  1.1300 0.0645  1.1400 0.0637  1.1500 0.0629  1.1600 0.0621  1.1700 0.0613  1.1800 0.0605  1.1900 0.0597  1.2000 0.0590  1.2100 0.0582  1.2200 0.0575  1.2300 0.0568  1.2400 0.0561  1.2500 0.0554  1.2600 0.0547  1.2700 0.0540  1.2800 0.0533  1.2900 0.0527  1.3000 0.0520  1.3100 0.0514  1.3200 0.0508  1.3300 0.0502  1.3400 0.0496  1.3500 0.0490  1.3600 0.0484  1.3700 0.0478  1.3800 0.0472  1.3900 0.0467  1.4000 0.0461  1.4100 0.0456  1.4200 0.0450  1.4300 0.0445  1.4400 0.0440  1.4500 0.0435  1.4600 0.0430  1.4700 0.0425  1.4800 0.0420  1.4900 0.0415  1.5000 0.0410  1.5100 0.0405  1.5200 0.0401  1.5300 0.0396  1.5400 0.0392  1.5500 0.0387  1.5600 0.0383  1.5700 0.0379  1.5800 0.0374  1.5900 0.0370  1.6000 0.0366  1.6100 0.0362  1.6200 0.0358  1.6300 0.0354  1.6400 0.0350  1.6500 0.0346  1.6600 0.0342  1.6700 0.0338  1.6800 0.0334  1.6900 0.0331  1.7000 0.0327  1.7100 0.0323  1.7200 0.0319  1.7300 0.0316  1.7400 0.0312  1.7500 0.0309  1.7600 0.0305  1.7700 0.0302  1.7800 0.0298  1.7900 0.0295  1.8000 0.0291  1.8100 0.0288  1.8200 0.0285  1.8300 0.0281  1.8400 0.0278  1.8500 0.0275  1.8600 0.0271  1.8700 0.0268  1.8800 0.0265  1.8900 0.0262  1.9000 0.0258  1.9100 0.0255  1.9200 0.0252  1.9300 0.0249  1.9400 0.0246  1.9500 0.0243  1.9600 0.0240  1.9700 0.0237  1.9800 0.0234  1.9900 0.0231  2.0000 0.0228  2.0100 0.0225  2.0200 0.0222  2.0300 0.0219  2.0400 0.0216  2.0500 0.0213  2.0600 0.0210  2.0700 0.0207  2.0800 0.0204  2.0900 0.0201  2.1000 0.0198  2.1100 0.0196  2.1200 0.0193  2.1300 0.0190  2.1400 0.0187  2.1500 0.0184  2.1600 0.0182  2.1700 0.0179  2.1800 0.0176  2.1900 0.0174  2.2000 0.0171  2.2100 0.0168  2.2200 0.0166  2.2300 0.0163  2.2400 0.0161  2.2500 0.0158  2.2600 0.0156  2.2700 0.0153  2.2800 0.0151  2.2900 0.0148  2.3000 0.0146  2.3100 0.0144  2.3200 0.0141  2.3300 0.0139  2.3400 0.0137  2.3500 0.0134  2.3600 0.0132  2.3700 0.0130  2.3800 0.0128  2.3900 0.0126  2.4000 0.0123  2.4100 0.0121  2.4200 0.0119  2.4300 0.0117  2.4400 0.0115  2.4500 0.0113  2.4600 0.0111  2.4700 0.0109  2.4800 0.0107  2.4900 0.0105  2.5000 0.0104  2.5100 0.0102  2.5200 0.0100  2.5300 0.0098  2.5400 0.0096  2.5500 0.0095  2.5600 0.0093  2.5700 0.0091  2.5800 0.0089  2.5900 0.0088  2.6000 0.0086  2.6100 0.0085  2.6200 0.0083  2.6300 0.0082  2.6400 0.0080  2.6500 0.0079  2.6600 0.0077  2.6700 0.0076  2.6800 0.0074  2.6900 0.0073  2.7000 0.0071  2.7100 0.0070  2.7200 0.0069  2.7300 0.0067  2.7400 0.0066  2.7500 0.0065  2.7600 0.0064  2.7700 0.0063  2.7800 0.0061  2.7900 0.0060  2.8000 0.0059  2.8100 0.0058  2.8200 0.0057  2.8300 0.0056  2.8400 0.0055  2.8500 0.0054  2.8600 0.0053  2.8700 0.0051  2.8800 0.0050  2.8900 0.0050  2.9000 0.0049  2.9100 0.0048  2.9200 0.0047  2.9300 0.0046  2.9400 0.0045  2.9500 0.0044  2.9600 0.0043  2.9700 0.0042  2.9800 0.0041  2.9900 0.0041  3.0000 0.0040   /
\color{Blue}
\plot -2.0000 0.0000  -1.9900 0.0000  -1.9800 0.0000  -1.9700 0.0000  -1.9600 0.0000  -1.9500 0.0000  -1.9400 0.0000  -1.9300 0.0000  -1.9200 0.0000  -1.9100 0.0000  -1.9000 0.0000  -1.8900 0.0000  -1.8800 0.0000  -1.8700 0.0000  -1.8600 0.0000  -1.8500 0.0000  -1.8400 0.0000  -1.8300 0.0000  -1.8200 0.0000  -1.8100 0.0000  -1.8000 0.0000  -1.7900 0.0000  -1.7800 0.0000  -1.7700 0.0000  -1.7600 0.0000  -1.7500 0.0000  -1.7400 0.0000  -1.7300 0.0000  -1.7200 0.0000  -1.7100 0.0000  -1.7000 0.0000  -1.6900 0.0000  -1.6800 0.0000  -1.6700 0.0000  -1.6600 0.0000  -1.6500 0.0000  -1.6400 0.0000  -1.6300 0.0000  -1.6200 0.0000  -1.6100 0.0000  -1.6000 0.0000  -1.5900 0.0000  -1.5800 0.0000  -1.5700 0.0000  -1.5600 0.0000  -1.5500 0.0000  -1.5400 0.0000  -1.5300 0.0000  -1.5200 0.0000  -1.5100 0.0000  -1.5000 0.0000  -1.4900 0.0000  -1.4800 0.0000  -1.4700 0.0000  -1.4600 0.0000  -1.4500 0.0000  -1.4400 0.0000  -1.4300 0.0000  -1.4200 0.0000  -1.4100 0.0000  -1.4000 0.0000  -1.3900 0.0000  -1.3800 0.0000  -1.3700 0.0000  -1.3600 0.0000  -1.3500 0.0000  -1.3400 0.0000  -1.3300 0.0000  -1.3200 0.0000  -1.3100 0.0000  -1.3000 0.0000  -1.2900 0.0000  -1.2800 0.0000  -1.2700 0.0000  -1.2600 0.0000  -1.2500 0.0000  -1.2400 0.0000  -1.2300 0.0000  -1.2200 0.0000  -1.2100 0.0000  -1.2000 0.0000  -1.1900 0.0000  -1.1800 0.0000  -1.1700 0.0000  -1.1600 0.0000  -1.1500 0.0000  -1.1400 0.0000  -1.1300 0.0000  -1.1200 0.0000  -1.1100 0.0000  -1.1000 0.0000  -1.0900 0.0000  -1.0800 0.0000  -1.0700 0.0001  -1.0600 0.0001  -1.0500 0.0001  -1.0400 0.0001  -1.0300 0.0001  -1.0200 0.0001  -1.0100 0.0001  -1.0000 0.0001  -0.9900 0.0001  -0.9800 0.0001  -0.9700 0.0001  -0.9600 0.0001  -0.9500 0.0001  -0.9400 0.0001  -0.9300 0.0001  -0.9200 0.0001  -0.9100 0.0001  -0.9000 0.0001  -0.8900 0.0001  -0.8800 0.0001  -0.8700 0.0001  -0.8600 0.0001  -0.8500 0.0001  -0.8400 0.0001  -0.8300 0.0001  -0.8200 0.0001  -0.8100 0.0001  -0.8000 0.0001  -0.7900 0.0001  -0.7800 0.0001  -0.7700 0.0001  -0.7600 0.0001  -0.7500 0.0001  -0.7400 0.0001  -0.7300 0.0001  -0.7200 0.0001  -0.7100 0.0001  -0.7000 0.0001  -0.6900 0.0001  -0.6800 0.0001  -0.6700 0.0001  -0.6600 0.0001  -0.6500 0.0001  -0.6400 0.0001  -0.6300 0.0001  -0.6200 0.0001  -0.6100 0.0001  -0.6000 0.0001  -0.5900 0.0001  -0.5800 0.0001  -0.5700 0.0001  -0.5600 0.0002  -0.5500 0.0002  -0.5400 0.0002  -0.5300 0.0002  -0.5200 0.0002  -0.5100 0.0002  -0.5000 0.0002  -0.4900 0.0002  -0.4800 0.0002  -0.4700 0.0002  -0.4600 0.0002  -0.4500 0.0002  -0.4400 0.0002  -0.4300 0.0002  -0.4200 0.0002  -0.4100 0.0002  -0.4000 0.0002  -0.3900 0.0002  -0.3800 0.0002  -0.3700 0.0002  -0.3600 0.0002  -0.3500 0.0002  -0.3400 0.0003  -0.3300 0.0003  -0.3200 0.0003  -0.3100 0.0003  -0.3000 0.0003  -0.2900 0.0003  -0.2800 0.0003  -0.2700 0.0003  -0.2600 0.0003  -0.2500 0.0003  -0.2400 0.0003  -0.2300 0.0003  -0.2200 0.0003  -0.2100 0.0003  -0.2000 0.0004  -0.1900 0.0004  -0.1800 0.0004  -0.1700 0.0004  -0.1600 0.0005  -0.1500 0.0015  -0.1400 0.0089  -0.1300 0.0704  -0.1200 0.5850  -0.1100 4.6562  -0.1000 23.3909  -0.0900 21.4715  -0.0800 4.2287  -0.0700 0.8616  -0.0600 0.4433  -0.0500 0.3827  -0.0400 0.3630  -0.0300 0.3488  -0.0200 0.3365  -0.0100 0.3257  0.0000 0.3161  0.0100 0.3079  0.0200 0.3007  0.0300 0.2946  0.0400 0.2894  0.0500 0.2848  0.0600 0.2808  0.0700 0.2773  0.0800 0.2741  0.0900 0.2711  0.1000 0.2683  0.1100 0.2656  0.1200 0.2628  0.1300 0.2601  0.1400 0.2573  0.1500 0.2545  0.1600 0.2516  0.1700 0.2486  0.1800 0.2455  0.1900 0.2424  0.2000 0.2392  0.2100 0.2359  0.2200 0.2326  0.2300 0.2292  0.2400 0.2259  0.2500 0.2225  0.2600 0.2191  0.2700 0.2156  0.2800 0.2122  0.2900 0.2089  0.3000 0.2055  0.3100 0.2021  0.3200 0.1988  0.3300 0.1956  0.3400 0.1923  0.3500 0.1892  0.3600 0.1860  0.3700 0.1830  0.3800 0.1799  0.3900 0.1770  0.4000 0.1741  0.4100 0.1712  0.4200 0.1685  0.4300 0.1658  0.4400 0.1631  0.4500 0.1605  0.4600 0.1580  0.4700 0.1555  0.4800 0.1531  0.4900 0.1508  0.5000 0.1485  0.5100 0.1462  0.5200 0.1440  0.5300 0.1419  0.5400 0.1398  0.5500 0.1377  0.5600 0.1357  0.5700 0.1338  0.5800 0.1318  0.5900 0.1300  0.6000 0.1281  0.6100 0.1263  0.6200 0.1245  0.6300 0.1228  0.6400 0.1211  0.6500 0.1194  0.6600 0.1177  0.6700 0.1161  0.6800 0.1145  0.6900 0.1129  0.7000 0.1114  0.7100 0.1099  0.7200 0.1084  0.7300 0.1069  0.7400 0.1055  0.7500 0.1041  0.7600 0.1027  0.7700 0.1013  0.7800 0.0999  0.7900 0.0986  0.8000 0.0973  0.8100 0.0960  0.8200 0.0948  0.8300 0.0936  0.8400 0.0923  0.8500 0.0912  0.8600 0.0900  0.8700 0.0888  0.8800 0.0877  0.8900 0.0866  0.9000 0.0855  0.9100 0.0844  0.9200 0.0834  0.9300 0.0824  0.9400 0.0814  0.9500 0.0804  0.9600 0.0794  0.9700 0.0784  0.9800 0.0775  0.9900 0.0766  1.0000 0.0756  1.0100 0.0748  1.0200 0.0739  1.0300 0.0730  1.0400 0.0721  1.0500 0.0713  1.0600 0.0705  1.0700 0.0697  1.0800 0.0689  1.0900 0.0681  1.1000 0.0673  1.1100 0.0665  1.1200 0.0658  1.1300 0.0650  1.1400 0.0643  1.1500 0.0636  1.1600 0.0628  1.1700 0.0621  1.1800 0.0614  1.1900 0.0607  1.2000 0.0601  1.2100 0.0594  1.2200 0.0587  1.2300 0.0581  1.2400 0.0574  1.2500 0.0568  1.2600 0.0561  1.2700 0.0555  1.2800 0.0549  1.2900 0.0543  1.3000 0.0536  1.3100 0.0530  1.3200 0.0524  1.3300 0.0518  1.3400 0.0512  1.3500 0.0507  1.3600 0.0501  1.3700 0.0495  1.3800 0.0489  1.3900 0.0483  1.4000 0.0478  1.4100 0.0472  1.4200 0.0466  1.4300 0.0461  1.4400 0.0455  1.4500 0.0450  1.4600 0.0444  1.4700 0.0438  1.4800 0.0433  1.4900 0.0427  1.5000 0.0422  1.5100 0.0417  1.5200 0.0411  1.5300 0.0406  1.5400 0.0400  1.5500 0.0395  1.5600 0.0390  1.5700 0.0384  1.5800 0.0379  1.5900 0.0374  1.6000 0.0368  1.6100 0.0363  1.6200 0.0358  1.6300 0.0353  1.6400 0.0348  1.6500 0.0342  1.6600 0.0337  1.6700 0.0332  1.6800 0.0327  1.6900 0.0322  1.7000 0.0317  1.7100 0.0312  1.7200 0.0307  1.7300 0.0302  1.7400 0.0297  1.7500 0.0292  1.7600 0.0288  1.7700 0.0283  1.7800 0.0278  1.7900 0.0273  1.8000 0.0269  1.8100 0.0264  1.8200 0.0260  1.8300 0.0255  1.8400 0.0251  1.8500 0.0246  1.8600 0.0242  1.8700 0.0238  1.8800 0.0234  1.8900 0.0229  1.9000 0.0225  1.9100 0.0221  1.9200 0.0217  1.9300 0.0213  1.9400 0.0210  1.9500 0.0206  1.9600 0.0202  1.9700 0.0198  1.9800 0.0195  1.9900 0.0191  2.0000 0.0188  2.0100 0.0184  2.0200 0.0181  2.0300 0.0178  2.0400 0.0174  2.0500 0.0171  2.0600 0.0168  2.0700 0.0165  2.0800 0.0162  2.0900 0.0159  2.1000 0.0156  2.1100 0.0153  2.1200 0.0150  2.1300 0.0148  2.1400 0.0145  2.1500 0.0142  2.1600 0.0140  2.1700 0.0137  2.1800 0.0135  2.1900 0.0132  2.2000 0.0130  2.2100 0.0128  2.2200 0.0126  2.2300 0.0123  2.2400 0.0121  2.2500 0.0119  2.2600 0.0117  2.2700 0.0115  2.2800 0.0113  2.2900 0.0111  2.3000 0.0109  2.3100 0.0107  2.3200 0.0106  2.3300 0.0104  2.3400 0.0102  2.3500 0.0101  2.3600 0.0099  2.3700 0.0097  2.3800 0.0096  2.3900 0.0094  2.4000 0.0093  2.4100 0.0091  2.4200 0.0090  2.4300 0.0089  2.4400 0.0087  2.4500 0.0086  2.4600 0.0085  2.4700 0.0084  2.4800 0.0082  2.4900 0.0081  2.5000 0.0080  2.5100 0.0079  2.5200 0.0078  2.5300 0.0077  2.5400 0.0076  2.5500 0.0075  2.5600 0.0074  2.5700 0.0073  2.5800 0.0072  2.5900 0.0071  2.6000 0.0070  2.6100 0.0069  2.6200 0.0068  2.6300 0.0068  2.6400 0.0067  2.6500 0.0066  2.6600 0.0065  2.6700 0.0064  2.6800 0.0064  2.6900 0.0063  2.7000 0.0062  2.7100 0.0062  2.7200 0.0061  2.7300 0.0060  2.7400 0.0060  2.7500 0.0059  2.7600 0.0058  2.7700 0.0058  2.7800 0.0057  2.7900 0.0057  2.8000 0.0056  2.8100 0.0055  2.8200 0.0055  2.8300 0.0054  2.8400 0.0054  2.8500 0.0053  2.8600 0.0053  2.8700 0.0052  2.8800 0.0052  2.8900 0.0051  2.9000 0.0051  2.9100 0.0051  2.9200 0.0050  2.9300 0.0050  2.9400 0.0049  2.9500 0.0049  2.9600 0.0048  2.9700 0.0048  2.9800 0.0048  2.9900 0.0047  3.0000 0.0047      /
\color{black}
%\put {\Large$\LA c_1 \RA_L$} at -0.95 0.75
\put {\large c) Across $\tau_2$} at 0.5 -12
\endpicture
\color{black}
\normalcolor
\end{subfigure}
\caption{(a)  The concentration of the outer polygon in the linked model
along a line segment with $\beta=-2$ crossing the $\lambda$ phase
boundary in the $c_1$-dominated phase, for system sizes $L=8$,
$12$, $16$, and $20$.  The location of the critical point is denoted by
a bullet.  (b) The variance of $\LA c_1 \RA$ across the $\tau_1$ phase
boundary along the line segment from $(0,5)$ to $(5,0)$.  The data develop
a sharp spike when the critical line is crossed, with minimal widening close
to the critical point.  The width decreases with increasing $L$.  In this graph
the data are plotted for system sizes $L=8$, $12$, $16$, and $20$.
(c) The variance of $\LA c_2 \RA$ as a function of $\beta$ along the
line segment from $(-2,-2)$ to $(-2,3)$.  With increases $L$ in
$\{8,12,16,20\}$ the data show an increasing spike which narrows
quickly in width and shifting to the left to a limiting location as
$L$ increases to infinity.}
\label{21}   % %ZXZ[21]
\color{black}
\normalcolor
\end{figure}

\color{black}
\normalcolor

In figure \ref{21}(a) the concentrations $\LA c_1 \RA$ are plotted along a line
segment as the $\lambda$ phase boundary is crossed into the $c_1$-dominated
phase in the linked model, for systems of sizes $L=8$, $12$, $16$ and $20$. 
The graphs show quick convergence towards a limiting curve with increasing
$L$, particularly in the $c_1$-dominated phase.  Finite size effects are most pronounced
close to the critical point, and is seen in the slight upturn of the plot
in figure \ref{18} when $\alpha$ approaches $\alpha_c$.  We accommodated for
this by fitting a linear function to data, while discarding data close to the critical point.

Figure \ref{21}(b) is a graph of the variance $\hbox{Var}(c_1)$ plotted along
a line segment from $(0,5)$ to $(5,0)$ crossing the $\tau_1$ phase boundary, for
$L=8$, $12$, $16$ and $20$.  These curves all spike sharply when $\tau_1$ is crossed
at $\alpha \approx 2.48$.  The heights of the spikes increase by almost doubling as
$L$ increments by $4$. Finite size effects can be seen in the slight broadening 
at the base of the spikes, and this broadening decreases with increasing $L$. 

The variance $\hbox{Var}(c_2)$ plotted along a vertical line from $(-2,-2)$ to 
$(-2,3)$ across the $\tau_2$ phase boundary is shown in figure \ref{21}(c) for
$L=8$, $12$, $16$ and $20$.  For each $L$ this shows a spike roughly twice
the height of the spikes in figure \ref{21}(b).  With increasing $L$ the spikes
move, towards the left, but with decreasing increments.  This shows that the location
of the $\tau_2$ phase boundary remains uncertain, but also that it should converge
to a limiting curve of first order transitions along which $\hbox{Var}(c_2)$ is divergent.

\section{Characterising the transitions}

The nature of transitions across phase boundaries in these models 
can also be explored by defining and tracking order parameters and 
metric observables as the system is taken through a transition.  Numerical 
estimates of order parameters were obtained by sampling the models using a 
Multiple Markov Chain Metropolis algorithm \cite{MRRTT53,GT95,TJvROW96} 
implementing BFACF elementary moves (see figure \ref{3}) \cite{ACF83,BF81} 
on each polygon.  In our implementation we sampled along 
$50$ Markov chains (performing $10^5$ iterations along each sequence) 
along lines in the $\alpha\beta$-plane.  Apart from tracking the lengths 
$n_1$ and $n_2$ of the polygons, these simulations also collected data 
on the number of nearest neighbour contacts $k_1$ and $k_2$ in each 
polygon (these are \textit{self-contacts}), as well as the
number of nearest neighbour contacts $k_m$ between polygons
(these are \textit{mutual contacts}). 

Let $w_L(n_1,n_2;k_1,k_2,k_m)$ be the number of conformations of the
unlinked or linked model in a square of side-length $L$, of lengths 
$n_1$ and $n_2$ respectively, with $k_1$ self-contacts in the first polygon, 
$k_2$ in the second, and $k_m$ mutual contacts.   The mean \textit{density of 
self-contacts} of the first polygon is defined by

\begin{equation}
\hspace{-0.2cm}\scalebox{0.85}{
$\displaystyle
\LA k_1 /L^2 \RA 
= 
\frac{\displaystyle
\sum_{{n_1,n_2}\atop{k_1,k_2,k_M}}
(k_1/L^2)\, u_L(n_1,n_2;k_1,k_2,k_M)\, e^{\alpha n_1}\,e^{\beta n_2} 
}{\displaystyle
\sum_{{n_1,n_2}\atop{k_1,k_2,k_M}}
 u_L(n_1,n_2;k_1,k_2,k_M)\, e^{\alpha n_1}\,e^{\beta n_2} 
}
.$
}
\label{39}  %ZXZ[39]
\end{equation}
The densities $\LA k_2/L^2 \RA$ of self-contacts in the second polygon,
and $\LA k_m/L^2 \RA$ of mutual contacts, are defined similarly. 

We sampled data on metric observables, namely the mean square radius of
gyration of the first polygon (a function of $(\alpha,\beta)$), $\LA r_1^2 \RA$, and  
the corresponding quantity for the second polygon, $\LA r_2^2 \RA$. Moreover
we look at the mean distance or separation between the centres-of-mass
of the two polygons, $\LA d_{cm}\RA$. 

In addition to these quantities, the correlations $Cor(k_1,k_2) 
= \LA k_1 k_2\RA - \LA k_1\RA \LA k_2 \RA$ (between self-contacts in the polygons), 
and $Cor(r_1^2,r_2^2)
= \LA r_1^2\,r_2^2 \RA - \LA r_1^2 \RA \LA r_2^2 \RA$
(between the square radii of gyration of the two polygons) were calculated.

\begin{figure}[h!]
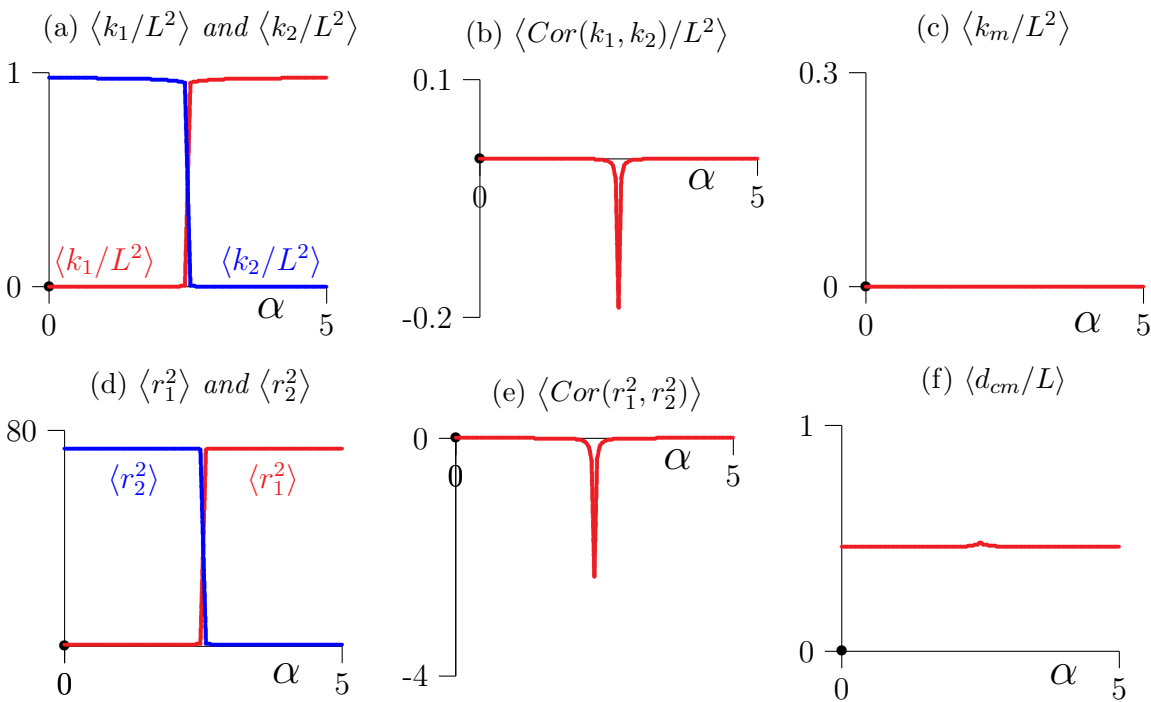
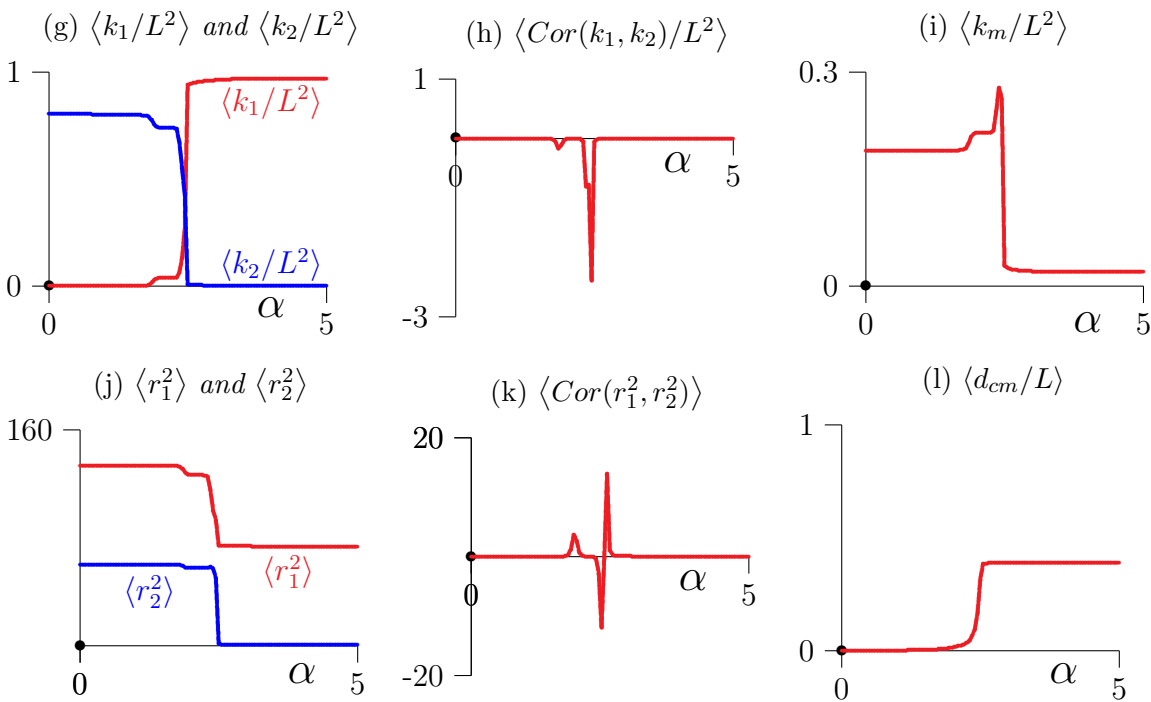

\U{Unlinked from $(0,5)$ to $(5,0)$ across $\tau$}: \vspace{1mm}

\begin{subfigure}{0.33\textwidth}
\normalcolor
\color{black}
	\caption{$\LA k_1/L^2 \RA$ and $\LA k_2/L^2 \RA$}
\beginpicture
\setcoordinatesystem units <21pt,81pt>
\axes{0}{0}{0}{0}{5}{1}{0}
\put {\Large$\alpha$} at 4.0 -0.10
\setplotsymbol ({\scalebox{0.3}{$\bullet$}})
\color{Red}
\plot 0.000 0.0000  0.050 0.0000  0.100 0.0000  0.150 0.0000  0.200 0.0000  0.250 0.0000  0.300 0.0000  0.350 0.0000  0.400 0.0000  0.450 0.0000  0.500 0.0000  0.550 0.0000  0.600 0.0000  0.650 0.0000  0.700 0.0000  0.750 0.0000  0.800 0.0000  0.850 0.0000  0.900 0.0000  0.950 0.0000  1.000 0.0000  1.050 0.0000  1.100 0.0000  1.150 0.0000  1.200 0.0000  1.250 0.0000  1.300 0.0000  1.350 0.0000  1.400 0.0000  1.450 0.0000  1.500 0.0000  1.550 0.0000  1.600 0.0000  1.650 0.0000  1.700 0.0001  1.750 0.0001  1.800 0.0001  1.850 0.0001  1.900 0.0001  1.950 0.0002  2.000 0.0002  2.050 0.0002  2.100 0.0003  2.150 0.0004  2.200 0.0005  2.250 0.0006  2.300 0.0009  2.350 0.0012  2.400 0.0018  2.450 0.0032  2.500 0.4753  2.550 0.9518  2.600 0.9558  2.650 0.9582  2.700 0.9600  2.750 0.9616  2.800 0.9628  2.850 0.9639  2.900 0.9650  2.950 0.9659  3.000 0.9666  3.050 0.9674  3.100 0.9680  3.150 0.9685  3.200 0.9691  3.250 0.9696  3.300 0.9700  3.350 0.9704  3.400 0.9708  3.450 0.9711  3.500 0.9713  3.550 0.9716  3.600 0.9718  3.650 0.9720  3.700 0.9722  3.750 0.9724  3.800 0.9726  3.850 0.9727  3.900 0.9728  3.950 0.9730  4.000 0.9731  4.050 0.9732  4.100 0.9733  4.150 0.9733  4.200 0.9734  4.250 0.9735  4.300 0.9735  4.350 0.9736  4.400 0.9736  4.450 0.9737  4.500 0.9737  4.550 0.9738  4.600 0.9738  4.650 0.9738  4.700 0.9738  4.750 0.9739  4.800 0.9739  4.850 0.9739  4.900 0.9739  4.950 0.9739  5.000 0.9740 /
\color{Blue}
\plot 0.000 0.9739  0.050 0.9739  0.100 0.9739  0.150 0.9739  0.200 0.9739  0.250 0.9738  0.300 0.9738  0.350 0.9738  0.400 0.9738  0.450 0.9737  0.500 0.9737  0.550 0.9737  0.600 0.9736  0.650 0.9736  0.700 0.9735  0.750 0.9735  0.800 0.9734  0.850 0.9734  0.900 0.9733  0.950 0.9732  1.000 0.9731  1.050 0.9730  1.100 0.9729  1.150 0.9727  1.200 0.9726  1.250 0.9725  1.300 0.9723  1.350 0.9721  1.400 0.9719  1.450 0.9716  1.500 0.9714  1.550 0.9711  1.600 0.9708  1.650 0.9704  1.700 0.9700  1.750 0.9696  1.800 0.9691  1.850 0.9686  1.900 0.9680  1.950 0.9674  2.000 0.9667  2.050 0.9659  2.100 0.9650  2.150 0.9640  2.200 0.9629  2.250 0.9616  2.300 0.9601  2.350 0.9582  2.400 0.9558  2.450 0.9518  2.500 0.4736  2.550 0.0031  2.600 0.0018  2.650 0.0012  2.700 0.0009  2.750 0.0006  2.800 0.0005  2.850 0.0004  2.900 0.0003  2.950 0.0002  3.000 0.0002  3.050 0.0002  3.100 0.0001  3.150 0.0001  3.200 0.0001  3.250 0.0001  3.300 0.0001  3.350 0.0000  3.400 0.0000  3.450 0.0000  3.500 0.0000  3.550 0.0000  3.600 0.0000  3.650 0.0000  3.700 0.0000  3.750 0.0000  3.800 0.0000  3.850 0.0000  3.900 0.0000  3.950 0.0000  4.000 0.0000  4.050 0.0000  4.100 0.0000  4.150 0.0000  4.200 0.0000  4.250 0.0000  4.300 0.0000  4.350 0.0000  4.400 0.0000  4.450 0.0000  4.500 0.0000  4.550 0.0000  4.600 0.0000  4.650 0.0000  4.700 0.0000  4.750 0.0000  4.800 0.0000  4.850 0.0000  4.900 0.0000  4.950 0.0000  5.000 0.0000 /
\color{black}
\put {\color{Red}$\LA k_1/L^2 \RA$\color{black}} at 1 0.12
\put {\color{Blue}$\LA k_2/L^2 \RA$\color{black}} at 4 0.12
\endpicture
\end{subfigure}
\begin{subfigure}{0.33\textwidth}
\normalcolor
\color{black}
	\caption{$\LA Cor(k_1,k_2)/L^2 \RA$}
\beginpicture
\setcoordinatesystem units <21pt,300pt>
\axes{0}{-0.2}{0}{0}{5}{0.1}{0}
\put {\Large$\alpha$} at 4.0 -0.025
\setplotsymbol ({\scalebox{0.3}{$\bullet$}})
\color{Red}
\plot 0.000 -0.0000  0.050 -0.0000  0.100 -0.0000  0.150 -0.0000  0.200 -0.0000  0.250 -0.0000  0.300 -0.0000  0.350 -0.0000  0.400 -0.0000  0.450 -0.0000  0.500 -0.0000  0.550 -0.0000  0.600 -0.0000  0.650 -0.0000  0.700 -0.0000  0.750 -0.0000  0.800 -0.0000  0.850 -0.0000  0.900 -0.0000  0.950 -0.0000  1.000 -0.0000  1.050 -0.0000  1.100 -0.0000  1.150 -0.0000  1.200 -0.0000  1.250 -0.0000  1.300 -0.0000  1.350 -0.0000  1.400 -0.0000  1.450 -0.0000  1.500 -0.0001  1.550 -0.0001  1.600 -0.0001  1.650 -0.0001  1.700 -0.0001  1.750 -0.0001  1.800 -0.0002  1.850 -0.0002  1.900 -0.0003  1.950 -0.0004  2.000 -0.0005  2.050 -0.0006  2.100 -0.0008  2.150 -0.0010  2.200 -0.0013  2.250 -0.0019  2.300 -0.0028  2.350 -0.0045  2.400 -0.0087  2.450 -0.0254  2.500 -0.1881  2.550 -0.0244  2.600 -0.0087  2.650 -0.0045  2.700 -0.0028  2.750 -0.0018  2.800 -0.0013  2.850 -0.0010  2.900 -0.0008  2.950 -0.0006  3.000 -0.0005  3.050 -0.0004  3.100 -0.0003  3.150 -0.0002  3.200 -0.0002  3.250 -0.0001  3.300 -0.0001  3.350 -0.0001  3.400 -0.0001  3.450 -0.0001  3.500 -0.0001  3.550 -0.0000  3.600 -0.0000  3.650 -0.0000  3.700 -0.0000  3.750 -0.0000  3.800 -0.0000  3.850 -0.0000  3.900 -0.0000  3.950 -0.0000  4.000 -0.0000  4.050 -0.0000  4.100 -0.0000  4.150 -0.0000  4.200 -0.0000  4.250 -0.0000  4.300 -0.0000  4.350 -0.0000  4.400 -0.0000  4.450 -0.0000  4.500 -0.0000  4.550 -0.0000  4.600 -0.0000  4.650 -0.0000  4.700 -0.0000  4.750 -0.0000  4.800 -0.0000  4.850 -0.0000  4.900 -0.0000  4.950 -0.0000  5.000 -0.0000  /
%\put {$\mathbf{\LA r_1^2 \RA}$} at -1 0.033
\endpicture
\end{subfigure}
\begin{subfigure}{0.33\textwidth}
\normalcolor
\color{black}
	\caption{$\LA k_m/L^2 \RA$}
\beginpicture
\setcoordinatesystem units <21pt,270pt>
\axes{0}{0}{0}{0}{5}{0.3}{0}
\put {\Large$\alpha$} at 4.0 -0.05
\setplotsymbol ({\scalebox{0.3}{$\bullet$}})
\color{Red}
\plot 0.000 0.0000  0.050 0.0000  0.100 0.0000  0.150 0.0000  0.200 0.0000  0.250 0.0000  0.300 0.0000  0.350 0.0000  0.400 0.0000  0.450 0.0000  0.500 0.0000  0.550 0.0000  0.600 0.0000  0.650 0.0000  0.700 0.0000  0.750 0.0000  0.800 0.0000  0.850 0.0000  0.900 0.0000  0.950 0.0000  1.000 0.0000  1.050 0.0000  1.100 0.0000  1.150 0.0000  1.200 0.0000  1.250 0.0000  1.300 0.0000  1.350 0.0000  1.400 0.0000  1.450 0.0000  1.500 0.0000  1.550 0.0000  1.600 0.0000  1.650 0.0000  1.700 0.0000  1.750 0.0000  1.800 0.0000  1.850 0.0000  1.900 0.0000  1.950 0.0000  2.000 0.0000  2.050 0.0000  2.100 0.0000  2.150 0.0000  2.200 0.0000  2.250 0.0000  2.300 0.0000  2.350 0.0000  2.400 0.0000  2.450 0.0000  2.500 0.0000  2.550 0.0000  2.600 0.0000  2.650 0.0000  2.700 0.0000  2.750 0.0000  2.800 0.0000  2.850 0.0000  2.900 0.0000  2.950 0.0000  3.000 0.0000  3.050 0.0000  3.100 0.0000  3.150 0.0000  3.200 0.0000  3.250 0.0000  3.300 0.0000  3.350 0.0000  3.400 0.0000  3.450 0.0000  3.500 0.0000  3.550 0.0000  3.600 0.0000  3.650 0.0000  3.700 0.0000  3.750 0.0000  3.800 0.0000  3.850 0.0000  3.900 0.0000  3.950 0.0000  4.000 0.0000  4.050 0.0000  4.100 0.0000  4.150 0.0000  4.200 0.0000  4.250 0.0000  4.300 0.0000  4.350 0.0000  4.400 0.0000  4.450 0.0000  4.500 0.0000  4.550 0.0000  4.600 0.0000  4.650 0.0000  4.700 0.0000  4.750 0.0000  4.800 0.0000  4.850 0.0000  4.900 0.0000  4.950 0.0000  5.000 0.0000  /
%\put {$\mathbf{\LA r_1^2 \RA}$} at -1 0.033
\endpicture
\end{subfigure}
\normalcolor
\color{black}

\vspace{3mm}

\begin{subfigure}{0.33\textwidth}
\normalcolor
\color{black}
	\caption{$\LA r_1^2\RA$ and $\LA r_2^2\RA$}
\beginpicture
\setcoordinatesystem units <21pt,1.0215pt>
\axescenter{0}{0}{0}{0}{5}{80}{0}
\put {\Large$\alpha$} at 4 -10
\setplotsymbol ({\scalebox{0.3}{$\bullet$}})
\color{Red}
\plot 0.000 0.5000  0.050 0.5000  0.100 0.5000  0.150 0.5000  0.200 0.5000  0.250 0.5000  0.300 0.5000  0.350 0.5000  0.400 0.5000  0.450 0.5000  0.500 0.5000  0.550 0.5001  0.600 0.5001  0.650 0.5002  0.700 0.5002  0.750 0.5002  0.800 0.5003  0.850 0.5003  0.900 0.5003  0.950 0.5004  1.000 0.5005  1.050 0.5007  1.100 0.5009  1.150 0.5011  1.200 0.5013  1.250 0.5014  1.300 0.5017  1.350 0.5021  1.400 0.5026  1.450 0.5033  1.500 0.5040  1.550 0.5050  1.600 0.5061  1.650 0.5077  1.700 0.5095  1.750 0.5115  1.800 0.5142  1.850 0.5170  1.900 0.5210  1.950 0.5265  2.000 0.5330  2.050 0.5418  2.100 0.5528  2.150 0.5660  2.200 0.5849  2.250 0.6108  2.300 0.6488  2.350 0.7071  2.400 0.8093  2.450 1.0399  2.500 37.4021  2.550 72.9540  2.600 73.0374  2.650 73.0757  2.700 73.0975  2.750 73.1124  2.800 73.1249  2.850 73.1323  2.900 73.1372  2.950 73.1412  3.000 73.1437  3.050 73.1460  3.100 73.1498  3.150 73.1512  3.200 73.1523  3.250 73.1531  3.300 73.1535  3.350 73.1541  3.400 73.1556  3.450 73.1559  3.500 73.1576  3.550 73.1577  3.600 73.1598  3.650 73.1599  3.700 73.1599  3.750 73.1600  3.800 73.1592  3.850 73.1593  3.900 73.1607  3.950 73.1608  4.000 73.1598  4.050 73.1598  4.100 73.1575  4.150 73.1574  4.200 73.1591  4.250 73.1591  4.300 73.1587  4.350 73.1587  4.400 73.1576  4.450 73.1576  4.500 73.1583  4.550 73.1583  4.600 73.1581  4.650 73.1580  4.700 73.1579  4.750 73.1578  4.800 73.1573  4.850 73.1573  4.900 73.1570  4.950 73.1570  5.000 73.1572  /
\color{Blue}
\plot 0.000 73.1534  0.050 73.1529  0.100 73.1529  0.150 73.1532  0.200 73.1532  0.250 73.1543  0.300 73.1543  0.350 73.1534  0.400 73.1534  0.450 73.1530  0.500 73.1530  0.550 73.1530  0.600 73.1530  0.650 73.1520  0.700 73.1520  0.750 73.1515  0.800 73.1515  0.850 73.1521  0.900 73.1521  0.950 73.1540  1.000 73.1540  1.050 73.1552  1.100 73.1552  1.150 73.1552  1.200 73.1552  1.250 73.1535  1.300 73.1534  1.350 73.1519  1.400 73.1517  1.450 73.1522  1.500 73.1520  1.550 73.1527  1.600 73.1524  1.650 73.1523  1.700 73.1517  1.750 73.1524  1.800 73.1513  1.850 73.1503  1.900 73.1487  1.950 73.1452  2.000 73.1425  2.050 73.1384  2.100 73.1337  2.150 73.1292  2.200 73.1211  2.250 73.1127  2.300 73.0971  2.350 73.0737  2.400 73.0320  2.450 72.9374  2.500 37.2451  2.550 1.0355  2.600 0.8079  2.650 0.7069  2.700 0.6487  2.750 0.6111  2.800 0.5855  2.850 0.5665  2.900 0.5522  2.950 0.5415  3.000 0.5332  3.050 0.5267  3.100 0.5212  3.150 0.5171  3.200 0.5137  3.250 0.5112  3.300 0.5092  3.350 0.5075  3.400 0.5063  3.450 0.5051  3.500 0.5043  3.550 0.5035  3.600 0.5029  3.650 0.5023  3.700 0.5019  3.750 0.5015  3.800 0.5012  3.850 0.5010  3.900 0.5009  3.950 0.5007  4.000 0.5006  4.050 0.5004  4.100 0.5004  4.150 0.5003  4.200 0.5003  4.250 0.5002  4.300 0.5002  4.350 0.5001  4.400 0.5001  4.450 0.5001  4.500 0.5001  4.550 0.5001  4.600 0.5001  4.650 0.5000  4.700 0.5000  4.750 0.5000  4.800 0.5000  4.850 0.5000  4.900 0.5000  4.950 0.5000  5.000 0.5000 /
\put {\color{Red}$\LA r^2_1 \RA$\color{black}} at 3.75 62
\put {\color{Blue}$\LA r^2_2 \RA$\color{black}} at 1.25 62
\endpicture
\end{subfigure}
\begin{subfigure}{0.33\textwidth}
\normalcolor
\color{black}
	\caption{$\LA Cor(r_1^2,r_2^2) \RA$}
\beginpicture
\setcoordinatesystem units <21pt,22.5pt>
\axescenter{0}{-4}{0}{0}{5}{0}{0}
\axis left ticks withvalues -4 / at -4 /
/
\put {\Large$\alpha$} at 4 -0.35
\setplotsymbol ({\scalebox{0.3}{$\bullet$}})
\color{Red}
\plot 0.000 -0.0000  0.050 -0.0000  0.100 -0.0000  0.150 -0.0000  0.200 -0.0000  0.250 0.0000  0.300 0.0000  0.350 0.0000  0.400 0.0000  0.450 0.0000  0.500 0.0000  0.550 -0.0000  0.600 -0.0000  0.650 -0.0000  0.700 -0.0001  0.750 -0.0001  0.800 -0.0001  0.850 -0.0001  0.900 -0.0001  0.950 -0.0001  1.000 -0.0002  1.050 -0.0002  1.100 -0.0002  1.150 -0.0002  1.200 -0.0003  1.250 -0.0003  1.300 -0.0003  1.350 -0.0005  1.400 -0.0006  1.450 -0.0007  1.500 -0.0008  1.550 -0.0011  1.600 -0.0014  1.650 -0.0019  1.700 -0.0024  1.750 -0.0025  1.800 -0.0031  1.850 -0.0036  1.900 -0.0045  1.950 -0.0057  2.000 -0.0072  2.050 -0.0098  2.100 -0.0128  2.150 -0.0167  2.200 -0.0226  2.250 -0.0314  2.300 -0.0447  2.350 -0.0744  2.400 -0.1362  2.450 -0.3776  2.500 -2.3301  2.550 -0.3427  2.600 -0.1291  2.650 -0.0693  2.700 -0.0441  2.750 -0.0296  2.800 -0.0206  2.850 -0.0152  2.900 -0.0113  2.950 -0.0087  3.000 -0.0070  3.050 -0.0055  3.100 -0.0044  3.150 -0.0035  3.200 -0.0027  3.250 -0.0022  3.300 -0.0018  3.350 -0.0015  3.400 -0.0013  3.450 -0.0010  3.500 -0.0009  3.550 -0.0007  3.600 -0.0006  3.650 -0.0005  3.700 -0.0004  3.750 -0.0003  3.800 -0.0002  3.850 -0.0002  3.900 -0.0002  3.950 -0.0001  4.000 -0.0001  4.050 -0.0001  4.100 -0.0001  4.150 -0.0001  4.200 -0.0000  4.250 -0.0000  4.300 -0.0000  4.350 -0.0000  4.400 -0.0000  4.450 -0.0000  4.500 -0.0000  4.550 -0.0000  4.600 -0.0000  4.650 -0.0000  4.700 -0.0000  4.750 -0.0000  4.800 -0.0000  4.850 -0.0000  4.900 -0.0000  4.950 -0.0000  5.000 -0.0000   /
\endpicture
\end{subfigure}
\begin{subfigure}{0.33\textwidth}
	\caption{$\LA d_{cm}/L \RA$}
\beginpicture
\normalcolor
\color{black}
\setcoordinatesystem units <21pt,85.5pt>
\axes{0}{0}{0}{0}{5}{1}{0}
\put {\Large$\alpha$} at 4 -0.1
\setplotsymbol ({\scalebox{0.3}{$\bullet$}})
\color{Red}
\plot 0.000 0.4610  0.050 0.4610  0.100 0.4610  0.150 0.4609  0.200 0.4609  0.250 0.4606  0.300 0.4605  0.350 0.4610  0.400 0.4610  0.450 0.4612  0.500 0.4612  0.550 0.4610  0.600 0.4611  0.650 0.4612  0.700 0.4612  0.750 0.4615  0.800 0.4615  0.850 0.4614  0.900 0.4614  0.950 0.4608  1.000 0.4609  1.050 0.4604  1.100 0.4604  1.150 0.4602  1.200 0.4603  1.250 0.4609  1.300 0.4609  1.350 0.4615  1.400 0.4615  1.450 0.4614  1.500 0.4614  1.550 0.4609  1.600 0.4609  1.650 0.4606  1.700 0.4606  1.750 0.4604  1.800 0.4605  1.850 0.4607  1.900 0.4609  1.950 0.4616  2.000 0.4619  2.050 0.4625  2.100 0.4630  2.150 0.4632  2.200 0.4640  2.250 0.4643  2.300 0.4655  2.350 0.4667  2.400 0.4691  2.450 0.4734  2.500 0.4814  2.550 0.4729  2.600 0.4703  2.650 0.4683  2.700 0.4670  2.750 0.4659  2.800 0.4645  2.850 0.4639  2.900 0.4637  2.950 0.4634  3.000 0.4633  3.050 0.4631  3.100 0.4625  3.150 0.4624  3.200 0.4621  3.250 0.4620  3.300 0.4621  3.350 0.4620  3.400 0.4618  3.450 0.4618  3.500 0.4613  3.550 0.4613  3.600 0.4606  3.650 0.4606  3.700 0.4607  3.750 0.4606  3.800 0.4609  3.850 0.4609  3.900 0.4605  3.950 0.4605  4.000 0.4609  4.050 0.4609  4.100 0.4616  4.150 0.4616  4.200 0.4610  4.250 0.4610  4.300 0.4612  4.350 0.4611  4.400 0.4613  4.450 0.4613  4.500 0.4610  4.550 0.4610  4.600 0.4610  4.650 0.4610  4.700 0.4611  4.750 0.4612  4.800 0.4613  4.850 0.4613  4.900 0.4613  4.950 0.4613  5.000 0.46135   /
\normalcolor
\color{black}
\endpicture
\end{subfigure}

\vspace{3mm}
\U{Linked from $(0,5)$ to $(5,0)$ across $\tau_1$}: \vspace{1mm}
\vspace{1mm}

\begin{subfigure}{0.33\textwidth}
\normalcolor
\color{black}
	\caption{$\LA k_1/L^2 \RA$ and $\LA k_2/L^2 \RA$}
\beginpicture
\setcoordinatesystem units <21pt,81pt>
\axes{0}{0}{0}{0}{5}{1}{0}
\put {\Large$\alpha$} at 4.0 -0.1
\setplotsymbol ({\scalebox{0.3}{$\bullet$}})
\color{Red}
\plot 0.000 0.0000  0.050 0.0000  0.100 0.0000  0.150 0.0000  0.200 0.0000  0.250 0.0000  0.300 0.0000  0.350 0.0000  0.400 0.0000  0.450 0.0000  0.500 0.0000  0.550 0.0000  0.600 0.0000  0.650 0.0000  0.700 0.0000  0.750 0.0000  0.800 0.0000  0.850 0.0000  0.900 0.0000  0.950 0.0001  1.000 0.0001  1.050 0.0001  1.100 0.0001  1.150 0.0001  1.200 0.0002  1.250 0.0002  1.300 0.0002  1.350 0.0003  1.400 0.0003  1.450 0.0004  1.500 0.0005  1.550 0.0006  1.600 0.0008  1.650 0.0010  1.700 0.0012  1.750 0.0019  1.800 0.0046  1.850 0.0147  1.900 0.0296  1.950 0.0359  2.000 0.0372  2.050 0.0374  2.100 0.0375  2.150 0.0375  2.200 0.0375  2.250 0.0376  2.300 0.0387  2.350 0.0601  2.400 0.1311  2.450 0.2744  2.500 0.9383  2.550 0.9453  2.600 0.9489  2.650 0.9515  2.700 0.9536  2.750 0.9555  2.800 0.9571  2.850 0.9585  2.900 0.9599  2.950 0.9610  3.000 0.9621  3.050 0.9630  3.100 0.9639  3.150 0.9646  3.200 0.9653  3.250 0.9658  3.300 0.9663  3.350 0.9668  3.400 0.9672  3.450 0.9675  3.500 0.9678  3.550 0.9681  3.600 0.9683  3.650 0.9685  3.700 0.9687  3.750 0.9688  3.800 0.9689  3.850 0.9691  3.900 0.9692  3.950 0.9693  4.000 0.9693  4.050 0.9694  4.100 0.9695  4.150 0.9695  4.200 0.9696  4.250 0.9696  4.300 0.9697  4.350 0.9697  4.400 0.9697  4.450 0.9698  4.500 0.9698  4.550 0.9698  4.600 0.9698  4.650 0.9698  4.700 0.9699  4.750 0.9699  4.800 0.9699  4.850 0.9699  4.900 0.9699  4.950 0.9699  5.000 0.9699  /
\color{Blue}
\plot 0.000 0.8032  0.050 0.8032  0.100 0.8032  0.150 0.8032  0.200 0.8031  0.250 0.8031  0.300 0.8031  0.350 0.8031  0.400 0.8030  0.450 0.8030  0.500 0.8030  0.550 0.8029  0.600 0.8029  0.650 0.8028  0.700 0.8028  0.750 0.8027  0.800 0.8026  0.850 0.8025  0.900 0.8025  0.950 0.8023  1.000 0.8022  1.050 0.8021  1.100 0.8020  1.150 0.8018  1.200 0.8016  1.250 0.8014  1.300 0.8011  1.350 0.8009  1.400 0.8005  1.450 0.8002  1.500 0.7998  1.550 0.7993  1.600 0.7987  1.650 0.7981  1.700 0.7972  1.750 0.7957  1.800 0.7910  1.850 0.7750  1.900 0.7520  1.950 0.7424  2.000 0.7404  2.050 0.7401  2.100 0.7400  2.150 0.7400  2.200 0.7399  2.250 0.7395  2.300 0.7346  2.350 0.6726  2.400 0.5493  2.450 0.4198  2.500 0.0074  2.550 0.0054  2.600 0.0046  2.650 0.0041  2.700 0.0037  2.750 0.0035  2.800 0.0033  2.850 0.0031  2.900 0.0030  2.950 0.0029  3.000 0.0028  3.050 0.0028  3.100 0.0027  3.150 0.0027  3.200 0.0026  3.250 0.0026  3.300 0.0026  3.350 0.0026  3.400 0.0026  3.450 0.0025  3.500 0.0025  3.550 0.0025  3.600 0.0025  3.650 0.0025  3.700 0.0025  3.750 0.0025  3.800 0.0025  3.850 0.0025  3.900 0.0025  3.950 0.0025  4.000 0.0025  4.050 0.0025  4.100 0.0025  4.150 0.0025  4.200 0.0025  4.250 0.0025  4.300 0.0025  4.350 0.0025  4.400 0.0025  4.450 0.0025  4.500 0.0025  4.550 0.0025  4.600 0.0025  4.650 0.0025  4.700 0.0025  4.750 0.0025  4.800 0.0025  4.850 0.0025  4.900 0.0025  4.950 0.0025  5.000 0.0025 /
\put {\color{Red}$\LA k_1/L^2 \RA$\color{black}} at 4 0.86
\put {\color{Blue}$\LA k_2/L^2 \RA$\color{black}} at 4 0.10
\endpicture
\color{black}
\end{subfigure}
\begin{subfigure}{0.33\textwidth}
\normalcolor
\color{black}
	\caption{$\LA Cor(k_1,k_2)/L^2 \RA$}
\beginpicture
\setcoordinatesystem units <21pt,22.5pt>
\axes{0}{-3}{0}{0}{5}{1}{0}
\put {\Large$\alpha$} at 4.0 -0.4
\setplotsymbol ({\scalebox{0.3}{$\bullet$}})
\color{Red}
\plot 0.000 -0.0000  0.050 -0.0000  0.100 -0.0000  0.150 -0.0000  0.200 -0.0000  0.250 -0.0000  0.300 -0.0000  0.350 -0.0000  0.400 -0.0000  0.450 -0.0000  0.500 -0.0000  0.550 -0.0000  0.600 -0.0000  0.650 -0.0000  0.700 -0.0001  0.750 -0.0001  0.800 -0.0001  0.850 -0.0001  0.900 -0.0001  0.950 -0.0001  1.000 -0.0002  1.050 -0.0002  1.100 -0.0003  1.150 -0.0003  1.200 -0.0004  1.250 -0.0005  1.300 -0.0006  1.350 -0.0007  1.400 -0.0008  1.450 -0.0010  1.500 -0.0012  1.550 -0.0015  1.600 -0.0019  1.650 -0.0026  1.700 -0.0045  1.750 -0.0140  1.800 -0.0621  1.850 -0.1790  1.900 -0.1309  1.950 -0.0319  2.000 -0.0057  2.050 -0.0010  2.100 -0.0002  2.150 -0.0001  2.200 -0.0003  2.250 -0.0036  2.300 -0.0616  2.350 -0.8210  2.400 -0.7777  2.450 -2.3988  2.500 -0.0570  2.550 -0.0159  2.600 -0.0079  2.650 -0.0048  2.700 -0.0033  2.750 -0.0025  2.800 -0.0020  2.850 -0.0016  2.900 -0.0013  2.950 -0.0011  3.000 -0.0009  3.050 -0.0007  3.100 -0.0006  3.150 -0.0005  3.200 -0.0004  3.250 -0.0003  3.300 -0.0003  3.350 -0.0002  3.400 -0.0002  3.450 -0.0001  3.500 -0.0001  3.550 -0.0001  3.600 -0.0001  3.650 -0.0001  3.700 -0.0001  3.750 -0.0000  3.800 -0.0000  3.850 -0.0000  3.900 -0.0000  3.950 -0.0000  4.000 -0.0000  4.050 -0.0000  4.100 -0.0000  4.150 -0.0000  4.200 -0.0000  4.250 -0.0000  4.300 -0.0000  4.350 -0.0000  4.400 -0.0000  4.450 -0.0000  4.500 -0.0000  4.550 -0.0000  4.600 -0.0000  4.650 -0.0000  4.700 -0.0000  4.750 -0.0000  4.800 -0.0000  4.850 -0.0000  4.900 -0.0000  4.950 -0.0000  5.000 -0.0000   /
\color{black}
%\put {\color{Red}$\LA k_1 \RA$\color{black}} at 1 0.05
\endpicture
\end{subfigure}
\begin{subfigure}{0.33\textwidth}
\normalcolor
\color{black}
	\caption{$\LA k_m/L^2 \RA$}
\beginpicture
\setcoordinatesystem units <21pt,270pt>
\axes{0}{0}{0}{0}{5}{0.3}{0}
\put {\Large$\alpha$} at 4.0 -0.05
\setplotsymbol ({\scalebox{0.3}{$\bullet$}})
\color{Red}
\plot 0.000 0.1895  0.050 0.1895  0.100 0.1895  0.150 0.1895  0.200 0.1895  0.250 0.1895  0.300 0.1895  0.350 0.1895  0.400 0.1895  0.450 0.1895  0.500 0.1895  0.550 0.1895  0.600 0.1895  0.650 0.1895  0.700 0.1895  0.750 0.1895  0.800 0.1895  0.850 0.1895  0.900 0.1895  0.950 0.1895  1.000 0.1895  1.050 0.1896  1.100 0.1896  1.150 0.1896  1.200 0.1896  1.250 0.1896  1.300 0.1897  1.350 0.1897  1.400 0.1897  1.450 0.1898  1.500 0.1899  1.550 0.1899  1.600 0.1901  1.650 0.1902  1.700 0.1904  1.750 0.1909  1.800 0.1927  1.850 0.1996  1.900 0.2096  1.950 0.2139  2.000 0.2148  2.050 0.2150  2.100 0.2150  2.150 0.2150  2.200 0.2150  2.250 0.2153  2.300 0.2177  2.350 0.2442  2.400 0.2786  2.450 0.2631  2.500 0.0286  2.550 0.0254  2.600 0.0239  2.650 0.0230  2.700 0.0224  2.750 0.0219  2.800 0.0215  2.850 0.0212  2.900 0.0210  2.950 0.0208  3.000 0.0206  3.050 0.0205  3.100 0.0204  3.150 0.0203  3.200 0.0203  3.250 0.0202  3.300 0.0202  3.350 0.0201  3.400 0.0201  3.450 0.0201  3.500 0.0201  3.550 0.0201  3.600 0.0200  3.650 0.0200  3.700 0.0200  3.750 0.0200  3.800 0.0200  3.850 0.0200  3.900 0.0200  3.950 0.0200  4.000 0.0200  4.050 0.0200  4.100 0.0200  4.150 0.0200  4.200 0.0200  4.250 0.0200  4.300 0.0200  4.350 0.0200  4.400 0.0200  4.450 0.0200  4.500 0.0200  4.550 0.0200  4.600 0.0200  4.650 0.0200  4.700 0.0200  4.750 0.0200  4.800 0.0200  4.850 0.0200  4.900 0.0200  4.950 0.0200  5.000 0.0200  /
%\put {$\mathbf{\LA r_1^2 \RA}$} at -1 0.033
\endpicture
\end{subfigure}
\normalcolor
\color{black}

\vspace{3mm}

\begin{subfigure}{0.33\textwidth}
\normalcolor
\color{black}
	\caption{$\LA r_1^2\RA$ and $\LA r_2^2\RA$}
\beginpicture
\setcoordinatesystem units <21pt,0.51075pt>
\axescenter{0}{0}{0}{0}{5}{160}{0}
\put {\Large$\alpha$} at 4 -20
\setplotsymbol ({\scalebox{0.3}{$\bullet$}})
\color{Red}
\plot 0.000 133.4651  0.050 133.4651  0.100 133.4651  0.150 133.4651  0.200 133.4651  0.250 133.4650  0.300 133.4650  0.350 133.4649  0.400 133.4649  0.450 133.4648  0.500 133.4646  0.550 133.4645  0.600 133.4642  0.650 133.4640  0.700 133.4637  0.750 133.4632  0.800 133.4627  0.850 133.4621  0.900 133.4613  0.950 133.4603  1.000 133.4590  1.050 133.4575  1.100 133.4556  1.150 133.4532  1.200 133.4503  1.250 133.4466  1.300 133.4420  1.350 133.4363  1.400 133.4292  1.450 133.4204  1.500 133.4093  1.550 133.3953  1.600 133.3775  1.650 133.3535  1.700 133.3135  1.750 133.2002  1.800 132.6931  1.850 130.7438  1.900 127.8508  1.950 126.6164  2.000 126.3555  2.050 126.3085  2.100 126.2995  2.150 126.2954  2.200 126.2819  2.250 126.1919  2.300 125.3057  2.350 115.2968  2.400 100.2395  2.450 92.7775  2.500 74.1035  2.550 73.9274  2.600 73.8581  2.650 73.8166  2.700 73.7856  2.750 73.7596  2.800 73.7371  2.850 73.7174  2.900 73.7003  2.950 73.6858  3.000 73.6735  3.050 73.6634  3.100 73.6551  3.150 73.6484  3.200 73.6431  3.250 73.6390  3.300 73.6358  3.350 73.6334  3.400 73.6316  3.450 73.6304  3.500 73.6295  3.550 73.6290  3.600 73.6286  3.650 73.6285  3.700 73.6285  3.750 73.6286  3.800 73.6288  3.850 73.6291  3.900 73.6293  3.950 73.6296  4.000 73.6299  4.050 73.6302  4.100 73.6305  4.150 73.6308  4.200 73.6311  4.250 73.6313  4.300 73.6316  4.350 73.6318  4.400 73.6321  4.450 73.6323  4.500 73.6325  4.550 73.6326  4.600 73.6328  4.650 73.6329  4.700 73.6331  4.750 73.6332  4.800 73.6333  4.850 73.6334  4.900 73.6335  4.950 73.6336  5.000 73.6337  /
\color{Blue}
\plot 0.000 59.9842  0.050 59.9842  0.100 59.9842  0.150 59.9842  0.200 59.9842  0.250 59.9842  0.300 59.9841  0.350 59.9841  0.400 59.9841  0.450 59.9840  0.500 59.9839  0.550 59.9838  0.600 59.9836  0.650 59.9834  0.700 59.9832  0.750 59.9829  0.800 59.9825  0.850 59.9819  0.900 59.9813  0.950 59.9805  1.000 59.9795  1.050 59.9782  1.100 59.9767  1.150 59.9747  1.200 59.9723  1.250 59.9693  1.300 59.9656  1.350 59.9610  1.400 59.9553  1.450 59.9482  1.500 59.9395  1.550 59.9286  1.600 59.9151  1.650 59.8978  1.700 59.8731  1.750 59.8210  1.800 59.6334  1.850 58.9626  1.900 57.9894  1.950 57.5798  2.000 57.4944  2.050 57.4793  2.100 57.4768  2.150 57.4767  2.200 57.4790  2.250 57.4935  2.300 57.6055  2.350 58.3349  2.400 57.0914  2.450 49.7069  2.500 1.3707  2.550 0.9914  2.600 0.8492  2.650 0.7658  2.700 0.7081  2.750 0.6648  2.800 0.6312  2.850 0.6047  2.900 0.5837  2.950 0.5669  3.000 0.5536  3.050 0.5430  3.100 0.5346  3.150 0.5278  3.200 0.5225  3.250 0.5182  3.300 0.5147  3.350 0.5119  3.400 0.5097  3.450 0.5079  3.500 0.5064  3.550 0.5052  3.600 0.5043  3.650 0.5035  3.700 0.5028  3.750 0.5023  3.800 0.5019  3.850 0.5015  3.900 0.5013  3.950 0.5010  4.000 0.5008  4.050 0.5007  4.100 0.5006  4.150 0.5005  4.200 0.5004  4.250 0.5003  4.300 0.5003  4.350 0.5002  4.400 0.5002  4.450 0.5001  4.500 0.5001  4.550 0.5001  4.600 0.5001  4.650 0.5001  4.700 0.5001  4.750 0.5000  4.800 0.5000  4.850 0.5000  4.900 0.5000  4.950 0.5000  5.000 0.5000 /
\put {\color{Red}$\LA r^2_1 \RA$\color{black}} at 3.75 55
\put {\color{Blue}$\LA r^2_2 \RA$\color{black}} at 1.25 40
\endpicture
\end{subfigure}
\begin{subfigure}{0.33\textwidth}
\normalcolor
\color{black}
	\caption{$\LA Cor(r_1^2,r_2^2) \RA$}
\beginpicture
\setcoordinatesystem units <21pt,2.25pt>
\axescenter{0}{-20}{0}{0}{5}{20}{0}
\axis left ticks withvalues -20 20 / at -20 20 /
/
\put {\Large$\alpha$} at 4 -4
\setplotsymbol ({\scalebox{0.3}{$\bullet$}})
\color{Red}
\plot 0.000 0.0094  0.050 0.0094  0.100 0.0094  0.150 0.0094  0.200 0.0094  0.250 0.0094  0.300 0.0094  0.350 0.0093  0.400 0.0093  0.450 0.0093  0.500 0.0093  0.550 0.0093  0.600 0.0093  0.650 0.0093  0.700 0.0093  0.750 0.0093  0.800 0.0093  0.850 0.0093  0.900 0.0094  0.950 0.0094  1.000 0.0094  1.050 0.0095  1.100 0.0095  1.150 0.0096  1.200 0.0098  1.250 0.0100  1.300 0.0103  1.350 0.0106  1.400 0.0112  1.450 0.0119  1.500 0.0129  1.550 0.0144  1.600 0.0169  1.650 0.0238  1.700 0.0558  1.750 0.2415  1.800 1.2239  1.850 3.6602  1.900 2.7137  1.950 0.6674  2.000 0.1210  2.050 0.0211  2.100 0.0016  2.150 -0.0111  2.200 -0.0620  2.250 -0.3741  2.300 -2.8200  2.350 -12.0263  2.400 -0.2003  2.450 13.8875  2.500 0.9629  2.550 0.2771  2.600 0.1540  2.650 0.1101  2.700 0.0877  2.750 0.0730  2.800 0.0617  2.850 0.0521  2.900 0.0439  2.950 0.0368  3.000 0.0307  3.050 0.0255  3.100 0.0211  3.150 0.0174  3.200 0.0143  3.250 0.0117  3.300 0.0096  3.350 0.0079  3.400 0.0065  3.450 0.0053  3.500 0.0043  3.550 0.0035  3.600 0.0029  3.650 0.0024  3.700 0.0019  3.750 0.0016  3.800 0.0013  3.850 0.0011  3.900 0.0009  3.950 0.0007  4.000 0.0006  4.050 0.0005  4.100 0.0004  4.150 0.0003  4.200 0.0003  4.250 0.0002  4.300 0.0002  4.350 0.0001  4.400 0.0001  4.450 0.0001  4.500 0.0001  4.550 0.0001  4.600 0.0001  4.650 0.0000  4.700 0.0000  4.750 0.0000  4.800 0.0000  4.850 0.0000  4.900 0.0000  4.950 0.0000  5.000 0.0000   /
\endpicture
\end{subfigure}
\begin{subfigure}{0.33\textwidth}
	\caption{$\LA d_{cm}/L \RA$}
\beginpicture
\normalcolor
\color{black}
\setcoordinatesystem units <21pt,85.5pt>
\axes{0}{0}{0}{0}{5}{1}{0}
\put {\Large$\alpha$} at 4 -0.1
\setplotsymbol ({\scalebox{0.3}{$\bullet$}})
\color{Red}
\plot 0.000 0.0010  0.050 0.0010  0.100 0.0010  0.150 0.0010  0.200 0.0010  0.250 0.0010  0.300 0.0010  0.350 0.0010  0.400 0.0010  0.450 0.0010  0.500 0.0010  0.550 0.0010  0.600 0.0010  0.650 0.0010  0.700 0.0011  0.750 0.0011  0.800 0.0011  0.850 0.0012  0.900 0.0012  0.950 0.0012  1.000 0.0013  1.050 0.0014  1.100 0.0014  1.150 0.0015  1.200 0.0017  1.250 0.0018  1.300 0.0020  1.350 0.0022  1.400 0.0024  1.450 0.0027  1.500 0.0031  1.550 0.0036  1.600 0.0041  1.650 0.0047  1.700 0.0055  1.750 0.0064  1.800 0.0075  1.850 0.0088  1.900 0.0103  1.950 0.0120  2.000 0.0142  2.050 0.0167  2.100 0.0197  2.150 0.0234  2.200 0.0279  2.250 0.0337  2.300 0.0415  2.350 0.0669  2.400 0.0917  2.450 0.1664  2.500 0.3156  2.550 0.3840  2.600 0.3864  2.650 0.3872  2.700 0.3882  2.750 0.3886  2.800 0.3891  2.850 0.3893  2.900 0.3898  2.950 0.3899  3.000 0.3904  3.050 0.3904  3.100 0.3903  3.150 0.3903  3.200 0.3904  3.250 0.3904  3.300 0.3905  3.350 0.3905  3.400 0.3904  3.450 0.3904  3.500 0.3910  3.550 0.3910  3.600 0.3906  3.650 0.3906  3.700 0.3901  3.750 0.3901  3.800 0.3902  3.850 0.3902  3.900 0.3903  3.950 0.3903  4.000 0.3901  4.050 0.3901  4.100 0.3906  4.150 0.3906  4.200 0.3909  4.250 0.3909  4.300 0.3902  4.350 0.3902  4.400 0.3896  4.450 0.3896  4.500 0.3897  4.550 0.3897  4.600 0.3893  4.650 0.3893  4.700 0.3887  4.750 0.3887  4.800 0.3890  4.850 0.3890  4.900 0.3890  4.950 0.3890  5.000 0.3890   /
\normalcolor
\color{black}
\endpicture
\end{subfigure}
\caption{Across the $\tau$ and $\tau_1$ phase boundaries in the $\alpha\beta$-plane.
Graphs for the unlinked model are shown in (a)-(f), and for the linked model
are shown in (g)-(l).  Consistent with the results in figure \ref{10}(a)-(c)
and \ref{17}(a)-(c), these results show first order transitions as the 
$\tau$ and $\tau_1$ phase boundaries are crossed in the two models respectively.}
\label{22}   % %ZXZ[22]
\color{black}
\normalcolor
\end{figure}

In figure \ref{22} we show thermodynamic and metric data estimates
along the line segment with endpoints $(0,5)$ and $(5,0)$ in the phase 
diagram for the unlinked model (figure \ref{10}) and the linked model
(figure \ref{17}).  This line crosses the $\tau$ phase boundary
in the unlinked model, and the $\tau_1$ phase boundary in the linked
model.  The results for the unlinked case are shown in figures
\ref{22}(a)-(f), while the results for the linked model are shown in
figures \ref{22}(g)-(l).

The results in both models are consistent with a first order transition
as the phase boundaries ($\tau$ or $\tau_1$, respectively) are crossed.
There are sharp discontinuities in the mean number of self-contacts
in both models, as well as in the metric quantities.  The correlations
show sharp peaks at the locations of the transitions.  Figure \ref{22}(j) indicates
that the outer polygon in the linked model is inflated by the inner polygon
in the $c_2$-dominated phase, and expands in the $c_1$-dominated 
phase to compress the inner polygon to a very small size. 

\begin{figure}[h!]
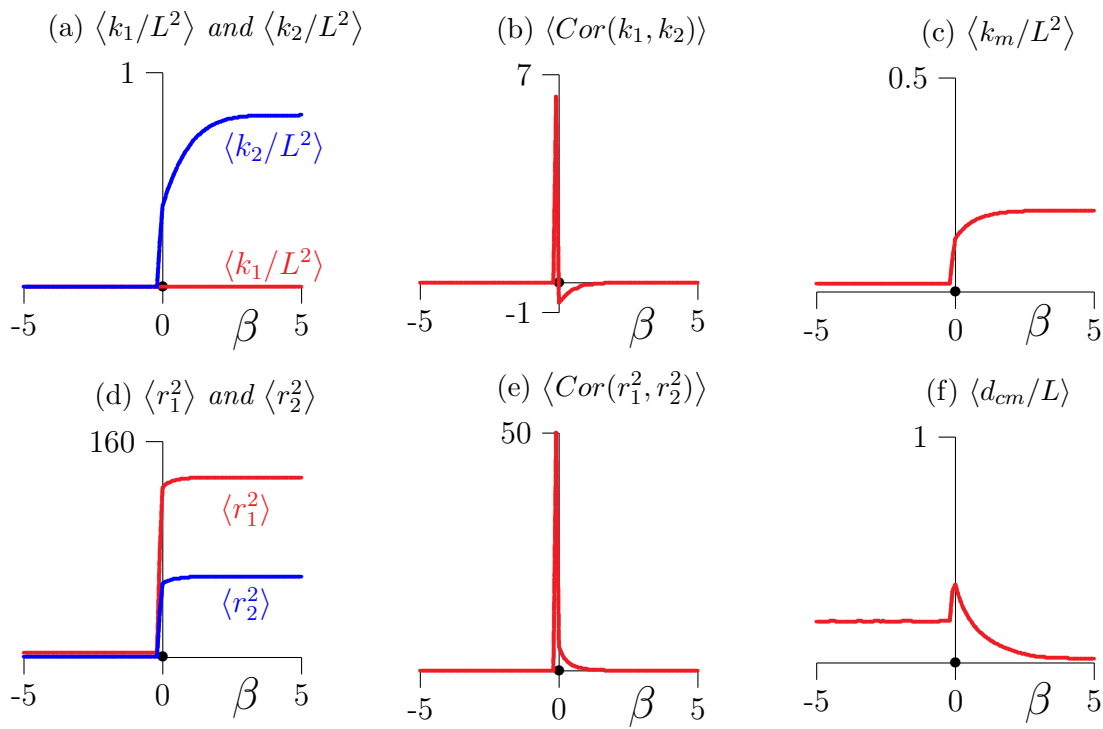

\U{Linked from $(-2,-5)$ to $(-2,5)$ across $\tau_2$}: \vspace{1mm}

\begin{subfigure}{0.33\textwidth}
\normalcolor
\color{black}
	\caption{$\LA k_1/L^2 \RA$ and $\LA k_2/L^2 \RA$}
\beginpicture
\setcoordinatesystem units <10.5pt,81pt>
\axescenter{-5}{0}{0}{0}{5}{1}{0}
\put {\Large$\beta$} at 3.0 -0.2
\setplotsymbol ({\scalebox{0.3}{$\bullet$}})
\color{Red}
\plot -5.000 0.0006  -4.900 0.0006  -4.800 0.0006  -4.700 0.0006  -4.600 0.0006  -4.500 0.0006  -4.400 0.0006  -4.300 0.0006  -4.200 0.0006  -4.100 0.0007  -4.000 0.0007  -3.900 0.0006  -3.800 0.0006  -3.700 0.0006  -3.600 0.0006  -3.500 0.0006  -3.400 0.0006  -3.300 0.0006  -3.200 0.0006  -3.100 0.0006  -3.000 0.0006  -2.900 0.0006  -2.800 0.0006  -2.700 0.0006  -2.600 0.0006  -2.500 0.0006  -2.400 0.0006  -2.300 0.0006  -2.200 0.0006  -2.100 0.0006  -2.000 0.0006  -1.900 0.0006  -1.800 0.0006  -1.700 0.0006  -1.600 0.0006  -1.500 0.0006  -1.400 0.0006  -1.300 0.0006  -1.200 0.0006  -1.100 0.0006  -1.000 0.0006  -0.900 0.0006  -0.800 0.0006  -0.700 0.0006  -0.600 0.0006  -0.500 0.0006  -0.400 0.0006  -0.300 0.0006  -0.200 0.0006  -0.100 0.0010  0.000 0.0009  0.100 0.0007  0.200 0.0005  0.300 0.0004  0.400 0.0003  0.500 0.0003  0.600 0.0002  0.700 0.0002  0.800 0.0001  0.900 0.0001  1.000 0.0001  1.100 0.0001  1.200 0.0001  1.300 0.0000  1.400 0.0000  1.500 0.0000  1.600 0.0000  1.700 0.0000  1.800 0.0000  1.900 0.0000  2.000 0.0000  2.100 0.0000  2.200 0.0000  2.300 0.0000  2.400 0.0000  2.500 0.0000  2.600 0.0000  2.700 0.0000  2.800 0.0000  2.900 0.0000  3.000 0.0000  3.100 0.0000  3.200 0.0000  3.300 0.0000  3.400 0.0000  3.500 0.0000  3.600 0.0000  3.700 0.0000  3.800 0.0000  3.900 0.0000  4.000 0.0000  4.100 0.0000  4.200 0.0000  4.300 0.0000  4.400 0.0000  4.500 0.0000  4.600 0.0000  4.700 0.0000  4.800 0.0000  4.900 0.0000  5.000 0.0000 /
\color{Blue}
\plot -5.000 0.0000  -4.900 0.0000  -4.800 0.0000  -4.700 0.0000  -4.600 0.0000  -4.500 0.0000  -4.400 0.0000  -4.300 0.0000  -4.200 0.0000  -4.100 0.0000  -4.000 0.0000  -3.900 0.0000  -3.800 0.0000  -3.700 0.0000  -3.600 0.0000  -3.500 0.0000  -3.400 0.0000  -3.300 0.0000  -3.200 0.0000  -3.100 0.0000  -3.000 0.0000  -2.900 0.0000  -2.800 0.0000  -2.700 0.0000  -2.600 0.0000  -2.500 0.0000  -2.400 0.0000  -2.300 0.0000  -2.200 0.0000  -2.100 0.0000  -2.000 0.0000  -1.900 0.0000  -1.800 0.0000  -1.700 0.0000  -1.600 0.0000  -1.500 0.0000  -1.400 0.0000  -1.300 0.0000  -1.200 0.0000  -1.100 0.0000  -1.000 0.0000  -0.900 0.0000  -0.800 0.0000  -0.700 0.0000  -0.600 0.0000  -0.500 0.0001  -0.400 0.0001  -0.300 0.0001  -0.200 0.0001  -0.100 0.2089  0.000 0.3754  0.100 0.4169  0.200 0.4556  0.300 0.4906  0.400 0.5227  0.500 0.5521  0.600 0.5787  0.700 0.6029  0.800 0.6253  0.900 0.6454  1.000 0.6635  1.100 0.6799  1.200 0.6944  1.300 0.7076  1.400 0.7193  1.500 0.7298  1.600 0.7392  1.700 0.7476  1.800 0.7550  1.900 0.7615  2.000 0.7673  2.100 0.7723  2.200 0.7766  2.300 0.7803  2.400 0.7836  2.500 0.7864  2.600 0.7887  2.700 0.7907  2.800 0.7923  2.900 0.7937  3.000 0.7949  3.100 0.7959  3.200 0.7967  3.300 0.7974  3.400 0.7979  3.500 0.7984  3.600 0.7988  3.700 0.7991  3.800 0.7994  3.900 0.7996  4.000 0.7997  4.100 0.7999  4.200 0.8000  4.300 0.8001  4.400 0.8002  4.500 0.8003  4.600 0.8003  4.700 0.8004  4.800 0.8004  4.900 0.8004  5.000 0.8005 /
\color{black}
\put {\color{Red}$\LA k_1/L^2 \RA$\color{black}} at 4 0.10
\put {\color{Blue}$\LA k_2/L^2 \RA$\color{black}} at 4 0.65
\endpicture
\end{subfigure}
\begin{subfigure}{0.33\textwidth}
\normalcolor
\color{black}
	\caption{$\LA Cor(k_1,k_2) \RA$}
\beginpicture
\setcoordinatesystem units <10.5pt,11.25pt>
\axes{-5}{-1}{0}{0}{5}{7}{0}
\put {\Large$\beta$} at 3.0 -1.5
\setplotsymbol ({\scalebox{0.3}{$\bullet$}})
\color{Red}
\plot -5.000 0.0000  -4.900 -0.0000  -4.800 -0.0000  -4.700 -0.0000  -4.600 -0.0000  -4.500 -0.0000  -4.400 -0.0000  -4.300 -0.0000  -4.200 -0.0000  -4.100 -0.0000  -4.000 -0.0000  -3.900 -0.0000  -3.800 -0.0000  -3.700 -0.0000  -3.600 -0.0000  -3.500 -0.0000  -3.400 -0.0000  -3.300 -0.0000  -3.200 -0.0000  -3.100 -0.0000  -3.000 -0.0000  -2.900 -0.0000  -2.800 -0.0000  -2.700 0.0000  -2.600 0.0000  -2.500 0.0000  -2.400 0.0001  -2.300 0.0001  -2.200 0.0001  -2.100 0.0001  -2.000 0.0001  -1.900 0.0000  -1.800 0.0001  -1.700 0.0001  -1.600 0.0001  -1.500 0.0001  -1.400 0.0002  -1.300 0.0003  -1.200 0.0004  -1.100 0.0003  -1.000 0.0004  -0.900 0.0003  -0.800 0.0003  -0.700 0.0006  -0.600 0.0007  -0.500 0.0009  -0.400 0.0011  -0.300 0.0018  -0.200 0.0023  -0.100 6.2542  0.000 -0.6825  0.100 -0.5667  0.200 -0.4813  0.300 -0.3607  0.400 -0.3074  0.500 -0.2236  0.600 -0.2129  0.700 -0.1756  0.800 -0.1110  0.900 -0.0794  1.000 -0.0823  1.100 -0.0715  1.200 -0.0595  1.300 -0.0454  1.400 -0.0364  1.500 -0.0268  1.600 -0.0206  1.700 -0.0154  1.800 -0.0110  1.900 -0.0094  2.000 -0.0074  2.100 -0.0058  2.200 -0.0057  2.300 -0.0043  2.400 -0.0032  2.500 -0.0026  2.600 -0.0013  2.700 -0.0012  2.800 -0.0010  2.900 -0.0009  3.000 -0.0015  3.100 -0.0011  3.200 -0.0009  3.300 -0.0007  3.400 -0.0006  3.500 -0.0005  3.600 -0.0004  3.700 -0.0003  3.800 -0.0002  3.900 -0.0001  4.000 -0.0002  4.100 -0.0001  4.200 -0.0001  4.300 -0.0001  4.400 -0.0001  4.500 -0.0001  4.600 -0.0001  4.700 -0.0001  4.800 -0.0000  4.900 -0.0000  5.000 0.0000  /
\endpicture
\end{subfigure}
\begin{subfigure}{0.33\textwidth}
\normalcolor
\color{black}
	\caption{$\LA k_m/L^2 \RA$}
\beginpicture
\setcoordinatesystem units <10.5pt,162pt>
\axescenter{-5}{0}{0}{0}{5}{0.5}{0}
\put {\Large$\beta$} at 3.0 -0.075
\setplotsymbol ({\scalebox{0.3}{$\bullet$}})
\color{Red}
\plot -5.000 0.0191  -4.900 0.0191  -4.800 0.0191  -4.700 0.0191  -4.600 0.0191  -4.500 0.0191  -4.400 0.0191  -4.300 0.0191  -4.200 0.0191  -4.100 0.0191  -4.000 0.0191  -3.900 0.0191  -3.800 0.0191  -3.700 0.0191  -3.600 0.0191  -3.500 0.0191  -3.400 0.0191  -3.300 0.0191  -3.200 0.0191  -3.100 0.0191  -3.000 0.0191  -2.900 0.0191  -2.800 0.0191  -2.700 0.0191  -2.600 0.0191  -2.500 0.0191  -2.400 0.0191  -2.300 0.0191  -2.200 0.0191  -2.100 0.0191  -2.000 0.0191  -1.900 0.0191  -1.800 0.0191  -1.700 0.0191  -1.600 0.0191  -1.500 0.0191  -1.400 0.0191  -1.300 0.0191  -1.200 0.0191  -1.100 0.0191  -1.000 0.0191  -0.900 0.0191  -0.800 0.0192  -0.700 0.0192  -0.600 0.0192  -0.500 0.0192  -0.400 0.0192  -0.300 0.0192  -0.200 0.0193  -0.100 0.0783  0.000 0.1235  0.100 0.1330  0.200 0.1411  0.300 0.1479  0.400 0.1538  0.500 0.1588  0.600 0.1631  0.700 0.1668  0.800 0.1700  0.900 0.1728  1.000 0.1751  1.100 0.1772  1.200 0.1789  1.300 0.1805  1.400 0.1818  1.500 0.1829  1.600 0.1839  1.700 0.1848  1.800 0.1855  1.900 0.1861  2.000 0.1867  2.100 0.1871  2.200 0.1875  2.300 0.1879  2.400 0.1882  2.500 0.1884  2.600 0.1886  2.700 0.1888  2.800 0.1889  2.900 0.1890  3.000 0.1891  3.100 0.1892  3.200 0.1893  3.300 0.1893  3.400 0.1894  3.500 0.1894  3.600 0.1894  3.700 0.1894  3.800 0.1895  3.900 0.1895  4.000 0.1895  4.100 0.1895  4.200 0.1895  4.300 0.1895  4.400 0.1895  4.500 0.1895  4.600 0.1895  4.700 0.1895  4.800 0.1895  4.900 0.1895  5.000 0.1895  /
\endpicture
\end{subfigure}
\normalcolor
\color{black}

\vspace{2mm}

\begin{subfigure}{0.33\textwidth}
\normalcolor
\color{black}
	\caption{$\LA r_1^2\RA$ and $\LA r_2^2\RA$}
\beginpicture
\setcoordinatesystem units <10.5pt,0.51075pt>
\axescenter{-5}{0}{0}{0}{5}{160}{0}
\put {\Large$\beta$} at 3 -30
\setplotsymbol ({\scalebox{0.3}{$\bullet$}})
\color{Red}
\plot -5.000 3.5180  -4.900 3.5174  -4.800 3.5174  -4.700 3.5181  -4.600 3.5181  -4.500 3.5218  -4.400 3.5218  -4.300 3.5219  -4.200 3.5219  -4.100 3.5191  -4.000 3.5191  -3.900 3.5180  -3.800 3.5180  -3.700 3.5204  -3.600 3.5204  -3.500 3.5221  -3.400 3.5221  -3.300 3.5185  -3.200 3.5185  -3.100 3.5184  -3.000 3.5184  -2.900 3.5208  -2.800 3.5209  -2.700 3.5209  -2.600 3.5210  -2.500 3.5196  -2.400 3.5197  -2.300 3.5188  -2.200 3.5189  -2.100 3.5190  -2.000 3.5192  -1.900 3.5217  -1.800 3.5221  -1.700 3.5233  -1.600 3.5238  -1.500 3.5233  -1.400 3.5240  -1.300 3.5239  -1.200 3.5250  -1.100 3.5253  -1.000 3.5269  -0.900 3.5298  -0.800 3.5323  -0.700 3.5364  -0.600 3.5404  -0.500 3.5448  -0.400 3.5511  -0.300 3.5589  -0.200 3.5698  -0.100 79.3987  0.000 126.0579  0.100 127.9451  0.200 129.2992  0.300 130.2518  0.400 130.9727  0.500 131.5013  0.600 131.8957  0.700 132.2049  0.800 132.4552  0.900 132.6458  1.000 132.7977  1.100 132.9188  1.200 133.0184  1.300 133.0975  1.400 133.1657  1.500 133.2205  1.600 133.2654  1.700 133.3029  1.800 133.3336  1.900 133.3598  2.000 133.3830  2.100 133.4010  2.200 133.4152  2.300 133.4279  2.400 133.4387  2.500 133.4475  2.600 133.4545  2.700 133.4605  2.800 133.4652  2.900 133.4694  3.000 133.4732  3.100 133.4760  3.200 133.4784  3.300 133.4803  3.400 133.4819  3.500 133.4832  3.600 133.4841  3.700 133.4850  3.800 133.4857  3.900 133.4863  4.000 133.4867  4.100 133.4871  4.200 133.4872  4.300 133.4875  4.400 133.4877  4.500 133.4878  4.600 133.4880  4.700 133.4881  4.800 133.4882  4.900 133.4883  5.000 133.4883  /
\color{Blue}
\plot -5.000 0.5000  -4.900 0.5000  -4.800 0.5000  -4.700 0.5000  -4.600 0.5000  -4.500 0.5000  -4.400 0.5000  -4.300 0.5000  -4.200 0.5000  -4.100 0.5000  -4.000 0.5000  -3.900 0.5000  -3.800 0.5000  -3.700 0.5000  -3.600 0.5000  -3.500 0.5000  -3.400 0.5000  -3.300 0.5000  -3.200 0.5000  -3.100 0.5000  -3.000 0.5000  -2.900 0.5001  -2.800 0.5001  -2.700 0.5001  -2.600 0.5001  -2.500 0.5002  -2.400 0.5002  -2.300 0.5002  -2.200 0.5003  -2.100 0.5003  -2.000 0.5004  -1.900 0.5005  -1.800 0.5006  -1.700 0.5008  -1.600 0.5009  -1.500 0.5011  -1.400 0.5014  -1.300 0.5017  -1.200 0.5020  -1.100 0.5025  -1.000 0.5030  -0.900 0.5037  -0.800 0.5046  -0.700 0.5057  -0.600 0.5071  -0.500 0.5088  -0.400 0.5109  -0.300 0.5139  -0.200 0.5175  -0.100 33.6622  0.000 54.5557  0.100 55.7887  0.200 56.7052  0.300 57.3794  0.400 57.9165  0.500 58.3262  0.600 58.6400  0.700 58.8919  0.800 59.1014  0.900 59.2664  1.000 59.4001  1.100 59.5078  1.200 59.6014  1.300 59.6746  1.400 59.7374  1.500 59.7879  1.600 59.8300  1.700 59.8626  1.800 59.8892  1.900 59.9114  2.000 59.9318  2.100 59.9460  2.200 59.9567  2.300 59.9662  2.400 59.9745  2.500 59.9807  2.600 59.9854  2.700 59.9893  2.800 59.9915  2.900 59.9941  3.000 59.9964  3.100 59.9981  3.200 59.9998  3.300 60.0009  3.400 60.0016  3.500 60.0023  3.600 60.0021  3.700 60.0026  3.800 60.0029  3.900 60.0032  4.000 60.0033  4.100 60.0035  4.200 60.0026  4.300 60.0027  4.400 60.0025  4.500 60.0026  4.600 60.0028  4.700 60.0028  4.800 60.0028  4.900 60.0028  5.000 60.0029 /
\put {\color{Red}$\LA r^2_1 \RA$\color{black}} at 3 110
\put {\color{Blue}$\LA r^2_2 \RA$\color{black}} at 3 40
\endpicture
\end{subfigure}
\begin{subfigure}{0.33\textwidth}
\normalcolor
\color{black}
	\caption{$\LA Cor(r_1^2,r_2^2) \RA$}
\beginpicture
\setcoordinatesystem units <10.5pt,1.8pt>
\axescenter{-5}{0}{0}{0}{5}{50}{0}
\put {\Large$\beta$} at 3 -7.5
\setplotsymbol ({\scalebox{0.3}{$\bullet$}})
\color{Red}
\plot -5.000 0.0000  -4.900 0.0000  -4.800 0.0000  -4.700 0.0000  -4.600 0.0000  -4.500 0.0000  -4.400 0.0000  -4.300 0.0000  -4.200 0.0000  -4.100 0.0000  -4.000 0.0000  -3.900 0.0000  -3.800 0.0000  -3.700 0.0000  -3.600 0.0000  -3.500 0.0000  -3.400 0.0000  -3.300 0.0000  -3.200 0.0000  -3.100 0.0000  -3.000 0.0001  -2.900 0.0001  -2.800 0.0001  -2.700 0.0001  -2.600 0.0001  -2.500 0.0002  -2.400 0.0002  -2.300 0.0003  -2.200 0.0003  -2.100 0.0004  -2.000 0.0005  -1.900 0.0006  -1.800 0.0008  -1.700 0.0009  -1.600 0.0011  -1.500 0.0014  -1.400 0.0017  -1.300 0.0020  -1.200 0.0025  -1.100 0.0030  -1.000 0.0037  -0.900 0.0047  -0.800 0.0059  -0.700 0.0074  -0.600 0.0093  -0.500 0.0116  -0.400 0.0148  -0.300 0.0196  -0.200 0.0261  -0.100 50.  0.000 5.1404  0.100 3.6702  0.200 2.6100  0.300 1.9234  0.400 1.4635  0.500 1.1332  0.600 0.8644  0.700 0.6702  0.800 0.5314  0.900 0.4247  1.000 0.3412  1.100 0.2771  1.200 0.2248  1.300 0.1848  1.400 0.1509  1.500 0.1247  1.600 0.1024  1.700 0.0856  1.800 0.0714  1.900 0.0597  2.000 0.0495  2.100 0.0413  2.200 0.0355  2.300 0.0298  2.400 0.0251  2.500 0.0212  2.600 0.0179  2.700 0.0153  2.800 0.0133  2.900 0.0115  3.000 0.0099  3.100 0.0087  3.200 0.0077  3.300 0.0069  3.400 0.0062  3.500 0.0057  3.600 0.0053  3.700 0.0049  3.800 0.0046  3.900 0.0044  4.000 0.0041  4.100 0.0040  4.200 0.0039  4.300 0.0038  4.400 0.0037  4.500 0.0036  4.600 0.0035  4.700 0.0035  4.800 0.0035  4.900 0.0034  5.000 0.0034    /
\endpicture
\end{subfigure}
\begin{subfigure}{0.33\textwidth}
	\caption{$\LA d_{cm}/L \RA$}
\beginpicture
\normalcolor
\color{black}
\setcoordinatesystem units <10.5pt,85.5pt>
\axescenter{-5}{0}{0}{0}{5}{1}{0}
\put {\Large$\beta$} at 3 -0.175
\setplotsymbol ({\scalebox{0.3}{$\bullet$}})
\color{Red}
\plot -5.000 0.1833  -4.900 0.1826  -4.800 0.1826  -4.700 0.1833  -4.600 0.1833  -4.500 0.1850  -4.400 0.1850  -4.300 0.1841  -4.200 0.1841  -4.100 0.1831  -4.000 0.1831  -3.900 0.1830  -3.800 0.1830  -3.700 0.1835  -3.600 0.1835  -3.500 0.1846  -3.400 0.1846  -3.300 0.1835  -3.200 0.1835  -3.100 0.1833  -3.000 0.1833  -2.900 0.1844  -2.800 0.1844  -2.700 0.1841  -2.600 0.1842  -2.500 0.1834  -2.400 0.1834  -2.300 0.1832  -2.200 0.1832  -2.100 0.1833  -2.000 0.1833  -1.900 0.1842  -1.800 0.1842  -1.700 0.1844  -1.600 0.1844  -1.500 0.1839  -1.400 0.1839  -1.300 0.1834  -1.200 0.1834  -1.100 0.1832  -1.000 0.1832  -0.900 0.1834  -0.800 0.1835  -0.700 0.1840  -0.600 0.1841  -0.500 0.1846  -0.400 0.1848  -0.300 0.1845  -0.200 0.1849  -0.100 0.3170  0.000 0.3452  0.100 0.3044  0.200 0.2691  0.300 0.2397  0.400 0.2164  0.500 0.1949  0.600 0.1774  0.700 0.1618  0.800 0.1473  0.900 0.1348  1.000 0.1239  1.100 0.1142  1.200 0.1053  1.300 0.0972  1.400 0.0900  1.500 0.0834  1.600 0.0772  1.700 0.0716  1.800 0.0661  1.900 0.0612  2.000 0.0564  2.100 0.0522  2.200 0.0482  2.300 0.0446  2.400 0.0412  2.500 0.0381  2.600 0.0354  2.700 0.0330  2.800 0.0310  2.900 0.0291  3.000 0.0275  3.100 0.0262  3.200 0.0250  3.300 0.0240  3.400 0.0232  3.500 0.0225  3.600 0.0220  3.700 0.0215  3.800 0.0211  3.900 0.0208  4.000 0.0205  4.100 0.0203  4.200 0.0202  4.300 0.0200  4.400 0.0199  4.500 0.0198  4.600 0.0197  4.700 0.0196  4.800 0.0196  4.900 0.0195  5.000 0.0195   /
\normalcolor
\color{black}
\endpicture
\end{subfigure}
\caption{Across the $\tau_2$ phase boundary in the $\alpha\beta$-plane
(linked mode).  Consistent  with the results in figure \ref{17}(d)-(f), these 
results show a  first order phase transition with sharp peaks in the
correlation functions as the phase is crossed.}
\label{23}   % %ZXZ[23]
\color{black}
\normalcolor
\end{figure}

In figure \ref{23} the estimated mean quantities are plotted as the
linked model is taken through the $\tau_2$ phase boundary.  The results
are consistent with those in figure \ref{17}(d)-(f), showing a strong first 
order character as the inner polygon inflates the outer polygon when the
model is taken from the empty phase into the $c_2$-dominated phase
in figure \ref{16}. The mean number of self-contacts in the inner polygon
changes discontinuously at the phase boundary, as the mean square
radii of gyration of both polygons change discontinuously.

\begin{figure}[t!]
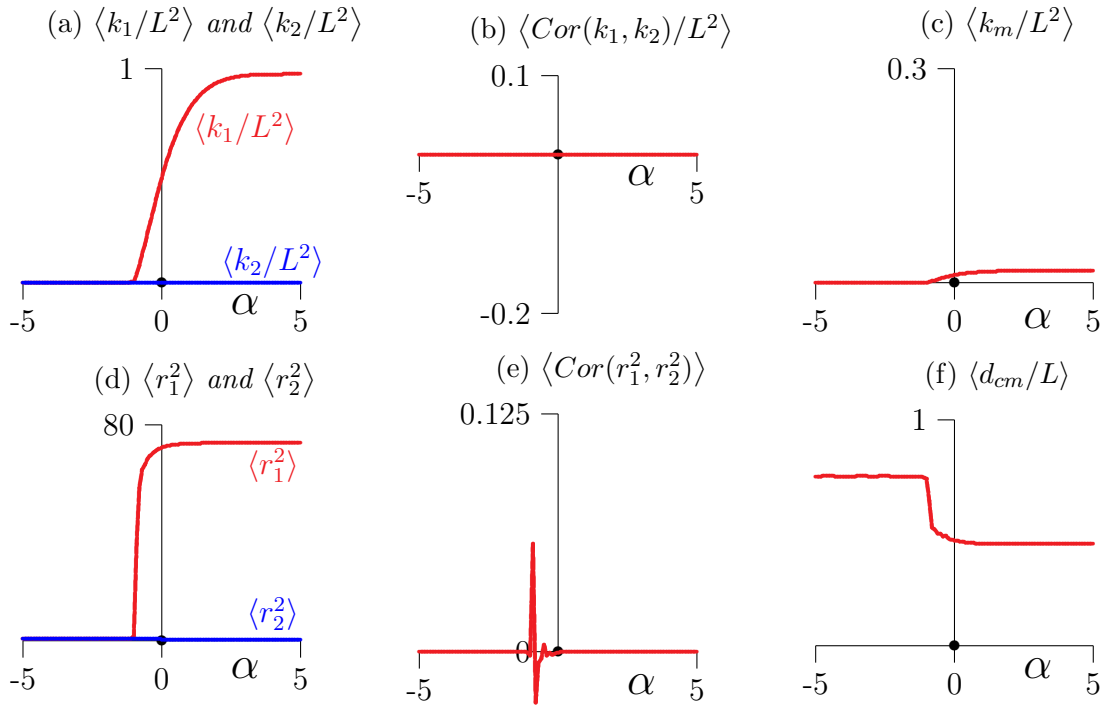
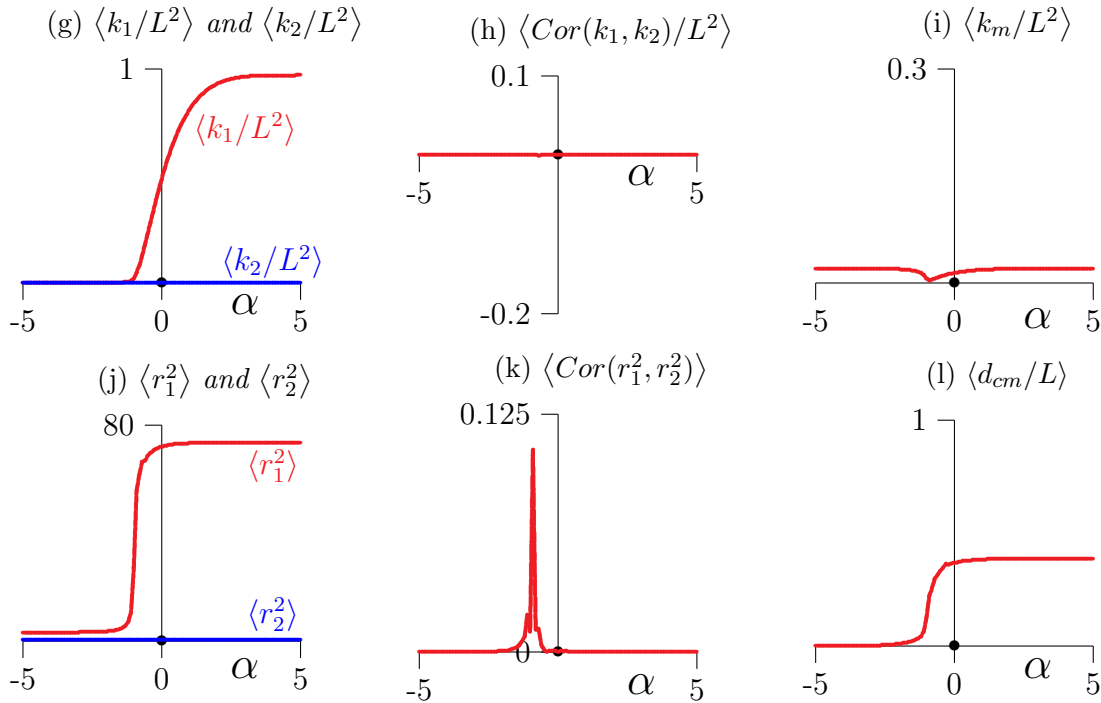

%\hrule\vspace{2mm}
\U{Unlinked from $(-5,-2)$ to $(5,-2)$ across $\lambda_1$ or $\lambda_2$}: \vspace{1mm}

\begin{subfigure}{0.33\textwidth}
\normalcolor
\color{black}
	\caption{$\LA k_1/L^2 \RA$ and $\LA k_2/L^2 \RA$}
\beginpicture
\setcoordinatesystem units <10.5pt,81pt>
\axescenter{-5}{0}{0}{0}{5}{1}{0}
\put {\Large$\alpha$} at 3.0 -0.10
\setplotsymbol ({\scalebox{0.3}{$\bullet$}})
\color{Red}
\plot -5.000 0.0000  -4.900 0.0000  -4.800 0.0000  -4.700 0.0000  -4.600 0.0000  -4.500 0.0000  -4.400 0.0000  -4.300 0.0000  -4.200 0.0000  -4.100 0.0000  -4.000 0.0000  -3.900 0.0000  -3.800 0.0000  -3.700 0.0000  -3.600 0.0000  -3.500 0.0000  -3.400 0.0000  -3.300 0.0000  -3.200 0.0000  -3.100 0.0000  -3.000 0.0000  -2.900 0.0000  -2.800 0.0000  -2.700 0.0000  -2.600 0.0000  -2.500 0.0000  -2.400 0.0001  -2.300 0.0001  -2.200 0.0001  -2.100 0.0001  -2.000 0.0001  -1.900 0.0002  -1.800 0.0002  -1.700 0.0002  -1.600 0.0003  -1.500 0.0004  -1.400 0.0005  -1.300 0.0007  -1.200 0.0010  -1.100 0.0014  -1.000 0.0027  -0.900 0.0387  -0.800 0.0797  -0.700 0.1320  -0.600 0.1766  -0.500 0.2360  -0.400 0.2841  -0.300 0.3356  -0.200 0.3897  -0.100 0.4392  0.000 0.4868  0.100 0.5317  0.200 0.5738  0.300 0.6136  0.400 0.6491  0.500 0.6826  0.600 0.7130  0.700 0.7413  0.800 0.7668  0.900 0.7902  1.000 0.8114  1.100 0.8303  1.200 0.8477  1.300 0.8631  1.400 0.8771  1.500 0.8895  1.600 0.9006  1.700 0.9104  1.800 0.9191  1.900 0.9269  2.000 0.9336  2.100 0.9395  2.200 0.9446  2.300 0.9491  2.400 0.9530  2.500 0.9563  2.600 0.9592  2.700 0.9616  2.800 0.9637  2.900 0.9654  3.000 0.9669  3.100 0.9681  3.200 0.9691  3.300 0.9699  3.400 0.9706  3.500 0.9712  3.600 0.9717  3.700 0.9721  3.800 0.9724  3.900 0.9726  4.000 0.9729  4.100 0.9730  4.200 0.9732  4.300 0.9733  4.400 0.9734  4.500 0.9735  4.600 0.9736  4.700 0.9736  4.800 0.9736  4.900 0.9737  5.000 0.9737 /
\color{Blue}
\plot -5.000 0.0001  -4.900 0.0001  -4.800 0.0001  -4.700 0.0001  -4.600 0.0001  -4.500 0.0001  -4.400 0.0001  -4.300 0.0001  -4.200 0.0001  -4.100 0.0001  -4.000 0.0001  -3.900 0.0001  -3.800 0.0001  -3.700 0.0001  -3.600 0.0001  -3.500 0.0001  -3.400 0.0001  -3.300 0.0001  -3.200 0.0001  -3.100 0.0001  -3.000 0.0001  -2.900 0.0001  -2.800 0.0001  -2.700 0.0001  -2.600 0.0001  -2.500 0.0001  -2.400 0.0001  -2.300 0.0001  -2.200 0.0001  -2.100 0.0001  -2.000 0.0001  -1.900 0.0001  -1.800 0.0001  -1.700 0.0001  -1.600 0.0001  -1.500 0.0001  -1.400 0.0001  -1.300 0.0001  -1.200 0.0001  -1.100 0.0001  -1.000 0.0001  -0.900 0.0001  -0.800 0.0001  -0.700 0.0001  -0.600 0.0001  -0.500 0.0000  -0.400 0.0000  -0.300 0.0000  -0.200 0.0000  -0.100 0.0000  0.000 0.0000  0.100 0.0000  0.200 0.0000  0.300 0.0000  0.400 0.0000  0.500 0.0000  0.600 0.0000  0.700 0.0000  0.800 0.0000  0.900 0.0000  1.000 0.0000  1.100 0.0000  1.200 0.0000  1.300 0.0000  1.400 0.0000  1.500 0.0000  1.600 0.0000  1.700 0.0000  1.800 0.0000  1.900 0.0000  2.000 0.0000  2.100 0.0000  2.200 0.0000  2.300 0.0000  2.400 0.0000  2.500 0.0000  2.600 0.0000  2.700 0.0000  2.800 0.0000  2.900 0.0000  3.000 0.0000  3.100 0.0000  3.200 0.0000  3.300 0.0000  3.400 0.0000  3.500 0.0000  3.600 0.0000  3.700 0.0000  3.800 0.0000  3.900 0.0000  4.000 0.0000  4.100 0.0000  4.200 0.0000  4.300 0.0000  4.400 0.0000  4.500 0.0000  4.600 0.0000  4.700 0.0000  4.800 0.0000  4.900 0.0000  5.000 0.0000 /
\color{black}
\put {\color{Red}$\LA k_1/L^2 \RA$\color{black}} at 3 0.72
\put {\color{Blue}$\LA k_2/L^2 \RA$\color{black}} at 4 0.1
\endpicture
\end{subfigure}
\begin{subfigure}{0.33\textwidth}
\normalcolor
\color{black}
	\caption{$\LA Cor(k_1,k_2)/L^2 \RA$}
\beginpicture
\setcoordinatesystem units <10.5pt,300pt>
\axes{-5}{-0.2}{0}{0}{5}{0.1}{0}
\put {\Large$\alpha$} at 3.0 -0.025
\setplotsymbol ({\scalebox{0.3}{$\bullet$}})
\color{Red}
\plot -5.000 0.0000  -4.900 -0.0000  -4.800 -0.0000  -4.700 -0.0000  -4.600 -0.0000  -4.500 -0.0000  -4.400 -0.0000  -4.300 -0.0000  -4.200 -0.0000  -4.100 -0.0000  -4.000 -0.0000  -3.900 -0.0000  -3.800 -0.0000  -3.700 0.0000  -3.600 0.0000  -3.500 0.0000  -3.400 0.0000  -3.300 0.0000  -3.200 0.0000  -3.100 -0.0000  -3.000 -0.0000  -2.900 -0.0000  -2.800 -0.0000  -2.700 -0.0000  -2.600 -0.0000  -2.500 -0.0000  -2.400 -0.0000  -2.300 -0.0000  -2.200 -0.0000  -2.100 -0.0000  -2.000 -0.0000  -1.900 -0.0000  -1.800 -0.0000  -1.700 -0.0000  -1.600 -0.0000  -1.500 -0.0000  -1.400 -0.0000  -1.300 -0.0000  -1.200 -0.0000  -1.100 -0.0000  -1.000 -0.0000  -0.900 -0.0004  -0.800 -0.0003  -0.700 0.0003  -0.600 -0.0000  -0.500 -0.0001  -0.400 -0.0001  -0.300 0.0001  -0.200 0.0000  -0.100 -0.0000  0.000 -0.0001  0.100 -0.0000  0.200 -0.0000  0.300 -0.0000  0.400 -0.0000  0.500 -0.0000  0.600 0.0000  0.700 -0.0000  0.800 -0.0000  0.900 -0.0000  1.000 -0.0000  1.100 -0.0000  1.200 -0.0000  1.300 -0.0000  1.400 -0.0000  1.500 -0.0000  1.600 -0.0000  1.700 -0.0000  1.800 -0.0000  1.900 -0.0000  2.000 -0.0000  2.100 -0.0000  2.200 -0.0000  2.300 -0.0000  2.400 -0.0000  2.500 -0.0000  2.600 -0.0000  2.700 -0.0000  2.800 -0.0000  2.900 -0.0000  3.000 -0.0000  3.100 -0.0000  3.200 -0.0000  3.300 -0.0000  3.400 0.0000  3.500 0.0000  3.600 0.0000  3.700 0.0000  3.800 0.0000  3.900 0.0000  4.000 0.0000  4.100 0.0000  4.200 0.0000  4.300 0.0000  4.400 0.0000  4.500 0.0000  4.600 -0.0000  4.700 -0.0000  4.800 -0.0000  4.900 -0.0000  5.000 0.0000  /
%\put {$\mathbf{\LA r_1^2 \RA}$} at -1 0.033
\endpicture
\end{subfigure}
\begin{subfigure}{0.33\textwidth}
\normalcolor
\color{black}
	\caption{$\LA k_m/L^2 \RA$}
\beginpicture
\setcoordinatesystem units <10.5pt,270pt>
\axescenter{-5}{0}{0}{0}{5}{0.3}{0}
\put {\Large$\alpha$} at 3.0 -0.05
\setplotsymbol ({\scalebox{0.3}{$\bullet$}})
\color{Red}
\plot -5.000 0.0000  -4.900 0.0000  -4.800 0.0000  -4.700 0.0000  -4.600 0.0000  -4.500 0.0000  -4.400 0.0000  -4.300 0.0000  -4.200 0.0000  -4.100 0.0000  -4.000 0.0000  -3.900 0.0000  -3.800 0.0000  -3.700 0.0000  -3.600 0.0000  -3.500 0.0000  -3.400 0.0000  -3.300 0.0000  -3.200 0.0000  -3.100 0.0000  -3.000 0.0000  -2.900 0.0000  -2.800 0.0000  -2.700 0.0000  -2.600 0.0000  -2.500 0.0000  -2.400 0.0000  -2.300 0.0000  -2.200 0.0000  -2.100 0.0000  -2.000 0.0000  -1.900 0.0000  -1.800 0.0000  -1.700 0.0000  -1.600 0.0000  -1.500 0.0000  -1.400 0.0000  -1.300 0.0000  -1.200 0.0000  -1.100 0.0000  -1.000 0.0000  -0.900 0.0004  -0.800 0.0021  -0.700 0.0031  -0.600 0.0045  -0.500 0.0057  -0.400 0.0067  -0.300 0.0075  -0.200 0.0087  -0.100 0.0096  0.000 0.0104  0.100 0.0111  0.200 0.0117  0.300 0.0123  0.400 0.0128  0.500 0.0133  0.600 0.0137  0.700 0.0141  0.800 0.0144  0.900 0.0147  1.000 0.0150  1.100 0.0152  1.200 0.0154  1.300 0.0155  1.400 0.0156  1.500 0.0158  1.600 0.0159  1.700 0.0160  1.800 0.0161  1.900 0.0161  2.000 0.0162  2.100 0.0163  2.200 0.0163  2.300 0.0163  2.400 0.0164  2.500 0.0164  2.600 0.0164  2.700 0.0164  2.800 0.0164  2.900 0.0164  3.000 0.0165  3.100 0.0165  3.200 0.0165  3.300 0.0165  3.400 0.0165  3.500 0.0165  3.600 0.0165  3.700 0.0165  3.800 0.0165  3.900 0.0165  4.000 0.0165  4.100 0.0165  4.200 0.0165  4.300 0.0165  4.400 0.0165  4.500 0.0165  4.600 0.0165  4.700 0.0165  4.800 0.0165  4.900 0.0165  5.000 0.0165   /
%\put {$\mathbf{\LA r_1^2 \RA}$} at -1 0.033
\endpicture
\end{subfigure}
\normalcolor
\color{black}

\vspace{2mm}

\begin{subfigure}{0.33\textwidth}
\normalcolor
\color{black}
	\caption{$\LA r_1^2\RA$ and $\LA r_2^2\RA$}
\beginpicture
\setcoordinatesystem units <10.5pt,1.0215pt>
\axescenter{-5}{0}{0}{0}{5}{80}{0}
\put {\Large$\alpha$} at 3 -10
\setplotsymbol ({\scalebox{0.3}{$\bullet$}})
\color{Red}
\plot -5.000 0.5001  -4.900 0.5001  -4.800 0.5001  -4.700 0.5001  -4.600 0.5001  -4.500 0.5001  -4.400 0.5002  -4.300 0.5002  -4.200 0.5002  -4.100 0.5003  -4.000 0.5003  -3.900 0.5004  -3.800 0.5005  -3.700 0.5006  -3.600 0.5008  -3.500 0.5009  -3.400 0.5011  -3.300 0.5013  -3.200 0.5016  -3.100 0.5021  -3.000 0.5025  -2.900 0.5032  -2.800 0.5039  -2.700 0.5047  -2.600 0.5058  -2.500 0.5070  -2.400 0.5086  -2.300 0.5109  -2.200 0.5135  -2.100 0.5168  -2.000 0.5210  -1.900 0.5262  -1.800 0.5332  -1.700 0.5423  -1.600 0.5546  -1.500 0.5720  -1.400 0.5970  -1.300 0.6365  -1.200 0.7054  -1.100 0.8594  -1.000 1.4617  -0.900 35.9304  -0.800 56.8032  -0.700 63.2082  -0.600 65.2417  -0.500 67.2853  -0.400 68.6802  -0.300 69.5417  -0.200 70.4798  -0.100 71.0642  0.000 71.4984  0.100 71.8331  0.200 72.1039  0.300 72.3259  0.400 72.4881  0.500 72.6219  0.600 72.7408  0.700 72.8330  0.800 72.9061  0.900 72.9661  1.000 73.0154  1.100 73.0568  1.200 73.0874  1.300 73.1140  1.400 73.1315  1.500 73.1489  1.600 73.1620  1.700 73.1721  1.800 73.1807  1.900 73.1851  2.000 73.1891  2.100 73.1924  2.200 73.1949  2.300 73.1955  2.400 73.1943  2.500 73.1948  2.600 73.1915  2.700 73.1908  2.800 73.1904  2.900 73.1897  3.000 73.1907  3.100 73.1905  3.200 73.1897  3.300 73.1895  3.400 73.1887  3.500 73.1885  3.600 73.1862  3.700 73.1862  3.800 73.1852  3.900 73.1853  4.000 73.1861  4.100 73.1862  4.200 73.1845  4.300 73.1844  4.400 73.1845  4.500 73.1845  4.600 73.1866  4.700 73.1866  4.800 73.1881  4.900 73.1880  5.000 73.1892   /
\color{Blue}
\plot -5.000 0.5207  -4.900 0.5207  -4.800 0.5207  -4.700 0.5207  -4.600 0.5207  -4.500 0.5209  -4.400 0.5209  -4.300 0.5211  -4.200 0.5211  -4.100 0.5210  -4.000 0.5210  -3.900 0.5205  -3.800 0.5205  -3.700 0.5208  -3.600 0.5208  -3.500 0.5211  -3.400 0.5211  -3.300 0.5207  -3.200 0.5207  -3.100 0.5206  -3.000 0.5206  -2.900 0.5208  -2.800 0.5208  -2.700 0.5208  -2.600 0.5208  -2.500 0.5206  -2.400 0.5206  -2.300 0.5208  -2.200 0.5208  -2.100 0.5211  -2.000 0.5211  -1.900 0.5211  -1.800 0.5211  -1.700 0.5208  -1.600 0.5208  -1.500 0.5204  -1.400 0.5203  -1.300 0.5204  -1.200 0.5203  -1.100 0.5201  -1.000 0.5198  -0.900 0.5226  -0.800 0.5144  -0.700 0.5117  -0.600 0.5095  -0.500 0.5070  -0.400 0.5062  -0.300 0.5049  -0.200 0.5049  -0.100 0.5042  0.000 0.5035  0.100 0.5026  0.200 0.5022  0.300 0.5019  0.400 0.5016  0.500 0.5013  0.600 0.5012  0.700 0.5011  0.800 0.5010  0.900 0.5008  1.000 0.5006  1.100 0.5005  1.200 0.5004  1.300 0.5003  1.400 0.5003  1.500 0.5002  1.600 0.5002  1.700 0.5001  1.800 0.5001  1.900 0.5001  2.000 0.5001  2.100 0.5001  2.200 0.5001  2.300 0.5000  2.400 0.5000  2.500 0.5000  2.600 0.5000  2.700 0.5000  2.800 0.5000  2.900 0.5000  3.000 0.5000  3.100 0.5000  3.200 0.5000  3.300 0.5000  3.400 0.5000  3.500 0.5000  3.600 0.5000  3.700 0.5000  3.800 0.5000  3.900 0.5000  4.000 0.5000  4.100 0.5000  4.200 0.5000  4.300 0.5000  4.400 0.5000  4.500 0.5000  4.600 0.5000  4.700 0.5000  4.800 0.5000  4.900 0.5000  5.000 0.5000  /
\put {\color{Red}$\LA r^2_1 \RA$\color{black}} at 4 65
\put {\color{Blue}$\LA r^2_2 \RA$\color{black}} at 4 10
\endpicture
\end{subfigure}
\begin{subfigure}{0.33\textwidth}
\normalcolor
\color{black}
	\caption{$\LA Cor(r_1^2,r_2^2) \RA$}
\beginpicture
\setcoordinatesystem units <10.5pt,720pt>
\axes{-5}{0}{0}{0}{5}{0.125}{0}
\put {\Large$\alpha$} at 3 -0.017
\setplotsymbol ({\scalebox{0.3}{$\bullet$}})
\color{Red}
\plot -5.000 0.0000  -4.900 -0.0000  -4.800 -0.0000  -4.700 -0.0000  -4.600 -0.0000  -4.500 -0.0000  -4.400 -0.0000  -4.300 -0.0000  -4.200 -0.0000  -4.100 -0.0000  -4.000 -0.0000  -3.900 -0.0000  -3.800 -0.0000  -3.700 0.0000  -3.600 0.0000  -3.500 0.0000  -3.400 0.0000  -3.300 0.0000  -3.200 0.0000  -3.100 -0.0000  -3.000 -0.0000  -2.900 -0.0000  -2.800 -0.0000  -2.700 -0.0000  -2.600 -0.0000  -2.500 -0.0000  -2.400 -0.0000  -2.300 -0.0000  -2.200 -0.0000  -2.100 -0.0000  -2.000 -0.0000  -1.900 -0.0001  -1.800 -0.0001  -1.700 -0.0001  -1.600 -0.0001  -1.500 -0.0001  -1.400 -0.0001  -1.300 -0.0001  -1.200 -0.0001  -1.100 -0.0004  -1.000 -0.0020  -0.900 0.0567  -0.800 -0.0269  -0.700 -0.0058  -0.600 -0.0036  -0.500 0.0035  -0.400 -0.0021  -0.300 -0.0008  -0.200 -0.0019  -0.100 -0.0009  0.000 -0.0008  0.100 -0.0003  0.200 -0.0002  0.300 0.0000  0.400 -0.0003  0.500 -0.0003  0.600 -0.0003  0.700 -0.0001  0.800 -0.0001  0.900 -0.0001  1.000 -0.0001  1.100 -0.0001  1.200 -0.0002  1.300 -0.0001  1.400 -0.0000  1.500 -0.0000  1.600 -0.0000  1.700 -0.0000  1.800 -0.0000  1.900 -0.0000  2.000 -0.0000  2.100 -0.0000  2.200 -0.0000  2.300 -0.0000  2.400 -0.0000  2.500 -0.0000  2.600 0.0000  2.700 0.0000  2.800 -0.0000  2.900 -0.0000  3.000 -0.0000  3.100 -0.0000  3.200 -0.0000  3.300 -0.0000  3.400 -0.0000  3.500 -0.0000  3.600 0.0000  3.700 0.0000  3.800 0.0000  3.900 0.0000  4.000 0.0000  4.100 0.0000  4.200 -0.0000  4.300 -0.0000  4.400 -0.0000  4.500 -0.0000  4.600 0.0000  4.700 0.0000  4.800 0.0000  4.900 0.0000  5.000 0.0000   /
\endpicture
\end{subfigure}
\begin{subfigure}{0.33\textwidth}
	\caption{$\LA d_{cm}/L \RA$}
\beginpicture
\normalcolor
\color{black}
\setcoordinatesystem units <10.5pt,85.5pt>
\axescenter{-5}{0}{0}{0}{5}{1}{0}
\put {\Large$\alpha$} at 3 -0.1
\setplotsymbol ({\scalebox{0.3}{$\bullet$}})
\color{Red}
\plot -5.000 0.7490  -4.900 0.7496  -4.800 0.7496  -4.700 0.7506  -4.600 0.7506  -4.500 0.7505  -4.400 0.7505  -4.300 0.7494  -4.200 0.7494  -4.100 0.7485  -4.000 0.7485  -3.900 0.7486  -3.800 0.7486  -3.700 0.7493  -3.600 0.7493  -3.500 0.7500  -3.400 0.7500  -3.300 0.7504  -3.200 0.7504  -3.100 0.7498  -3.000 0.7498  -2.900 0.7490  -2.800 0.7490  -2.700 0.7499  -2.600 0.7499  -2.500 0.7504  -2.400 0.7504  -2.300 0.7496  -2.200 0.7496  -2.100 0.7498  -2.000 0.7498  -1.900 0.7492  -1.800 0.7491  -1.700 0.7488  -1.600 0.7485  -1.500 0.7491  -1.400 0.7487  -1.300 0.7481  -1.200 0.7469  -1.100 0.7442  -1.000 0.7371  -0.900 0.6409  -0.800 0.5228  -0.700 0.5114  -0.600 0.4962  -0.500 0.4931  -0.400 0.4830  -0.300 0.4866  -0.200 0.4724  -0.100 0.4687  0.000 0.4660  0.100 0.4650  0.200 0.4619  0.300 0.4608  0.400 0.4568  0.500 0.4562  0.600 0.4543  0.700 0.4534  0.800 0.4524  0.900 0.4525  1.000 0.4516  1.100 0.4513  1.200 0.4514  1.300 0.4514  1.400 0.4518  1.500 0.4518  1.600 0.4520  1.700 0.4518  1.800 0.4515  1.900 0.4517  2.000 0.4516  2.100 0.4514  2.200 0.4508  2.300 0.4509  2.400 0.4509  2.500 0.4507  2.600 0.4515  2.700 0.4515  2.800 0.4513  2.900 0.4513  3.000 0.4507  3.100 0.4505  3.200 0.4508  3.300 0.4507  3.400 0.4509  3.500 0.4508  3.600 0.4513  3.700 0.4513  3.800 0.4513  3.900 0.4512  4.000 0.4510  4.100 0.4510  4.200 0.4515  4.300 0.4515  4.400 0.4514  4.500 0.4514  4.600 0.4507  4.700 0.4507  4.800 0.4502  4.900 0.4502  5.000 0.4498   /
\normalcolor
\color{black}
\endpicture
\end{subfigure}

\vspace{2mm}
\U{Linked from $(-5,-2)$ to $(5,-2)$ across $\lambda$}: \vspace{1mm}
\vspace{1mm}

\begin{subfigure}{0.33\textwidth}
\normalcolor
\color{black}
	\caption{$\LA k_1/L^2 \RA$ and $\LA k_2/L^2 \RA$}
\beginpicture
\setcoordinatesystem units <10.5pt,81pt>
\axescenter{-5}{0}{0}{0}{5}{1}{0}
\put {\Large$\alpha$} at 3.0 -0.10
\setplotsymbol ({\scalebox{0.3}{$\bullet$}})
\color{Red}
\plot -5.000 0.0000  -4.900 0.0000  -4.800 0.0000  -4.700 0.0000  -4.600 0.0000  -4.500 0.0000  -4.400 0.0000  -4.300 0.0000  -4.200 0.0000  -4.100 0.0000  -4.000 0.0000  -3.900 0.0000  -3.800 0.0000  -3.700 0.0000  -3.600 0.0000  -3.500 0.0000  -3.400 0.0000  -3.300 0.0000  -3.200 0.0001  -3.100 0.0001  -3.000 0.0001  -2.900 0.0001  -2.800 0.0001  -2.700 0.0002  -2.600 0.0002  -2.500 0.0002  -2.400 0.0003  -2.300 0.0003  -2.200 0.0004  -2.100 0.0005  -2.000 0.0006  -1.900 0.0008  -1.800 0.0010  -1.700 0.0012  -1.600 0.0016  -1.500 0.0020  -1.400 0.0026  -1.300 0.0035  -1.200 0.0050  -1.100 0.0085  -1.000 0.0207  -0.900 0.0490  -0.800 0.0800  -0.700 0.1270  -0.600 0.1778  -0.500 0.2300  -0.400 0.2821  -0.300 0.3340  -0.200 0.3863  -0.100 0.4364  0.000 0.4836  0.100 0.5284  0.200 0.5708  0.300 0.6101  0.400 0.6457  0.500 0.6789  0.600 0.7099  0.700 0.7379  0.800 0.7637  0.900 0.7869  1.000 0.8079  1.100 0.8270  1.200 0.8443  1.300 0.8599  1.400 0.8737  1.500 0.8861  1.600 0.8971  1.700 0.9069  1.800 0.9156  1.900 0.9234  2.000 0.9302  2.100 0.9361  2.200 0.9413  2.300 0.9458  2.400 0.9496  2.500 0.9530  2.600 0.9558  2.700 0.9582  2.800 0.9603  2.900 0.9620  3.000 0.9634  3.100 0.9647  3.200 0.9657  3.300 0.9665  3.400 0.9672  3.500 0.9677  3.600 0.9682  3.700 0.9686  3.800 0.9690  3.900 0.9692  4.000 0.9694  4.100 0.9696  4.200 0.9698  4.300 0.9699  4.400 0.9700  4.500 0.9700  4.600 0.9701  4.700 0.9702  4.800 0.9702  4.900 0.9703  5.000 0.9703 /
\color{Blue}
\plot -5.000 0.0000  -4.900 0.0000  -4.800 0.0000  -4.700 0.0000  -4.600 0.0000  -4.500 0.0000  -4.400 0.0000  -4.300 0.0000  -4.200 0.0000  -4.100 0.0000  -4.000 0.0000  -3.900 0.0000  -3.800 0.0000  -3.700 0.0000  -3.600 0.0000  -3.500 0.0000  -3.400 0.0000  -3.300 0.0000  -3.200 0.0000  -3.100 0.0000  -3.000 0.0000  -2.900 0.0000  -2.800 0.0000  -2.700 0.0000  -2.600 0.0000  -2.500 0.0000  -2.400 0.0000  -2.300 0.0000  -2.200 0.0000  -2.100 0.0000  -2.000 0.0000  -1.900 0.0000  -1.800 0.0000  -1.700 0.0000  -1.600 0.0000  -1.500 0.0000  -1.400 0.0000  -1.300 0.0000  -1.200 0.0000  -1.100 0.0000  -1.000 0.0001  -0.900 0.0001  -0.800 0.0001  -0.700 0.0001  -0.600 0.0000  -0.500 0.0000  -0.400 0.0000  -0.300 0.0000  -0.200 0.0000  -0.100 0.0000  0.000 0.0000  0.100 0.0000  0.200 0.0000  0.300 0.0000  0.400 0.0000  0.500 0.0000  0.600 0.0000  0.700 0.0000  0.800 0.0000  0.900 0.0000  1.000 0.0000  1.100 0.0000  1.200 0.0000  1.300 0.0000  1.400 0.0000  1.500 0.0000  1.600 0.0000  1.700 0.0000  1.800 0.0000  1.900 0.0000  2.000 0.0000  2.100 0.0000  2.200 0.0000  2.300 0.0000  2.400 0.0000  2.500 0.0000  2.600 0.0000  2.700 0.0000  2.800 0.0000  2.900 0.0000  3.000 0.0000  3.100 0.0000  3.200 0.0000  3.300 0.0000  3.400 0.0000  3.500 0.0000  3.600 0.0000  3.700 0.0000  3.800 0.0000  3.900 0.0000  4.000 0.0000  4.100 0.0000  4.200 0.0000  4.300 0.0000  4.400 0.0000  4.500 0.0000  4.600 0.0000  4.700 0.0000  4.800 0.0000  4.900 0.0000  5.000 0.0000 /
\color{black}
\put {\color{Red}$\LA k_1/L^2 \RA$\color{black}} at 3 0.72
\put {\color{Blue}$\LA k_2/L^2 \RA$\color{black}} at 4 0.1
\endpicture
\end{subfigure}
\begin{subfigure}{0.33\textwidth}
\normalcolor
\color{black}
	\caption{$\LA Cor(k_1,k_2)/L^2 \RA$}
\beginpicture
\setcoordinatesystem units <10.5pt,300pt>
\axes{-5}{-0.2}{0}{0}{5}{0.1}{0}
\put {\Large$\alpha$} at 3.0 -0.025
\setplotsymbol ({\scalebox{0.3}{$\bullet$}})
\color{Red}
\plot -5.000 0.0000  -4.900 -0.0000  -4.800 -0.0000  -4.700 -0.0000  -4.600 -0.0000  -4.500 0.0000  -4.400 0.0000  -4.300 -0.0000  -4.200 -0.0000  -4.100 -0.0000  -4.000 -0.0000  -3.900 -0.0000  -3.800 -0.0000  -3.700 -0.0000  -3.600 -0.0000  -3.500 -0.0000  -3.400 -0.0000  -3.300 -0.0000  -3.200 -0.0000  -3.100 -0.0000  -3.000 -0.0000  -2.900 -0.0000  -2.800 -0.0000  -2.700 0.0000  -2.600 0.0000  -2.500 0.0000  -2.400 0.0000  -2.300 0.0000  -2.200 0.0000  -2.100 0.0000  -2.000 0.0000  -1.900 0.0000  -1.800 0.0000  -1.700 0.0000  -1.600 0.0000  -1.500 0.0000  -1.400 0.0000  -1.300 0.0000  -1.200 0.0000  -1.100 0.0000  -1.000 -0.0000  -0.900 -0.0001  -0.800 -0.0000  -0.700 -0.0004  -0.600 -0.0002  -0.500 0.0001  -0.400 -0.0000  -0.300 0.0001  -0.200 -0.0000  -0.100 -0.0000  0.000 -0.0000  0.100 -0.0000  0.200 -0.0000  0.300 -0.0000  0.400 -0.0000  0.500 -0.0000  0.600 -0.0000  0.700 -0.0000  0.800 -0.0000  0.900 -0.0000  1.000 -0.0000  1.100 -0.0000  1.200 -0.0000  1.300 -0.0000  1.400 -0.0000  1.500 0.0000  1.600 -0.0000  1.700 -0.0000  1.800 -0.0000  1.900 -0.0000  2.000 -0.0000  2.100 -0.0000  2.200 -0.0000  2.300 -0.0000  2.400 -0.0000  2.500 -0.0000  2.600 -0.0000  2.700 -0.0000  2.800 -0.0000  2.900 -0.0000  3.000 -0.0000  3.100 -0.0000  3.200 -0.0000  3.300 -0.0000  3.400 -0.0000  3.500 -0.0000  3.600 0.0000  3.700 0.0000  3.800 0.0000  3.900 0.0000  4.000 0.0000  4.100 0.0000  4.200 0.0000  4.300 0.0000  4.400 0.0000  4.500 0.0000  4.600 0.0000  4.700 0.0000  4.800 0.0000  4.900 0.0000  5.000 0.0000  /
%\put {$\mathbf{\LA r_1^2 \RA}$} at -1 0.033
\endpicture
\end{subfigure}
\begin{subfigure}{0.33\textwidth}
\normalcolor
\color{black}
	\caption{$\LA k_m/L^2 \RA$}
\beginpicture
\setcoordinatesystem units <10.5pt,270pt>
\axescenter{-5}{0}{0}{0}{5}{0.3}{0}
\put {\Large$\alpha$} at 3.0 -0.05
\setplotsymbol ({\scalebox{0.3}{$\bullet$}})
\color{Red}
\plot -5.000 0.0200  -4.900 0.0200  -4.800 0.0200  -4.700 0.0200  -4.600 0.0200  -4.500 0.0200  -4.400 0.0200  -4.300 0.0200  -4.200 0.0200  -4.100 0.0200  -4.000 0.0200  -3.900 0.0200  -3.800 0.0200  -3.700 0.0200  -3.600 0.0200  -3.500 0.0200  -3.400 0.0200  -3.300 0.0199  -3.200 0.0199  -3.100 0.0199  -3.000 0.0199  -2.900 0.0199  -2.800 0.0198  -2.700 0.0198  -2.600 0.0197  -2.500 0.0197  -2.400 0.0196  -2.300 0.0195  -2.200 0.0194  -2.100 0.0193  -2.000 0.0191  -1.900 0.0189  -1.800 0.0186  -1.700 0.0183  -1.600 0.0178  -1.500 0.0172  -1.400 0.0164  -1.300 0.0152  -1.200 0.0136  -1.100 0.0109  -1.000 0.0063  -0.900 0.0035  -0.800 0.0047  -0.700 0.0062  -0.600 0.0075  -0.500 0.0089  -0.400 0.0101  -0.300 0.0109  -0.200 0.0121  -0.100 0.0130  0.000 0.0138  0.100 0.0145  0.200 0.0151  0.300 0.0158  0.400 0.0162  0.500 0.0167  0.600 0.0171  0.700 0.0175  0.800 0.0178  0.900 0.0181  1.000 0.0183  1.100 0.0185  1.200 0.0187  1.300 0.0189  1.400 0.0190  1.500 0.0192  1.600 0.0193  1.700 0.0194  1.800 0.0195  1.900 0.0195  2.000 0.0196  2.100 0.0197  2.200 0.0197  2.300 0.0197  2.400 0.0198  2.500 0.0198  2.600 0.0198  2.700 0.0199  2.800 0.0199  2.900 0.0199  3.000 0.0199  3.100 0.0199  3.200 0.0199  3.300 0.0199  3.400 0.0199  3.500 0.0199  3.600 0.0199  3.700 0.0199  3.800 0.0199  3.900 0.0199  4.000 0.0199  4.100 0.0199  4.200 0.0199  4.300 0.0199  4.400 0.0199  4.500 0.0199  4.600 0.0200  4.700 0.0200  4.800 0.0200  4.900 0.0200  5.000 0.0200  /
%\put {$\mathbf{\LA r_1^2 \RA}$} at -1 0.033
\endpicture
\end{subfigure}
\normalcolor
\color{black}

\vspace{2mm}

\begin{subfigure}{0.33\textwidth}
\normalcolor
\color{black}
	\caption{$\LA r_1^2\RA$ and $\LA r_2^2\RA$}
\beginpicture
\setcoordinatesystem units <10.5pt,1.0215pt>
\axescenter{-5}{0}{0}{0}{5}{80}{0}
\put {\Large$\alpha$} at 3 -10
\setplotsymbol ({\scalebox{0.3}{$\bullet$}})
\color{Red}
\plot -5.000 3.1678  -4.900 3.1680  -4.800 3.1682  -4.700 3.1685  -4.600 3.1689  -4.500 3.1691  -4.400 3.1696  -4.300 3.1700  -4.200 3.1707  -4.100 3.1717  -4.000 3.1728  -3.900 3.1744  -3.800 3.1761  -3.700 3.1778  -3.600 3.1803  -3.500 3.1823  -3.400 3.1858  -3.300 3.1900  -3.200 3.1951  -3.100 3.2021  -3.000 3.2100  -2.900 3.2201  -2.800 3.2322  -2.700 3.2460  -2.600 3.2642  -2.500 3.2863  -2.400 3.3143  -2.300 3.3502  -2.200 3.3940  -2.100 3.4508  -2.000 3.5210  -1.900 3.6109  -1.800 3.7285  -1.700 3.8859  -1.600 4.0979  -1.500 4.4065  -1.400 4.8580  -1.300 5.5896  -1.200 6.9447  -1.100 10.5698  -1.000 27.0323  -0.900 55.5869  -0.800 61.5900  -0.700 66.3186  -0.600 66.9257  -0.500 68.4568  -0.400 69.6222  -0.300 70.5017  -0.200 71.1189  -0.100 71.6273  0.000 71.9868  0.100 72.2721  0.200 72.5015  0.300 72.6842  0.400 72.8323  0.500 72.9466  0.600 73.0402  0.700 73.1234  0.800 73.1807  0.900 73.2341  1.000 73.2722  1.100 73.3075  1.200 73.3360  1.300 73.3583  1.400 73.3763  1.500 73.3905  1.600 73.4001  1.700 73.4090  1.800 73.4157  1.900 73.4193  2.000 73.4193  2.100 73.4205  2.200 73.4224  2.300 73.4221  2.400 73.4214  2.500 73.4205  2.600 73.4202  2.700 73.4191  2.800 73.4181  2.900 73.4170  3.000 73.4165  3.100 73.4156  3.200 73.4167  3.300 73.4159  3.400 73.4133  3.500 73.4127  3.600 73.4101  3.700 73.4097  3.800 73.4086  3.900 73.4084  4.000 73.4083  4.100 73.4081  4.200 73.4085  4.300 73.4084  4.400 73.4103  4.500 73.4103  4.600 73.4115  4.700 73.4115  4.800 73.4116  4.900 73.4116  5.000 73.4113  /
\color{Blue}
\plot -5.000 0.5000  -4.900 0.5000  -4.800 0.5000  -4.700 0.5000  -4.600 0.5000  -4.500 0.5000  -4.400 0.5000  -4.300 0.5000  -4.200 0.5000  -4.100 0.5000  -4.000 0.5000  -3.900 0.5000  -3.800 0.5000  -3.700 0.5000  -3.600 0.5000  -3.500 0.5000  -3.400 0.5000  -3.300 0.5000  -3.200 0.5000  -3.100 0.5000  -3.000 0.5000  -2.900 0.5001  -2.800 0.5001  -2.700 0.5001  -2.600 0.5001  -2.500 0.5001  -2.400 0.5002  -2.300 0.5002  -2.200 0.5003  -2.100 0.5004  -2.000 0.5004  -1.900 0.5005  -1.800 0.5006  -1.700 0.5009  -1.600 0.5011  -1.500 0.5016  -1.400 0.5021  -1.300 0.5028  -1.200 0.5040  -1.100 0.5052  -1.000 0.5089  -0.900 0.5163  -0.800 0.5131  -0.700 0.5122  -0.600 0.5084  -0.500 0.5052  -0.400 0.5053  -0.300 0.5042  -0.200 0.5043  -0.100 0.5037  0.000 0.5033  0.100 0.5027  0.200 0.5022  0.300 0.5020  0.400 0.5014  0.500 0.5011  0.600 0.5010  0.700 0.5008  0.800 0.5007  0.900 0.5006  1.000 0.5005  1.100 0.5004  1.200 0.5003  1.300 0.5003  1.400 0.5003  1.500 0.5002  1.600 0.5001  1.700 0.5001  1.800 0.5001  1.900 0.5001  2.000 0.5001  2.100 0.5001  2.200 0.5000  2.300 0.5000  2.400 0.5000  2.500 0.5000  2.600 0.5000  2.700 0.5000  2.800 0.5000  2.900 0.5000  3.000 0.5000  3.100 0.5000  3.200 0.5000  3.300 0.5000  3.400 0.5000  3.500 0.5000  3.600 0.5000  3.700 0.5000  3.800 0.5000  3.900 0.5000  4.000 0.5000  4.100 0.5000  4.200 0.5000  4.300 0.5000  4.400 0.5000  4.500 0.5000  4.600 0.5000  4.700 0.5000  4.800 0.5000  4.900 0.5000  5.000 0.5000  /
\put {\color{Red}$\LA r^2_1 \RA$\color{black}} at 4 65
\put {\color{Blue}$\LA r^2_2 \RA$\color{black}} at 4 10
\endpicture
\end{subfigure}
\begin{subfigure}{0.33\textwidth}
\normalcolor
\color{black}
	\caption{$\LA Cor(r_1^2,r_2^2) \RA$}
\beginpicture
\setcoordinatesystem units <10.5pt,720pt>
\axes{-5}{0}{0}{0}{5}{0.125}{0}
\put {\Large$\alpha$} at 3 -0.017
\setplotsymbol ({\scalebox{0.3}{$\bullet$}})
\color{Red}
\plot -5.000 0.0000  -4.900 0.0000  -4.800 0.0000  -4.700 0.0000  -4.600 0.0000  -4.500 0.0000  -4.400 0.0000  -4.300 0.0000  -4.200 0.0000  -4.100 0.0000  -4.000 0.0000  -3.900 0.0000  -3.800 0.0000  -3.700 0.0000  -3.600 0.0000  -3.500 0.0000  -3.400 0.0000  -3.300 0.0000  -3.200 0.0000  -3.100 0.0000  -3.000 0.0000  -2.900 0.0001  -2.800 0.0001  -2.700 0.0001  -2.600 0.0001  -2.500 0.0002  -2.400 0.0002  -2.300 0.0002  -2.200 0.0003  -2.100 0.0004  -2.000 0.0006  -1.900 0.0006  -1.800 0.0008  -1.700 0.0012  -1.600 0.0016  -1.500 0.0023  -1.400 0.0033  -1.300 0.0051  -1.200 0.0079  -1.100 0.0195  -1.000 0.0108  -0.900 0.1062  -0.800 0.0114  -0.700 0.0124  -0.600 0.0039  -0.500 0.0009  -0.400 -0.0002  -0.300 0.0007  -0.200 0.0006  -0.100 0.0007  0.000 -0.0001  0.100 0.0004  0.200 0.0003  0.300 0.0004  0.400 0.0001  0.500 0.0001  0.600 0.0002  0.700 0.0001  0.800 0.0000  0.900 -0.0000  1.000 0.0000  1.100 0.0000  1.200 0.0000  1.300 0.0000  1.400 0.0000  1.500 0.0000  1.600 -0.0000  1.700 0.0000  1.800 -0.0000  1.900 -0.0000  2.000 0.0000  2.100 0.0000  2.200 0.0000  2.300 0.0000  2.400 0.0000  2.500 0.0000  2.600 -0.0000  2.700 -0.0000  2.800 -0.0000  2.900 -0.0000  3.000 0.0000  3.100 0.0000  3.200 0.0000  3.300 0.0000  3.400 -0.0000  3.500 -0.0000  3.600 0.0000  3.700 0.0000  3.800 0.0000  3.900 0.0000  4.000 0.0000  4.100 0.0000  4.200 0.0000  4.300 0.0000  4.400 0.0000  4.500 0.0000  4.600 0.0000  4.700 0.0000  4.800 0.0000  4.900 0.0000  5.000 0.0000   /
\endpicture
\end{subfigure}
\begin{subfigure}{0.33\textwidth}
	\caption{$\LA d_{cm}/L \RA$}
\beginpicture
\normalcolor
\color{black}
\setcoordinatesystem units <10.5pt,85.5pt>
\axescenter{-5}{0}{0}{0}{5}{1}{0}
\put {\Large$\alpha$} at 3 -0.1
\setplotsymbol ({\scalebox{0.3}{$\bullet$}})
\color{Red}
\plot -5.000 0.0000  -4.900 0.0000  -4.800 0.0000  -4.700 0.0000  -4.600 0.0001  -4.500 0.0001  -4.400 0.0001  -4.300 0.0001  -4.200 0.0001  -4.100 0.0002  -4.000 0.0002  -3.900 0.0002  -3.800 0.0003  -3.700 0.0004  -3.600 0.0004  -3.500 0.0005  -3.400 0.0006  -3.300 0.0007  -3.200 0.0009  -3.100 0.0011  -3.000 0.0014  -2.900 0.0017  -2.800 0.0020  -2.700 0.0024  -2.600 0.0030  -2.500 0.0036  -2.400 0.0043  -2.300 0.0053  -2.200 0.0063  -2.100 0.0077  -2.000 0.0092  -1.900 0.0110  -1.800 0.0131  -1.700 0.0156  -1.600 0.0186  -1.500 0.0222  -1.400 0.0267  -1.300 0.0328  -1.200 0.0423  -1.100 0.0617  -1.000 0.1256  -0.900 0.2207  -0.800 0.2558  -0.700 0.2975  -0.600 0.3146  -0.500 0.3331  -0.400 0.3469  -0.300 0.3628  -0.200 0.3594  -0.100 0.3648  0.000 0.3676  0.100 0.3703  0.200 0.3730  0.300 0.3757  0.400 0.3761  0.500 0.3775  0.600 0.3790  0.700 0.3797  0.800 0.3817  0.900 0.3824  1.000 0.3828  1.100 0.3836  1.200 0.3838  1.300 0.3839  1.400 0.3839  1.500 0.3840  1.600 0.3840  1.700 0.3841  1.800 0.3843  1.900 0.3845  2.000 0.3856  2.100 0.3858  2.200 0.3862  2.300 0.3862  2.400 0.3860  2.500 0.3860  2.600 0.3857  2.700 0.3858  2.800 0.3861  2.900 0.3862  3.000 0.3860  3.100 0.3861  3.200 0.3852  3.300 0.3853  3.400 0.3859  3.500 0.3860  3.600 0.3868  3.700 0.3868  3.800 0.3873  3.900 0.3873  4.000 0.3873  4.100 0.3873  4.200 0.3872  4.300 0.3872  4.400 0.3871  4.500 0.3870  4.600 0.3867  4.700 0.3867  4.800 0.3865  4.900 0.3865  5.000 0.3867   /
\normalcolor
\color{black}
\endpicture
\end{subfigure}
\caption{Across the $\lambda_1$ (or $\lambda_2$) and $\lambda$ 
phase boundaries in the $\alpha\beta$-plane.  Graphs for the unlinked model 
are shown in (a)-(f), and for the linked model are shown in (g)-(l).  Consistent 
with the results in figure \ref{11}(a)-(c) and \ref{17}(g)-(i), these results show 
continuous transitions as these phase boundaries are crossed in the two models 
respectively.}
\label{24}   % %ZXZ[24]
\color{black}
\normalcolor
\end{figure}

In figure \ref{24} we show changes in the order parameters as $\lambda_1$ 
(or $\lambda_2$) phase boundary is crossed in the unlinked model, and 
the similar changes as the $\lambda$ boundary is crossed in the linked model.  
Our results in the previous sections show that these are continuous phase 
boundaries, and the results in figure \ref{24} are consistent with this.  The mean 
density of self-crossings in figure \ref{24}(a) (across the $\lambda_1$
phase boundary), and figure \ref{24}(g) (across the $\lambda$ phase
boundary) show very similar behaviour in the two models.  Crossing
these boundaries into the $c_1$-dominated phase causes the first
polygon in the unlinked model, and the outside polygon in the
linked model, to expand into the dense phase occupying almost all
sites in the confining square.  The second polygon (in the unlinked model) 
and the inside polygon (in the linked model) are both squeezed to minimal
length by the expanding polygon, and this is consistent with the metric
data reported in the figure.  The mean distance between the centres-of-mass
in the unlinked model decreases from the mean distance in the empty
phase to a smaller distance as one polygon increases in length.  In 
the linked model, this distance is zero in the empty phase, but becomes 
positive when the outer polygon expands into the $c_1$-dominated phase
with centre-of-mass near the center of the confining square,
while the inner polygon has almost minimal length and explores the
inside of the square.  These observations differ from those shown 
in figure \ref{23}, where a first order transition is seen along the 
$\tau_2$ phase boundary in the linked model.

\section{Conclusions}

We determined the phase diagrams of two square lattice models of pairs of ring
polymers in the dense phase.  The placements of the two polygons 
in a confining square were chosen in two distinct topological ways, in one case
they were unlinked (or splittable) in the plane, and in the second case, linked (or
unsplittable) in the plane.  The phase diagrams include, in addition to an empty phase, 
a phase dominated by the first polygon (it is dense inside the square),
and a phase dominated by the second polygon; see figures
\ref{9} and \ref{16}.  These phases are separated by boundaries whose
nature depends on the topology of the model.   In the unlinked model,
there are two lines of continuous transitions ($\lambda_1$ and $\lambda_2$)
separating the empty phase from the two dense phases, while the two dense 
phases are separated by a line of first order transitions ($\tau$).  In contrast, 
the linked model has two curves of first order transitions ($\tau_1$ and $\tau_2$), 
where $\tau_1$ separates a phase in which the outer polygon is dense from 
a phase in which the inner polygon is dense, and $\tau_2$ separates
the empty phase from the dense inner polygon phase.  A line of continuous 
transitions $\lambda$ separates the empty phase from the dense outer 
polygon phase.

In the case of both models we were able to determine
critical exponents along the critical curves.   Along the curves of first
order transition the exponents are given in equations \ref{19} (for
$\tau$), \ref{32} ($\tau_1$) and \ref{35} ($\tau_2$).  In each case
the result is consistent with the expected value 
$2-\alpha_s=2-\alpha_s^\prime=1$ reported in equation \Ref{5}.  We were similarly
able to estimate these exponents along the lines of continuous transitions.
The results are shown in table \ref{tablealpha}.

\begin{table}[h!]
\caption{Estimates of $2-\alpha_s$ and $2-\alpha_s^\prime$ across phase boundaries}
%\begin{indented}
%\lineup
%\item[]
\begin{tabular}{c| @{}*{3}{c} | @{}*{3}{c} }
%\br            
  &\multicolumn{3}{c}{Unlinked model} &\multicolumn{3}{c}{Linked model} \cr                  
  & $\tau$ & $\lambda_1$ & $\lambda_2$ & $\tau_1$ & $\tau_2$ & $\lambda$ \cr 
\hline
 & & & & & \cr
 $2-\alpha_s             $ &\; $1.003(3)$ & $0$               & $0$               &\;  $1.02(4)$ & $1.06(12)$ & $0$ \cr
 $2-\alpha^\prime_s$ &\;  $1.003(3)$ & $1.55(3)$  & $1.55(3)$        &\;  $1.02(4)$ & $1.06(12)$ & $1.57(6)$ \cr
\end{tabular}
%\end{indented}
\label{tablealpha}   %ZXZ[tablealpha]
\end{table}

The concentration gap associated with the first order transitions may also be
estimated.  For example, for the critical line $\tau$ in the unlinked model
the free energy crossing $\tau$ is given by equation \Ref{17}.  Taking
derivatives on either side of $\tau$ and subtracting to determine the
gap gives $H_\tau\!\svv_{\alpha=2.5} \approx 1.95$.  In the same
way, by equations \Ref{27} and \Ref{29},
$H_{\tau_1}\!\svv_{\alpha=2.5} \approx 1.55$ and
$H_{\tau_2}\!\svv_{\alpha=-2.0} \approx 0.70$.

Multicritical scaling around the multicritical points were more difficult
to determine, in particular in the linked model.  Our estimates of the critical exponents
are less secure in the linked model.   We show our best estimates in table \ref{tablebeta}.

\begin{table}[h!]
\caption{Estimates of $2-\alpha_t$ and $2-\alpha_u$ between $\tau$ and $\lambda$ boundaries}
%\begin{indented}
%\lineup
%\item[]
\begin{tabular}{c| @{}*{2}{c} | @{}*{2}{c} }
%\br            
  &\multicolumn{2}{c}{Unlinked model} &\multicolumn{2}{c}{Linked model} \cr                  
  & $\tau$-$\lambda_1$ & $\tau$-$\lambda_2$ & $\tau_1$-$\lambda$ & $\tau_2$-$\lambda$ \cr 
\hline
 & & & & \cr
 $2-\alpha_t             $ &\;  $1.55(3)$ & $1.55(3)$ &\quad $1.57(6)$ & $0$ \cr
 $2-\alpha_u             $ &\; $1.55(3)$ & $1.55(3)$ &\quad $1.17(7)$ & $0$ \cr
 $\phi                       $ &\;  $1.00(4)$ & $1.00(4)$ &\quad $1.3(2)$ & $1$ \cr
\end{tabular}
%\end{indented}
\label{tablebeta}   %ZXZ[tablebeta]
\end{table}

\section*{Acknowledgements} EJJvR acknowledges financial support 
from NSERC (Canada) in the form of Discovery Grant RGPIN-2019-06303. 
EJJvR was also grateful to the Department of Physics and Astronomy at
the University of Padova for the financial support during a visit in 2019. 

\section*{References}
\bibliographystyle{plain}
\bibliography{latticelinks}

\begin{thebibliography}{10}

\bibitem{ACF83}
C~Aragao~de Carvalho, S~Caracciolo, and J~Fr{\"o}hlich.
\newblock Polymers and $g\phi^4$-theory in four dimensions.
\newblock {\em Nucl Phys B}, 215:209--248, 1983.

\bibitem{BO12}
M~Baiesi and E~Orlandini.
\newblock Universal properties of knotted polymer rings.
\newblock {\em Phys Rev E}, 86:031805, 2012.

\bibitem{BF81}
B~Berg and D~Foerster.
\newblock Random paths and random surfaces on a digital computer.
\newblock {\em Phys Lett B}, 106:323--326, 1981.

\bibitem{CJvR20}
S~Campbell and EJ~Janse~van Rensburg.
\newblock Parallel {PERM}.
\newblock {\em J Phys A: Math Theo}, 2020.

\bibitem{deG79}
P-G de~Gennes.
\newblock {\em Scaling Concepts in Polymer Physics}.
\newblock Cornell, 1979.

\bibitem{deG84}
P-G de~Gennes.
\newblock Tight knots.
\newblock {\em Macromol}, 17:703--704, 1984.

\bibitem{D62}
M~Delbr{\"u}ck.
\newblock Knotting problems in biology.
\newblock {\em Proc Symp Appl Math}, 14:55--63, 1962.

\bibitem{GJvR18}
F~Gassoumov and EJ~Janse~van Rensburg.
\newblock Osmotic pressure of confined square lattice self-avoiding walks.
\newblock {\em J Phys A: Math Theor}, 52:025004, 2018.

\bibitem{GT95}
CJ~Geyer and EA~Thompson.
\newblock Annealing {Markov} chain {Monte} {Carlo} with applications to
  ancestral inference.
\newblock {\em J Amer Stat Assoc}, 90:909--920, 1995.

\bibitem{GFR96}
AY~Grosberg, A~Feigel, and Y~Rabin.
\newblock Flory-type theory of a knotted ring polymer.
\newblock {\em Phys Rev E}, 54:6618--6622, 1996.

\bibitem{H61A}
JM~Hammersley.
\newblock The number of polygons on a lattice.
\newblock {\em Proc Camb Phil Soc}, 57:516--523, 1961.

\bibitem{HM54}
JM~Hammersley and KW~Morton.
\newblock Poor man's {Monte} {Carlo}.
\newblock {\em J Roy Stat Soc Ser B (Meth)}, 16:23--38, 1954.

\bibitem{HW62A}
JM~Hammersley and DJA Welsh.
\newblock Further results on the rate of convergence to the connective constant
  of the hypercubical lattice.
\newblock {\em Quart J Math}, 13:108--110, 1962.

\bibitem{JvR99}
EJ~Janse~van Rensburg.
\newblock Composite models of polygons.
\newblock {\em J Phys A: Math Gen}, 32:4351--4372, 1999.

\bibitem{JvR02}
EJ~Janse~van Rensburg.
\newblock The probability of knotting in lattice polygons.
\newblock {\em Contemp Math}, 304:125--136, 2002.

\bibitem{JvR07}
EJ~Janse~van Rensburg.
\newblock Squeezing knots.
\newblock {\em J Stat Mech: Theo Expr}, 2007:P03001, 2007.

\bibitem{JvR19}
EJ~Janse~van Rensburg.
\newblock Osmotic pressure of compressed lattice knots.
\newblock {\em Phys Rev E}, 100:012501, 2019.

\bibitem{JOTW07b}
EJ~Janse~van Rensburg, E~Orlandini, MC~Tesi, and SG~Whittington.
\newblock Knot probability of polygons subjected to a force: a {Monte} {Carlo}
  study.
\newblock {\em J Phys A: Math Theo}, 41:025003, 2008.

\bibitem{JOTW07}
EJ~Janse~van Rensburg, E~Orlandini, MC~Tesi, and SG~Whittington.
\newblock Knotting in stretched polygons.
\newblock {\em J Phys A: Math Theo}, 41:015003, 2008.

\bibitem{JvRR11A}
EJ~Janse~van Rensburg and A~Rechnitzer.
\newblock Generalized atmospheric sampling of knotted polygons.
\newblock {\em J Knot Theo Ram}, 20:1145--1171, 2011.

\bibitem{JvRW91}
EJ~Janse~van Rensburg and SG~Whittington.
\newblock The {BFACF} algorithm and knotted polygons.
\newblock {\em J Phys A: Math Gen}, 24:5553--5567, 1991.

\bibitem{KM91}
K~Koniaris and M~Muthukumar.
\newblock Knottedness in ring polymers.
\newblock {\em Phys Rev Lett}, 66:2211--2214, 1991.

\bibitem{LS84}
ID~Lawrie and S~Sarlbach.
\newblock Tricriticality.
\newblock In {C Domb and JL Lebowitz}, editor, {\em Phase Transitions and
  Critical Phenomena}, volume~9, pages 65--161. Academic Press, 1984.

\bibitem{M95}
N~Madras.
\newblock Critical behaviour of self-avoiding walks that cross a square.
\newblock {\em J Phys A: Math Gen}, 28:1535--1547, 1995.

\bibitem{Maple17}
{Maple 17}.
\newblock {Waterloo Maple, Inc}.

\bibitem{MLY11}
R~Matthews, AA~Louis, and JM~Yeomans.
\newblock Confinement of knotted polymers in a slit.
\newblock {\em Mol Phys}, 109:1289--1295, 2011.

\bibitem{MRRTT53}
N~Metropolis, AW~{Rosenbluth}, MN~{Rosenbluth}, AH~Teller, and E~Teller.
\newblock Equation of state calculations by fast computing machines.
\newblock {\em J Chem Phys}, 21:1087--1092, 1953.

\bibitem{MW84}
JPJ Michels and FW~Wiegel.
\newblock Probability of knots in a polymer ring.
\newblock {\em Phys Lett A}, 90:381--384, 1982.

\bibitem{OSV04}
E~Orlandini, AL~Stella, and C~Vanderzande.
\newblock Loose, flat knots in collapsed polymers.
\newblock {\em J Stat Phys}, 115:681--700, 2004.

\bibitem{RJvR08}
A~Rechnitzer and EJ~Janse~van Rensburg.
\newblock Generalized atmospheric {Rosenbluth} methods ({GARM}).
\newblock {\em J Phys A: Math Theo}, 41:442002, 2008.

\bibitem{RCV93}
VV~Rybenkov, NR~Cozzarelli, and AV~Vologodskii.
\newblock Probability of {DNA} knotting and the effective diameter of the {DNA}
  double helix.
\newblock {\em Proc Nat Acad Sci}, 90:5307--5311, 1993.

\bibitem{SW93}
SY~Shaw and JC~Wang.
\newblock Knotting of a {DNA} chain during ring closure.
\newblock {\em Science}, 260:533--536, 1993.

\bibitem{TJvROSW94}
MC~Tesi, EJ~Janse~van Rensburg, E~Orlandini, DW~Sumners, and SG~Whittington.
\newblock Knotting and supercoiling in circular {DNA}: a model incorporating
  the effect of added salt.
\newblock {\em Phys Rev E}, 49:868--872, 1994.

\bibitem{TJvROW96}
MC~Tesi, EJ~Janse~van Rensburg, E~Orlandini, and SG~Whittington.
\newblock {Monte} {Carlo} study of the interacting self-avoiding walk model in
  three dimensions.
\newblock {\em J Stat Phys}, 82:155--181, 1996.

\bibitem{V95}
C~Vanderzande.
\newblock On knots in a model for the adsorption of ring polymers.
\newblock {\em J Phys A: Math Gen}, 28:3681--3700, 1995.

\end{thebibliography}

\end{document}